\documentclass[english, reprint, aps, prb, twocolumn, superscriptaddress, floatfix]{revtex4}
\usepackage{babel}
\usepackage{graphicx}
\usepackage{latexsym}
\usepackage{amsmath}
\usepackage{graphics}
\usepackage{amssymb}
\usepackage{layout}
\usepackage{verbatim}
\usepackage{amsfonts,epsfig}
\usepackage{slashed}
\usepackage{feynmp}
\usepackage{float}
\usepackage{hyperref}
\usepackage[usenames]{color}

\newcommand{\beq}{\begin{equation}}
\newcommand{\eeq}{\end{equation}}
\newcommand{\bea}{\begin{eqnarray}}
\newcommand{\eea}{\end{eqnarray}}
\newcommand{\nn}{\nonumber}
\newcommand{\tr}{\hbox{Tr}}

\newcommand{\eup}{E_{\uparrow}(k)}
\newcommand{\edo}{E_{\downarrow}(k)}

\newcommand{\bk}{{\vec k}}

\newcommand{\bp}{{\vec p}}

\newcommand{\bq}{{\vec q}}

\begin{document}

\title{Effective field theory, three-loop perturbative expansion, and their experimental implications in graphene many-body effects}
\author{Edwin Barnes}
\affiliation{Condensed Matter Theory Center, Department of Physics, University of Maryland, College Park, Maryland 20742, USA}
\affiliation{Joint Quantum Institute, University of Maryland, College Park, Maryland 20742, USA}
\author{E.~H. Hwang}
\affiliation{Condensed Matter Theory Center, Department of Physics, University of Maryland, College Park, Maryland 20742, USA}
\affiliation{SKKU Advanced Institute of Nanotechnology and Department of Physics, Sungkyunkwan University, Suwon, 440-746, Korea}
\author{R.~E. Throckmorton}
\affiliation{Condensed Matter Theory Center, Department of Physics, University of Maryland, College Park, Maryland 20742, USA}
\author{S. Das Sarma}
\affiliation{Condensed Matter Theory Center, Department of Physics, University of Maryland, College Park, Maryland 20742, USA}
\affiliation{Joint Quantum Institute, University of Maryland, College Park, Maryland 20742, USA}

\begin{abstract}
Many-body electron-electron interaction effects are theoretically considered in monolayer graphene from a continuum effective field-theoretic perspective by going beyond the standard leading-order single-loop perturbative renormalization group (RG) analysis. Given that the effective (bare) coupling constant (i.e. the fine structure constant) in graphene is of order unity, which is neither small to justify a perturbative expansion nor large enough for strong-coupling theories to be applicable, the problem is a difficult one, with some similarity to 2+1-dimensional strong-coupling quantum electrodynamics (QED). In this work, we take a systematic and comprehensive analytical approach in theoretically studying graphene many-body effects, primarily at the Dirac point (i.e., in undoped, intrinsic graphene), by going up to three loops in the diagrammatic expansion to both ascertain the validity of a perturbative expansion in the coupling constant and to develop an RG theory that can be used to estimate the actual quantitative renormalization effect to higher-order accuracy. Electron-electron interactions are expected to play an important role in intrinsic graphene due to the absence of screening at the Dirac (charge neutrality) point, potentially leading to strong deviations from the Fermi liquid description around the charge neutrality point where the graphene Fermi velocity should manifest an ultraviolet logarithmic divergence because of the linear band dispersion. While no direct signatures for non-Fermi liquid behavior at the Dirac point have yet been observed experimentally, there is ample evidence for the interaction-induced renormalization of the graphene velocity as the Dirac point is approached by lowering the carrier density. We provide a critical comparison between theory and experiment, using both higher-order diagrammatic and RPA (i.e., infinite-order bubble diagrams) calculations, emphasizing future directions for a deeper understanding of the graphene effective field theory. We find that while the one-loop RG analysis gives reasonable quantitative agreement with the experimental data, both for graphene in vacuum and graphene on substrates, particularly when dynamical screening effects and finite carrier density effects are incorporated in the theory through the random phase approximation, the two-loop analysis reveals an interacting strong-coupling critical point in graphene suspended in vacuum signifying either a quantum phase transition or a breakdown of the weak-coupling renormalization group approach. By adapting a version of Dyson's argument for the breakdown of the QED perturbative expansion to the case of graphene, we show that in contrast to QED where the asymptotic perturbative series in the coupling constant converges to at least 137 orders (and possibly to much higher order) before diverging in higher orders, the graphene perturbative series in the coupling constant may manifest asymptotic divergence already in the first or second order in the coupling constant, favoring the conclusion that perturbation theory may be inadequate, particularly for graphene suspended in vacuum. We propose future experiments and theoretical directions to make further progress on this important and difficult problem. The question of convergence of the asymptotic perturbative expansion for graphene many-body effects is discussed critically in the context of the available experimental results and our theoretical calculations.
\end{abstract}

\maketitle

\section{Introduction}\label{sec:intro}

Graphene, or more precisely monolayer graphene, is a 2d honeycomb lattice of carbon atoms with a single-particle band structure\cite{Wallace_PR47} which in the long-wavelength limit is linear and chiral: $E_\pm(q)=\pm v_F q$, where $\pm$ refer to the two chiral linear (conduction/valence) bands, $q$ is the 2d wavenumber (momentum), and $v_F$ the so-called graphene (Fermi) velocity. (We use $\hbar=1$ throughout.) Graphene is the most studied topic in physics during the 2005-2013 time period, and the subject is well reviewed in the literature.\cite{CastroNeto_RMP09, DasSarma_RMP11} Our interest in the current work is a theoretical study of electron-electron interaction effects on graphene single-particle properties, specifically the linear band dispersion, a subject which was originally studied in the literature\cite{Semenoff_PRL84,Gonzalez_NPB94,Gonzalez_PRB99,Gonzalez_PRB01} long before the current interest in graphene exploded following the experimental work of Geim and Novoselov.\cite{Novoselov_Science04,Novoselov_Nature05} The specific issues we want to explore in great detail in the current work include the role and importance of graphene many-body effects, the renormalization of ultraviolet divergences, and the ability of a perturbative RG analysis to accurately capture the signatures of electron-electron interactions in real graphene systems. We investigate these questions by going beyond the leading-order one-loop RG analysis, which is necessitated by the fact that, unlike in QED where the interaction strength is given by the vacuum fine structure constant, $e^2/\hbar c\sim 1/137$, the effective coupling constant defining the interaction strength in graphene is of order unity due to the smallness of the graphene Fermi velocity, $v_F\approx c/300$, defining its chiral linear dispersion. A related question to be discussed in this paper in reasonable depth is how well (or poorly) experimental data and theoretical many-body calculations agree with each other with respect to the interaction-induced renormalization of the graphene Fermi velocity, a question of fundamental importance in understanding the quantitative aspects of graphene effective field theory.

Intrinsic (i.e. undoped) graphene,\cite{Hwang_PRL07b} our primary interest in the current work, is a seemingly simple system of two chiral, linear electron-hole single-particle bands with the valence band completely full, the conduction band completely empty, and the Fermi level pinned precisely at the Dirac point (considered to be at zero energy with the valence/conduction band states occupying negative/positive energy states). The system is thus a chiral, gapless semiconductor with a massless linear electron-hole energy dispersion characterized by the Fermi velocity, $v_F$. The long-wavelength single-particle theory is precisely described by the chiral, massless 2+1-dimensional Dirac-Weyl equation. The relevant ``spin" index for this effective theory is, in fact, the pseudo-spin index arising from the two sublattices of the 2d honeycomb lattice with pseudospin-momentum locking leading to the long-wavelength description being the massless 2d chiral Dirac equation. In addition to the pseudospin index, graphene also has the usual spin index with a two-fold spin degeneracy and a valley index (arising from the band structure effect at $K$, $K'$ points of the Brillouin zone), adding an additional valley degeneracy factor of 2 (i.e., graphene has a four-fold single-particle ground state degeneracy).

The bare (i.e., unrenormalized) value of the Fermi velocity $v_F$ is given by the basic tight-binding band structure calculation for intrinsic graphene to be roughly $v_F\sim 10^8\hbox{cm/s}\approx c/300$. This immediately leads to a bare coupling constant defined by
\beq
\alpha\equiv\frac{e^2}{\kappa v_F}=2.2/\kappa,
\eeq
where $\kappa$ is the background lattice dielectric constant arising from the substrate on which the graphene layer resides. For suspended graphene in vacuum, $\kappa=1$, and thus $\alpha\approx2.2$ for suspended graphene, whereas $\alpha\approx0.9 (0.4)$ for graphene on SiO${}_2$ (BN) substrates, which are two common substrate materials used in the experimental studies of graphene. It is worth noting that these values of $\alpha$ for graphene on substrates represent typical rather than precise values; the effective dielectric constant $\kappa$ can vary substantially between different experiments. For example, the effective $\kappa$ for graphene on a BN substrate has been found to range from 2.5 to 8 depending on the specific experimental setup.\cite{Wang_NP12,Yu_PNAS13} We further note that the standard definition of the dimensionless interaction strength in a solid state electronic material is the so-called (dimensionless) Wigner-Seitz radius, universally referred to as the $r_s$-parameter\cite{FetterWalecka,Mahan} and defined as $r_s=\langle PE\rangle/\langle KE\rangle$, where $\langle PE\rangle$, $\langle KE\rangle$ refer respectively to the average ground state Coulomb potential energy and the average kinetic energy. For an ordinary parabolic dispersion 2d (3d) electron gas system, $r_s$ depends on the 2d (3d) electron density $n$ as $r_s\sim n^{-1/2} (n^{-1/3})$. In graphene, however, the linear dispersion makes $r_s$ independent of carrier density, and it is easy to show that $r_s=\alpha= e^2/(\kappa v_F)$ for graphene, which is independent of density. This density-independent constant interaction strength of graphene sets it apart from regular metals and semiconductors in terms of electron-electron interaction effects, although the interaction itself is the standard long-range $1/r$ Coulomb interaction between two electrons in all of these systems.

In addition to the chiral, linear energy dispersion and the density-independent coupling constant, graphene many-body theories require an explicit ultraviolet momentum (i.e., short-distance) cutoff because of the linear energy dispersion. Such a momentum cutoff is unnecessary in the usual 2d or 3d electron gas problem where the only divergence one needs to worry about is the long-distance (i.e., low momentum) infrared divergence associated with the long-range nature of the Coulomb interaction, which leads to a well known infrared logarithmic divergence in the self-energy at leading order (i.e., Hartree-Fock) in the bare Coulomb interaction. It has been known for more than fifty years that the infrared log divergence of the Hartree-Fock self-energy is regularized simply by carrying out an expansion in the dynamically screened Coulomb interaction, which is equivalent to the random phase approximation (RPA) in which the infinite series of bubble diagrams is summed up for the reducible polarizability and the self-energy. The ultraviolet divergence in the graphene self-energy is, however, present in both Hartree-Fock and RPA theories, i.e., whether the perturbative expansion is carried out in the bare interaction or in the dynamically screened Coulomb interaction, introducing a $\alpha\log(k_c/k)$ (or equivalently $\alpha\log(E_c/E)$) renormalization of the graphene Fermi velocity akin to the corresponding logarithmic mass renormalization in QED. Using weak-coupling RG (and noting the $\alpha\propto v_F^{-1}$ relationship between the bare coupling constant and the bare graphene velocity), the running of the renormalized coupling and the renormalized Fermi velocity then imply a logarithmic divergence of the graphene Fermi velocity as the Dirac point is approached. One of the key questions we explore in the current work is how the higher-order corrections in the coupling contant $\alpha$ modify the leading-order velocity (and polarizability) renormalization in graphene. This question is substantive since the bare coupling constant in graphene is not small ($\alpha\sim1$), and as such there is no a priori reason to trust a weak-coupling leading-order perturbation theory.

The issue of higher-order corrections to graphene many-body effects is not just of abstract theoretical interest since a number of recent experimental studies\cite{Elias_NP11,Chae_PRL12,Yu_PNAS13} claim to have directly observed the expected logarithmic renormalization of the graphene Fermi velocity, obtaining, in fact, reasonable quantitative agreement with the leading-order theoretical results. In 3+1-dimensional QED, of course, the celebrated quantitative agreement between electron g-2 experiments and high-order perturbative diagrammatic calculations\cite{Aoyama_PRL12a,Aoyama_PRL12b} has now been established up to an astounding twelve decimal places, which is amazing, but understandable, since $\alpha\sim1/137$ for QED, and thus, the perturbation theory, although asymptotic in nature, should give ``correct" convergent results (at least) up to 137 decimal places, in principle, before diverging at order 137 or above. In graphene, however, $\alpha\sim1$, and therefore, it is mysterious why theory and experiment should agree at all quantitatively. Even a qualitative agreement between theory and experiment in graphene necessitates (at least) estimating the higher-order diagrams at some reasonable level of quantitative accuracy so that one has some faith in believing that the agreement between experimental data and the leading-order theory is not just simply the magic of data fitting with enough free (and adjustable) parameters (or just a pure lucky coincidence), but is a real advance in our fundamental understanding of the quantitative aspects of the graphene effective field theory.

Our goal in the paper is to systematically go beyond the leading-order perturbative theory to calculate the ultraviolet divergent contributions to the graphene self-energy and polarizability function so that we can make some concrete quantitative statements about velocity and charge renormalization in graphene using a perturbative expansion in the coupling constant. Our goal is ambitious and difficult, but is clearly necessary given that the coupling constant in relevant graphene experiments is simply not perturbatively small for one to be restricted to just the simple leading-order theory. We provide complete technical details for our complex calculations so that others may check our results and go beyond our calculations as necessary.

The rest of this paper is organized as follows: In Sec. II we provide a background mentioning earlier work in the literature on graphene many-body effects so that our extensive calculations are set in the proper context; in Sec. III we provide our detailed higher-order two and three-loop perturbative self-energy and polarization results for the Coulomb interaction, discussing the corresponding RG analysis and questions of renormalizability; in Sec. IV we critically discuss the important issue of the comparison between theory and experiment; in Sec. V we discuss our results and open questions along with the issue of the renormalizability of the theory for a hypothetical zero-range electron-electron interaction, contrasting it with the Coulomb interaction results given in Sec. III; we conclude in Sec. VI. Two appendices provide many technical details.

\section{Background}\label{sec:background}

There has been substantial earlier work in the literature on graphene many-body effects, which we briefly summarize in the current section in order to put our work in the proper context. This also seems to motivate our work, establishing its necessity in spite of the substantial body of existing work on this topic in the literature.

The existing many-body theoretic work on monolayer graphene divides itself naturally into a number of different categories arising from the complex and difficult nature of the theoretical issues involved in the problem, where many different approaches from complementary viewpoints can be useful. The first category, and our work fits firmly into this category, is the most obvious approach to the problem, spiritually following the classic perturbative field theoretic approach pioneered by Tomonaga, Schwinger, Feynman, and Dyson more than sixty years ago.\cite{Tomonaga_PTP46,Koba_PTP47,Schwinger_PR48a,Schwinger_PR48b,Dyson_PR48,Dyson_PR49,Feynman_PR49,Feynman_PR50} Although it was already known in the early 1950s\cite{Dyson_PR52} that the QED perturbative series is only asymptotic and eventually diverges at very high orders (order $\gtrsim 137$), the perturbation series analysis of QED has remained the most quantitatively successful theory ever developed anywhere, with an astonishing 1 part in $10^{12}$ type agreement achieved between the perturbative results (up to $O(\alpha^6)$) and precision measurements.\cite{Aoyama_PRL12a,Aoyama_PRL12b} For graphene, the leading-order (i.e., $O(\alpha)$) perturbative calculation coupled with a weak-coupling RG analysis strictly at the Dirac point was already carried out in the 1990s.\cite{Gonzalez_NPB94} This leading-order weak-coupling theory is equivalent to the many-body Hartree-Fock approximation, i.e., the exchange self-energy calculation. As such, the leading-order (in the bare coupling) theory fails\cite{Hwang_PRL07b} at finite carrier density (i.e., finite chemical potential) when the Fermi energy is no longer at the Dirac point since it incorrectly predicts the presence of an infrared divergence (induced by the long-range Coulomb interaction) at the Fermi energy. This pathological feature of the leading-order theory was corrected in Ref.~[\onlinecite{DasSarma_PRB07}] by summing the infinite series of bubble diagrams so as to dynamically screen the bare Coulomb interaction in the RPA theory. Similar analyses based on RPA were also carried out by other groups,\cite{Polini_SSC07} reaching the same conclusions as Ref.~[\onlinecite{DasSarma_PRB07}]. Such an RPA theory is devoid of the pathological infrared divergence of the leading-order Hartree-Fock theory, and manifests only the ultraviolet $\log(E_c/E_F)$ divergence inherent in the QED nature of the graphene problem. Thus, as the Dirac point is approached (i.e., $E_F\to0$), the ultraviolet logarithmic divergence in the renormalized Fermi velocity becomes apparent and leads to a running coupling constant $\alpha$ ($\propto v_F^{-1}$) which flows to zero logarithmically, thus justifying the applicability of the weak-coupling perturbative expansion in a heuristic manner. Thus, RPA, which we will discuss in some detail in Sec. IV, remains a powerful quantitative tool for studying graphene many-body effects and comparing with experimental data, and indeed there is a substantial theoretical literature on the use of RPA for calculating graphene many-body renormalization.\cite{Hwang_PRL07a,Hwang_PRB07a,DasSarma_PRB07,Hwang_PRB07b,Barlas_PRL07,Hwang_PRL07b,Polini_SSC07,Adam_PNAS07,Hwang_PE08,Hwang_PRB08a,Polini_PRB08,Tse_PRL08,Hwang_PRB09a,LeBlanc_PRB11,DasSarma_PRB13}

There is, however, an important shortcoming of both Hartree-Fock and RPA theories which arises from the fact that the bare interaction strength in graphene (i.e., $\alpha$) is by no means small, and therefore, a leading-order perturbation theory in $\alpha$ is questionable. One can argue that RPA, in fact, becomes exact in the limit in which the ground state degeneracy $N$ diverges, where real graphene has $N=2$, corresponding to the valley degeneracy. From this point of view, the RPA theory developed in Refs.~[\onlinecite{DasSarma_PRB07,Polini_SSC07}] can be thought of as a non-perturbative theory that coincides precisely with the leading-order $1/N$ expansion. Several other works have employed the $1/N$ expansion as well, focusing primarily on the case of intrinsic graphene,\cite{Vafek_PRL07,Son_PRB07,Drut_PRB08,Foster_PRB08} and a subset of these works studied the influence of the critical point at infinite coupling on the large-$N$ theory.\cite{Son_PRB07,Drut_PRB08} In particular, the infinite-coupling critical point was predicted to give rise to a modification of the graphene dispersion from $E=v_Fk$ to $E\sim k^z$ with $z<1$. Experiments typically report a renormalization of the graphene velocity rather than a nontrivial new dynamical exponent $z<1$, so the importance of the infinite-coupling critical point has not been validated experimentally.

In addition to the extensive literature on the weak-coupling perturbative expansion (including RPA which is still a weak-coupling theory except for the perturbative expansion being in the dynamically screened Coulomb interaction rather than the bare interaction) and large-$N$ theories, there is an extensive literature on graphene many-body effects being studied purely from the perspective of a strong-coupling theory.\cite{Juricic_PRB09,Araki_PRB10,Araki_AP10,Semenoff_PS11,Araki_PRB12,Drut_PRL09,Drut_PRB09a,Drut_PRB09b,Drut_arXiv13,Armour_PRB10,Armour_PRB11,Wang_NJP12,Gamayun_PRB10,Giuliani_PRB09,Giuliani_PRB10,Giuliani_AHP10,Giuliani_AP12} Typically, three different approaches are used: (i) direct numerical work using lattice QCD type calculations;\cite{Drut_PRL09,Drut_PRB09a,Drut_PRB09b,Drut_arXiv13,Armour_PRB10,Armour_PRB11} (ii) using some sort of Schwinger-Dyson theory,\cite{Khveshchenko_PRL01,Khveshchenko_JPCM09,Gamayun_PRB10,Wang_NJP12,Gonzalez_PRB12,Gonzalez_JHEP12} which involves solving an integral equation built from some infinite subset of (usually ladder-type) diagrams; and (iii) mapping the strong-coupling problem onto some known strong-coupling field theoretic model (e.g. the Gross-Neveu model\cite{Gross_PRD74}) through some ad-hoc approximations.\cite{Herbut_PRL06,Juricic_PRB09} All of these strong-coupling theories predict the existence of a strong-coupling fixed point with the opening of an energy gap at the Dirac point (i.e., the massless theory achieves a mass through the spontaneous breaking of chiral symmetry) provided the coupling constant is larger than some critical value, $\alpha_c$. There is no consensus on the value of $\alpha_c$ in the literature, with $\alpha_c\sim1-10$ being quoted by different authors. We mention that $\alpha>2.2$ is unphysical for graphene since $\kappa\ge1$ always. It must be emphasized that the strong-coupling theories are not yet validated by any experimental observations since no one has reported the observation of any gap in monolayer graphene at the Dirac point on any substrates or in vacuum. Thus, in spite of their theoretical importance and elegance, the strong-coupling graphene many-body theories are all on shaky empirical grounds. This is in stark contrast to the perturbative theories and RPA, which have been quite successful in capturing the signatures of electron-electron interactions observed in experiments, as will be discussed at length in Sec.~\ref{sec:experiment}.

Our current work makes an attempt to bridge the gap between the existing weak-coupling leading-order (i.e., $O(\alpha)$) graphene many-body theories and the strong-coupling theories by going to $O(\alpha^2)$ in the perturbation theory exactly and to $O(\alpha^3)$ approximately. Given that the most common values of the graphene coupling strength ($\alpha\approx0.4-2.2$) range around unity, it is important to go beyond $O(\alpha)$ theories at least to make sure that there is no manifest problem in the perturbation series in higher orders. We calculate the polarizability and the self-energy at the Dirac point exactly to $O(\alpha^2)$, and the self-energy approximately to $O(\alpha^3)$. These results are presented in detail in the next section.

There have been a few earlier theoretical attempts to go to higher orders in graphene many-body perturbation theories, often obtaining conflicting (and partial) results. For example, Ref.~[\onlinecite{Kotov_PRB08}] and Ref.~[\onlinecite{Sodemann_PRB12}] calculate the polarizability to $O(\alpha^2)$, finding different results. Our completely independent calculations agree with Ref.~[\onlinecite{Sodemann_PRB12}], but not with Ref.~[\onlinecite{Kotov_PRB08}]. The self-energy correction to $O(\alpha^2)$ has earlier been calculated by Ref.~[\onlinecite{Mishchenko_PRL07}] and by Ref.~[\onlinecite{Vafek_PRB08}]. We find that our $O(\alpha^2)$ self-energy results disagree with Ref.~[\onlinecite{Vafek_PRB08}], albeit in a fairly minor way. The $O(\alpha^2)$ results of Ref.~[\onlinecite{Mishchenko_PRL07}] are incomplete in that they only include the zero-energy limit of the self-energy correction. Our partial $O(\alpha^3)$ self-energy calculations are provided mainly to show explicitly that higher-order ultraviolet logarithmic terms (e.g. $\log^2(E_c/E)$) indeed arise in graphene at third order as is necessary for the renormalizability of the theory at higher orders. Our estimated $O(\alpha^3)$ perturbative self-energy corrections indicate that the graphene perturbative series is probably asymptotic only up to $O(\alpha)$ terms, and as such, RPA may very well be the best quantitative theory we can have, and going to $O(\alpha^2)$ or $O(\alpha^3)$ may make the agreement between theory and experiment actually worse since the expansion may have already started diverging at $O(\alpha^2)$!

Given the highly technically demanding nature of our $O(\alpha^2)$ and $O(\alpha^3)$ perturbative results and the fact that the existing higher-order results in the literature disagree with each other, we have decided to provide {\it all} the technical details of our theory so that others can check our results for consistency.

\section{Perturbation theory with bare Coulomb interaction}\label{sec:perttheory}

It may be instructive to carry out a simple dimensional analysis to explain why the graphene many-body problem in condensed matter physics resembles a field-theoretic problem with an ultraviolet logarithmic divergence in contrast to the usual situation (i.e. parabolic electronic energy band dispersion) where such large momentum singularities are considered to be fundamentally absent in solid state systems by virtue of the actual physical existence of a large momentum lattice cutoff. This dimensional analysis also serves to distinguish the graphene condensed matter effective field theory from its relativistic quantum field-theoretic analogs, 3+1- and 2+1-dimensional QED.

First, it must be emphasized that the Coulomb interaction for graphene is a true $1/r$-type Coulomb interaction (in contrast to the $\log r$ Coulomb interaction of the purely 2d world) as in 3d systems since graphene is a 2d membrane in the 3d world, and the electric field and potential lines exist in the 3d world. Thus, the corresponding momentum space interaction in graphene is $1/q$, and not $1/q^2$, as it is in both 3+1d and 2+1d QED. Second, the electron propagator in graphene scales as $1/q$ (as in QED) for large momentum since the energy dispersion is linear in momentum, distinguishing it from the usual $1/q^2$ scaling at large momentum of the ordinary parabolic-dispersion electron propagator. This distinguishes graphene from typical solid state systems. Graphene thus shares features of both QED (in energy dispersion) and solid state physics (2d membrane in a 3d world with a true ultraviolet cutoff, $q_c\sim1/a$, where $a$ is the graphene lattice constant associated with the existence of a physical honeycomb lattice comprised of carbon atoms). In addition, the bare Coulomb interaction in graphene is not retarded since it is nonrelativistic by virtue of the graphene velocity $v_F$ being only $\sim c/300$, in contrast to QED.

The fact that the Coulomb interaction for graphene scales as $1/q$ in momentum space instead of $1/q^2$ is one of two important differences between 2d graphene and 2+1d QED (the other important difference being the nonrelativistic nature of the interaction due to $v_F\ll c$). This difference leads to the fact that the electron self-energy in graphene is logarithmically divergent in the ultraviolet (large momentum) regime, unlike 2+1d QED, which is a superrenormalizable theory. On the other hand, the ultraviolet behavior of the self-energy in graphene coincides with that in 3+1d QED because the difference in the Coulomb interactions ($1/q$ versus $1/q^2$) is compensated by the difference in the number of spatial dimensions. Technically speaking, the interaction in graphene is marginal from an RG perspective rather than irrelevant (as it is in 2+1d QED). Thus, graphene effective field theory has the same logarithmic ultraviolet divergence with a running effective coupling constant as in 3+1d QED. Graphene effective field theory is thus a 2+1d theory that shares several common features with 3+1d QED.

What about the corresponding Schr\"odinger-like parabolic energy dispersion situation common to most solid state systems (where no one ever worries about logarithmic ultraviolet divergences or running coupling constants)? For a parabolic electron energy dispersion, the large momentum structure of the self-energy integrals (for the physical Coulomb interaction going as $1/q$ or $1/q^2$ in 2d or 3d respectively) goes as the integral of $dq/q^2$ in both 2d and 3d, thus rendering the corresponding effective field theory superrenormalizable, with the Coulomb interaction being irrelevant and the ultraviolet cutoff a non-issue. If the 2d Coulomb interaction is assumed to be $1/q^2$ (i.e., logarithmic electron-electron interactions in real space instead of the physical $1/r$ interaction), then both graphene with its linear energy dispersion and ordinary solid state systems with parabolic dispersion become superrenormalizable field theories in two dimensions with no ultraviolet divergences.

Thus, 2d graphene field theory, because of the linear energy dispersion and the $1/r$-Coulomb interaction (both are necessary), suffers from exactly the same ultraviolet self-energy divergence as one faces in the usual 3+1d QED, except that the effective coupling constant is of order unity rather than $\alpha_{QED}\sim1/137$ because of the fact that the graphene velocity is $v_F\approx c/300$. We note that since the one-loop self-energy correction enhances the graphene effective velocity as one approaches the Dirac point (i.e., at lower momentum or energy scales), it seems that the weak-coupling theory is always applicable very close to the Dirac point since the effective coupling approaches $\alpha_{QED}$ at the Dirac point. This argument (which has been repeatedly made in the literature), while technically correct, is completely impractical since there is absolutely no justification in constructing the graphene RG flow starting with the one-loop self-energy calculation since the measured effective coupling (determined via the velocity) at experimentally relevant energy scales (corresponding to carrier densities on the order of $n\sim10^{12}\hbox{cm}^{-2}$) is not small but instead of order one. Because of the logarithmically slow running of the effective coupling, one would need to further reduce the energy scale or carrier density by roughly six orders of magnitude to achieve a factor of ten reduction in the coupling, a task which is far beyond current experimental capabilities. Thus, even the fact that the effective running coupling decreases with decreasing energy in the graphene effective field theory may very well be an artifact of the one-loop approximation which cannot be justified for an effective coupling of order unity. It might very well be that the graphene theory at experimentally relevant energy scales is more QCD-like rather than QED-like, and becomes asymptotically free at high energy with the effective coupling (velocity) going to infinity (zero) at the Dirac point. This is the weak coupling versus strong coupling conundrum at the heart of our work.  We want to study the graphene effective field theory at higher loop orders to see the extent to which a weak-coupling perturbative RG makes any sense at all for a seemingly strong-coupling problem with $\alpha\sim1$. Thus, graphene is truly a strong-coupling QED problem where experiments (so far) tend to indicate a weak-coupling behavior!

It may also be worthwhile to mention that for a hypothetical short-range Coulomb interaction (i.e. a constant in momentum space), the same dimensional analysis indicates that the self-energy terms will have power-law divergences that scale with the order of perturbation theory, leading to a horribly complicated and possibly nonrenormalizable interacting theory (see the appendix for details).  We emphasize that later in the paper we make these dimensional arguments rigorous by systematically carrying out the dimensional analysis to infinite loop order, showing that the above conclusions remain valid to all loop orders in the graphene effective field theory, but the perturbation series in loops may only be asymptotic to the first or the second loop orders for the actual physical values of the graphene effective coupling constant.

Finally, what is the role of the nonrelativistic Coulomb interaction in graphene (in QED of course the interaction is manifestly relativistic)?  It turns out that the nonrelativistic (i.e. nonretarded) form of the Coulomb interaction plays no important role in graphene except to assert that if the weak-coupling theory remains valid with the graphene velocity increasing logarithmically to arbitrarily low energy, then the theory must eventually be cut off at an astronomically low momentum scale where $v_F\sim c$, and relativistic effects come into the theory at these exponentially ($e^{-300}\hbox{cm}^{-1}$!) small momentum scales. Thus, the effective coupling constant reaches $\alpha\sim1/137$, and the graphene velocity becomes $c$, with the RG flow saturating at that point.

We mention that the screening of the Coulomb interaction (i.e. the insertion of polarization diagrams in the interaction lines) does not change the above dimensional analysis at the ultraviolet scale since screening is not operational at high momenta. Screening does serve a very important purpose however; it serves to eliminate the infrared divergence associated with the bare Coulomb interaction at any finite doping. Thus, the graphene theory at any finite carrier density (again, a complication not arising in QED) must use the screened Coulomb interaction and not the bare Coulomb interaction in order to avoid logarithmically divergent infrared singularities arising from the long-range nature of the bare Coulomb interaction. Screening, however, plays no role in the ultraviolet divergence and in the running coupling constant of the graphene field theory since the logarithmic divergence structure of the theory remains unaffected by screening, with only the subleading terms being affected quantitatively.

In this section, we compute first, second, and third-order corrections to the electron self-energy and vacuum polarization using an effective field theory diagrammatic expansion that is perturbative in the strength of the bare Coulomb interaction, which is quantified by the effective graphene fine structure constant $\alpha$. The results are then used to determine the renormalized polarization, the Fermi velocity, and the running of the effective coupling strength. We are particularly interested in testing the reliability of perturbation theory in light of the fact that the expansion parameter, $\alpha$, is of order unity in real graphene experiments.

First-order results for the Fermi velocity and vacuum polarization have already been shown to give good qualitative (and perhaps quantitative) agreement with experiments.\cite{Elias_NP11,Sodemann_PRB12} We find that while second-order corrections improve the agreement between theory and experiment in the case of the vacuum polarization,\cite{Sodemann_PRB12} second-order corrections to the Fermi velocity lead to a strong-coupling critical point at $\alpha=\alpha_c\approx0.78$, suggesting either a phase transition or a breakdown of perturbation theory when it is applied to experimental setups with $\alpha>\alpha_c$, such as graphene suspended in vacuum. We find support for the conclusion that perturbation theory is failing for $\alpha>\alpha_c$ by estimating the order where the perturbative asymptotic series in $\alpha$ begins to diverge from the true result, finding that this happens around first or second order. This evidence is also consistent with the fact that no indication of a phase transition has been seen in any experimental measurements of many-body effects in single-layer graphene.

We further provide in this section a general analysis of the divergence structure of the graphene effective field theory. We show that only logarithmic ultraviolet divergences can arise in higher-order corrections to the electron self-energy, with higher powers of logarithms arising at third order and above. Only the linear-log divergences contribute to the renormalization of the Fermi velocity and effective coupling, while all the higher-power log divergences can be determined from the linear-log terms through a set of recursion relations, which we derive explicitly. These recursion relations reveal that if graphene is to remain renormalizable at higher orders, it must be the case that $\log^2$ divergences arise at third order, while higher-power logarithmic divergences cannot appear at this order. We verify this explicitly by computing third-order diagrams which are expected to be the most divergent, namely those which contain divergent self-energy subdiagrams.

Although several groups have already reported results for the second-order polarizability\cite{Kotov_PRB08,Sodemann_PRB12}, the electron self-energy, and the velocity renormalization,\cite{Mishchenko_PRL07,Vafek_PRB08} these results generally conflict with one another and with our own findings as we discussed in the previous section. Because of these disagreements, we felt it necessary to provide the full details of our calculations in the main text and to point out the specific places where our results agree or disagree with previous findings as they occur. On the other hand, the first-order results are well established in the literature; we include a detailed review of these results as well for the sake of completeness and because these results are used in the higher-order calculations. We have attempted to keep the calculation of each diagram as self-contained as possible for the sake of readability.

\subsection{Conventions, Feynman rules and useful identities}

We follow the conventions of Ref.~[\onlinecite{Son_PRB07}] and work with the Euclidean action,
\bea
S\!\!\!&=&\!\!\!{-}\sum_{a=1}^N\int dt d^2x({\bar\psi}_a\gamma^0\partial_0\psi_a{+}v_F{\bar\psi}_a\gamma^i\partial_i\psi_a{+}A_0{\bar\psi}_a\gamma^0\psi_a)
\nn\\&&\!\!\!+{1\over2g^2}\int dtd^3x(\partial_iA_0)^2.
\eea
The fields ${\bar\psi}_a$ are four-component fermion fields describing electrons and holes, with $a$ labeling the fermion species. The number of species, $N$, is equal to 2 in real graphene, corresponding to the spin degeneracy. $v_F$ denotes the Fermi velocity. The $\gamma$'s are Dirac matrices satisfying the Euclidean Clifford algebra, $\{\gamma^\mu,\gamma^\nu\}=2\delta^{\mu\nu}$, which will we choose as
\beq
\gamma^0=\sigma_3\otimes\sigma_3, \qquad \gamma^i=\sigma_i\otimes I,
\eeq
where the $\sigma_i$ are Pauli matrices. A useful identity involving the gamma matrices is
\beq
\tr\{\gamma^\mu\gamma^\nu\gamma^\rho\gamma^\sigma\}
=4(\delta^{\mu\nu}\delta^{\rho\sigma}-\delta^{\mu\rho}\delta^{\nu\sigma}+\delta^{\mu\sigma}\delta^{\nu\rho}).\label{gammaid}
\eeq
The coupling $g^2$ is given by
\beq
g^2={2\over1+\epsilon}{e^2\over\epsilon_0}=\frac{4\pi e^2}{\kappa},
\eeq
where $e$ is the electric charge, $\epsilon_0$ is the vacuum permeability, and $\epsilon$ and $\kappa$ are two different definitions of the dielectric constant of the substrate (SI and cgs units, respectively). Combining these quantities with the Fermi velocity, we can define an effective fine structure constant for graphene:
\beq
\alpha\equiv\frac{g^2}{4\pi v_F}=\frac{e^2}{2\pi(1+\epsilon)\epsilon_0 v_F}=\frac{e^2}{\kappa v_F}.\label{defofalpha}
\eeq
We find it convenient to make use of the quasirelativistic notation,
\beq
\slashed p=\gamma^0p_0+v_F\vec{\gamma}\cdot\vec{p},\qquad p^2=p_0^2+v_F^2|\vec{p}|^2.
\eeq
The free fermion propagator is
\beq
G_0(p)=\frac{i}{\slashed p}=\frac{i\slashed p}{p^2},
\eeq
the effective propagator for the Coulomb interaction is
\beq
D_0(p)=g^2\int\frac{dp_z}{2\pi}\frac{1}{p_z^2+|\vec{p}|^2}=\frac{g^2}{2|\vec{p}|},
\eeq
and the interaction vertex is $i\gamma^0$. Every closed fermion loop contributes an overall minus sign to the value of the diagram.

\subsection{One-loop electron self-energy}\label{sec:oneloopSE}

\begin{figure}
\includegraphics[width=0.5\columnwidth]{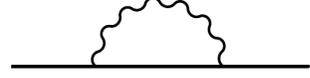}
\caption[Self-energy]{One-loop correction to the electron self-energy.}
\label{fig:oneloopselfenergy}
\end{figure}
Both for the sake of completeness and to make our conventions more transparent, we begin by rederiving the result for the one-loop electron self-energy, Fig.~\ref{fig:oneloopselfenergy}, which was originally computed in Ref.~[\onlinecite{Gonzalez_NPB94}]. The diagram evaluates to
\bea
\Sigma_1(q)&=&-\int\frac{d^3k}{(2\pi)^3}\gamma^0 G_0(k+q)\gamma^0 D_0(k)
\nn\\&=&-\frac{ig^2}{2}\int\frac{d^3k}{(2\pi)^3}\gamma^0\frac{\slashed k+\slashed q}{(k+q)^2}\gamma^0\frac{1}{|\vec{k}|}
\nn\\&=&-\frac{ig^2}{2}\int \frac{d^2k}{(2\pi)^2}\frac{1}{|\vec{k}|}\gamma^0\Upsilon(q,\vec{k})\gamma^0,
\eea
\bea
\Upsilon(q,\vec{k})&=&\int\frac{dk_0}{2\pi}\frac{\slashed k+\slashed q}{(k+q)^2}\nn\\&=&\int\frac{dk_0}{2\pi}\left[\frac{v_F(\vec{k}+\vec{q})\cdot\vec{\gamma}}{k_0^2+v_F^2|\vec{k}+\vec{q}|^2}+\frac{k_0\gamma^0}{k_0^2+v_F^2|\vec{k}+\vec{q}|^2}\right]
\nn\\&=&\frac{(\vec{k}+\vec{q})\cdot\vec{\gamma}}{2|\vec{k}+\vec{q}|},\label{expressionforY}
\eea
\beq
\Sigma_1(q)=\frac{ig^2}{4}\int\frac{d^2k}{(2\pi)^2}\frac{1}{|\vec{k}|}\frac{(\vec{k}+\vec{q})\cdot\vec{\gamma}}{|\vec{k}+\vec{q}|}.
\eeq
In order to perform the remaining integrals, we choose the coordinate system such that $\vec{q}=(|\vec{q}|,0)$ and adopt the transformation to elliptical coordinates used in Ref.~[\onlinecite{Sodemann_PRB12}]:
\bea
&&k_x=\frac{|\vec{q}|}{2}(\cosh\mu\cos\nu-1),\qquad k_y=\frac{|\vec{q}|}{2}\sinh\mu\sin\nu,\nn\\&& d^2k=\frac{|\vec{q}|^2}{4}(\cosh^2\mu-\cos^2\nu)d\mu d\nu,
\eea
yielding
\bea
&&\!\!\!\!\!\!\Sigma_1(q)=
\nn\\&&\!\!\!\!\!\!\frac{i g^2|\vec{q}|}{32\pi^2}\gamma\cdot\int_0^{2\pi}d\nu\int_0^{\mu_{max}}d\mu(1+\cos\nu\cosh\mu,\sin\nu\sinh\mu).\nn\\&&\label{eqn12}
\eea
We have regulated the integral by including the cutoff, $\mu_{max}$. We would like to relate this cutoff to a more physical cutoff on the momentum: $\Lambda\ge|\vec{k}|$. The mapping to elliptical coordinates given above implies
\beq
|\vec{k}|=\frac{|\vec{q}|}{2}(\cosh\mu-\cos\nu),
\eeq
so that
\bea
&&\Lambda=\frac{|\vec{q}|}{2}(\cosh\mu_{max}-\cos\nu)\nn\\&&\Rightarrow\mu_{max}=\cosh^{-1}\left(\frac{2\Lambda}{|\vec{q}|}+\cos\nu\right).
\eea
The three integrals in (\ref{eqn12}) then evaluate to
\bea
&&\!\!\!\!\!\!{\int_0^{2\pi}}d\nu{\int_0^{\mu_{max}}}d\mu={\int_0^{2\pi}}d\nu \cosh^{{-}1}\left(\frac{2\Lambda}{|\vec{q}|}{+}\cos\nu\right)
\nn\\&&=2\pi\log(4\Lambda/|\vec{q}|){+}O(|\vec{q}|^2/\Lambda^2),\nn\\
&&\!\!\!\!\!\!{\int_0^{2\pi}}d\nu\cos\nu{\int_0^{\mu_{max}}}d\mu \cosh\mu
\nn\\&&={\int_0^{2\pi}}d\nu\cos\nu\sqrt{\left(\frac{2\Lambda}{|\vec{q}|}{+}\cos\nu\right)^2{-}1}=\pi{+}O(|\vec{q}|^2/\Lambda^2),\nn\\
&&\!\!\!\!\!\!{\int_0^{2\pi}}d\nu\sin\nu{\int_0^{\mu_{max}}}d\mu \sinh\mu
\nn\\&&={\int_0^{2\pi}}d\nu\sin\nu\left(\frac{2\Lambda}{|\vec{q}|}{+}\cos\nu\right)=0.
\eea
Plugging these results into (\ref{eqn12}), we obtain
\bea
\Sigma_1(q)\!\!\!&{=}&\!\!\!\frac{i g^2|\vec{q}|}{32\pi^2}\gamma^1\left[2\pi\log(\Lambda/|\vec{q}|){+}4\pi\log2{+}\pi\right]{+}O\left(\frac{|\vec{q}|^2}{\Lambda^2}\right)
\nn\\\!\!\!&\rightarrow&\!\!\!\frac{i g^2}{16\pi}\vec{q}\cdot\vec\gamma\left[\log(\Lambda/|\vec{q}|){+}2\log2{+}1/2\right]{+}O\left(\frac{|\vec{q}|^2}{\Lambda^2}\right).\nn\\&&
\eea
In the final step, we have reverted to a general coordinate system, i.e., $\vec{q}=(q_x,q_y)$. We therefore have
\beq
\Sigma_1(q)=\frac{i g^2}{16\pi}\vec{q}\cdot\vec\gamma\log(\Lambda/|\vec{q}|),\label{selfenergy}
\eeq
where the finite part has been absorbed into a redefinition of the ultraviolet cutoff $\Lambda$. Since the full two-point function is given by
\beq
\langle \psi(p)\bar\psi(0)\rangle=\frac{i}{\slashed p-i\Sigma(p)},
\eeq
the one-loop self-energy leads to a renormalization of the Fermi velocity:
\bea
v_q&\equiv&v_F^*(q)=v_F-i\frac{1}{|\vec{q}|^2}\tr[\vec q\cdot\vec\gamma\Sigma_1(q)]\nn\\&=&v_F+\frac{g^2}{16\pi}\log(\Lambda/|\vec{q}|)=v_F\left[1+\frac{\alpha}{4}\log(\Lambda/|\vec{q}|)\right].\nn\\\label{oneloopvq}
\eea

The one-loop velocity renormalization of Eq.~(\ref{oneloopvq}) can be inverted using the definition of the graphene coupling constant, Eq.~(\ref{defofalpha}), to express the coupling itself at a momentum $|\vec{q}|$ in terms of the bare coupling $\alpha$, which is to be interpreted as the coupling strength at the ultraviolet cutoff momentum scale $\Lambda$ in Eq.~(\ref{oneloopvq}):
\beq
\alpha_q= \frac{\alpha}{1 + \frac{\alpha}{4}\log (\Lambda/|\vec{q}|)}.\label{defofalphaq}
\eeq
Equation (\ref{defofalphaq}), sometimes referred to as the equation for the running coupling constant (i.e. a scale-dependent coupling constant), connects the effective interaction strength at a particular momentum scale $|\vec{q}|$ with that at the ultraviolet cutoff momentum $\Lambda$.  Since the ultraviolet momentum scale $\Lambda$ is arbitrary and unknown (physically, it is of the order of the inverse lattice spacing of graphene, but its precise value is arbitrary), it is preferable to eliminate the unknown parameter $\Lambda$ from the theory by considering the relationship between two momentum scales $|\vec{q}|$ and $|\vec{k}|$, which is easily done by writing down two equations similar to Eqs.~(\ref{oneloopvq}) and (\ref{defofalphaq}), but with $|\vec{q}|$ replaced by $|\vec{k}|$. It is then easy to eliminate the unknown cutoff scale $\Lambda$ from the equations to obtain the following relationship:
\beq
\frac{v_q}{v_k}=\frac{\alpha_k}{\alpha_q}=1 + \frac{\alpha_k}{4}\log(|\vec{k}|/|\vec{q}|).\label{vqfromvk}
\eeq
Equation (\ref{vqfromvk}) connects the physical velocity (and coupling strength) at one momentum $|\vec{k}|$ with that at another momentum $|\vec{q}|$ with no reference to the unknown ultraviolet scale $\Lambda$, and is thus the appropriate equation for the running coupling and the velocity renormalization in graphene up to one-loop interaction corrections.  We emphasize (this seems to have been completely missed in the graphene literature) that Eq.~(\ref{vqfromvk}) or its equivalent counterpart in higher loop orders, connecting the graphene velocity at one scale with that at another scale, is the correct formula to use in comparing theory and experiment in graphene, and not Eq.~(\ref{oneloopvq}) which has the unknown ultraviolet cutoff explicitly in the formula.  Of course, there is still the problem of the theory being explicitly done for undoped intrinsic graphene, whereas experiments are done as a function of density in doped extrinsic graphene (and not as a function of momentum in undoped graphene), but as discussed later on in Sec.~\ref{sec:experiment}, the substitution of the density-dependent Fermi momentum for $|\vec{q}|$ and $|\vec{k}|$ is justified by the RPA theory\cite{DasSarma_PRB07,Polini_SSC07} carried out for doped graphene (with the Fermi momentum being proportional to the square root of density, Eqs.~(\ref{oneloopvq})-(\ref{vqfromvk}) pick up an extra factor of 2 so that all the factors of 4 become factors of 8, and the momenta $|\vec{k}|$ and $|\vec{q}|$ are replaced by two different densities, $n_1$ and $n_2$).  Then, Eq.~(\ref{vqfromvk}) connects the graphene velocity or coupling strength at two distinct carrier densities by the following equation:
\beq
\frac{v_1}{v_2}=\frac{\alpha_2}{\alpha_1}=1 + \frac{\alpha_2}{8}\log(n_2/n_1),\label{v1fromv2}
\eeq
where $n_1$ and $n_2$ are two distinct carrier densities with Fermi velocities $v_1$ and $v_2$ respectively.

The existence of the logarithm in Eqs.~(\ref{oneloopvq})-(\ref{v1fromv2}) is entirely due to the linear graphene dispersion, which leads to properties at the ultraviolet scale (very high momentum or very high density) being reflected in the corresponding velocity or running coupling at very low momentum or carrier density, something that usually does not happen in ordinary Fermi liquids with parabolic energy dispersion.  We emphasize that our discussion above connecting the infrared and the ultraviolet scales in graphene did not refer at all to an RG flow or beta function, although the same results can also be obtained by constructing the graphene effective beta function and integrating it. We do not think that it is necessary to appeal to an RG analysis to discuss the results derived in Eqs.~(\ref{oneloopvq})-(\ref{v1fromv2}); in fact, we believe that the graphene effective field theory is a beautiful example of the innate simplicity of the RG analysis developed by Ken Wilson which was designed specifically to handle large log divergences in a theory through the systematic use of a momentum cutoff and then eliminating the cutoff in terms of effective physical quantities calculated at physical scales. Invoking the RG terminology to discuss the graphene one-loop calculation does not in any way give us any deeper insight than the simple analysis given above; the two are in fact completely equivalent since they deal with the logarithmic divergence in the theory arising from the linear dispersion, leading to the influence of the ultraviolet cutoff scale showing up in the infrared through a logarithmic divergence without the explicit presence of the arbitrary ultraviolet cutoff itself showing up in the theory.  In the rest of this paper, we will critically investigate the extent to which the same remains true in higher-order calculations, which is necessitated by the graphene coupling constant being of $O(1)$ in real laboratory systems.

\subsection{Vacuum polarization function bubble diagram}\label{sec:oneloopvacpol}

\begin{figure}
\includegraphics[width=0.5\columnwidth]{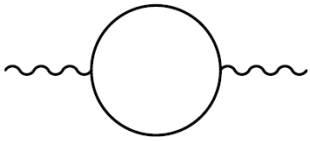}
\caption[Bubble]{One-loop bubble correction to the vacuum polarization.}
\label{fig:vacpolbubble}
\end{figure}
We also review the calculation of the vacuum polarization bubble diagram\cite{Gonzalez_NPB94} shown in Fig.~\ref{fig:vacpolbubble} as this diagram will enter into our two-loop calculations later on. This diagram translates to the expression
\bea
\Pi_B(q)&=&N\int\frac{d^3k}{(2\pi)^3}\tr\left[\gamma^0G_0(k)\gamma^0G_0(k+q)\right]\nn\\&=&-N\int\frac{d^3k}{(2\pi)^3}\tr\left[\gamma^0\frac{\slashed k}{k^2}\gamma^0\frac{\slashed k+\slashed q}{(k+q)^2}\right].
\eea
The trace over gamma matrices evaluates to
\beq
\tr\left[\gamma^0\gamma^\mu\gamma^0\gamma^\nu\right]=4(2\delta^{0\mu}\delta^{0\nu}-\delta^{\mu\nu}),
\eeq
leading to
\beq
\Pi_B(q){=}{-}4N\int\frac{d^3k}{(2\pi)^3}\frac{k_0(k_0{+}q_0){-}v_F^2\vec{k}\cdot(\vec{k}{+}\vec{q})}{(k_0^2{+}v_F^2|\vec{k}|^2)[(k_0{+}q_0)^2{+}v_F^2|\vec{k}{+}\vec{q}|^2]}.
\eeq
Performing the integral over $k_0$, we find
\bea
&&\Pi_B(q){=}\nn\\&&{-}2Nv_F\int\frac{d^2k}{(2\pi)^2}\frac{|\vec{k}|{+}|\vec{k}{+}\vec{q}|}{v_F^2(|\vec{k}|{+}|\vec{k}{+}\vec{q}|)^2{+}q_0^2}\left[1{-}\frac{\vec{k}\cdot(\vec{k}{+}\vec{q})}{|\vec{k}||\vec{k}{+}\vec{q}|}\right].\nn\\&&
\eea
We again choose the coordinate system such that $\vec{q}=(|\vec{q}|,0)$ and transform to elliptical coordinates:
\bea
&&k_x=\frac{|\vec{q}|}{2}(\cosh\mu\cos\nu-1),\qquad k_y=\frac{|\vec{q}|}{2}\sinh\mu\sin\nu,\nn\\&& d^2k=\frac{|\vec{q}|^2}{4}(\cosh^2\mu-\cos^2\nu)d\mu d\nu,
\eea
yielding
\beq
\Pi_B(q)=-\frac{Ng^2v_F|\vec{q}|^3}{4\pi^2}\int_0^\infty d\mu\int_0^{2\pi}d\nu\frac{\cosh\mu\sin^2\nu}{q_0^2+v_F^2|\vec{q}|^2\cosh^2\mu}.
\eeq
The integrals are trivial, and we obtain
\beq
\Pi_B(q)=-\frac{N}{8}\frac{|\vec{q}|^2}{\sqrt{q_0^2+v_F^2|\vec{q}|^2}}.
\eeq
Performing the analytic continuation, $q_0\to-i\omega$, this becomes the well known result
\beq
\Pi_B(q)=-\frac{N|\vec{q}|}{8v_F}\frac{1}{\sqrt{1-y^2}},\label{oneloopbubble}
\eeq
where we have defined
\beq
y\equiv \frac{\omega}{v_F|\vec{q}|}.
\eeq

\subsection{One-loop vertex diagram}

\begin{figure}
\includegraphics[width=0.5\columnwidth]{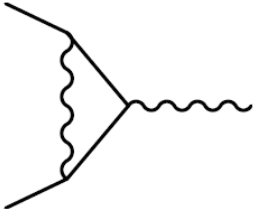}
\caption[Self-energy]{One-loop correction to the vertex function.}
\label{fig:vertex}
\end{figure}
Here, we will consider the remaining one-loop diagram, namely the vertex diagram shown in Fig.~(\ref{fig:vertex}). We will not compute this diagram explicitly since the result is not needed for our higher-order calculations of the vacuum polarization and electron self-energy. However, we do wish to demonstrate that this diagram has no ultraviolet divergences as part of our general analysis of the divergence structure of the graphene effective field theory. The vertex diagram evaluates to
\beq
V_1(p,q)=\frac{ig^2}{2}\int\frac{d^3k}{(2\pi)^3}\gamma^0\frac{\slashed p+\slashed k}{(p+k)^2}\gamma^0\frac{\slashed q+\slashed k}{(q+k)^2}\gamma^0\frac{1}{|\vec{k}|}.
\eeq
This integral has a potential logarithmic divergence which can be isolated by setting $p=q=0$ in the integrand. We then obtain
\bea
V_{1,div}\!\!\!&=&\!\!\!\frac{ig^2}{2}\int\frac{d^3k}{(2\pi)^3}\gamma^0\frac{\slashed k}{k^2}\gamma^0\frac{\slashed k}{k^2}\gamma^0\frac{1}{|\vec{k}|}
\nn\\\!\!\!&=&\!\!\!\frac{ig^2}{2}\int\frac{d^3k}{(2\pi)^3}\frac{1}{|\vec{k}|k^4}(k_0^2\gamma^0-2v_Fk_0\vec{k}\cdot\vec\gamma-v_F^2|\vec{k}|^2\gamma^0)
\nn\\&&=0.
\eea
The final expression vanishes due to the integration over $k_0$, demonstrating that the vertex function is ultraviolet finite. We will comment further on this result in our general discussion of ultraviolet divergences later on in Sec.~\ref{sec:divergencestructure}.

\subsection{Self-energy correction to vacuum polarization function}\label{sec:vacpolSE}

\subsubsection{Computing the diagram}

\begin{figure}
\includegraphics[width=0.5\columnwidth]{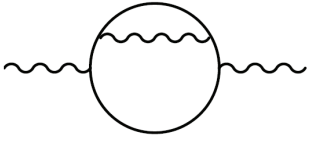}
\caption[Self-energy]{Two-loop self-energy correction to the vacuum polarization.}
\label{fig:vacpolselfenergy}
\end{figure}

In this section and the next, we compute the two-loop corrections to the vacuum polarization function. The results can be used to calculate corrections to the dielectric function and to test the conjecture that all ultraviolet divergences can be absorbed into the renormalized Fermi velocity and electron field strength. These corrections were previously computed in Refs.~[\onlinecite{Kotov_PRB08}] and [\onlinecite{Sodemann_PRB12}], with conflicting results. Our calculation of the first two-loop correction, given in this section, finds agreement with Ref.~[\onlinecite{Sodemann_PRB12}].

The diagram corresponding to the self-energy correction to the vacuum polarization function is shown in Fig.~\ref{fig:vacpolselfenergy} and evaluates to the expression
\bea
&&\!\!\!\!\!\!\Pi_{SE}(q)=
\nn\\&&\!\!\!\!\!\!N\int\frac{d^3k}{(2\pi)^3}\tr\left[\gamma^0G_0(k)\gamma^0G_0(k+q)\Sigma_1(k+q)G_0(k+q)\right]
\nn\\&&\!\!\!\!\!\!=\frac{Ng^2}{16\pi}\int\frac{d^3k}{(2\pi)^3}\tr\left[\gamma^0\frac{\slashed k}{k^2}\gamma^0\frac{\slashed k{+}\slashed q}{(k{+}q)^2}(\vec{k}{+}\vec{q})\cdot\vec{\gamma}\frac{\slashed k{+}\slashed q}{(k{+}q)^2}\right]
\nn\\&&\qquad\qquad\times\log(\Lambda/|\vec{k}{+}\vec{q}|)
\eea
We begin by focusing on the trace of gamma matrices, which can be expressed as
\bea
&&\tr\left[\gamma^0\gamma^\mu\gamma^0\gamma^\nu\gamma^i\gamma^\lambda\right]
\nn\\&&=\delta_{\mu0}\tr\left[\gamma^0\gamma^\nu\gamma^i\gamma^\lambda\right]
-(1-\delta_{\mu0})\tr\left[\gamma^\mu\gamma^\nu\gamma^i\gamma^\lambda\right]\nn\\
&&=4\delta_{\mu0}(\delta_{\nu0}\delta_{i\lambda}+\delta_{\lambda0}\delta_{i\nu})
\nn\\&&\qquad-4(1-\delta_{\mu0})(\delta_{\mu\nu}\delta_{i\lambda}-\delta_{\mu i}\delta_{\nu\lambda}+\delta_{\mu\lambda}\delta_{i\nu}).
\eea
where $\mu,\nu,\lambda=0,1,2$ and $i=1,2$. To arrive at the second line above, we made use of the identity in Eq.~(\ref{gammaid}). The expression for $\Pi_{SE}(q)$ therefore breaks up into a sum of five contributions:
\bea
T_1\!\!\!&=&\!\!\!\frac{Ng^2v_F}{4\pi}\int\frac{d^3k}{(2\pi)^3}\frac{k_0(k_0+q_0)|\vec{k}+\vec{q}|^2}{k^2(k+q)^4}\log(\Lambda/|\vec{k}+\vec{q}|),\nn\\
T_2\!\!\!&=&\!\!\!T_1,\nn\\
T_3\!\!\!&=&\!\!\!-\frac{Ng^2v_F^3}{4\pi}\int\frac{d^3k}{(2\pi)^3}\frac{\vec{k}\cdot(\vec{k}+\vec{q})|\vec{k}+\vec{q}|^2}{k^2(k+q)^4}\log(\Lambda/|\vec{k}+\vec{q}|),\nn\\
T_4\!\!\!&=&\!\!\!\frac{Ng^2v_F}{4\pi}\int\frac{d^3k}{(2\pi)^3}\frac{\vec{k}\cdot(\vec{k}+\vec{q})}{k^2(k+q)^2}\log(\Lambda/|\vec{k}+\vec{q}|)),\nn\\
T_5\!\!\!&=&\!\!\!T_3.
\eea
The function $\Pi_{SE}(q)$ is then given by
\beq
\Pi_{SE}(q)=2T_1+2T_3+T_4.
\eeq
The $k_0$ integrals appearing in the above approximations can be evaluated directly:
\bea
I_1&\equiv&\int\frac{dk_0}{2\pi}\frac{k_0(k_0+q_0)}{(k_0^2+v_F^2|\vec{k}|^2)[(k_0+q_0)^2+v_F^2|\vec{k}+\vec{q}|^2]^2}
\nn\\&=&\frac{1}{4v_F|\vec{k}+\vec{q}|}\frac{v_F^2(|\vec{k}|+|\vec{k}+\vec{q}|)^2-q_0^2}{[v_F^2(|\vec{k}|+|\vec{k}+\vec{q}|)^2+q_0^2]^2},\nn\\
I_3&\equiv&\int\frac{dk_0}{2\pi}\frac{1}{(k_0^2+v_F^2|\vec{k}|^2)[(k_0+q_0)^2+v_F^2|\vec{k}+\vec{q}|^2]^2}
\nn\\&=&\frac{v_F^2(|\vec{k}|+|\vec{k}+\vec{q}|)^2(|\vec{k}|+2|\vec{k}+\vec{q}|)+|\vec{k}|q_0^2}{4v_F^3|\vec{k}||\vec{k}+\vec{q}|^3[v_F^2(|\vec{k}|+|\vec{k}+\vec{q}|)^2+q_0^2]^2},\nn\\
I_4&\equiv&\int\frac{dk_0}{2\pi}\frac{1}{(k_0^2+v_F^2|\vec{k}|^2)[(k_0+q_0)^2+v_F^2|\vec{k}+\vec{q}|^2]}
\nn\\&=&\frac{|\vec{k}|+|\vec{k}+\vec{q}|}{2v_F|\vec{k}||\vec{k}+\vec{q}|[v_F^2(|\vec{k}|+|\vec{k}+\vec{q}|)^2+q_0^2]},
\eea
yielding
\bea
\Pi_{SE}(q)&=&{-}\frac{Ng^2v_F}{4\pi}\int\frac{d^2k}{(2\pi)^2}\bigg[-2|\vec{k}+\vec{q}|^2I_1
\nn\\&&+(2v_F^2|\vec{k}{+}\vec{q}|^2I_3{-}I_4)\vec{k}\cdot(\vec{k}{+}\vec{q})  \bigg]\log(\Lambda/|\vec{k}{+}\vec{q}|)\nn\\
&=&-\frac{Ng^2}{8\pi}\int\frac{d^2k}{(2\pi)^2}\frac{\vec{k}\cdot(\vec{k}+\vec{q})-|\vec{k}||\vec{k}+\vec{q}|}{|\vec{k}|}\nn\\&&\times\frac{[v_F^2(|\vec{k}|{+}|\vec{k}{+}\vec{q}|)^2{-}q_0^2]}
{[v_F^2(|\vec{k}|{+}|\vec{k}{+}\vec{q}|)^2{+}q_0^2]^2}\log(\Lambda/|\vec{k}{+}\vec{q}|).
\eea
In order to perform the remaining integrals, we first change variables according to $\vec{k}\to-\vec{k}-\vec{q}$, and we choose the coordinate system such that $\vec{q}=(|\vec{q}|,0)$. We then transform to elliptical coordinates defined by
\bea
&&k_x=\frac{|\vec{q}|}{2}(\cosh\mu\cos\nu-1),\qquad k_y=\frac{|\vec{q}|}{2}\sinh\mu\sin\nu,\nn\\&& d^2k=\frac{|\vec{q}|^2}{4}(\cosh^2\mu-\cos^2\nu)d\mu d\nu,
\eea
to obtain
\bea
\Pi_{SE}(q)\!\!\!&{=}&\!\!\!\frac{Ng^2|\vec{q}|^3}{128\pi^3}\int_0^\infty d\mu\int_0^{2\pi}d\nu \log\left[\frac{2\Lambda}{|\vec{q}|(\cosh\mu{-}\cos\nu)}\right]
\nn\\&&\!\!\!\times
\sin^2\nu(\cosh\mu{-}\cos\nu)\frac{v_F^2|\vec{q}|^2\cosh^2\mu{-}q_0^2}{[v_F^2|\vec{q}|^2\cosh^2\mu{+}q_0^2]^2}.\label{Paintegral1}\nn\\&&
\eea
We first focus on the term proportional to $\log(\Lambda/|\vec{q}|)$:
\bea
&&\!\!\!\!\!\!\!\frac{Ng^2|\vec{q}|^3}{128\pi^3}\int_0^\infty d\mu\int_0^{2\pi}d\nu\log(\Lambda/|\vec{q}|)\sin^2\nu(\cosh\mu-\cos\nu)
\nn\\&&\qquad\qquad\times\frac{v_F^2|\vec{q}|^2\cosh^2\mu-q_0^2}{[v_F^2|\vec{q}|^2\cosh^2\mu+q_0^2]^2}
\nn\\&&\!\!\!\!\!\!\!=\frac{Ng^2|\vec{q}|^3}{128\pi^2}\log(\Lambda/|\vec{q}|)\int_0^\infty d\mu\cosh\mu\frac{v_F^2|\vec{q}|^2\cosh^2\mu-q_0^2}{[v_F^2|\vec{q}|^2\cosh^2\mu+q_0^2]^2}
\nn\\&&\!\!\!\!\!\!\!=\frac{Ng^2|\vec{q}|}{256\pi v_F^2}\log(\Lambda/|\vec{q}|)\frac{1}{[1+(q_0/v_F|\vec{q}|)^2]^{3/2}}
\nn\\&&\!\!\!\!\!\!\!=\frac{N\alpha|\vec{q}|}{64 v_F}\log(\Lambda/|\vec{q}|)\frac{1}{[1-(\omega/v_F|\vec{q}|)^2]^{3/2}},\label{logterm}
\eea
where we have performed the analytic continuation, $q_0\to-i\omega$. Note that this analytic continuation must be performed after the integration over $\mu$ and $\nu$; otherwise the final expression in Eq.~(\ref{logterm}) only holds if $\omega<v_F|\vec{q}|$. The above result for the coefficient of the logarithmic divergence, Eq.~(\ref{logterm}), agrees with that obtained in Ref.~[\onlinecite{Sodemann_PRB12}]. It should also be noted that an apparent factor of 2 difference comes from different definitions of the fermion degeneracy factor, $N$. In that reference, real graphene corresponds to $N=4$, whereas here it corresponds to $N=2$. It is also worth noting that the factor $\sin^2\nu$ appearing in the integrand of Eq.~(\ref{Paintegral1}) is missing in Eq.~(35) of Ref.~[\onlinecite{Sodemann_PRB12}]. However, this appears to be merely a typo---this factor effectively produces an additional factor of 1/2 which appears to have been included in the final result, Eq.~(19) of that paper.

Returning to the full expression for the self-energy correction to the vacuum polarization, we may write
\beq
\Pi_{SE}(q)=\frac{N\alpha|\vec{q}|}{32\pi v_F}\left[\frac{\pi}{2}\frac{1}{(1-y^2)^{3/2}}\log(\Lambda/|\vec{q}|)+I_a(y)\right],\label{fullPa}
\eeq
with
\bea
&&I_a(y)\equiv \frac{1}{\pi}\int d\mu d\nu\log\left[\frac{2}{\cosh\mu-\cos\nu}\right]\nn\\&&\qquad\times\sin^2\nu(\cosh\mu-\cos\nu)\frac{\cosh^2\mu+y^2}{(\cosh^2\mu-y^2)^2},
\eea
and
\beq
y\equiv\frac{\omega}{v_F|\vec{q}|}.
\eeq
$I_a(x)$ can be expressed analytically as\cite{Sodemann_PRB12}
\bea
I_a(x)\!\!\!&=&\!\!\!\frac{1}{3}\frac{1+2x^2}{1-x^2}-\frac{x}{6}\frac{5-2x^2}{1-x^2}\log\left(\frac{1-x}{1+x}\right)
\nn\\&&\!\!\!-\frac{\pi}{12}\frac{3-12\log2+6x^2-4x^4}{(1-x^2)^{3/2}}
\nn\\&&\!\!\!-\frac{i}{(1-x^2)^{3/2}}\bigg[\frac{\pi^2}{4}-\hbox{Li}_2(x+i\sqrt{1-x^2})
\nn\\&&\!\!\!+\hbox{Li}_2(-x-i\sqrt{1-x^2})+\frac{i\pi}{2}\log(x+i\sqrt{1-x^2})\bigg],\label{Iaexplicit}\nn\\&&
\eea
where $\hbox{Li}_2(z)$ denotes the dilogarithm function. In conclusion, the result for $\Pi_{SE}(q)$, Eq.~(\ref{fullPa}), is equal to the result obtained in Ref.~[\onlinecite{Sodemann_PRB12}]. Ref.~[\onlinecite{Kotov_PRB08}] also computes the self-energy correction to the vacuum polarization, but considers only the static limit, $y=0$. In that reference, the constant term, $\frac{2}{\pi}I_a(0)$, is absorbed into the definition of $\Lambda$. However, as pointed out in Ref.~[\onlinecite{Sodemann_PRB12}], absorbing this constant into $\Lambda$ is inconsistent with the original definition of $\Lambda$, which was set by the one-loop electron self-energy calculation given above. We therefore follow Ref.~[\onlinecite{Sodemann_PRB12}] in retaining this constant contribution.

\subsubsection{Renormalization}

In the one-loop calculations, we have seen that only the electron self-energy is ultraviolet divergent, meaning that only the Fermi velocity is renormalized to first order, and not, for example, the electric charge as well. It is expected that this trend will persist to higher orders. This can already be checked at second order using the result we have just obtained for the self-energy correction to the vacuum polarization. In particular, this correction should combine with the first-order bubble diagram contribution in such a way that the $\log(\Lambda/|\vec{q}|)$ divergence is naturally absorbed into a renormalization of the Fermi velocity, i.e., the bare velocity $v_F$ is effectively replaced by the one-loop expression for $v_q$, Eq.~(\ref{oneloopvq}), in the one-loop vacuum polarization.\cite{Sodemann_PRB12} Here, we verify explicitly that this is indeed the case.

Recall that the bubble diagram contribution is
\beq
\Pi_B(q)=-\frac{N|\vec{q}|}{8v_F}\frac{1}{\sqrt{1-y^2}},
\eeq
while the self-energy contribution is
\beq
\Pi_{SE}(q)=\frac{N\alpha|\vec{q}|}{64v_F}\frac{1}{(1-y^2)^{3/2}}\log(\Lambda/|\vec{q}|)+\frac{N\alpha|\vec{q}|}{32\pi v_F}I_a(y).
\eeq
The net contribution to the vacuum polarization function from the self-energy diagram receives an extra factor of 2 due to the symmetry of the diagram. We therefore consider
\bea
&&\!\!\!\!\!\!\Pi_B(q)+2\Pi_{SE}(q)=
\nn\\&&\!\!\!\!\!\!-\frac{N|\vec{q}|}{8v_F}\bigg[\frac{1}{\sqrt{1-y^2}}-\frac{\alpha}{4}\frac{1}{(1-y^2)^{3/2}}\log(\Lambda/|\vec{q}|) -\frac{\alpha}{2\pi}I_a(y)\bigg].\label{renorm}\nn\\&&
\eea

First consider the static limit, $y=0$.\cite{Kotov_PRB08} Renormalization amounts to the second (divergent) term in Eq.~(\ref{renorm}) getting absorbed into the first such that $v_F$ is replaced by $v_q$:
\bea
&&\!\!\!\!\!\!\!\!\!\!1-\frac{\alpha}{4}\log(\Lambda/|\vec{q}|)\to \frac{v_F}{v_q}
\nn\\&&\!\!\!\!\!\!\!\!\!\!\Rightarrow\Pi_B(\vec{q},\omega=0)+2\Pi_{SE}(\vec{q},\omega=0)\to-\frac{N|\vec{q}|}{8v_q}+\ldots
\eea
This replacement seems to assume that $\frac{\alpha}{4}\log(\Lambda/|\vec{q}|)$ can be treated as a small quantity, so that
\beq
\frac{1}{1+\frac{\alpha}{4}\log(\Lambda/|\vec{q}|)}\approx 1-\frac{\alpha}{4}\log(\Lambda/|\vec{q}|).
\eeq
However, we have already assumed that the ultraviolet cutoff satisfies $\Lambda\gg|\vec{q}|$, so one cannot assume that this is a small quantity. This replacement is more properly justified by systematically expanding the bare coupling $\alpha$ in terms of the renormalized coupling
\beq
\alpha_q\equiv \frac{g^2}{4\pi v_q}=\frac{\alpha}{1+(\alpha/4)\log(\Lambda/|\vec{q}|)}.
\eeq
This expansion is easily constructed up to second order:
\beq
\alpha=\alpha_q+\frac{\alpha_q^2}{4}\log(\Lambda/|\vec{q}|)+O(\alpha_q^3).
\eeq
This expansion is sensible and works even for large $\log(\Lambda/|\vec{q}|)$ since in this limit $\alpha_q^2\log(\Lambda/|\vec{q}|)\sim\alpha_q$ is still small. Using this expansion, we can justify the above approximation:
\bea
&&\frac{1}{v_F}\left(1-\frac{\alpha}{4}\log(\Lambda/|\vec{q}|)\right)=\frac{4\pi\alpha}{g^2}\left(1-\frac{\alpha}{4}\log(\Lambda/|\vec{q}|)\right)\nn\\
&&=\frac{4\pi\alpha_q}{g^2}\left(1{+}\frac{\alpha_q}{4}\log(\Lambda/|\vec{q}|){+}O(\alpha_q^2)\right)\nn\\&&\qquad\qquad\times\left(1{-}\frac{\alpha_q}{4}\log(\Lambda/|\vec{q}|){+}O(\alpha_q^2)\right)\nn\\
&&=\frac{4\pi\alpha_q}{g^2}\left(1+O(\alpha_q^2)\right)=\frac{1}{v_q}+O(\alpha_q^3).
\eea

The basic procedure works even if we do not restrict attention to the static limit.\cite{Sodemann_PRB12} In general, renormalization amounts to throwing away the log term and replacing all occurrences of $v_F$ by $v_q$:
\beq
\Pi_B(q)+2\Pi_{SE}(q)\to-\frac{N|\vec{q}|}{8v_q}\left[\frac{1}{\sqrt{1-x^2}}-\frac{\alpha_q}{2\pi}I_a(x)\right],\label{renormalized}
\eeq
where
\beq
x\equiv \frac{\omega}{v_q|\vec{q}|}.
\eeq
To check explicitly that this is the correct prescription, let us first define the ``small" quantity
\beq
\xi_1\equiv\frac{\alpha}{4}\log(\Lambda/|\vec{q}|),
\eeq
so that the variables $y$ and $x$ are related by
\beq
y=x[1+\xi_1]+O(\alpha^2).
\eeq
We first focus on the disappearance of the log term. Begin by expressing the $y$-dependence of $\Pi_B$ in terms of $x$:
\bea
&&[1-y^2]^{-1/2}\approx[1-x^2-2\xi_1x^2+O(\alpha^2)]^{-1/2}\nn\\&&\approx(1-x^2)^{-1/2}\left[1+\xi_1\frac{x^2}{1-x^2}\right]+O(\alpha^2).
\eea
Similarly, the $y$-dependence of the log term in $\Pi_{SE}$ can be expressed as
\beq
-\xi_1[1-y^2]^{-3/2}\approx-\xi_1[1-x^2]^{-3/2}+O(\alpha^2).
\eeq
Combining these two results gives
\bea
&&[1-y^2]^{-1/2}-\xi_1[1-y^2]^{-3/2}\nn\\&&\approx(1-x^2)^{-1/2}\left[1+\xi_1\frac{x^2}{1-x^2}-\xi_1\frac{1}{1-x^2}\right]+O(\alpha^2)\nn\\&&=(1-x^2)^{-1/2}[1-\xi_1]+O(\alpha^2).
\eea
Finally, using that
\beq
\frac{1}{v_F}[1-\xi_1]\approx\frac{1}{v_q},
\eeq
we see that the log term effectively disappears and $v_F$ is replaced by $v_q$ to arrive at Eq.~(\ref{renormalized}).

A similar analysis can be performed for the finite term in Eq.~(\ref{renorm}). However, this is now completely trivial since the finite term is proportional to $\alpha$:
\beq
\alpha I_a(y)\approx \alpha I_a(x)+O(\alpha^2)\approx \alpha(1-\xi_1)I_a(x)+O(\alpha^2).
\eeq
Again the $(1-\xi_1)$ factor combines with the overall $1/v_F$ in Eq.~(\ref{renorm}) to produce $1/v_q$.

In summary, the renormalization prescription of absorbing ultraviolet divergences into the Fermi velocity appears to be self-consistent in both the static and non-static regimes. Note that this result already implies that the second two-loop correction to the vacuum polarization, namely the vertex correction, is ultraviolet finite since the self-energy correction has fully accounted for the renormalization of the Fermi velocity in the expression for the one-loop vacuum polarization. The finiteness of the vertex correction is verified explicitly in the next section.

\subsection{Vertex correction to vacuum polarization function}

\begin{figure}
\includegraphics[width=0.5\columnwidth]{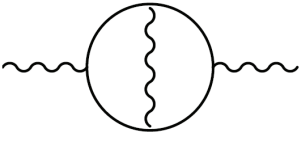}
\caption[Self-energy]{Two-loop vertex correction to the vacuum polarization.}
\label{fig:vacpolvertex}
\end{figure}

The diagram corresponding to the two-loop vertex correction to the vacuum polarization is shown in Fig.~\ref{fig:vacpolvertex}. This diagram evaluates to
\bea
\Pi_V(q)\!\!\!&=&\!\!\!-N\int\frac{d^3k}{(2\pi)^3}\int\frac{d^3p}{(2\pi)^3}D_0(k-p)
\nn\\&&\!\!\!\times\tr\left[\gamma^0G_0(k)\gamma^0G_0(p)\gamma^0G_0(p+q)\gamma^0G_0(k+q)\right]\nn\\
\!\!\!&=&\!\!\!-\frac{g^2N}{2}\int\frac{d^3k}{(2\pi)^3}\int\frac{d^3p}{(2\pi)^3}\frac{1}{|\vec{k}-\vec{p}|}
\nn\\&&\!\!\!\times\tr\left[\gamma^0\frac{\slashed k}{k^2}\gamma^0\frac{\slashed p}{p^2}\gamma^0\frac{\slashed p+\slashed q}{(p+q)^2}\gamma^0\frac{\slashed k+\slashed q}{(k+q)^2}\right].
\eea
The $\gamma$-matrix trace in the integrand is
\bea
&&\tr\left[\gamma^0\gamma^\mu\gamma^0\gamma^\nu\gamma^0\gamma^\rho\gamma^0\gamma^\sigma\right]=
\nn\\&&4\delta_{\mu0}\delta_{\rho0}\left(2\delta_{\nu0}\delta_{\sigma0}-\delta_{\nu\sigma}\right)
\nn\\&&-4\delta_{\mu0}\delta_{\rho\ne0}\left(\delta_{\nu0}\delta_{\rho\sigma}+\delta_{\sigma0}\delta_{\nu\rho}\right)
\nn\\
&&-4\delta_{\mu\ne0}\delta_{\rho0}\left(\delta_{\mu\nu}\delta_{\sigma0}+\delta_{\mu\sigma}\delta_{\nu0}\right)
\nn\\&&+4\delta_{\mu\ne0}\delta_{\rho\ne0}\left(\delta_{\mu\nu}\delta_{\rho\sigma}-\delta_{\mu\rho}\delta_{\nu\sigma}+\delta_{\mu\sigma}\delta_{\nu\rho}\right).
\eea
We therefore have nine contributions:
\bea
&&\!\!\!\!\!\!\!R_1{\equiv}{-}4g^2N{\int}\frac{d^3k}{(2\pi)^3}{\int}\frac{d^3p}{(2\pi)^3}\frac{1}{|\vec{k}{-}\vec{p}|}\frac{k_0p_0(p_0{+}q_0)(k_0{+}q_0)}{k^2p^2(p{+}q)^2(k{+}q)^2},\nn\\
&&\!\!\!\!\!\!\!R_2{\equiv}2g^2N{\int}\frac{d^3k}{(2\pi)^3}{\int}\frac{d^3p}{(2\pi)^3}\frac{1}{|\vec{k}{-}\vec{p}|}\frac{k_0(p_0{+}q_0)p\cdot(k{+}q)}{k^2p^2(p{+}q)^2(k{+}q)^2},\nn\\
&&\!\!\!\!\!\!\!R_3{\equiv}2g^2Nv_F^2{\int}\frac{d^3k}{(2\pi)^3}{\int}\frac{d^3p}{(2\pi)^3}\frac{1}{|\vec{k}{-}\vec{p}|}\frac{k_0p_0(\vec{p}{+}\vec{q})\cdot(\vec{k}{+}\vec{q})}{k^2p^2(p{+}q)^2(k{+}q)^2},\nn\\
&&\!\!\!\!\!\!\!R_4{\equiv}2g^2Nv_F^2{\int}\frac{d^3k}{(2\pi)^3}{\int}\frac{d^3p}{(2\pi)^3}\frac{1}{|\vec{k}{-}\vec{p}|}\frac{k_0(k_0{+}q_0)\vec{p}\cdot(\vec{p}{+}\vec{q})}{k^2p^2(p{+}q)^2(k{+}q)^2},\nn\\
&&\!\!\!\!\!\!\!R_5{\equiv}2g^2Nv_F^2{\int}\frac{d^3k}{(2\pi)^3}{\int}\frac{d^3p}{(2\pi)^3}\frac{1}{|\vec{k}{-}\vec{p}|}\frac{(p_0{+}q_0)(k_0{+}q_0)\vec{k}\cdot\vec{p}}{k^2p^2(p{+}q)^2(k{+}q)^2},\nn\\
&&\!\!\!\!\!\!\!R_6{\equiv}2g^2Nv_F^2{\int}\frac{d^3k}{(2\pi)^3}{\int}\frac{d^3p}{(2\pi)^3}\frac{1}{|\vec{k}{-}\vec{p}|}\frac{p_0(p_0{+}q_0)\vec{k}\cdot(\vec{k}{+}\vec{q})}{k^2p^2(p{+}q)^2(k{+}q)^2},\nn\\
&&\!\!\!\!\!\!\!R_7{\equiv}{-}2g^2Nv_F^4{\int}\frac{d^3k}{(2\pi)^3}{\int}\frac{d^3p}{(2\pi)^3}\frac{1}{|\vec{k}{-}\vec{p}|}\frac{\vec{k}\cdot\vec{p}(\vec{p}{+}\vec{q})\cdot(\vec{k}{+}\vec{q})}{k^2p^2(p{+}q)^2(k{+}q)^2},\nn\\
&&\!\!\!\!\!\!\!R_8{\equiv}2g^2Nv_F^2{\int}\frac{d^3k}{(2\pi)^3}{\int}\frac{d^3p}{(2\pi)^3}\frac{1}{|\vec{k}{-}\vec{p}|}\frac{\vec{k}\cdot(\vec{p}{+}\vec{q})p\cdot(k{+}q)}{k^2p^2(p{+}q)^2(k{+}q)^2},\nn\\
&&\!\!\!\!\!\!\!R_9{\equiv}{-}2g^2Nv_F^4{\int}\frac{d^3k}{(2\pi)^3}{\int}\frac{d^3p}{(2\pi)^3}\frac{1}{|\vec{k}{-}\vec{p}|}\frac{\vec{k}\cdot(\vec{k}{+}\vec{q})\vec{p}\cdot(\vec{p}{+}\vec{q})}{k^2p^2(p{+}q)^2(k{+}q)^2},\nn\\
\eea
As before, we first perform the integrations over $k_0$ and $p_0$. There are three basic integrals over these frequencies that appear in the $R_i$:
\bea
Q_1^u&\equiv&\int\frac{du_0}{2\pi}\frac{u_0(u_0+q_0)}{(u_0^2+v_F^2|\vec{u}|^2)[(u_0+q_0)^2+v_F^2|\vec{u}+\vec{q}|^2]}
\nn\\&&=\frac{v_F(|\vec{u}|+|\vec{u}+\vec{q}|)}{2[v_F^2(|\vec{u}|+|\vec{u}+\vec{q}|)^2+q_0^2]},\nn\\
\eea
\bea
Q_2^u&\equiv&\int\frac{du_0}{2\pi}\frac{u_0}{(u_0^2+v_F^2|\vec{u}|^2)[(u_0+q_0)^2+v_F^2|\vec{u}+\vec{q}|^2]}
\nn\\&&=-\frac{q_0}{2v_F|\vec{u}+\vec{q}|[v_F^2(|\vec{u}|+|\vec{u}+\vec{q}|)^2+q_0^2]},\nn\\
\eea
\bea
Q_3^u&\equiv&\int\frac{du_0}{2\pi}\frac{1}{(u_0^2+v_F^2|\vec{u}|^2)[(u_0+q_0)^2+v_F^2|\vec{u}+\vec{q}|^2]}
\nn\\&&=\frac{|\vec{u}|+|\vec{u}+\vec{q}|}{2v_F|\vec{u}||\vec{u}+\vec{q}|[v_F^2(|\vec{u}|+|\vec{u}+\vec{q}|)^2+q_0^2]}.
\eea
The $R_i$ can be expressed in terms of these three functions:
\bea
&&\!\!\!\!\!\!R_1{\equiv}{-}4g^2N{\int}\frac{d^2k}{(2\pi)^2}{\int}\frac{d^2p}{(2\pi)^2}\frac{1}{|\vec{k}{-}\vec{p}|}Q_1^kQ_1^p,\nn\\
&&\!\!\!\!\!\!R_2{\equiv}2g^2N{\int}\frac{d^2k}{(2\pi)^2}{\int}\frac{d^2p}{(2\pi)^2}\frac{1}{|\vec{k}{-}\vec{p}|}\bigg[Q_1^kQ_1^p\nn\\&&\qquad{+}v_F^2\vec{p}\cdot(\vec{k}{+}\vec{q})Q_2^k(Q_2^p{+}q_0Q_3^p)\bigg],\nn\\
&&\!\!\!\!\!\!R_3{\equiv}2g^2Nv_F^2{\int}\frac{d^2k}{(2\pi)^2}{\int}\frac{d^2p}{(2\pi)^2}\frac{1}{|\vec{k}{-}\vec{p}|}(\vec{p}{+}\vec{q})\cdot(\vec{k}{+}\vec{q})Q_2^kQ_2^p,\nn\\
&&\!\!\!\!\!\!R_4{\equiv}2g^2Nv_F^2{\int}\frac{d^2k}{(2\pi)^2}{\int}\frac{d^2p}{(2\pi)^2}\frac{1}{|\vec{k}{-}\vec{p}|}\vec{p}\cdot(\vec{p}{+}\vec{q})Q_1^kQ_3^p,\nn\\
&&\!\!\!\!\!\!R_5{\equiv}2g^2Nv_F^2{\int}\frac{d^2k}{(2\pi)^2}{\int}\frac{d^2p}{(2\pi)^2}\frac{1}{|\vec{k}{-}\vec{p}|}\vec{k}\cdot\vec{p}(Q_2^k{+}q_0Q_3^k)\nn\\&&\qquad\times(Q_2^p{+}q_0Q_3^p),\nn\\
&&\!\!\!\!\!\!R_6{\equiv}2g^2Nv_F^2{\int}\frac{d^2k}{(2\pi)^2}{\int}\frac{d^2p}{(2\pi)^2}\frac{1}{|\vec{k}{-}\vec{p}|}\vec{k}\cdot(\vec{k}{+}\vec{q})Q_3^kQ_1^p,\nn\\
&&\!\!\!\!\!\!R_7{\equiv}{-}2g^2Nv_F^4{\int}\frac{d^2k}{(2\pi)^2}{\int}\frac{d^2p}{(2\pi)^2}\frac{1}{|\vec{k}{-}\vec{p}|}\vec{k}\cdot\vec{p}\nn\\&&\qquad\times(\vec{p}{+}\vec{q})\cdot(\vec{k}{+}\vec{q})Q_3^kQ_3^p,\nn\\
&&\!\!\!\!\!\!R_8{\equiv}2g^2Nv_F^2{\int}\frac{d^2k}{(2\pi)^2}{\int}\frac{d^2p}{(2\pi)^2}\frac{1}{|\vec{k}{-}\vec{p}|}\vec{k}\cdot(\vec{p}{+}\vec{q})\nn\\&&\qquad\times\left[(Q_2^k{+}q_0Q_3^k)Q_2^p{+}v_F^2\vec{p}\cdot(\vec{k}{+}\vec{q})Q_3^kQ_3^p\right],\nn\\
&&\!\!\!\!\!\!R_9{\equiv}{-}2g^2Nv_F^4{\int}\frac{d^2k}{(2\pi)^2}{\int}\frac{d^2p}{(2\pi)^2}\frac{1}{|\vec{k}{-}\vec{p}|}\vec{k}\cdot(\vec{k}{+}\vec{q})\nn\\&&\qquad\times\vec{p}\cdot(\vec{p}{+}\vec{q})Q_3^kQ_3^p.
\eea
Adding together these contributions, we find
\bea
&&\!\!\!\!\!\!\Pi_V(q){=}{-}2g^2N\int\frac{d^2k}{(2\pi)^2}\int\frac{d^2p}{(2\pi)^2}\frac{1}{|\vec{k}{-}\vec{p}|}\bigg\{Q_1^kQ_1^p
\nn\\&&\!\!\!\!\!\!
{-}v_F^2\left[\vec{p}{\cdot}(\vec{p}{+}\vec{q})Q_1^kQ_3^p{+}\vec{k}{\cdot}(\vec{k}{+}\vec{q})Q_3^kQ_1^p\right]\nn\\
&&\!\!\!\!\!\!{-}v_F^2\left[\vec{k}{\cdot}\vec{p}{+}\vec{k}{\cdot}(\vec{p}{+}\vec{q}){+}\vec{p}{\cdot}(\vec{k}{+}\vec{q}){+}(\vec{p}{+}\vec{q}){\cdot}(\vec{k}{+}\vec{q})\right]Q_2^kQ_2^p\nn\\
&&\!\!\!\!\!\!{-}v_F^2q_0\left[\vec{k}{\cdot}\vec{p}{+}\vec{p}{\cdot}(\vec{k}{+}\vec{q})\right]Q_2^kQ_3^p
{-}v_F^2q_0\left[\vec{k}{\cdot}\vec{p}{+}\vec{k}{\cdot}(\vec{p}{+}\vec{q})\right]Q_3^kQ_2^p\nn\\
&&\!\!\!\!\!\!{+}v_F^2\Big[{-}q_0^2\vec{k}{\cdot}\vec{p}{+}v_F^2\vec{k}{\cdot}\vec{p}(\vec{p}{+}\vec{q}){\cdot}(\vec{k}{+}\vec{q})
{+}v_F^2\vec{k}{\cdot}(\vec{k}{+}\vec{q})\vec{p}{\cdot}(\vec{p}{+}\vec{q})\nn\\&&{-}v_F^2\vec{k}{\cdot}(\vec{p}{+}\vec{q})\vec{p}{\cdot}(\vec{k}{+}\vec{q})\Big]Q_3^kQ_3^p\bigg\}.
\eea
First collect all the terms proportional to $q_0^2$:
\bea
&&\!\!\!\!\!\!{-}v_F^2\left[\vec{k}{\cdot}\vec{p}+\vec{k}{\cdot}(\vec{p}{+}\vec{q}){+}\vec{p}{\cdot}(\vec{k}{+}\vec{q}){+}(\vec{p}{+}\vec{q}){\cdot}(\vec{k}{+}\vec{q})\right]Q_2^kQ_2^p\nn\\
&&\!\!\!\!\!\!{-}v_F^2q_0\left[\vec{k}{\cdot}\vec{p}{+}\vec{p}{\cdot}(\vec{k}{+}\vec{q})\right]Q_2^kQ_3^p
{-}v_F^2q_0\left[\vec{k}{\cdot}\vec{p}{+}\vec{k}{\cdot}(\vec{p}{+}\vec{q})\right]Q_3^kQ_2^p\nn\\&&{-}v_F^2q_0^2\vec{k}{\cdot}\vec{p}Q_3^kQ_3^p\nn\\
&&\!\!\!\!\!\!{=}{-}\frac{q_0^2}{4}\frac{1}{[v_F^2(|\vec{k}|{+}|\vec{k}{+}\vec{q}|)^2{+}q_0^2][v_F^2(|\vec{p}|{+}|\vec{p}{+}\vec{q}|)^2{+}q_0^2]}
\nn\\&&\!\!\!\!\!\!\times\left(\frac{\vec{k}{\cdot}\vec{p}}{|\vec{k}||\vec{p}|}{-}\frac{\vec{p}{\cdot}(\vec{k}{+}\vec{q})}{|\vec{k}{+}\vec{q}||\vec{p}|}
{-}\frac{\vec{k}{\cdot}(\vec{p}{+}\vec{q})}{|\vec{k}||\vec{p}{+}\vec{q}|}{+}\frac{(\vec{k}{+}\vec{q}){\cdot}(\vec{p}{+}\vec{q})}{|\vec{k}{+}\vec{q}||\vec{p}{+}\vec{q}|}\right).\nn\\&&
\eea
The remaining terms are proportional to $v_F^2$ and sum to
\bea
&&Q_1^kQ_1^p
{-}v_F^2\left[\vec{p}{\cdot}(\vec{p}{+}\vec{q})Q_1^kQ_3^p{+}\vec{k}{\cdot}(\vec{k}{+}\vec{q})Q_3^kQ_1^p\right]\nn\\&&{+}v_F^4\Big[\vec{k}{\cdot}\vec{p}(\vec{p}{+}\vec{q}){\cdot}(\vec{k}{+}\vec{q})
{+}\vec{k}{\cdot}(\vec{k}{+}\vec{q})\vec{p}{\cdot}(\vec{p}{+}\vec{q})\nn\\&&{-}\vec{k}{\cdot}(\vec{p}{+}\vec{q})\vec{p}{\cdot}(\vec{k}{+}\vec{q})\Big]Q_3^kQ_3^p\nn\\
&&{=}\frac{v_F^2}{4}\frac{(|\vec{k}|{+}|\vec{k}{+}\vec{q}|)(|\vec{p}|{+}|\vec{p}{+}\vec{q}|)}{[v_F^2(|\vec{k}|{+}|\vec{k}{+}\vec{q}|)^2{+}q_0^2][v_F^2(|\vec{p}|{+}|\vec{p}{+}\vec{q}|)^2{+}q_0^2]}
\nn\\&&\times\bigg\{(|\vec{k}|^2{+}\vec{k}{\cdot}\vec{q}{-}|\vec{k}||\vec{k}{+}\vec{q}|)(|\vec{p}|^2{+}\vec{p}{\cdot}\vec{q}{-}|\vec{p}||\vec{p}{+}\vec{q}|)
\nn\\&&{+}\vec{k}{\cdot}\vec{p}|\vec{q}|^2{-}(\vec{k}{\cdot}\vec{q})(\vec{p}{\cdot}\vec{q})\bigg\}\frac{1}{|\vec{k}||\vec{p}||\vec{k}{+}\vec{q}||\vec{p}{+}\vec{q}|}.
\eea
Combining these two results gives
\bea
&&\Pi_V(q){=}{-}\frac{g^2N}{2}{\int}\frac{d^2k}{(2\pi)^2}{\int}\frac{d^2p}{(2\pi)^2}
\frac{1}{|\vec{k}{-}\vec{p}|}\nn\\&&\times\frac{1}{[v_F^2(|\vec{k}|{+}|\vec{k}{+}\vec{q}|)^2{+}q_0^2][v_F^2(|\vec{p}|{+}|\vec{p}{+}\vec{q}|)^2{+}q_0^2]}\nn\\
&&\times \bigg\{{-}q_0^2\left(\frac{\vec{k}{\cdot}\vec{p}}{|\vec{k}||\vec{p}|}{-}\frac{\vec{p}{\cdot}(\vec{k}{+}\vec{q})}{|\vec{k}{+}\vec{q}||\vec{p}|}
{-}\frac{\vec{k}{\cdot}(\vec{p}{+}\vec{q})}{|\vec{k}||\vec{p}{+}\vec{q}|}{+}\frac{(\vec{k}{+}\vec{q}){\cdot}(\vec{p}{+}\vec{q})}{|\vec{k}{+}\vec{q}||\vec{p}{+}\vec{q}|}\right)
\nn\\&&{+}v_F^2\frac{(|\vec{k}|{+}|\vec{k}{+}\vec{q}|)(|\vec{p}|{+}|\vec{p}{+}\vec{q}|)}{|\vec{k}||\vec{p}||\vec{k}{+}\vec{q}||\vec{p}{+}\vec{q}|}
\Big[\vec{k}{\cdot}\vec{p}|\vec{q}|^2{-}(\vec{k}{\cdot}\vec{q})(\vec{p}{\cdot}\vec{q})\nn\\&&{+}(|\vec{k}|^2{+}\vec{k}{\cdot}\vec{q}{-}|\vec{k}||\vec{k}{+}\vec{q}|)(|\vec{p}|^2{+}\vec{p}{\cdot}\vec{q}{-}|\vec{p}||\vec{p}{+}\vec{q}|)
\Big]\bigg\}.\nn\\&&
\eea
This result agrees with Eq.~(34) of Ref.~[\onlinecite{Sodemann_PRB12}]. Note that an apparent discrepancy by a factor of 2 is accounted for by the different definitions of the fermion degeneracy, $N$. To proceed further, we again make use of elliptic coordinates:
\bea
&&k_x{=}\frac{|\vec{q}|}{2}(\cosh\mu\cos\nu{-}1),\qquad k_y{=}\frac{|\vec{q}|}{2}\sinh\mu\sin\nu,\nn\\
&&p_x{=}\frac{|\vec{q}|}{2}(\cosh\mu'\cos\nu'{-}1),\qquad p_y{=}\frac{|\vec{q}|}{2}\sinh\mu'\sin\nu',\nn\\
&&d^2kd^2p{=}\nn\\&&\frac{|\vec{q}|^4}{16}(\cosh^2\mu{-}\cos^2\nu)(\cosh^2\mu'{-}\cos^2\nu')d\mu d\nu d\mu' d\nu',\nn\\&&
\eea
and set $\vec{q}{=}(|\vec{q}|,0)$. The Coulomb propagator becomes
\bea
&&\!\!\!\!\!\!\frac{1}{|\vec{k}{-}\vec{p}|}{=}
\nn\\&&\!\!\!\!\!\!\frac{2}{|\vec{q}|\sqrt{\cosh(\mu{+}\mu'){-}\cos(\nu{+}\nu')}\sqrt{\cosh(\mu{-}\mu'){-}\cos(\nu{-}\nu')}},\nn\\&&
\eea
the additional overall factor in the integrand becomes
\bea
&&\frac{1}{[v_F^2(|\vec{k}|{+}|\vec{k}{+}\vec{q}|)^2{+}q_0^2][v_F^2(|\vec{p}|{+}|\vec{p}{+}\vec{q}|)^2{+}q_0^2]}
\nn\\&&{=}\frac{1}{(v_F^2|\vec{q}|^2\cosh^2\mu{+}q_0^2)(v_F^2|\vec{q}|^2\cosh^2\mu'{+}q_0^2)},
\eea
the factor proportional to $q_0^2$ reads
\bea
&&\!\!\!\!\!\!{-}q_0^2\left(\frac{\vec{k}{\cdot}\vec{p}}{|\vec{k}||\vec{p}|}{-}\frac{\vec{p}{\cdot}(\vec{k}{+}\vec{q})}{|\vec{k}{+}\vec{q}||\vec{p}|}
{-}\frac{\vec{k}{\cdot}(\vec{p}{+}\vec{q})}{|\vec{k}||\vec{p}{+}\vec{q}|}{+}\frac{(\vec{k}{+}\vec{q}){\cdot}(\vec{p}{+}\vec{q})}{|\vec{k}{+}\vec{q}||\vec{p}{+}\vec{q}|}\right)
\nn\\&&\!\!\!\!\!\!{=}{-}4q_0^2\frac{\sin\nu\sin\nu'}{(\cosh^2\mu{-}\cos^2\nu)(\cosh^2\mu'{-}\cos^2\nu')}
\nn\\&&\!\!\!\!\!\!\times\left(\cosh\mu\cosh\mu'\sin\nu\sin\nu'{+}\cos\nu\cos\nu'\sinh\mu\sinh\mu')\right),\nn\\&&
\eea
and the factor multiplying $v_F^2$ is
\bea
&&v_F^2\frac{(|\vec{k}|{+}|\vec{k}{+}\vec{q}|)(|\vec{p}|{+}|\vec{p}{+}\vec{q}|)}{|\vec{k}||\vec{p}||\vec{k}{+}\vec{q}||\vec{p}{+}\vec{q}|}
\Big[\vec{k}{\cdot}\vec{p}|\vec{q}|^2{-}(\vec{k}{\cdot}\vec{q})(\vec{p}{\cdot}\vec{q})\nn\\&&{+}(|\vec{k}|^2{+}\vec{k}{\cdot}\vec{q}{-}|\vec{k}||\vec{k}{+}\vec{q}|)(|\vec{p}|^2{+}\vec{p}{\cdot}\vec{q}{-}|\vec{p}||\vec{p}{+}\vec{q}|)
\Big]
\nn\\&&{=}4v_F^2|\vec{q}|^2\frac{\cosh\mu\cosh\mu'\sin\nu\sin\nu'}{(\cosh^2\mu{-}\cos^2\nu)(\cosh^2\mu'{-}\cos^2\nu')}\nn\\&&\times\left(\sin\nu\sin\nu'{+}\sinh\mu\sinh\mu'\right).
\eea
Combining all these results gives
\begin{widetext}
\bea
\Pi_V(q)&{=}&{-}\frac{N|\vec{q}|\alpha}{16\pi^3v_F}{\int}\frac{d\mu d\mu'd\nu d\nu'}{\sqrt{\cosh(\mu{+}\mu'){-}\cos(\nu{+}\nu')}\sqrt{\cosh(\mu{-}\mu'){-}\cos(\nu{-}\nu')}}
\frac{\cosh\mu\cosh\mu'\sin\nu\sin\nu'}{(\cosh^2\mu{-}y^2)(\cosh^2\mu'{-}y^2)}\nn\\
&&\times[(\sin\nu\sin\nu'{+}\sinh\mu\sinh\mu'){+}y^2(\sin\nu\sin\nu'{+}\tanh\mu\tanh\mu'\cos\nu\cos\nu')],\label{Pb4int}
\eea
\end{widetext}
with $y{=}iq_0/(v_F|\vec{q}|)$. This result differs from Eq.~(37) of Ref.~[\onlinecite{Sodemann_PRB12}]: in that reference, the factor multiplying $y^2$ in the numerator contains a term $\cosh\mu\cosh\mu'$, whereas in the above result, this term is replaced by $\sin\nu\sin\nu'$.

The integrals over $\nu$ and $\nu'$ in Eq.~(\ref{Pb4int}) can be performed exactly. To do these integrals, it helps to first define
\beq
\sigma\equiv \nu{+}\nu',\qquad \tau\equiv\nu{-}\nu',\qquad d\nu d\nu'{=}1/2d\sigma d\tau.
\eeq
Since
\bea
\sin\nu\sin\nu'&{=}&\sin[(\sigma{+}\tau)/2]\sin[(\sigma{-}\tau)/2]\nn\\&&{=}{-}1/2(\cos\sigma{-}\cos\tau),\nn\\
\cos\nu\cos\nu'&{=}&\cos[(\sigma{+}\tau)/2]\cos[(\sigma{-}\tau)/2]\nn\\&&{=}1/2(\cos\sigma{+}\cos\tau),
\eea
and
\bea
&&{\int_0^{2\pi}}d\nu{\int_0^{2\pi}}d\nu' f(\nu{+}\nu',\nu{-}\nu')\nn\\&&{=}\left[{\int_0^{2\pi}}d\sigma{\int_0^\sigma} d\tau{+}{\int_{2\pi}^{4\pi}}d\sigma{\int_0^{4\pi{-}\sigma}}d\tau\right]f(\sigma,\tau)\nn\\&&{=}{\int_0^{2\pi}}d\sigma{\int_0^{2\pi}}d\tau f(\sigma,\tau),
\eea
if $f(\pm\sigma{+}2\pi,\pm\tau{+}2\pi){=}f(\sigma,\tau)$, the integrals over $\nu$ and $\nu'$ amount to computing the following three integrals:
\bea
&&\!\!\!\!\!\!{\cal I}_1{=}{\int_0^{2\pi}}d\nu{\int_{0}^{2\pi}}d\nu'\frac{\sin\nu\sin\nu'}{\sqrt{\cosh(\mu{+}\mu'){-}\cos(\nu{+}\nu')}}
\nn\\&&\qquad\times\frac{1}{\sqrt{\cosh(\mu{-}\mu'){-}\cos(\nu{-}\nu')}}
\nn\\&&\!\!\!\!\!\!
{=}{-}\frac{1}{2}\sqrt{w_{+}w_{-}}{\int_0^{2\pi}}d\sigma{\int_0^{2\pi}}d\tau\frac{\cos\sigma{-}\cos\tau}{\sqrt{1{-}w_{+}\cos\sigma}\sqrt{1{-}w_{-}\cos\tau}},\nn\\
\eea
\bea
&&\!\!\!\!\!\!{\cal I}_2{=}{\int_0^{2\pi}}d\nu{\int_{0}^{2\pi}}d\nu'\frac{\sin^2\nu\sin^2\nu'}{\sqrt{\cosh(\mu{+}\mu'){-}\cos(\nu{+}\nu')}}
\nn\\&&\qquad\times\frac{1}{\sqrt{\cosh(\mu{-}\mu'){-}\cos(\nu{-}\nu')}}
\nn\\
&&\!\!\!\!\!\!{=}\frac{1}{4}\sqrt{w_{+}w_{-}}{\int_0^{2\pi}}d\sigma{\int_0^{2\pi}}d\tau\frac{(\cos\sigma{-}\cos\tau)^2}{\sqrt{1{-}w_{+}\cos\sigma}\sqrt{1{-}w_{-}\cos\tau}},\nn\\
\nn\\
\eea
\bea
&&\!\!\!\!\!\!{\cal I}_3{=}{\int_0^{2\pi}}d\nu{\int_{0}^{2\pi}}d\nu'\frac{\sin\nu\sin\nu'\cos\nu\cos\nu'}{\sqrt{\cosh(\mu{+}\mu'){-}\cos(\nu{+}\nu')}}
\nn\\&&\qquad\times\frac{1}{\sqrt{\cosh(\mu{-}\mu'){-}\cos(\nu{-}\nu')}}
\nn\\
&&\!\!\!\!\!\!{=}{-}\frac{1}{4}\sqrt{w_{+}w_{-}}{\int_0^{2\pi}}d\sigma{\int_0^{2\pi}}d\tau\frac{\cos^2\sigma{-}\cos^2\tau}{\sqrt{1{-}w_{+}\cos\sigma}\sqrt{1{-}w_{-}\cos\tau}},\nn\\&&
\eea
with $w_\pm\equiv\hbox{sech}(\mu\pm\mu')$. These integrals can be performed with the help of the following results:
\bea
{\cal J}_0(w)&{=}&{\int_0^{2\pi}}du\frac{1}{\sqrt{1{-}w\cos u}}{=}\frac{4}{\sqrt{1{+}w}}\hbox{K}\left(\frac{2w}{1{+}w}\right),\nn\\
\eea
\bea
{\cal J}_1(w)&{=}&{\int_0^{2\pi}}du\frac{\cos u}{\sqrt{1{-}w\cos u}}\nn\\&{=}&{-}\frac{4\sqrt{1{+}w}}{w}\hbox{E}\left(\frac{2w}{1{+}w}\right)
{+}\frac{4}{w\sqrt{1{+}w}}\hbox{K}\left(\frac{2w}{1{+}w}\right),\nn\\
\eea
\bea
{\cal J}_2(w)&{=}&{\int_0^{2\pi}}du\frac{\cos^2u}{\sqrt{1{-}w\cos u}}\nn\\&{=}&
{-}\frac{8\sqrt{1{+}w}}{3w^2}\hbox{E}\left(\frac{2w}{1{+}w}\right)
{+}\frac{4(2{+}w^2)}{3w^2\sqrt{1{+}w}}\hbox{K}\left(\frac{2w}{1{+}w}\right),\nn\\&&
\eea
where $K$ and $E$ are complete elliptic integrals of the first and second kind:
\beq
K(z){=}{\int_0^{\pi/2}}\frac{d\theta}{\sqrt{1{-}z\sin^2\theta}},\quad E(z){=}{\int_0^{\pi/2}}d\theta\sqrt{1{-}z\sin^2\theta}.
\eeq
In terms of the ${\cal J}_i$, the ${\cal I}_i$ are
\bea
{\cal I}_1&{=}&{-}\frac{1}{2}\sqrt{w_{+}w_{-}}\left[{\cal J}_1(w_{+}){\cal J}_0(w_{-}){-}{\cal J}_0(w_{+}){\cal J}_1(w_{-})\right],\nn\\
{\cal I}_2&{=}&\frac{1}{4}\sqrt{w_{+}w_{-}}\Big[{\cal J}_2(w_{+}){\cal J}_0(w_{-}){-}2{\cal J}_1(w_{+}){\cal J}_1(w_{-})\nn\\&&\qquad{+}{\cal J}_0(w_{+}){\cal J}_2(w_{-})\Big],\nn\\
{\cal I}_3&{=}&{-}\frac{1}{4}\sqrt{w_{+}w_{-}}\left[{\cal J}_2(w_{+}){\cal J}_0(w_{-}){-}{\cal J}_0(w_{+}){\cal J}_2(w_{-})\right].\nn\\&&
\eea
In terms of the ${\cal I}_i$, the vertex correction reads
\bea
&&\Pi_V(q){=}{-}\frac{N|\vec{q}|\alpha}{16\pi^3v_F}{\int} d\mu d\mu'\frac{\cosh\mu\cosh\mu'}{(\cosh^2\mu{-}y^2)(\cosh^2\mu'{-}y^2)}
\nn\\&&\times\left[{\cal I}_1\sinh\mu\sinh\mu'{+}(1{+}y^2){\cal I}_2{+}y^2{\cal I}_3\tanh\mu\tanh\mu'\right].\nn\\&&
\eea
We can make the following coordinate transformation:
\beq
a\equiv \mu{+}\mu',\quad b\equiv \mu{-}\mu',\quad w_{+}{=}\hbox{sech}(a),\quad w_{-}{=}\hbox{sech}(b).
\eeq
The symmetries $\mu\leftrightarrow\mu'$ and $\mu'\leftrightarrow{-}\mu'$ of the integrand translate to $b\leftrightarrow{-}b$ and $a\leftrightarrow b$, allowing us to express the integration limits as
\bea
&&{\int_0^\infty} d\mu{\int_0^\infty} d\mu'f(\mu,\mu'){=}\frac{1}{2}{\int_0^\infty} da{\int_{{-}a}^a} db f'(a,b)\nn\\&&{=}{\int_0^\infty} da{\int_0^a} db f'(a,b){=}\frac{1}{2}{\int_0^\infty} da{\int_0^\infty} db f'(a,b).
\eea
We then have
\beq
\Pi_V(q){=}{-}\frac{N|\vec{q}|\alpha}{8v_F}I_b(y),
\eeq
with
\bea
I_b(y)&\equiv&\frac{1}{2\pi^3}{\int_0^1}{\int_0^1} dw_{+} dw_{-}\frac{w_{+}{+}w_{-}}{\sqrt{1{-}w_{+}^2}\sqrt{1{-}w_{-}^2}} \nn\\&&\times\frac{1}{4w_{+}^2w_{-}^2y^4{-}4w_{+}w_{-}(1{+}w_{+}w_{-})y^2{+}(w_{+}{+}w_{-})^2}
\nn\\&&\times\bigg[\frac{w_{-}{-}w_{+}}{2w_{+}w_{-}}{\cal I}_1(w_{+},w_{-}){+}(1{+}y^2){\cal I}_2(w_{+},w_{-})
\nn\\&&{+}y^2\frac{w_{-}{-}w_{+}}{w_{+}{+}w_{-}}{\cal I}_3(w_{+},w_{-})\bigg].\label{defIb}
\eea
We have checked numerically that this double integral appears to agree with the plot shown in Fig.~2a of Ref.~[\onlinecite{Sodemann_PRB12}] despite the differences occurring at intermediate steps of the respective calculations and the different forms of the final expressions. We therefore conclude that our final result for the dielectric function coincides with that obtained in Ref.~[\onlinecite{Sodemann_PRB12}], where it is shown that the second-order corrections lead to an improvement in comparing the theoretical results to experimentally measured values. In the next section, we will see that this improvement does not persist in the case of the Fermi velocity, at least for graphene suspended in vacuum.

\subsection{Two-loop corrections to electron self-energy and a strong-coupling quantum critical point}\label{sec:twoloopSE}

We now move on to our calculation of the second-order corrections to the electron self-energy. Our results differ from previous results reported in Refs.~[\onlinecite{Mishchenko_PRL07}] and [\onlinecite{Vafek_PRB08}]. In the course of the calculations, we indicate the specific points at which these differences arise. Our results for the electron self-energy are used to compute the renormalized Fermi velocity and the running of the effective coupling to second-order, where we find a critical point $\alpha_c$ in the RG flow, signifying either a breakdown of perturbation theory or a quantum phase transition.

\subsubsection{Two-loop rainbow correction to self-energy}\label{sec:twolooprainbow}

\begin{figure}
\includegraphics[width=0.5\columnwidth]{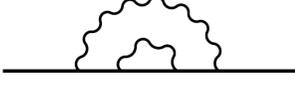}
\caption[Self-energy]{Two-loop self-energy correction to the electron self-energy.}
\label{fig:selfenergy2b}
\end{figure}

The first potential two-loop correction to the electron self-energy is shown in Fig.~\ref{fig:selfenergy2b}.
This diagram evaluates to
\bea
\Sigma_{2b}(q)&=&-\int\frac{d^3k}{(2\pi)^3}D_0(q-k)\gamma^0G_0(k)\Sigma_1(k)G_0(k)\gamma^0\nn\\
&=&\frac{ig^4}{32\pi}\int\frac{d^3k}{(2\pi)^3}\frac{1}{|\vec{q}-\vec{k}|}\log(\Lambda/|\vec{k}|)\gamma^0\frac{\slashed k}{k^2}\vec{k}\cdot\vec\gamma\frac{\slashed k}{k^2}\gamma^0.\nn\\&&
\eea
Straightforward algebra reveals that the integral over $k_0$ vanishes identically:
\bea
&&\int\frac{dk_0}{2\pi}\frac{1}{k^4}\gamma^0(k_0\gamma^0+v_F\vec{k}\cdot\vec\gamma)\vec{k}\cdot\vec\gamma(k_0\gamma^0+v_F\vec{k}\cdot\vec\gamma)\gamma^0\nn\\&&
=\vec{k}\cdot\vec\gamma\int\frac{dk_0}{2\pi}\frac{k_0^2-v_F^2|\vec{k}|^2}{(k_0^2+v_F^2|\vec{k}|^2)^2}=0.
\eea
Therefore, the full contribution vanishes identically:
\beq
\Sigma_{2b}(q)=0.\label{selfenergy2b}
\eeq

\subsubsection{Two-loop vertex correction to self-energy}
\begin{figure}
\includegraphics[width=0.5\columnwidth]{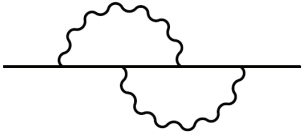}
\caption[Self-energy]{Two-loop vertex correction to the electron self-energy.}
\label{fig:selfenergy2a}
\end{figure}

{\it Reduction to a quadruple integral:-}
The second two-loop correction to the electron self-energy is shown in Fig.~\ref{fig:selfenergy2a}.
This diagram has the value
\bea
\Sigma_{2a}(q)&{=}&{\int}\frac{d^3k}{(2\pi)^3}{\int}\frac{d^3p}{(2\pi)^3}D_0(q{-}k)D_0(q{-}p)\nn\\&&\qquad\qquad\times\gamma^0G_0(k)\gamma^0G_0(k{+}p{-}q)\gamma^0G_0(p)\gamma^0\nn\\
&{=}&\frac{{-}ig^4}{4}{\int}\frac{d^3k}{(2\pi)^3}{\int}\frac{d^3p}{(2\pi)^3}\frac{1}{|\vec{q}{-}\vec{k}|}\frac{1}{|\vec{q}{-}\vec{p}|}\nn\\&&\qquad\qquad\times\gamma^0\frac{\slashed k}{k^2}
\gamma^0\frac{\slashed k{+}\slashed p{-}\slashed q}{(k{+}p{-}q)^2}\gamma^0\frac{\slashed p}{p^2}\gamma^0.
\eea
The product of gamma matrices in the integrand can be expanded as
\bea
&&\gamma^0\gamma^\mu\gamma^0\gamma^\nu\gamma^0\gamma^\rho\gamma^0=\delta^{\mu0}\delta^{\nu0}\delta^{\rho0}\gamma^0{-}\delta^{\mu0}\delta^{\nu0}(1{-}\delta^{\rho0})\gamma^\rho \nn\\&&{-}\delta^{\mu0}(1{-}\delta^{\nu0})\delta^{\rho0}\gamma^\nu {-}\delta^{\mu0}(1{-}\delta^{\nu0})(1{-}\delta^{\rho0})\gamma^\nu\gamma^\rho\gamma^0 \nn\\&&{-}(1{-}\delta^{\mu0})\delta^{\nu0}\delta^{\rho0}\gamma^\mu {-}(1{-}\delta^{\mu0})\delta^{\nu0}(1{-}\delta^{\rho0})\gamma^\mu\gamma^\rho\gamma^0\nn\\&&{-}(1{-}\delta^{\mu0})(1{-}\delta^{\nu0})\delta^{\rho0}\gamma^\mu\gamma^\nu\gamma^0 \nn\\&&{+}(1{-}\delta^{\mu0})(1{-}\delta^{\nu0})(1{-}\delta^{\rho0})\gamma^\mu\gamma^\nu\gamma^\rho,
\eea
which leads to
\bea
&&\Sigma_{2a}(q){=}\nn\\&&\frac{{-}ig^4}{4}{\int}\frac{d^3k}{(2\pi)^3}{\int}\frac{d^3p}{(2\pi)^3}\frac{1}{|\vec{q}{-}\vec{k}|}\frac{1}{|\vec{q}{-}\vec{p}|}\frac{1}{k^2p^2(k{+}p{-}q)^2}\nn\\&&\times\bigg\{k_0p_0(k_0{+}p_0{-}q_0)\gamma^0 {-}v_Fk_0(k_0{+}p_0{-}q_0)\vec{p}{\cdot}\vec\gamma\nn\\&&{-}v_Fk_0p_0(\vec{k}{+}\vec{p}{-}\vec{q}){\cdot}\vec\gamma{-}v_F^2k_0\gamma^0(\vec{k}{+}\vec{p}{-}\vec{q}){\cdot}\vec\gamma \vec{p}{\cdot}\vec\gamma\nn\\&&{-}v_Fp_0(k_0{+}p_0{-}q_0)\vec{k}{\cdot}\vec\gamma  {-}v_F^2(k_0{+}p_0{-}q_0)\gamma^0 \vec{k}{\cdot}\vec\gamma \vec{p}{\cdot} \vec\gamma\nn\\&&{-}v_F^2p_0\gamma^0\vec{k}{\cdot}\vec\gamma(\vec{k}{+}\vec{p}{-}\vec{q}){\cdot}\vec\gamma{+}v_F^3\vec{k}{\cdot}\vec\gamma(\vec{k}{+}\vec{p}{-}\vec{q}){\cdot}\vec\gamma \vec{p}{\cdot}\vec\gamma\bigg\}.\nn\\&&
\eea
As usual, we first perform the integrals over the energies $k_0$ and $p_0$:
\bea
B_1&\equiv&{\int}\frac{dk_0}{2\pi}{\int}\frac{dp_0}{2\pi}\frac{k_0p_0(k_0{+}p_0{-}q_0)}{M(k,p)} \nn\\&{=}&\frac{1}{4}\frac{q_0}{q_0^2{+}v_F^2(|\vec{k}|{+}|\vec{p}|{+}|\vec{k}{+}\vec{p}{-}\vec{q}|)^2},\nn\\
\eea
\bea
B_2&\equiv&{\int}\frac{dk_0}{2\pi}{\int}\frac{dp_0}{2\pi}\frac{k_0(k_0{+}p_0{-}q_0)}{M(k,p)} \nn\\&{=}&\frac{1}{4}\frac{|\vec{k}|{+}|\vec{p}|{+}|\vec{k}{+}\vec{p}{-}\vec{q}|}{|\vec{p}|[q_0^2{+}v_F^2(|\vec{k}|{+}|\vec{p}|{+}|\vec{k}{+}\vec{p}{-}\vec{q}|)^2]},\nn\\
\eea
\bea
B_3&\equiv&{\int}\frac{dk_0}{2\pi}{\int}\frac{dp_0}{2\pi}\frac{k_0p_0}{M(k,p)} \nn\\&{=}&{-}\frac{1}{4}\frac{|\vec{k}|{+}|\vec{p}|{+}|\vec{k}{+}\vec{p}{-}\vec{q}|}{|\vec{k}{+}\vec{p}{-}\vec{q}|[q_0^2{+}v_F^2(|\vec{k}|{+}|\vec{p}|{+}|\vec{k}{+}\vec{p}{-}\vec{q}|)^2]},\nn\\
\eea
\bea
B_4&\equiv&{\int}\frac{dk_0}{2\pi}{\int}\frac{dp_0}{2\pi}\frac{k_0}{M(k,p)} \nn\\&{=}&\frac{1}{4v_F^2}\frac{q_0}{|\vec{p}||\vec{k}{+}\vec{p}{-}\vec{q}|[q_0^2{+}v_F^2(|\vec{k}|{+}|\vec{p}|{+}|\vec{k}{+}\vec{p}{-}\vec{q}|)^2]},\nn\\
\eea
\bea
B_5&\equiv&{\int}\frac{dk_0}{2\pi}{\int}\frac{dp_0}{2\pi}\frac{p_0(k_0{+}p_0{-}q_0)}{M(k,p)} \nn\\&{=}&\frac{1}{4}\frac{|\vec{k}|{+}|\vec{p}|{+}|\vec{k}{+}\vec{p}{-}\vec{q}|}{|\vec{k}|[q_0^2{+}v_F^2(|\vec{k}|{+}|\vec{p}|{+}|\vec{k}{+}\vec{p}{-}\vec{q}|)^2]},\nn\\
\eea
\bea
B_6&\equiv&{\int}\frac{dk_0}{2\pi}{\int}\frac{dp_0}{2\pi}\frac{k_0{+}p_0{-}q_0}{M(k,p)} \nn\\&{=}&{-}\frac{1}{4v_F^2}\frac{q_0}{|\vec{k}||\vec{p}|[q_0^2{+}v_F^2(|\vec{k}|{+}|\vec{p}|{+}|\vec{k}{+}\vec{p}{-}\vec{q}|)^2]},\nn\\
\eea
\bea
B_7&\equiv&{\int}\frac{dk_0}{2\pi}{\int}\frac{dp_0}{2\pi}\frac{p_0}{M(k,p)} \nn\\&{=}&\frac{1}{4v_F^2}\frac{q_0}{|\vec{k}||\vec{k}{+}\vec{p}{-}\vec{q}|[q_0^2{+}v_F^2(|\vec{k}|{+}|\vec{p}|{+}|\vec{k}{+}\vec{p}{-}\vec{q}|)^2]},\nn\\
\eea
\bea
B_8&\equiv&{\int}\frac{dk_0}{2\pi}{\int}\frac{dp_0}{2\pi}\frac{1}{M(k,p)} \nn\\&{=}&\frac{1}{4v_F^2}\frac{|\vec{k}|{+}|\vec{p}|{+}|\vec{k}{+}\vec{p}{-}\vec{q}|}{|\vec{k}||\vec{p}||\vec{k}{+}\vec{p}{-}\vec{q}|[q_0^2{+}v_F^2(|\vec{k}|{+}|\vec{p}|{+}|\vec{k}{+}\vec{p}{-}\vec{q}|)^2]},\nn\\&&
\eea
with
\bea
&&M(k,p)\equiv\nn\\&&(k_0^2{+}v_F^2|\vec{k}|^2)(p_0^2{+}v_F^2|\vec{p}|^2)[(k_0{+}p_0{-}q_0)^2{+}v_F^2|\vec{k}{+}\vec{p}{-}\vec{q}|^2].\nn\\&&
\eea
We then have
\bea
&&\Sigma_{2a}(q){=}\frac{{-}ig^4}{4}{\int}\frac{d^2k}{(2\pi)^2}{\int}\frac{d^2p}{(2\pi)^2}\frac{1}{|\vec{q}{-}\vec{k}|}\frac{1}{|\vec{q}{-}\vec{p}|}\nn\\&&\times\bigg\{B_1\gamma^0{-}B_2v_F\vec{p}{\cdot}\vec\gamma {-}B_3v_F(\vec{k}{+}\vec{p}{-}\vec{q}){\cdot}\vec\gamma\nn\\&&{-}B_4v_F^2\gamma^0(\vec{k}{+}\vec{p}{-}\vec{q}){\cdot}\vec\gamma\vec{p}{\cdot}\vec\gamma{-}B_5v_F\vec{k}{\cdot}\vec\gamma {-}B_6v_F^2\gamma^0\vec{k}{\cdot}\vec\gamma\vec{p}{\cdot}\vec\gamma\nn\\&&{-}B_7v_F^2\gamma^0\vec{k}{\cdot}\vec\gamma(\vec{k}{+}\vec{p}{-}\vec{q}){\cdot}\vec\gamma {+}B_8v_F^3\vec{k}{\cdot}\vec\gamma(\vec{k}{+}\vec{p}{-}\vec{q}){\cdot}\vec\gamma\vec{p}{\cdot}\vec\gamma\bigg\}.\nn\\&&
\eea
The $\gamma^0$ component of this is
\bea
&&{1\over4}\tr[\gamma^0\Sigma_{2a}(q)]{=}\nn\\&&\frac{{-}ig^4}{16}{\int}\frac{d^2k}{(2\pi)^2}{\int}\frac{d^2p}{(2\pi)^2}\frac{1}{|\vec{q}{-}\vec{k}|}\frac{1}{|\vec{q}{-}\vec{p}|}\big\{4B_1 \nn\\&&{-}4B_4v_F^2\vec{p}{\cdot}(\vec{k}{+}\vec{p}{-}\vec{q}){-}4B_6v_F^2\vec{k}{\cdot}\vec{p}{-}4B_7v_F^2\vec{k}{\cdot}(\vec{k}{+}\vec{p}{-}\vec{q})\big\}\nn\\
&&{=}\frac{{-}ig^4q_0}{16}{\int}\frac{d^2k}{(2\pi)^2}{\int}\frac{d^2p}{(2\pi)^2}\frac{1}{|\vec{q}{-}\vec{k}|}\frac{1}{|\vec{q}{-}\vec{p}|}\nn\\&&\times\frac{1}{q_0^2{+}v_F^2(|\vec{k}|{+}|\vec{p}|{+}|\vec{k}{+}\vec{p}{-}\vec{q}|)^2} \bigg\{1{-}\nn\\&& \frac{\vec{p}}{|\vec{p}|}{\cdot}\frac{\vec{k}{+}\vec{p}{-}\vec{q}}{|\vec{k}{+}\vec{p}{-}\vec{q}|}{+}\frac{\vec{k}}{|\vec{k}|}{\cdot}\frac{\vec{p}}{|\vec{p}|} {-}\frac{\vec{k}}{|\vec{k}|}{\cdot}\frac{\vec{k}{+}\vec{p}{-}\vec{q}}{|\vec{k}{+}\vec{p}{-}\vec{q}|}\bigg\},\label{temporal2loop}
\eea
while the spatial components are
\bea
&&{1\over4}\tr[\gamma^i\Sigma_{2a}(q)]{=}\nn\\&&\frac{{-}ig^4}{16}{\int}\frac{d^2k}{(2\pi)^2}{\int}\frac{d^2p}{(2\pi)^2}\frac{1}{|\vec{q}{-}\vec{k}|}\frac{1}{|\vec{q}{-}\vec{p}|}\Big\{{-}4B_2v_Fp_i\nn\\&&{-}4B_3v_F(k_i{+}p_i{-}q_i) {-}4B_5v_Fk_i{+}4B_8v_F^3\Big[k_i\vec{p}{\cdot}(\vec{k}{+}\vec{p}{-}\vec{q})\nn\\&&{-}(k_i{+}p_i{-}q_i)\vec{k}{\cdot}\vec{p}{+}p_i\vec{k}{\cdot}(\vec{k}{+}\vec{p}{-}\vec{q})\Big]\Big\}\nn\\
&&{=}\frac{{-}ig^4v_F}{16}{\int}\frac{d^2k}{(2\pi)^2}{\int}\frac{d^2p}{(2\pi)^2}\frac{1}{|\vec{q}{-}\vec{k}|}\frac{1}{|\vec{q}{-}\vec{p}|} \nn\\&&\times\frac{|\vec{k}|{+}|\vec{p}|{+}|\vec{k}{+}\vec{p}{-}\vec{q}|}{q_0^2{+}v_F^2(|\vec{k}|{+}|\vec{p}|{+}|\vec{k}{+}\vec{p}{-}\vec{q}|)^2}\bigg\{{-}\frac{k_i}{|\vec{k}|}{-}\frac{p_i}{|\vec{p}|} \nn\\&&{+}\frac{k_i{+}p_i{-}q_i}{|\vec{k}{+}\vec{p}{-}\vec{q}|} {+}\frac{k_i(|\vec{p}|^2{-}\vec{p}{\cdot}\vec{q}){+}p_i(|\vec{k}|^2{-}\vec{k}{\cdot}\vec{q}){+}q_i\vec{k}{\cdot}\vec{p}}{|\vec{k}||\vec{p}||\vec{k}{+}\vec{p}{-}\vec{q}|}\bigg\}.\nn\\&&
\eea
If we choose the coordinates such that $\vec{q}{=}(|\vec{q}|,0)$, then it becomes apparent that the terms of the integrand which are proportional to $k_y$ or $p_y$ are odd functions of these variables, implying that these terms vanish upon integration. We may then make the replacement $k_i\to q_i\vec{k}{\cdot}\vec{q}/|\vec{q}|^2$, and similarly for $p_i$. The total integral is therefore proportional to $q_i$:
\bea
&&{1\over4}\tr[\gamma^i\Sigma_{2a}(q)]{=}\frac{{-}ig^4q_iv_F}{16|\vec{q}|^2}{\int}\frac{d^2k}{(2\pi)^2}{\int}\frac{d^2p}{(2\pi)^2}\frac{1}{|\vec{q}{-}\vec{k}|}\frac{1}{|\vec{q}{-}\vec{p}|} \nn\\&&\times\frac{|\vec{k}|{+}|\vec{p}|{+}|\vec{k}{+}\vec{p}{-}\vec{q}|}{q_0^2{+}v_F^2(|\vec{k}|{+}|\vec{p}|{+}|\vec{k}{+}\vec{p}{-}\vec{q}|)^2}\bigg\{{-}\frac{\vec{k}{\cdot}\vec{q}}{|\vec{k}|}{-}\frac{\vec{p}{\cdot}\vec{q}}{|\vec{p}|}
{+}\frac{(\vec{k}{+}\vec{p}{-}\vec{q}){\cdot}\vec{q}}{|\vec{k}{+}\vec{p}{-}\vec{q}|}
\nn\\&&{+}\frac{|\vec{p}|^2\vec{k}{\cdot}\vec{q}{+}|\vec{k}|^2\vec{p}{\cdot}\vec{q}{+}|\vec{q}|^2\vec{k}{\cdot}\vec{p}{-}2(\vec{k}{\cdot}\vec{q})(\vec{p}{\cdot}\vec{q})}{|\vec{k}||\vec{p}||\vec{k}{+}\vec{p}{-}\vec{q}|}\bigg\}.\label{spatial2loop}\nn\\&&
\eea

{\it Extracting the divergence in the temporal part:-}
To extract the divergent logarithm term in the temporal part of the two{-}loop self{-}energy, Eq.~(\ref{temporal2loop}), we must examine the behavior of the quadruple integral in the region $|\vec{k}|,|\vec{p}|\gg|\vec{q}|$. In this regime, the integral reduces to
\bea
\!\!\!&&{1\over4}\tr[\gamma^0\Sigma_{2a}(q)]=
\nn\\\!\!\!&&\frac{{-}ig^4q_0}{16}{\int}\frac{d^2k}{(2\pi)^2}{\int}\frac{d^2p}{(2\pi)^2}\frac{1}{|\vec{k}|}\frac{1}{|\vec{p}|}\frac{1}{q_0^2{+}v_F^2(|\vec{k}|{+}|\vec{p}|{+}|\vec{k}{+}\vec{p}|)^2} \nn\\\!\!\!&&\times\bigg\{1{-}\frac{\vec{p}{\cdot}(\vec{k}{+}\vec{p})}{|\vec{p}||\vec{k}{+}\vec{p}|} {-}\frac{\vec{k}{\cdot}(\vec{k}{+}\vec{p})}{|\vec{k}||\vec{k}{+}\vec{p}|}{+}\frac{\vec{k}{\cdot}\vec{p}}{|\vec{k}||\vec{p}|}\bigg\},
\eea
where we have discarded terms which become odd under the change of variable $\vec{k}\to{-}\vec{k}$, $\vec{p}\to{-}\vec{p}$ in the limit of large $|\vec{k}|, |\vec{p}|$. It helps to consider each of the four terms above separately:
\bea
\Psi_1&{=}&\frac{{-}ig^4q_0}{16}{\int}\frac{d^2k}{(2\pi)^2}{\int}\frac{d^2p}{(2\pi)^2}\frac{S(k,p)}{|\vec{k}||\vec{p}|},
\nn\\\Psi_2&{=}&\frac{ig^4q_0}{16}{\int}\frac{d^2k}{(2\pi)^2}{\int}\frac{d^2p}{(2\pi)^2}\frac{\vec{p}{\cdot}(\vec{k}{+}\vec{p})S(k,p)}{|\vec{k}||\vec{p}|^2|\vec{k}{+}\vec{p}|},
\nn\\\Psi_3&{=}&\frac{ig^4q_0}{16}{\int}\frac{d^2k}{(2\pi)^2}{\int}\frac{d^2p}{(2\pi)^2}\frac{\vec{k}{\cdot}(\vec{k}{+}\vec{p})S(k,p)}{|\vec{k}|^2|\vec{p}||\vec{k}{+}\vec{p}|},
\nn\\\Psi_4&{=}&\frac{{-}ig^4q_0}{16}{\int}\frac{d^2k}{(2\pi)^2}{\int}\frac{d^2p}{(2\pi)^2}\frac{\vec{k}{\cdot}\vec{p}S(k,p)}{|\vec{k}|^2|\vec{p}|^2},
\eea
with
\beq
S(k,p)\equiv\frac{1}{q_0^2{+}v_F^2(|\vec{k}|{+}|\vec{p}|{+}|\vec{k}{+}\vec{p}|)^2}.
\eeq
It is clear that $\Psi_2{=}\Psi_3$, so we need to only compute one of these integrals. In each of $\Psi_1,\Psi_3,\Psi_4$, we first perform the integration over $\vec{p}$. This is facilitated by choosing the coordinate system such that $\vec{k}{=}(|\vec{k}|,0)$ and then switching to elliptic coordinates:
\beq
p_x{=}\frac{|\vec{k}|}{2}(\cosh\mu\cos\nu{-}1),\quad p_y{=}\frac{|\vec{k}|}{2}\sinh\mu\sin\nu,
\eeq
for which
\bea
&&|\vec{p}|{=}\frac{|\vec{k}|}{2}(\cosh\mu{-}\cos\nu),\quad |\vec{k}{+}\vec{p}|{=}\frac{|\vec{k}|}{2}(\cosh\mu{+}\cos\nu),\nn\\&& |\vec{k}|{+}|\vec{p}|{+}|\vec{k}{+}\vec{p}|{=}|\vec{k}|(\cosh\mu{+}1),\nn\\
&&d^2p{=}\frac{|\vec{k}|^2}{4}(\cosh^2\mu{-}\cos^2\nu)d\mu d\nu{=}|\vec{p}||\vec{k}{+}\vec{p}|d\mu d\nu.
\eea
We then have
\bea
&&\Psi_1=\nn\\&&\frac{{-}ig^4q_0}{128\pi^2}{\int}\frac{d^2k}{(2\pi)^2}{\int_0^\infty} d\mu{\int_0^{2\pi}} d\nu\frac{\cos\nu{+}\cosh\mu}{q_0^2{+}v_F^2|\vec{k}|^2(1{+}\cosh\mu)^2},\nn\\
&&\Psi_3=\nn\\&&\frac{ig^4q_0}{128\pi^2}{\int}\frac{d^2k}{(2\pi)^2}{\int_0^\infty} d\mu{\int_0^{2\pi}} d\nu\frac{1{+}\cos\nu\cosh\mu}{q_0^2{+}v_F^2|\vec{k}|^2(1{+}\cosh\mu)^2},\nn\\
&&\Psi_4=\nn\\&&\frac{{-}ig^4q_0}{128\pi^2}{\int}\frac{d^2k}{(2\pi)^2}{\int_0^\infty} d\mu{\int_0^{2\pi}} d\nu\frac{\cos\nu{+}\cosh\mu}{q_0^2{+}v_F^2|\vec{k}|^2(1{+}\cosh\mu)^2}\nn\\&&\qquad\times\frac{\cosh\mu\cos\nu{-}1}{\cosh\mu{-}\cos\nu}.
\eea
Next, we restore $\vec{k}$ to being a general vector, $\vec{k}{=}(|\vec{k}|,0)\to(k_x,k_y)$, and switch to polar coordinates $|\vec{k}|$ and $\theta_k$. The integrations over $\theta_k$ are trivial, while the integrations over $\nu$ can be performed exactly, with the result
\bea
\Psi_1&{=}&\frac{{-}ig^4q_0}{128\pi^2}{\int_0^\infty} d\mu{\int_0^{\Lambda}}d|\vec{k}|\frac{|\vec{k}|\cosh\mu}{q_0^2{+}v_F^2|\vec{k}|^2(1{+}\cosh\mu)^2},\nn\\
\Psi_3&{=}&\frac{ig^4q_0}{128\pi^2}{\int_0^\infty} d\mu{\int_0^{\Lambda}}d|\vec{k}|\frac{|\vec{k}|}{q_0^2{+}v_F^2|\vec{k}|^2(1{+}\cosh\mu)^2},\nn\\
\Psi_4&{=}&\frac{ig^4q_0}{128\pi^2}{\int_0^\infty} d\mu{\int_0^{\Lambda}}d|\vec{k}|\frac{|\vec{k}|e^{{-}2\mu}}{q_0^2{+}v_F^2|\vec{k}|^2(1{+}\cosh\mu)^2}.\nn\\&&
\eea
The integration over $|\vec{k}|$ is the same in each case and evaluates to
\bea
&&{\int_0^{\Lambda}}d|\vec{k}|\frac{|\vec{k}|}{q_0^2{+}v_F^2|\vec{k}|^2(1{+}\cosh\mu)^2}=\nn\\&&\qquad\qquad\frac{\log\left[1{+}\frac{v_F^2\Lambda^2}{q_0^2}(1{+}\cosh\mu)^2\right]}{2v_F^2(1{+}\cosh\mu)^2}.
\eea
Plugging this result into the above expressions for $\Psi_1,\Psi_3,\Psi_4$, assuming $v_F\Lambda\gg|q_0|$, and keeping only the term proportional to $\log(v_F\Lambda/|q_0|)$ yields
\bea
\Psi_1&{=}&\frac{{-}ig^4q_0}{128\pi^2v_F^2}\left({\int_0^\infty} d\mu\frac{\cosh\mu}{(1{+}\cosh\mu)^2}\right)\log(v_F\Lambda/|q_0|)
\nn\\&&{=}\frac{{-}ig^4q_0}{192\pi^2v_F^2}\log(v_F\Lambda/|q_0|),\nn\\
\Psi_3&{=}&\frac{ig^4q_0}{128\pi^2v_F^2}\left({\int_0^\infty} d\mu\frac{1}{(1{+}\cosh\mu)^2}\right)\log(v_F\Lambda/|q_0|)
\nn\\&&{=}\frac{ig^4q_0}{384\pi^2v_F^2}\log(v_F\Lambda/|q_0|),\nn\\
\Psi_4&{=}&\frac{ig^4q_0}{128\pi^2v_F^2}\left({\int_0^\infty} d\mu\frac{e^{{-}2\mu}}{(1{+}\cosh\mu)^2}\right)\log(v_F\Lambda/|q_0|)
\nn\\&&{=}\frac{ig^4q_0}{32\pi^2v_F^2}(\log2{-}2/3)\log(v_F\Lambda/|q_0|).
\eea
We see that $\Psi_1{+}2\Psi_3{=}0$, so that the divergence of $\tr[\gamma^0\Sigma_{2a}(q)]$ comes solely from $\Psi_4$. We may rewrite $\Psi_4$ in the following way:
\bea
&&\Psi_4=\frac{ig^4q_0}{32\pi^2v_F^2}(\log2{-}2/3)\log(v_F\Lambda/|q_0|)\nn\\&&\qquad\to\frac{ig^4q_0}{32\pi^2v_F^2}(\log2{-}2/3)\log(\Lambda/|\vec{q}|).
\eea
Here, we have restored the renormalization scale $|\vec{q}|$ in the argument of the log divergence with the expectation that this dependence on $|\vec{q}|$ arises from the integration region we have neglected, namely the $|\vec{k}|\lesssim|\vec{q}|$ region. In particular, it must be the case that this region produces a term of the form $\log(|q_0|/(v_F|\vec{q}|))$ as follows from two simple observations regarding the integral in Eq.~(\ref{temporal2loop}). The first observation is that this integral can be rewritten as a dimensionless function of two dimensionless parameters, $v_F\Lambda/|q_0|$ and $v_F|\vec{q}|/|q_0|$. The above calculation shows that the large $|\vec{k}|$ portion of the integral depends only on the former parameter, while the small $|\vec{k}|$ portion depends only on the latter. The second observation is that Eq.~(\ref{temporal2loop}) is finite in the static limit $q_0\to0$, implying that the apparent divergence of $\log(v_F\Lambda/|q_0|)$ in this limit must be canceled by a similar term coming from the small $|\vec{k}|$ region. The only possible term that would cancel this divergence is $\log(|q_0|/(v_F|\vec{q}|))$, leaving behind $\log(\Lambda/|\vec{q}|)$. We therefore arrive at the following expression for the divergent term in the temporal part of $\Sigma_{2a}(q)$:
\bea
\frac{1}{4}\tr[\gamma^0\Sigma_{2a}(q)]&{=}&\frac{ig^4q_0}{32\pi^2v_F^2}(\log2{-}2/3)\log(\Lambda/|\vec{q}|)\nn\\&{=}&i\frac{3\log2{-}2}{6}\alpha^2q_0\log(\Lambda/|\vec{q}|).
\eea

{\it Extracting the divergence in the spatial part:-}
To extract the divergent logarithm term in the spatial part of the two{-}loop self{-}energy, Eq.~(\ref{spatial2loop}), we must examine the behavior of the quadruple integral in the region $|\vec{k}|,|\vec{p}|\gg|\vec{q}|$. It helps to first redefine $\vec{k}\to\vec{k}{+}\vec{q}$:
\bea
&&\!\!\!\!\!{1\over4}\tr[\gamma^i\Sigma_{2a}(q)]{=}\frac{{-}ig^4q_iv_F}{16|\vec{q}|^2}{\int}\frac{d^2k}{(2\pi)^2}{\int}\frac{d^2p}{(2\pi)^2}\frac{1}{|\vec{k}|}\frac{1}{|\vec{q}{-}\vec{p}|} \nn\\&&\!\!\!\!\!\times\frac{|\vec{k}{+}\vec{q}|{+}|\vec{p}|{+}|\vec{k}{+}\vec{p}|}{q_0^2{+}v_F^2(|\vec{k}{+}\vec{q}|{+}|\vec{p}|{+}|\vec{k}{+}\vec{p}|)^2}\bigg\{{-}\frac{(\vec{k}{+}\vec{q}){\cdot}\vec{q}}{|\vec{k}{+}\vec{q}|} {-}\frac{\vec{p}{\cdot}\vec{q}}{|\vec{p}|}{+}\frac{(\vec{k}{+}\vec{p}){\cdot}\vec{q}}{|\vec{k}{+}\vec{p}|}
\nn\\&&\!\!\!\!\!{+}\frac{|\vec{p}|^2(\vec{k}{+}\vec{q}){\cdot}\vec{q}{+}|\vec{k}{+}\vec{q}|^2\vec{p}{\cdot}\vec{q}{+}|\vec{q}|^2\vec{k}{\cdot}\vec{p}{-}2(\vec{k}{\cdot}\vec{q})(\vec{p}{\cdot}\vec{q}) {-}|\vec{q}|^2\vec{p}{\cdot}\vec{q}}{|\vec{k}{+}\vec{q}||\vec{p}||\vec{k}{+}\vec{p}|}\bigg\}.\nn\\&&
\eea
We then make the following expansions in the large momentum limit:
\bea
&&|\vec{k}{+}\vec{q}|{\approx}|\vec{k}|{+}\frac{\vec{k}{\cdot}\vec{q}}{|\vec{k}|},\; \frac{1}{|\vec{k}{+}\vec{q}|}{\approx}\frac{1}{|\vec{k}|}{-}\frac{\vec{k}{\cdot}\vec{q}}{|\vec{k}|^3},\; \frac{1}{|\vec{p}{-}\vec{q}|}{\approx}\frac{1}{|\vec{p}|}{+}\frac{\vec{p}{\cdot}\vec{q}}{|\vec{p}|^3},\nn\\
&&\frac{|\vec{k}{+}\vec{q}|{+}|\vec{p}|{+}|\vec{k}{+}\vec{p}|}{q_0^2{+}v_F^2(|\vec{k}{+}\vec{q}|{+}|\vec{p}|{+}|\vec{k}{+}\vec{p}|)^2}{\approx} \frac{1}{q_0^2{+}v_F^2(|\vec{k}|{+}|\vec{p}|{+}|\vec{k}{+}\vec{p}|)^2}\nn\\&&\times\bigg[ |\vec{k}|{+}|\vec{p}|{+}|\vec{k}{+}\vec{p}|{+} \frac{q_0^2{-}v_F^2(|\vec{k}|{+}|\vec{p}|{+}|\vec{k}{+}\vec{p}|)^2}{q_0^2{+}v_F^2(|\vec{k}|{+}|\vec{p}|{+}|\vec{k}{+}\vec{p}|)^2}\frac{\vec{k}{\cdot}\vec{q}}{|\vec{k}|}\bigg] \nn\\&&\qquad\qquad\equiv Q(\vec{k},\vec{p}){+}R(\vec{k},\vec{p})\frac{\vec{k}{\cdot}\vec{q}}{|\vec{k}|}.
\eea
We then have
\bea
&&{1\over4}\tr[\gamma^i\Sigma_{2a}(q)]{=}\frac{{-}ig^4q_iv_F}{16|\vec{q}|^2}{\int}\frac{d^2k}{(2\pi)^2}{\int}\frac{d^2p}{(2\pi)^2}\frac{1}{|\vec{k}||\vec{p}|} \nn\\&&\times\left(1{+}\frac{\vec{p}{\cdot}\vec{q}}{|\vec{p}|^2}\right)\left[Q(\vec{k},\vec{p}){+}R(\vec{k},\vec{p})\frac{\vec{k}{\cdot}\vec{q}}{|\vec{k}|}\right]\bigg\{ {-}\frac{\vec{p}{\cdot}\vec{q}}{|\vec{p}|} {+}\frac{(\vec{k}{+}\vec{p}){\cdot}\vec{q}}{|\vec{k}{+}\vec{p}|}
\nn\\&&{-}\frac{(\vec{k}{+}\vec{q}){\cdot}\vec{q}}{|\vec{k}|}\left(1{-}\frac{\vec{k}{\cdot}\vec{q}}{|\vec{k}|^2}\right)
{+}\frac{|\vec{p}|(\vec{k}{+}\vec{q}){\cdot}\vec{q}}{|\vec{k}||\vec{k}{+}\vec{p}|}\left(1{-}\frac{\vec{k}{\cdot}\vec{q}}{|\vec{k}|^2}\right)
\nn\\&&{+}\frac{|\vec{k}|\vec{p}{\cdot}\vec{q}}{|\vec{p}||\vec{k}{+}\vec{p}|}\left(1{+}\frac{\vec{k}{\cdot}\vec{q}}{|\vec{k}|^2}\right)
{+}\frac{|\vec{q}|^2\vec{k}{\cdot}\vec{p}}{|\vec{k}||\vec{p}||\vec{k}{+}\vec{p}|}\left(1{-}\frac{\vec{k}{\cdot}\vec{q}}{|\vec{k}|^2}\right) \nn\\&&{-}2\frac{(\vec{k}{\cdot}\vec{q})(\vec{p}{\cdot}\vec{q})}{|\vec{k}||\vec{p}||\vec{k}{+}\vec{p}|}\left(1{-}\frac{\vec{k}{\cdot}\vec{q}}{|\vec{k}|^2}\right) {-}\frac{|\vec{q}|^2\vec{p}{\cdot}\vec{q}}{|\vec{k}||\vec{p}||\vec{k}{+}\vec{p}|}\left(1{-}\frac{\vec{k}{\cdot}\vec{q}}{|\vec{k}|^2}\right)\bigg\}.\label{spatial2loop2}\nn\\&&
\eea
The terms that scale as the inverse fourth power in the momenta $\vec{k},\vec{p}$ give rise to a logarithmic divergence. These are the terms we are interested in. There are also terms in Eq.~(\ref{spatial2loop2}) which scale as the inverse third power and so would seem to produce a linear divergence. However these terms vanish identically as can be seen by performing a coordinate transformation $\vec{k}\to{-}\vec{k}$, $\vec{p}\to{-}\vec{p}$. We isolate each of the terms which contribute to the logarithmic divergence in the following series of integrals:
\beq
\Xi_1{=}{\int}\frac{d^2k}{(2\pi)^2}{\int}\frac{d^2p}{(2\pi)^2}\frac{Q(\vec{k},\vec{p})}{|\vec{k}||\vec{p}|}\left(\frac{(\vec{k}{\cdot}\vec{q})^2}{|\vec{k}|^3}{-}\frac{|\vec{q}|^2}{|\vec{k}|}\right),
\eeq
\beq
\Xi_2{=}{-}{\int}\frac{d^2k}{(2\pi)^2}{\int}\frac{d^2p}{(2\pi)^2}\frac{Q(\vec{k},\vec{p})}{|\vec{k}||\vec{p}|}\frac{|\vec{p}|}{|\vec{k}{+}\vec{p}|}\left(\frac{(\vec{k}{\cdot}\vec{q})^2}{|\vec{k}|^3}{-}\frac{|\vec{q}|^2}{|\vec{k}|}\right),
\eeq
\beq
\Xi_3{=}{\int}\frac{d^2k}{(2\pi)^2}{\int}\frac{d^2p}{(2\pi)^2}\frac{Q(\vec{k},\vec{p})}{|\vec{k}||\vec{p}|}\frac{(\vec{k}{\cdot}\vec{q})(\vec{p}{\cdot}\vec{q})}{|\vec{k}||\vec{p}||\vec{k}{+}\vec{p}|},
\eeq
\beq
\Xi_4{=}{\int}\frac{d^2k}{(2\pi)^2}{\int}\frac{d^2p}{(2\pi)^2}\frac{Q(\vec{k},\vec{p})}{|\vec{k}||\vec{p}|}\frac{|\vec{q}|^2(\vec{k}{\cdot}\vec{p})}{|\vec{k}||\vec{p}||\vec{k}{+}\vec{p}|},
\eeq
\beq
\Xi_5{=}{-}2{\int}\frac{d^2k}{(2\pi)^2}{\int}\frac{d^2p}{(2\pi)^2}\frac{Q(\vec{k},\vec{p})}{|\vec{k}||\vec{p}|}\frac{(\vec{k}{\cdot}\vec{q})(\vec{p}{\cdot}\vec{q})}{|\vec{k}||\vec{p}||\vec{k}{+}\vec{p}|}{=}{-}2\Xi_3,
\eeq
\beq
\Xi_6{=}{-}{\int}\frac{d^2k}{(2\pi)^2}{\int}\frac{d^2p}{(2\pi)^2}\frac{Q(\vec{k},\vec{p})}{|\vec{k}||\vec{p}|}\frac{(\vec{k}{\cdot}\vec{q})(\vec{p}{\cdot}\vec{q})}{|\vec{k}||\vec{p}|^2},
\eeq
\beq
\Xi_7{=}{-}{\int}\frac{d^2k}{(2\pi)^2}{\int}\frac{d^2p}{(2\pi)^2}\frac{Q(\vec{k},\vec{p})}{|\vec{k}||\vec{p}|}\frac{(\vec{p}{\cdot}\vec{q})^2}{|\vec{p}|^3},
\eeq
\beq
\Xi_8{=}{\int}\frac{d^2k}{(2\pi)^2}{\int}\frac{d^2p}{(2\pi)^2}\frac{Q(\vec{k},\vec{p})}{|\vec{k}||\vec{p}|}\frac{(\vec{p}{\cdot}\vec{q})(\vec{k}{+}\vec{p}){\cdot}\vec{q}}{|\vec{p}|^2|\vec{k}{+}\vec{p}|},
\eeq
\beq
\Xi_9{=}{\int}\frac{d^2k}{(2\pi)^2}{\int}\frac{d^2p}{(2\pi)^2}\frac{Q(\vec{k},\vec{p})}{|\vec{k}||\vec{p}|}\frac{(\vec{k}{\cdot}\vec{q})(\vec{p}{\cdot}\vec{q})}{|\vec{k}||\vec{p}||\vec{k}{+}\vec{p}|}{=}\Xi_3,
\eeq
\beq
\Xi_{10}{=}{\int}\frac{d^2k}{(2\pi)^2}{\int}\frac{d^2p}{(2\pi)^2}\frac{Q(\vec{k},\vec{p})}{|\vec{k}||\vec{p}|}\frac{|\vec{k}|(\vec{p}{\cdot}\vec{q})^2}{|\vec{p}|^3|\vec{k}{+}\vec{p}|},
\eeq
\beq
\Xi_{11}{=}{-}{\int}\frac{d^2k}{(2\pi)^2}{\int}\frac{d^2p}{(2\pi)^2}\frac{R(\vec{k},\vec{p})}{|\vec{k}||\vec{p}|}\frac{(\vec{k}{\cdot}\vec{q})^2}{|\vec{k}|^2},
\eeq
\beq
\Xi_{12}{=}{-}{\int}\frac{d^2k}{(2\pi)^2}{\int}\frac{d^2p}{(2\pi)^2}\frac{R(\vec{k},\vec{p})}{|\vec{k}||\vec{p}|}\frac{(\vec{k}{\cdot}\vec{q})(\vec{p}{\cdot}\vec{q})}{|\vec{k}||\vec{p}|},
\eeq
\beq
\Xi_{13}{=}{\int}\frac{d^2k}{(2\pi)^2}{\int}\frac{d^2p}{(2\pi)^2}\frac{R(\vec{k},\vec{p})}{|\vec{k}||\vec{p}|}\frac{(\vec{k}{\cdot}\vec{q})(\vec{k}{+}\vec{p}){\cdot}\vec{q}}{|\vec{k}||\vec{k}{+}\vec{p}|},
\eeq
\beq
\Xi_{14}{=}{\int}\frac{d^2k}{(2\pi)^2}{\int}\frac{d^2p}{(2\pi)^2}\frac{R(\vec{k},\vec{p})}{|\vec{k}||\vec{p}|}\frac{|\vec{p}|(\vec{k}{\cdot}\vec{q})^2}{|\vec{k}|^2|\vec{k}{+}\vec{p}|},
\eeq
\beq
\Xi_{15}{=}{\int}\frac{d^2k}{(2\pi)^2}{\int}\frac{d^2p}{(2\pi)^2}\frac{R(\vec{k},\vec{p})}{|\vec{k}||\vec{p}|}\frac{(\vec{k}{\cdot}\vec{q})(\vec{p}{\cdot}\vec{q})}{|\vec{p}||\vec{k}{+}\vec{p}|},
\eeq
In each of these integrals, we first choose the coordinate system such that $\vec{k}{=}(|\vec{k}|,0)$ and then replace $\vec{p}$ by elliptic coordinates:
\beq
p_x{=}\frac{|\vec{k}|}{2}(\cosh\mu\cos\nu{-}1),\quad p_y{=}\frac{|\vec{k}|}{2}\sinh\mu\sin\nu,
\eeq
for which
\bea
&&|\vec{p}|{=}\frac{|\vec{k}|}{2}(\cosh\mu{-}\cos\nu),\quad |\vec{k}{+}\vec{p}|{=}\frac{|\vec{k}|}{2}(\cosh\mu{+}\cos\nu),\nn\\&& |\vec{k}|{+}|\vec{p}|{+}|\vec{k}{+}\vec{p}|{=}|\vec{k}|(\cosh\mu{+}1),\nn\\
&&d^2p{=}\frac{|\vec{k}|^2}{4}(\cosh^2\mu{-}\cos^2\nu)d\mu d\nu{=}|\vec{k}||\vec{k}{+}\vec{p}|d\mu d\nu.
\eea
Under this coordinate transformation, we also have
\bea
Q(\vec{k},\vec{p})&{\to}&\frac{|\vec{k}|(\cosh\mu{+}1)}{q_0^2{+}v_F^2|\vec{k}|^2(\cosh\mu{+}1)^2}\equiv \widetilde Q(\mu,|\vec{k}|),\nn\\
R(\vec{k},\vec{p})&{\to}&\frac{q_0^2{-}v_F^2|\vec{k}|^2(\cosh\mu{+}1)^2}{\left[q_0^2{+}v_F^2|\vec{k}|^2(\cosh\mu{+}1)^2\right]^2}\equiv\widetilde R(\mu,|\vec{k}|).\nn\\&&
\eea
After making this coordinate transformation, we first perform the integrations over $\nu$, which can all be done exactly. We then restore $\vec{k}$ to a general vector by making the replacements
\beq
q_x\to\frac{\vec{k}{\cdot}\vec{q}}{|\vec{k}|},\qquad q_y\to \sqrt{|\vec{q}|^2{-}\frac{(\vec{k}{\cdot}\vec{q})^2}{|\vec{k}|^2}}.
\eeq
The integration over $\vec{k}$ is then expressed in terms of polar coordinates $|\vec{k}|$ and $\theta$, where $\theta$ is defined by $\vec{k}{\cdot}\vec{q}{=}|\vec{k}||\vec{q}|\cos\theta$. All the $\theta$ integrations are easily done, and we arrive at the following expressions for the integrals:
\beq
\Xi_1{=}{-}\frac{|\vec{q}|^2}{16\pi^2}{\int_0^\infty} d\mu{\int_0^\Lambda} d|\vec{k}| \widetilde Q(\mu,|\vec{k}|)\cosh\mu,
\eeq
\beq
\Xi_2{=}{-}\Xi_1,
\eeq
\beq
\Xi_3{=}{-}\frac{|\vec{q}|^2}{8\pi^2}{\int_0^\infty} d\mu{\int_0^\Lambda} d|\vec{k}| \widetilde Q(\mu,|\vec{k}|)e^{{-}\mu},
\eeq
\beq
\Xi_4{=}2\Xi_3,
\eeq
\beq
\Xi_5{=}{-}2\Xi_3,
\eeq
\beq
\Xi_6{=}\Xi_3,
\eeq
\beq
\Xi_7{=}{-}\frac{|\vec{q}|^2}{4\pi^2}{\int_0^\infty} d\mu{\int_0^\Lambda} d|\vec{k}| \widetilde Q(\mu,|\vec{k}|)[\coth\mu{-}1/2],
\eeq
\beq
\Xi_8{=}\frac{|\vec{q}|^2}{8\pi^2}{\int_0^\infty} d\mu{\int_0^\Lambda} d|\vec{k}| \widetilde Q(\mu,|\vec{k}|),
\eeq
\beq
\Xi_9{=}\Xi_3,
\eeq
\beq
\Xi_{10}{=}\frac{|\vec{q}|^2}{4\pi^2}{\int_0^\infty} d\mu{\int_0^\Lambda} d|\vec{k}| \widetilde Q(\mu,|\vec{k}|)\hbox{csch}\mu,
\eeq
\beq
\Xi_{11}{=}{-}\frac{|\vec{q}|^2}{16\pi^2}{\int_0^\infty} d\mu{\int_0^\Lambda} d|\vec{k}| \widetilde R(\mu,|\vec{k}|)|\vec{k}|\cosh\mu,
\eeq
\beq
\Xi_{12}{=}\frac{|\vec{q}|^2}{16\pi^2}{\int_0^\infty} d\mu{\int_0^\Lambda} d|\vec{k}| \widetilde R(\mu,|\vec{k}|)|\vec{k}|e^{{-}2\mu},
\eeq
\beq
\Xi_{13}{=}\frac{|\vec{q}|^2}{16\pi^2}{\int_0^\infty} d\mu{\int_0^\Lambda} d|\vec{k}| \widetilde R(\mu,|\vec{k}|)|\vec{k}|,
\eeq
\beq
\Xi_{14}{=}{-}\Xi_{11},
\eeq
\beq
\Xi_{15}{=}{-}\frac{|\vec{q}|^2}{8\pi^2}{\int_0^\infty} d\mu{\int_0^\Lambda} d|\vec{k}| \widetilde R(\mu,|\vec{k}|)|\vec{k}|e^{{-}\mu}.
\eeq

We separately combine the contributions containing $\widetilde Q$ and those containing $\widetilde R$ to obtain the following formula for the spatial part of the two{-}loop self{-}energy:
\bea
&&{1\over4}\tr[\gamma^i\Sigma_{2a}(q)]=\frac{{-}ig^4q_iv_F}{128\pi^2}\times\nn\\&&\times\bigg\{{\int_0^\infty} d\mu[2{-}3e^{{-}\mu}{-}2\coth\mu{+}2\hbox{csch}\mu]{\int_0^\Lambda} d|\vec{k}|\widetilde Q(\mu,|\vec{k}|)\nn\\&& {+}2{\int_0^\infty} d\mu e^{{-}\mu}\sinh^2(\mu/2){\int_0^\Lambda} d|\vec{k}||\vec{k}|\widetilde R(\mu,|\vec{k}|)\bigg\}.
\eea
The integrations over $|\vec{k}|$ can be performed exactly:
\beq
{\int_0^\Lambda} d|\vec{k}|\widetilde Q(\mu,|\vec{k}|){=}\frac{\log\left[1{+}\frac{v_F^2\Lambda^2}{q_0^2}(\cosh\mu{+}1)^2\right]}{2v_F^2(\cosh\mu{+}1)},
\eeq
\begin{widetext}
\beq
{\int_0^\Lambda} d|\vec{k}||\vec{k}|\widetilde R(\mu,|\vec{k}|)=\frac{2v_F^2\Lambda^2(\cosh\mu{+}1)^2{-}\left\{q_0^2{+}v_F^2\Lambda^2(\cosh\mu{+}1)^2\right\}\log\left[1{+}\frac{v_F^2\Lambda^2}{q_0^2}(\cosh\mu{+}1)^2\right]}{2v_F^2[q_0^2{+}v_F^2\Lambda^2(\cosh\mu{+}1)^2](\cosh\mu{+}1)^2}.
\eeq
\end{widetext}

We assume $v_F\Lambda/|q_0|\gg1$ for simplicity. We may then neglect the 1 in the argument of the logarithm and keep track of only the term proportional to $\log(v_F\Lambda/|q_0|)$:
\beq
{\int_0^\Lambda} d|\vec{k}|\widetilde Q(\mu,|\vec{k}|){\to}\frac{\log(v_F\Lambda/|q_0|)}{v_F^2(\cosh\mu{+}1)}{\to}\frac{\log(\Lambda/|\vec{q}|)}{v_F^2(\cosh\mu{+}1)},
\eeq
\beq
{\int_0^\Lambda} d|\vec{k}||\vec{k}|\widetilde R(\mu,|\vec{k}|){\to}{-}\frac{\log(v_F\Lambda/|q_0|)}{v_F^2(\cosh\mu{+}1)^2}{\to}{-}\frac{\log(\Lambda/|\vec{q}|)}{v_F^2(\cosh\mu{+}1)^2}.
\eeq
In the final step, we restore the renormalization scale $|\vec{q}|$ in the argument of the log divergence following the same reasoning as in the case of the temporal part of $\Sigma_{2a}(q)$ discussed above.

The remaining integrals over $\mu$ evaluate to
\bea
{\int_0^\infty} d\mu\frac{2{-}3e^{{-}\mu}{-}2\coth\mu{+}2\hbox{csch}\mu}{\cosh\mu{+}1}&{=}&4{-}6\log2,\nn\\
{-}{\int_0^\infty} d\mu\frac{e^{{-}\mu}\sinh^2(\mu/2)}{(\cosh\mu{+}1)^2}&{=}&4/3{-}2\log2.\nn\\&&
\eea
The result for the divergent term in the spatial part of the two{-}loop self{-}energy is then
\beq
\frac{1}{4}\tr[\gamma^i\Sigma_{2a}(q)]{=}i(\log2{-}2/3)\alpha^2v_Fq_i\log(\Lambda/|\vec{q}|).
\eeq
The full value of the diagram in Fig.~\ref{fig:selfenergy2a} is finally
\beq
\Sigma_{2a}(q){=}i(\log2{-}2/3)\alpha^2\left[\frac{1}{2}q_0\gamma^0{+}v_F\vec{q}{\cdot}\vec\gamma\right]\log(\Lambda/|\vec{q}|).\label{selfenergy2a}
\eeq
This result disagrees with both Refs.~[\onlinecite{Mishchenko_PRL07}] and [\onlinecite{Vafek_PRB08}]. It disagrees with Ref.~[\onlinecite{Mishchenko_PRL07}] because that reference
did not include the correction coming from the temporal part, although the expression for the spatial part obtained in that reference does agree with the present result. On the other hand, Ref.~[\onlinecite{Vafek_PRB08}] obtained a different result for the spatial part, although the result for the temporal part found in that reference does agree with the present result.

\subsubsection{Two{-}loop bubble correction to self{-}energy}

\begin{figure}
\includegraphics[width=0.5\columnwidth]{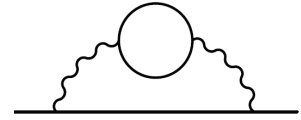}
\caption[Bubble]{Two-loop bubble correction to the electron self-energy.}
\label{fig:selfenergy2c}
\end{figure}

The two-loop bubble correction to the self-energy is shown in Fig.~\ref{fig:selfenergy2c}. This diagram evaluates to
\beq
\Sigma_{2c}(q)={-}{\int}\frac{d^3k}{(2\pi)^3}\gamma^0G_0(q{-}k)\gamma^0\Pi_B(k)[D_0(k)]^2,
\eeq
where $\Pi_B(q)$ is the result for the one{-}loop bubble diagram given in Eq.~(\ref{oneloopbubble}). We therefore have
\beq
\Sigma_{2c}(q)=\frac{iNg^4}{32}{\int}\frac{d^3k}{(2\pi)^3}\gamma^0\frac{\slashed q{-}\slashed k}{(q{-}k)^2\sqrt{k^2}}\gamma^0.
\eeq

{\it Extracting the divergence in the temporal part:-}
The temporal part of $\Sigma_{2c}(q)$ is
\bea
\frac{1}{4}\tr[\gamma^0\Sigma_{2c}(q)]&=&\frac{iNg^4}{128}{\int}\frac{d^3k}{(2\pi)^3}\frac{q_\mu{-}k_\mu}{(q{-}k)^2\sqrt{k^2}}\tr[\gamma^\mu\gamma^0]\nn\\
&=&\frac{iNg^4}{32}{\int}\frac{d^3k}{(2\pi)^3}\frac{q_0{-}k_0}{(q{-}k)^2\sqrt{k^2}}.\label{temporalSE2c}
\eea
To evaluate the integral over $k$, it helps to remove the square root in the denominator using the identity
\beq
\frac{1}{\sqrt{k^2}}=\frac{2}{\pi}{\int_0^\infty} \frac{dz}{z^2{+}k^2}.
\eeq
We then have
\beq
\frac{1}{4}\tr[\gamma^0\Sigma_{2c}(q)]=\frac{iNg^4}{16\pi}{\int_0^\infty} dz{\int}\frac{d^3k}{(2\pi)^3}\frac{q_0{-}k_0}{(q{-}k)^2(z^2{+}k^2)}.
\eeq
Next, we introduce a Feynman parameter via the identity
\beq
\frac{1}{AB}={\int_0^1}\frac{dw}{[wA{+}(1{-}w)B]^2},
\eeq
yielding
\bea
&&\!\!\!\!\!\!\frac{1}{4}\tr[\gamma^0\Sigma_{2c}(q)]=
\nn\\&&\!\!\!\!\!\!\frac{iNg^4}{16\pi}{\int_0^\infty} dz{\int_0^1}dw{\int}\frac{d^3k}{(2\pi)^3}\frac{q_0{-}k_0}{[w(q{-}k)^2{+}(1{-}w)(z^2{+}k^2)]^2}.\nn\\&&
\eea
We can simplify the denominator by changing variables to $\tilde k=k{-}wq$ to obtain
\bea
&&\!\!\!\!\!\!\frac{1}{4}\tr[\gamma^0\Sigma_{2c}(q)]=
\nn\\&&\!\!\!\!\!\!\frac{iNg^4}{16\pi}{\int_0^\infty} dz{\int_0^1}dw{\int}\frac{d^3\tilde k}{(2\pi)^3}\frac{(1{-}w)q_0{-}\tilde k_0}{[\tilde k^2{+}w(1{-}w)q^2{+}(1{-}w)z^2]^2}
\nn\\&&\!\!\!\!\!\!{=}\frac{iNg^4q_0}{16\pi}{\int_0^\infty} dz{\int_0^1}dw{\int}\frac{d^3\tilde k}{(2\pi)^3}\frac{1{-}w}{[\tilde k^2{+}(1{-}w)(wq^2{+}z^2)]^2}.\nn\\&&
\eea
In the last step, we discarded the $\tilde k_0$ in the numerator because that term is odd under a change of sign of $\tilde k$ and so vanishes identically. The integral over $\tilde k$ is now easily done using spherical coordinates, with the result
\bea
&&\!\!\!\!\!\!\frac{1}{4}\tr[\gamma^0\Sigma_{2c}(q)]=\frac{iNg^4q_0}{128\pi^2v_F^2}{\int_0^\infty}dz{\int_0^1}dw\frac{\sqrt{1{-}w}}{\sqrt{wq^2{+}z^2}}
\nn\\&&\!\!\!\!\!\!=\frac{iNg^4q_0}{128\pi^2v_F^2}{\int_0^\infty} dz\frac{1}{q^3}\left[{-}qz{+}(q^2{+}z^2)\tan^{{-}1}(q/z)\right].\nn\\&&
\eea
Note that here, $q=\sqrt{q_0^2{+}v_F^2|\vec{q}|^2}$. The integral over $z$ is of course ultraviolet divergent, and the upper integration limit must be replaced by a finite cutoff. Since $z$ has units of energy, the natural choice for this cutoff is $q\Lambda/|\vec{q}|$. The integral can be done exactly:
\bea
&&\!\!\!\!\!\!{\int_0^{q\Lambda/|\vec{q}|}} dz\frac{1}{q^3}\left[{-}qz{+}(q^2{+}z^2)\tan^{{-}1}(q/z)\right]
\nn\\&&\!\!\!\!\!\!=\frac{\Lambda}{6|\vec{q}|^3}\left[\pi\Lambda^2 {-}2\Lambda|\vec{q}|{+}3\pi|\vec{q}|^2{-}2(3|\vec{q}|^2{+}\Lambda^2)\tan^{{-}1}(\Lambda/|\vec{q}|)\right]
\nn\\&&\qquad{+}\frac{1}{3}\log(1{+}\Lambda^2/|\vec{q}|^2)
\nn\\&&=\frac{2}{3}\log(\Lambda/|\vec{q}|){+}\frac{8}{9}{+}O(|\vec{q}|^2/\Lambda^2).
\eea
We therefore obtain
\bea
\frac{1}{4}\tr[\gamma^0\Sigma_{2c}(q)]&=&\frac{iNg^4q_0}{128\pi^2v_F^2}\left[\frac{2}{3}\log(\Lambda/|\vec{q}|){+}\frac{8}{9}\right]\nn\\
&=&\frac{i\alpha^2q_0}{6}\left[\log(\Lambda/|\vec{q}|){+}4/3\right].
\eea

{\it Extracting the divergence in the spatial part:-}
The spatial part of $\Sigma_{2c}(q)$ is
\bea
\frac{1}{4}\tr[\gamma^i\Sigma_{2c}(q)]&{=}&\frac{iNg^4}{128}{\int}\frac{d^3k}{(2\pi)^3}\frac{q_\mu{-}k_\mu}{(q{-}k)^2\sqrt{k^2}}\tr[\gamma^i\gamma^0\gamma^\mu\gamma^0]\nn\\
&{=}&\frac{{-}iNg^4}{128}{\int}\frac{d^3k}{(2\pi)^3}\frac{q_\mu{-}k_\mu}{(q{-}k)^2\sqrt{k^2}}\tr[\gamma^i\gamma^\mu]\nn\\
&{=}&\frac{{-}iNg^4}{32}{\int}\frac{d^3k}{(2\pi)^3}\frac{q_i{-}k_i}{(q{-}k)^2\sqrt{k^2}}.
\eea
The final expression is almost identical to Eq.~(\ref{temporalSE2c}), and the same integration steps can be followed, with the result
\beq
\frac{1}{4}\tr[\gamma^i\Sigma_{2c}(q)]{=}\frac{{-}i\alpha^2v_Fq_i}{6}\left[\log(\Lambda/|\vec{q}|){+}4/3\right].
\eeq
Therefore we have
\beq
\Sigma_{2c}(q){=}\frac{i\alpha^2}{6}\left[q_0\gamma^0{-}v_F\vec{q}{\cdot}\vec\gamma\right]\left[\log(\Lambda/|\vec{q}|){+}4/3\right].\label{selfenergy2c}
\eeq
The spatial part agrees with that found in Ref.~[\onlinecite{Mishchenko_PRL07}]. However, that work does not appear to include the temporal part. Both the spatial and temporal parts agree with those found in Ref.~[\onlinecite{Vafek_PRB08}]. Combining Eqs.~(\ref{selfenergy2b}), (\ref{selfenergy2a}), (\ref{selfenergy2c}), the full two{-}loop self{-}energy is then
\bea
&&\!\!\!\!\!\!\Sigma_{2}(q){=}\Sigma_{2a}(q){+}\Sigma_{2b}(q){+}\Sigma_{2c}(q){=}
\nn\\&&\!\!\!\!\!\!i\alpha^2\left[\left(\frac{1}{2}\log2{-}\frac{1}{6}\right)q_0\gamma^0 {+}\left(\log2{-}\frac{5}{6}\right)v_F\vec{q}{\cdot}\vec\gamma\right]\log(\Lambda/|\vec{q}|).\label{2loopselfenergy}\nn\\&&
\eea
In this result, we have neglected all finite contributions to the electron self-energy as they do not affect the renormalization of the Fermi velocity, which is calculated in the next section. A proper justification for neglecting the finite terms is somewhat subtle and will be postponed until Sec.~\ref{sec:redefineLambda} since we must first derive some general properties of the divergence structure of the graphene field theory.

\subsubsection{Renormalization of the Fermi velocity to order $\alpha^2$}\label{sec:velrenorm}

We now use the result we have just obtained for the second-order electron self-energy to compute the corresponding correction to the renormalization of the Fermi velocity and the running of the effective coupling. For brevity of notation, let us define the following quantities:
\bea
\xi_1&{\equiv}&\frac{{-}i}{4v_F|\vec{q}|^2}\tr[\vec{q}{\cdot}\vec\gamma\Sigma_1(q)],\nn\\
\xi_{2t}&{\equiv}&\frac{{-}i}{4q_0}\tr[\gamma^0\Sigma_2(q)],\nn\\
\xi_{2x}&{\equiv}&\frac{{-}i}{4v_F|\vec{q}|^2}\tr[\vec{q}{\cdot}\vec\gamma\Sigma_2(q)].
\eea
The two{-}point function may then be expressed as
\bea
&&\!\!\!\!\!\!\langle\psi(p)\bar\psi(0)\rangle{=}i\left\{[1{+}\xi_{2t}]\gamma^0p_0{+}[1{+}\xi_1{+}\xi_{2x}]v_F\vec\gamma{\cdot}\vec{p}\right\}^{{-}1}
\nn\\&&\!\!\!\!\!\!{=}i[1{+}\xi_{2t}]^{{-}1}\left\{\gamma^0p_0{+}[1{+}\xi_1{+}\xi_{2x}{-}\xi_{2t}{+}O(\alpha^3)]v_F\vec\gamma{\cdot}\vec{p}\right\}^{{-}1}.\nn\\&&
\eea
This then implies that the second{-}order correction to the renormalization of the Fermi velocity reads
\beq
v_q{=}v_F\left[1{+}\xi_1{+}\xi_{2x}{-}\xi_{2t}\right].
\eeq
Using Eq.~(\ref{2loopselfenergy}), we have the following explicit forms of the parameters:
\bea
\xi_1&{=}&\frac{\alpha}{4}\log(\Lambda/|\vec{q}|),\nn\\
\xi_{2t}&{=}&\frac{3\log2{-}1}{6}\alpha^2\log(\Lambda/|\vec{q}|),\nn\\
\xi_{2x}&{=}&(\log2{-}5/6)\alpha^2\log(\Lambda/|\vec{q}|).\label{defofxi1xi2}
\eea
Combining these expressions, we obtain
\beq
v_q{=}v_F\left\{1{+}\left[\frac{\alpha}{4}{+}\left(\frac{1}{2}\log2{-}\frac{2}{3}\right)\alpha^2\right]\log(\Lambda/|\vec{q}|)\right\}.\label{velexp}
\eeq
This disagrees with the result quoted in Ref.~[\onlinecite{Mishchenko_PRL07}] because the contribution from the temporal part of the two{-}loop self{-}energy was not included in that reference. We therefore conclude that the velocity renormalization formula obtained in Ref.~[\onlinecite{Mishchenko_PRL07}] (Eq.~(12) of that reference) is incorrect. Eq.~(\ref{velexp}) also disagrees slightly with the result obtained in Ref.~[\onlinecite{Vafek_PRB08}] due to the discrepancy between the respective results for the self-energy correction $\Sigma_{2a}$, as discussed above.

We would like to convert the above expression for $v_q$ into a proper RG flow; this would allow us to compare the renormalized velocity at two different, arbitrary momentum scales. We begin by writing the renormalized coupling:
\beq
\alpha_q{=}\frac{\alpha}{1{+}(f_1\alpha{+}f_2\alpha^2)\log(\Lambda/|\vec{q}|)},
\eeq
with
\beq
f_1{=}\frac{1}{4},\qquad f_2{=}\frac{3\log2{-}4}{6}.
\eeq
This expression can be inverted to obtain a systematic expansion of the bare coupling $\alpha$ in terms of the renormalized coupling:
\bea
\alpha&{=}&\alpha_q{+}f_1\alpha_q^2\log(\Lambda/|\vec{q}|)\nn\\&&{+}\left[f_2\log(\Lambda/|\vec{q}|){+}f_1^2\log^2(\Lambda/|\vec{q}|)\right]\alpha_q^3{+}O(\alpha_q^4).\nn\\&&
\eea
Writing down a similar expansion, but with $q$ replaced by a different momentum $k$, and equating the two expansions allows us to obtain an expansion for $\alpha_k$ in terms of $\alpha_q$:
\bea
&&\!\!\!\!\!\!\!\alpha_k{-}\alpha_q=
\nn\\&&\!\!\!\!\!\!\!f_1\alpha_q^2\log(|\vec{k}|/|\vec{q}|){+}\Big[f_2\log(|\vec{k}|/|\vec{q}|){+}f_1^2\Big(\log^2(\Lambda/|\vec{q}|)
\nn\\&&\!\!\!\!\!\!\!-\log^2(\Lambda/|\vec{k}|){-}2\log(|\vec{k}|/|\vec{q}|)\log(\Lambda/|\vec{k}|)\Big)\Big]\alpha_q^3{+}O(\alpha_q^4).\nn\\&&
\eea
Dividing both sides by $|\vec{k}|{-}|\vec{q}|$ and taking the limit $|\vec{k}|\to|\vec{q}|$, we find
\beq
|\vec{q}|\frac{d\alpha_q}{d|\vec{q}|}{=}\alpha_q^2\left(f_1{+}f_2\alpha_q\right).
\eeq
Thus, we may define a beta function for the effective coupling:
\beq
\beta(\alpha_q){=}\alpha_q^2\left(f_1{+}f_2\alpha_q\right).
\eeq
Integrating the RG flow equation, we obtain
\beq
\log(|\vec{q}|/|\vec{k}|){=}\frac{\alpha_q{-}\alpha_k}{f_1\alpha_q\alpha_k}{+} \frac{f_2}{f_1^2}\log\left(\frac{\alpha_k(f_1{+}f_2\alpha_q)}{\alpha_q(f_1{+}f_2\alpha_k)}\right).
\eeq
Given the effective coupling $\alpha_k$ at moment scale $|\vec{k}|$, this equation allows us to predict the coupling at a different scale $|\vec{q}|$. We may convert this to a similar relation for the Fermi velocities:
\beq
\log(|\vec{q}|/|\vec{k}|){=}{-}\frac{4\pi}{f_1g^2}(v_q{-}v_k){+}\frac{f_2}{f_1^2}\log\left(\frac{g^2 f_2{+}4\pi f_1v_q}{g^2 f_2{+}4\pi f_1v_k}\right).\label{twoloopvq}
\eeq
When the two scales are close to each other, $|\vec{k}|\approx|\vec{q}|$, this relation can be approximately expressed as
\beq
v_q\approx v_k\left[1{+}(f_1\alpha_k{+}f_2\alpha_k^2)\log(|\vec{k}|/|\vec{q}|)\right].
\eeq

It is interesting to note that when $\hbox{sign}(f_1f_2){=}{-}1$, there is an interacting fixed point of the RG flow:
\beq
\alpha_c{=}{-}f_1/f_2.
\eeq
This is indeed the case for the values of $f_1$ and $f_2$ we obtained earlier:
\beq
\alpha_c{=}\frac{3}{8{-}6\log2}\approx 0.78.
\eeq
The existence of this critical point is one of the main results of the current work. This result means that if the flow begins at high energy/momentum at a value of $\alpha$ such that $\alpha<\alpha_c$, then the theory becomes weakly interacting in the infrared, as is the case when only the leading-order term in the beta function is retained. On the other hand, if $\alpha>\alpha_c$ initially, then the theory flows to a strongly-interacting fixed point in the infrared, indicating that either a phase transition has occurred, or that perturbation theory is failing at these larger values of the coupling.

It is important to note that the value of $\alpha_c$ we have obtained, 0.78, lies in the middle of the spectrum of values of $\alpha$ relevant for experiments, where graphene on a BN substrate corresponds to $\alpha\approx0.4$, while for graphene in vacuum, $\alpha\approx2.2$. This implies that if the critical point is physical, then a quantum phase transition should occur in suspended graphene. Since no such phase transition has been seen in experiments, this suggests that the critical point we have found is more likely unphysical and indicative of a breakdown of perturbation theory. Further evidence for such a breakdown will be given in the next section, where we perform a semiclassical analysis to estimate the order at which the perturbative series ceases to be asymptotic. A more complete discussion of comparisons to experiment will be given in Sec.~\ref{sec:experiment}.

\subsection{Estimating the breakdown of perturbation theory}

It is generally the case that a perturbative field theory expansion does not converge, and instead constitutes an asymptotic series approximation to a physical quantity, implying that any such expansion will inevitably begin to fail if calculations are carried out to sufficiently high order. In light of the critical point discovered in the previous section, we would like to estimate the order at which graphene perturbation theory begins to fail to check if it is in fact this failure which gives rise to the critical point and not a real phase transition.

We can obtain an estimate following Dyson's argument [\onlinecite{Dyson_PR52}] for the breakdown of perturbative QED. In that context, Dyson argued that the point in parameter space where the electric charge vanishes cannot be analytic since the theory in which the Coulomb interaction between electrons is taken to be attractive rather than repulsive does not have a stable ground state. The latter follows from the observation that the total energy can be made arbitrarily negative by producing a sufficiently large number of electron/positron pairs and grouping all the electrons in one region of space and all the positrons in another, distant region of space. If the number $N$ of electron/positron pairs is sufficiently large, the potential energy will dominate the kinetic energy of the particles since the kinetic energy scales like $N$, while the potential energy scales like $N^2$. This implies that there is a critical number $N_c$ of such pairs beyond which the pressure due to the zero-point motion of the electrons is insufficient to counteract the inward pull of the Coulomb force, leading to a collapse of the electron cloud. In terms of the real theory with a repulsive Coulomb interaction between electrons, this instability manifests itself as a failure of perturbation theory at higher orders where the number of virtual electrons appearing in a typical Feynman diagram is comparable to $N_c$.

Dyson's argument applies equally well to our effective graphene theory. We may therefore estimate the order at which perturbation theory breaks down by considering a gas of massless two-dimensional electrons with an attractive Coulomb interaction and determining the critical number of electrons $N_c$ beyond which the gas collapses on itself. This problem is very similar to that of gravitational collapse of a star, in which case $N_c$ corresponds to the Chandrasekhar limit.\cite{Chandrasekhar_AJ31,LevyLeblond_JMP69} Our approach will therefore be to adapt the standard semi-classical derivation of the Chandrasekhar limit to the case of graphene electrons. In the context of QED, this approach was recently advocated for in Ref.~[\onlinecite{Kolomeisky_arXiv13}]. To make the discussion as transparent as possible, we first review the gravitational problem.

\subsubsection{Review of the Chandrasekhar limit}

Consider a gas of massless neutral spin 1/2 fermions in three spatial dimensions interacting pairwise under the Newtonian gravitational potential,
\beq
U=-\frac{\beta}{r},
\eeq
where $r$ is the distance between two particles, and $\beta=Gm^2$. There exists a critical number of fermions, $N_c$, such that for $N<N_c$, the system is in a stable configuration in which the gravitational attraction is balanced by the outward pressure due to the Pauli exclusion principle, while for $N>N_c$, no such stable configuration exists, and the fermion gas implodes, forming a black hole. We wish to calculate $N_c$.

For simplicity, we assume that the fermions form a static ball of radius $r_0$, and that the number density can be characterized by $\rho(r)$, with
\beq
N=N(r_0)=4\pi\int_0^{r_0}dr r^2\rho(r).
\eeq
When the system is in equilibrium, it can be described semi-classically by the Lane-Emden equation:
\beq
r^2P'(r)+\beta N(r)\rho(r)=0,\label{laneemden}
\eeq
where $P(r)$ is the pressure of the gas at radius $r$ from the center. This equation is obtained simply from the condition that the gravitational attraction must balance the outward pressure on an infinitesimal volume element at each radius $r$ inside the gas. In order to solve this equation for $\rho(r)$, we must first determine the equation of state, $P(\rho)$. This can be determined by using the ultra-relativistic result for the relation between pressure and average momentum:
\beq
P=\frac{c\rho}{3}\langle p\rangle,
\eeq
where $c$ is the speed of light. Combining this with the expression for the density of a three-dimensional Fermi gas in terms of the Fermi momentum $p_F$,
\beq
\rho=\frac{p_F^3}{3\pi^2},
\eeq
we obtain
\beq
P=\frac{cp_F^3}{9\pi^2}\frac{4\pi\int_0^{p_F}dp p^3}{\frac{4\pi}{3}p_F^3}=\frac{c(3\pi^2)^{1/3}}{4}\rho^{4/3}.
\eeq
Using this equation of state, we can rearrange Eq.~(\ref{laneemden}) to read
\beq
x^2f^3(x)+\frac{d}{dx}\left(x^2\frac{d}{dx}f(x)\right)=0,\label{feqn}
\eeq
where we have defined dimensionless quantities $f$ and $x$ via
\beq
\rho=\rho_c f^3,
\eeq
and
\beq
x\equiv\left(\frac{\pi}{3}\right)^{1/6}\sqrt{\frac{\beta}{c}}\rho_c^{1/3}r.
\eeq
Here, $\rho_c$ is some characteristic density that we have introduced solely for the purpose of rendering $f(x)$ dimensionless. In terms of $f(x)$, the number of particles can be written as
\bea
N&=&\frac{\sqrt{3\pi}}{2}\left(\frac{c}{\beta}\right)^{3/2}\int_0^{x_0}dx x^2f^3(x)
\nn\\&=&-\frac{\sqrt{3\pi}}{2}\left(\frac{c}{\beta}\right)^{3/2}x_0^2f'(x_0).\label{Neqn}
\eea
From Eq.~(\ref{feqn}), it is straightforward to see that if $f(x)$ is a solution, $\lambda f(\lambda x)$ is also a solution for any real constant $\lambda$. Therefore, it suffices to solve Eq.~(\ref{feqn}) with initial conditions $f(0)=1$, $f'(0)=0$, since solutions with other values of $f(0)$ can be obtained from this solution by choosing $\lambda$ appropriately. These observations also make it clear that $N$ is independent of the initial conditions, implying that there is a unique value of $N$ corresponding to the critical value.  Solving Eq.~(\ref{feqn}) numerically and plugging the result into Eq.~(\ref{Neqn}) yields
\beq
N_c=3.09797\left(\frac{c}{\beta}\right)^{3/2}.
\eeq
The uniqueness of this result is due to the fact that we are working in the ultra-relativistic limit; in order to obtain stable configurations with $N<N_c$, we would need to move away from the ultra-relativistic regime. Since there cannot be stable solutions for $N>N_c$, we see that $N_c$ is a critical value beyond which the fermion gas collapses. This value of $N_c$ is the Chandrasekhar limit.\cite{Chandrasekhar_AJ31,LevyLeblond_JMP69} This expression was used in Ref.~[\onlinecite{Kolomeisky_arXiv13}] to estimate that perturbative QED breaks down at approximately the 5000th order.

\subsubsection{Breakdown of graphene perturbation theory}

We can apply the same steps for a gas of electrons (with four-fold degeneracy coming from spin and valley degrees of freedom) in two-dimensions with a repulsive Coulomb interaction, in which case we have $\beta\to e^2/\kappa$. The number of electrons is given by
\beq
N=N(r_0)=2\pi\int_0^{r_0}dr r n(r),
\eeq
where $n(r)$ is the number density. Since the form of the interaction is the same as in the previous subsection, the stability equation remains the same except that now the pressure $P(r)$ is a force per unit length:
\beq
r^2P'(r)+\beta N(r)n(r)=0.\label{laneemdenii}
\eeq
We can obtain the equation of state from the relations (again using the ultra-relativistic limit, but with $c\to v_F$)
\beq
P=\frac{v_Fn}{2}\langle p\rangle,\quad n=\frac{p_F^2}{\pi},
\eeq
with the result
\beq
P=\frac{v_F\sqrt{\pi}}{3}n^{3/2}.
\eeq
Plugging this into Eq.~(\ref{laneemdenii}) and defining
\beq
u\equiv 2\sqrt{\pi n_c}\frac{\beta}{v_F}r=2\alpha\sqrt{\pi n_c}r, \quad n(r)\equiv h^2(u)n_c,
\eeq
we find
\beq
uh^2(u)+\frac{d}{du}\left(u^2\frac{d}{du}h(u)\right)=0,\label{heqn}
\eeq
while the number of electrons is
\beq
N=\frac{1}{2\alpha^2}\int_0^{u_0}du u h^2(u)=-\frac{1}{2\alpha^2}u_0^2h'(u_0).\label{Neqnii}
\eeq
Solving Eq.~(\ref{heqn}) numerically and plugging the result into Eq.~(\ref{Neqnii}), we obtain the critical number of electrons:
\beq
N_c=0.70114/\alpha^{2}.
\eeq
For graphene suspended in vacuum with $\alpha=2.2$, we find $N_c=0.14$, while for $\alpha=0.5$, we obtain $N_c=2.8$. This behavior is more or less consistent with our perturbative RG analysis, assuming that the critical point we obtained earlier signifies the breakdown of graphene perturbation theory at second order. Of course, $\alpha$ measures the ratio of the potential energy of an electron pair relative to the kinetic energy, so it is not surprising that when $\alpha\sim1$, the vacuum is becoming unstable to electron/positron pair creation even with only one or two virtual electrons present when the Coulomb interaction is attractive. Given that the breakdown appears to happen at very low order, it should be possible to perform a more precise analysis since this corresponds to solving a few-body problem. We leave this to future work.

\subsection{Divergence structure of graphene effective field theory}\label{sec:divergencestructure}

The classic BPHZL theorem states that for a general renormalizable field theory, physical observables can be rendered finite by introducing a finite number of counterterms, one for each divergence appearing in superficially divergent amplitudes.\cite{Weinberg_PR60,Bogoliubov_AM57,Hepp_CMP66,Zimmermann_CMP69,Lowenstein_CMP75} This theorem immediately implies that an arbitrary amplitude in the graphene effective field theory can be made finite by renormalizing a finite number of parameters in the Lagrangian. Determining which and how many parameters must run with the coupling in order for the renormalization procedure to work requires a detailed analysis of the divergence structure of the theory. Below, we will show that there are three superficially divergent classes of diagrams, two of which potentially exhibit both linear and logarithmic divergences, while the third can have only logarithmic divergences. Nominally, this would suggest that we may need several different parameters to absorb these divergences if we also take into account the fact that the temporal and spatial components of the electron self-energy can diverge independently. However, there is an expectation that all these divergences can be removed to arbitrarily high order in the perturbative expansion by renormalizing only the Fermi velocity and electron field strength. Since we have already seen that the renormalization of the electron self-energy requires the renormalization of both these quantities, in order for these parameters to absorb all possible divergences in the graphene theory, it would have to be the case that all such divergences can be traced back to divergent self-energy corrections or possibly to divergent vertex corrections as well since the vertex can be renormalized through the field strength renormalization. We would like to demonstrate this explicitly.

\subsubsection{General diagram statistics}

The superficial degree of divergence of a Feynman diagram is defined to be the total power of momenta in the numerator of the integrand minus the power of momenta in the denominator. Defining the number of independent loops in the diagram to be $L$, and the number of internal photon (Coulomb) propagators and fermion propagators to be $P$ and $F$, respectively, the superficial degree of divergence, $D$, is easily seen to be
\beq
D=3L-F-P,\label{DfromLFP}
\eeq
in the case of Dirac fermions in 2+1 dimensions interacting via an effective 2d Coulomb interaction. For a general Feynman diagram in this theory, the number of loops can be expressed in terms of the number of vertices by observing that in the position space Feynman rules, every propagator contributes an integration over a distinct three-momentum, every vertex comes with an associated delta function, and an extra delta function imposes total momentum conservation. Therefore,
\beq
L=F+P-V+1.\label{LfromFPV}
\eeq
A general diagram with $E$ external lines must also obey the topological identity
\beq
2(F+P)+E=3V.
\eeq
The above three relations together imply the following simple result for the superficial degree of divergence of a diagram with $E$ external legs:
\beq
D=3-E.\label{DfromE}
\eeq
This result immediately implies that diagrams contributing to either the electron self-energy or the vacuum polarization have $D=1$, while diagrams contributing to the vertex function (with two external electron lines and one external photon) have $D=0$, as do diagrams with three external photon lines (e.g., the triangle diagram). All other diagrams in the theory have $D<0$, and are thus superficially convergent. Diagrams with three external photon lines vanish identically due to charge-conjugation invariance (under which the gauge field acquires a minus sign).\cite{Furry_PR37} We therefore have three types of diagrams whose divergence structure will dictate the renormalization procedure.

\subsubsection{Absence of linear divergences in electron self-energy}

Using the simple relation between the superficial degree of divergence and the number of external lines, Eq.~(\ref{DfromE}), we have just seen that diagrams contributing to the electron self-energy have $D=1$. This result can also be derived another way by making further observations about the structure of diagrams contributing to the electron self-energy. These additional observations will allow us to show that all such diagrams never exhibit linear divergences, instead possessing at most logarithmic divergences.

We begin by noticing that for diagrams contributing to the electron self-energy, every vertex is connected to one photon propagator while every photon propagator joins two distinct vertices, so that we have
\beq
V=2P,
\eeq
which implies
\beq
L=F-P+1.\label{LfromFP}
\eeq
Furthermore, it is not hard to convince oneself that the additional relation also holds for this class of diagrams:
\beq
L=P.
\eeq
This relation can be seen by imagining that every diagram contributing to the self-energy is built up by starting from the fermion propagator and adding photon propagators and fermion loops, one at a time. Since every additional photon propagator either creates a new loop or joins a new fermion loop to the diagram, we see that $L=P$. Using Eqs. (\ref{LfromFP}) and (\ref{DfromLFP}), this result immediately implies
\beq
F=2P-1,
\eeq
and
\beq
D=1.
\eeq
The former relation means that every diagram contributing to the electron self-energy contains an odd number of fermion propagators, while the latter means that every such diagram superficially has at most a linear divergence, as we have already seen. However, since every diagram contains an odd number of fermion propagators, it is easy to show that this linear divergence always cancels, leaving at most a logarithmic divergence. This can be seen by noting that when one expands the integrand of a diagrammatic contribution in the limit where all loop momenta are taken to be large, the leading term is independent of the external momentum $q$. Since the number of fermion propagators is odd, this leading term will be odd under the operation in which all the signs of the loop momenta are flipped. Since the measures of the loop integrations are invariant under this operation, as are the Coulomb propagators, this leading term must vanish, eliminating the possibility of a linear divergence. The absence of linear divergences in diagrams contributing to the electron self-energy means that these diagrams are at most logarithmically divergent. Diagrams containing divergent subdiagrams will generally diverge like the power of a logarithm.

\subsubsection{Linear divergences in corrections to vacuum polarization}

In the previous subsection, we showed that there are no linear divergences arising from diagrams that contribute to the electron self-energy. Therefore, if it is true that all divergences are due to divergent self-energy or vertex corrections, then we would expect that no linear divergences appear anywhere in the theory. We have seen that linear divergences also potentially arise in diagrams that contribute to the vacuum polarization function, i.e., diagrams with two external photon lines (which have $D=1$), as well as diagrams which contain these diagrams as subdiagrams. The fact that linear divergences do not arise in contributions to the electron self-energy strongly implies that no linear divergences are present in any subdiagram, and since all vacuum polarization corrections appear as subdiagrams in some electron self-energy contributions, this suggests that linear divergences do not arise in any vacuum polarization corrections either. However, it is still possible that a certain symmetry results in an exact cancelation of self-energy diagrams that contain linearly divergent vacuum polarization subdiagrams even if these subdiagrams do not themselves vanish identically. Therefore, it is worth taking a closer look at this issue.

We have seen in Sec~\ref{sec:oneloopvacpol} that the one-loop vacuum polarization has neither linear nor logarithmic divergences, while in Sec.~\ref{sec:vacpolSE} we showed one of the two-loop corrections contained a logarithmic divergence coming from the appearance of the one-loop electron self-energy as a subdiagram. It was further shown that this logarithmic divergence could be absorbed into a redefinition of the Fermi velocity, demonstrating the absence of charge renormalization up to second order. In the course of these calculations, however, it was not made clear why linear divergences did not appear in either the one-loop or two-loop corrections to the vacuum polarization, nor is it clear whether such divergences can arise in higher-order corrections. Let us first consider why there was no linear divergence in the one-loop vacuum polarization. This diagram has the value,
\beq
\Pi_B(q)=-N\int\frac{d^3k}{(2\pi)^3}\tr\left[\gamma^0\frac{\slashed k}{k^2}\gamma^0\frac{\slashed k+\slashed q}{(k+q)^2}\right].
\eeq
The leading linear divergence can be extracted by taking the limit $q\to0$. In this limit, the integral can be shown to vanish because of the integration over the loop frequency $k_0$:
\beq
\int_{-\infty}^{\infty}dk_0\frac{k_0^2-v_F^2|\vec{k}|^2}{k^4}=0.
\eeq
This identity also ensures the absence of a linear divergence in the two-loop vertex correction to the vacuum polarization. The above identity is one of a family of identities:
\beq
\int_{-\infty}^{\infty}dk_0\frac{(3-2m)k_0^2+v_F^2|\vec{k}|^2}{k^{2m}}=0,
\eeq
where $m\ge2$ is an integer. It can also be shown that the absence of a linear divergence in the two-loop self-energy correction to the vacuum polarization is due to the identity with $m=3$. Thus, these identities ensure the absence of linear divergences in first and second-order perturbation theory. This family of integral identities is also responsible for the fact that the one-loop vertex diagram is finite. Furthermore, it leads to the vanishing of the rainbow diagram corrections to the electron self-energy. One of these identities also gives rise to the vanishing of a third-order correction to the electron self-energy, as shown in Appendix~\ref{app:threelooprainbowSE}. These identities also play an important role for non-renormalization theorems in the context of ladder diagram expansions.\cite{Gonzalez_JHEP12} It is not clear whether the appearance of these identities is due to a hidden symmetry of the theory or whether these identities continue to ensure the cancelation of linear divergences at higher orders in perturbation theory. We will thus leave it as a conjecture that it is these identities that allow all divergences to be attributed to electron self-energy or vertex corrections.

\subsubsection{Recursion relation for coefficients of divergences}\label{sec:recursionrels}

In this section, we show that only linear-log divergences in electron self-energy diagrams contribute to the Fermi velocity renormalization (and to the running of the effective coupling). We also show that higher-power log divergences are completely determined by the linear-log terms and explicitly derive the recursion relations that govern this dependence.

The fact that only logarithmic divergences appear in the expansion of the electron self-energy means that we may write the following expansion for the Fermi velocity and effective coupling:
\beq
\frac{v_\mu}{v_F}=\frac{\alpha}{\alpha_\mu}=1+\sum_{n=1}^\infty F_n(\alpha)\log^n(\Lambda/\mu),\label{vmuexpand}
\eeq
where we have introduced the shorthand notation $\mu=|\vec{q}|$ for the renormalization scale, so that $v_\mu$ and $\alpha_\mu$ denote the renormalized Fermi velocity and coupling, respectively. This expression can be inverted to obtain an expansion for the bare coupling in terms of the renormalized coupling:
\beq
\alpha=\alpha_\mu+\sum_{n=1}^\infty G_n(\alpha_\mu)\log^n(\Lambda/\mu),\label{bareexpand}
\eeq
where $G_n$ can be calculated systematically as a power series in $\alpha_\mu$, with the coefficients given by combinations of the coefficients appearing in similar expansions of the $F_n$. In particular, one finds
\bea
G_1(\alpha_\mu)&=&\alpha_\mu F_1(\alpha_\mu),\label{G1fromF1}\\
G_2(\alpha_\mu)&=&G_{23}\alpha_\mu^3+G_{24}\alpha_\mu^4+G_{25}\alpha_\mu^5+\ldots
\nn\\&=&F_{11}^2\alpha_\mu^3+(3F_{11}F_{12}{+}F_{23})\alpha_\mu^4
\nn\\&& +(2F_{12}^2{+}4F_{11}F_{13}{+}F_{24})\alpha_\mu^5+\ldots,
\eea
where e.g.,
\beq
F_1(\alpha)=F_{11}\alpha+F_{12}\alpha^2+F_{13}\alpha^3+\ldots
\eeq
In terms of the notation introduced earlier, $f_1$, $f_2$, for the coefficients of the first- and second-order terms in the Fermi velocity expansion, we have simply $F_{11}=f_1$, $F_{12}=f_2$.

The fact that the bare coupling, $\alpha$, is independent of the renormalization scale $\mu$ means that we can differentiate Eq.~(\ref{bareexpand}) with respect to $\mu$ and then multiply by $\mu$ everywhere to obtain
\bea
0&=&\mu\alpha_\mu'-G_1+\mu\alpha_\mu' \frac{dG_1}{d\alpha_\mu}\log(\Lambda/\mu)-2G_2\log(\Lambda/\mu)\nn\\&&\!\!\!\!\!\!+\mu \alpha_\mu' \frac{dG_2}{d\alpha_\mu}\log^2(\Lambda/\mu)-3G_3\log^2(\Lambda/\mu)+\ldots,
\eea
where the prime denotes differentiation with respect to $\mu$. Since the renormalized coupling, $\alpha_\mu$, is independent of the cutoff, it must be the case that the coefficients of each power of $\log(\Lambda/\mu)$ vanish separately, leading to the following set of relations:
\bea
\mu\alpha_\mu'&=&G_1(\alpha_\mu),\\
G_2(\alpha_\mu)&=&\frac{1}{2}\mu\alpha_\mu' \frac{dG_1}{d\alpha_\mu},\\
G_3(\alpha_\mu)&=&\frac{1}{3}\mu\alpha_\mu' \frac{dG_2}{d\alpha_\mu},\\
&\vdots&\nn\\
G_n(\alpha_\mu)&=&\frac{1}{n}\mu\alpha_\mu'\frac{dG_{n-1}}{d\alpha_\mu}.
\eea
We see that the beta function for the coupling, $\beta(\alpha_\mu)$, is determined solely by the coefficient of the $\log(\Lambda/\mu)$ term in the expansion of the bare coupling, Eq.~(\ref{bareexpand}), which in turn is determined by the coefficient of the $\log(\Lambda/\mu)$ term in the expansion of the Fermi velocity in terms of the bare coupling, Eqs. (\ref{vmuexpand}),(\ref{G1fromF1}). The coefficients of all higher powers of $\log(\Lambda/\mu)$ are fully determined by $\beta(\alpha_\mu)$:
\bea
G_1(\alpha_\mu)\!\!\!&{=}&\!\!\!\beta(\alpha_\mu){=}F_{11}\alpha_\mu^2{+}F_{12}\alpha_\mu^3{+}F_{13}\alpha_\mu^4{+}\ldots,\\
G_n(\alpha_\mu)\!\!\!&{=}&\!\!\!\frac{\beta(\alpha_\mu)}{n}\frac{dG_{n-1}}{d\alpha_\mu},\qquad n>1.
\eea
These relations between the different $G_n$ imply relations among the coefficients appearing in the $F_n$, i.e. the coefficients appearing in the perturbative expansion of the Fermi velocity. Going up to fifth order in perturbation theory, we find explicitly:
\bea
&&\hbox{From }G_2:\nn\\
&&F_{22}=0,\; F_{23}=-\frac{1}{2}F_{11}F_{12}, \; F_{24}=-\frac{1}{2}F_{12}^2-F_{11}F_{13},
\nn\\&& F_{25}=-\frac{3}{2}(F_{12}F_{13}+F_{11}F_{14}),\label{FrelsfromG2}\\
&&\hbox{From }G_3:\nn\\
&&F_{33}=0,\quad F_{34}=-\frac{5}{3}F_{11}^2F_{12}-4F_{11}F_{23},\label{FrelsfromG3}
\\&&F_{35}=-\frac{25}{6}F_{11}F_{12}^2-4F_{11}^2F_{13}-5F_{12}F_{23}-5F_{11}F_{24},\nn\\
&&\hbox{From }G_4:\nn\\
&&F_{44}=0,\quad F_{45}=-\frac{43}{12}F_{11}^3F_{12}-10F_{11}^2F_{23}-5F_{11}F_{34}.\nn\\
\eea
For example, the second equation in (\ref{FrelsfromG2}) shows that the coefficient of the $\log^2(\Lambda/\mu)$ term in the third-order result for the Fermi velocity is completely determined by the first and second-order coefficients of the $\log(\Lambda/\mu)$ term. In Sec.~\ref{sec:threeloop} below, we will verify explicitly that such terms arise at third order as required by the recursion relation. The first equation in (\ref{FrelsfromG2}) implies that there is no $\log^2(\Lambda/\mu)$ divergence term at order $\alpha^2$, as we have already seen in our explicit calculation of the two-loop Fermi velocity renormalization. In the next subsection, we will see that the condition $F_{22}=0$ is in fact required by the standard theorem on the renormalization scheme independence of the first two terms of the beta function for the effective coupling. The first equation of (\ref{FrelsfromG3}) states that no $\log^3(\Lambda/\mu)$ divergence terms can arise at third-order. We will check this explicitly in Sec.~\ref{sec:threeloop} below.

\subsubsection{Redefining $\Lambda$ at higher orders in $\alpha$ and scheme dependence}\label{sec:redefineLambda}

We now finally return to the question of why we can neglect the finite part of the second{-}order self{-}energy. We raised this issue briefly at the end of Sec.~\ref{sec:twoloopSE}, but postponed addressing it until after we presented some general properties of the divergence structure of the perturbation theory. In particular, it may not be clear why we can neglect the finite contributions since we have already given a precise definition of the ultraviolet cutoff $\Lambda$ when we computed the first{-}order self{-}energy, $\Sigma_1(q)$. Once this definition is made, we must take care to maintain consistency with this definition, and it may not be immediately obvious why we can absorb a new constant into $\Lambda$ when we compute $\Sigma_2(q)$. To be more precise, suppose that we have not absorbed any constants into $\Lambda$ at all, and consider the renormalized Fermi velocity to second order in $\alpha$:
\beq
v_q{=}v_F\left\{1{+}\alpha f_1[\log(\Lambda/|\vec{q}|){+}C_1]{+}\alpha^2f_2[\log(\Lambda/|\vec{q}|){+}C_2]\right\}.
\eeq
Here, $C_1$ and $C_2$ are the finite parts of $\Sigma_1$ and $\Sigma_2$, respectively, and we have found above that
\beq
f_1{=}\frac{1}{4},\qquad f_2{=}\frac{3\log2{-}4}{6}.
\eeq
When we computed $\Sigma_1$ in Sec.~\ref{sec:oneloopSE}, we absorbed $C_1$ into the definition of $\Lambda$. If we then absorb $C_2$ into $\Lambda$, it would appear that we have an inconsistency since the $\Lambda$'s appearing in $\Sigma_1$ and $\Sigma_2$ would not be the same.

In order to understand how to resolve this apparent discrepancy, it helps to first make a few observations about the scheme dependence of the renormalization procedure. Suppose that we have computed the self{-}energy to all orders to obtain a renormalized velocity of the form
\bea
&&\!\!\!\!\!\!\!\!\frac{v_q}{v_F}{=}1{+}\widetilde{C}(\alpha){+}F_1(\alpha)\log(\Lambda/|\vec{q}|){+}F_2(\alpha)\log^2(\Lambda/|\vec{q}|)){+}\ldots,\nn\\&&
\eea
where $\widetilde{C}(\alpha)$ is the all{-}orders finite contribution, where we know that
\beq
F_1(\alpha){=}f_1\alpha{+}f_2\alpha^2{+}O(\alpha^3),
\eeq
and where we have used the result that the self{-}energy and hence the Fermi velocity can be expressed in terms of a power series in $\log(\Lambda/|\vec{q}|)$. Because of the appearance of higher{-}power logarithms, if we change the definition of $\Lambda$, then the coefficients $F_n$ of the logarithms will get modified. As we explained in the previous subsection, the beta function $\beta(\alpha_q)$ is determined by $F_1$, so that it too will be modified by a redefinition of $\Lambda$. However, since the higher{-}power logarithms occur only at third order and higher, the first two terms of $\beta(\alpha_q){=}\alpha_qF_1(\alpha_q)$ will remain invariant. This is a reflection of the well known scheme independence of the two lowest order terms in the beta function. Here, we see that it was necessarily the case that higher powers of divergent logarithms appear only at third order and above since otherwise the standard scheme{-}independence theorem would be violated.

When we choose to expand the Fermi velocity as in Eq.~(\ref{vmuexpand}) (i.e., where the finite term has been completely absorbed into the cutoff), then we are supposing that we first redistributed the finite part according to
\beq
\frac{v_q}{v_F}{=}1{+}\sum_{n{=}1}^\infty F_n(\alpha)[\log(\Lambda/|\vec{q}|){+}C(\alpha)]^n,
\eeq
in the process fixing the definitions of the $F_n$ to all orders (where the $F_n$ in this scheme obey the recursion relations derived in the previous subsection), and then we absorbed the all{-}orders constant $C(\alpha)$ into $\Lambda$. This quantity also has a power{-}series expansion in $\alpha$:
\beq
C(\alpha){=}C_1{+}\alpha(C_2{-}C_1){+}O(\alpha^2),
\eeq
where $C_1$ and $C_2$ were defined above. It is then apparent that the cutoff $\Lambda$ naturally gets redefined at every order in $\alpha$, and that it is not the finite contribution at order $\alpha$ that gets absorbed into the $\Lambda$ appearing in the self{-}energy at that order, but rather a combination of all the finite contributions occurring up to and including that order. What this means effectively is that we simply ignore the finite contributions appearing in the self{-}energy at every order in $\alpha$.

\subsection{Three-loop corrections to electron self-energy}\label{sec:threeloop}

\begin{figure}
\qquad\includegraphics[width=\columnwidth]{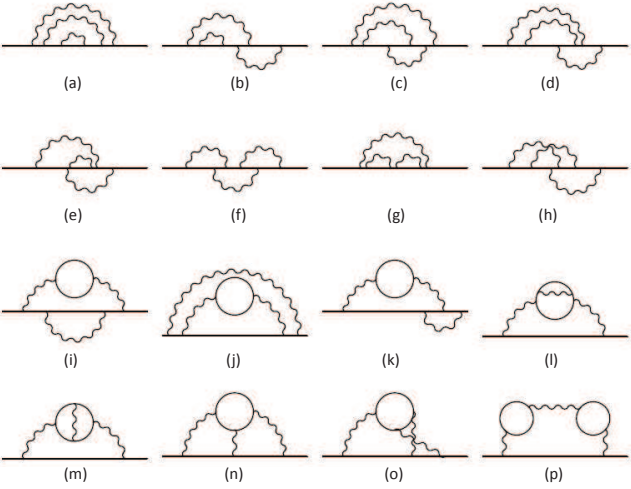}
\caption{Distinct diagrams contributing to third-order electron self-energy.}\label{fig:3loopdiagrams}
\end{figure}
There are 16 distinct diagrams contributing to the electron self-energy at third order; these are displayed in Fig.~\ref{fig:3loopdiagrams}. Note that diagrams (b), (d), (k), and (l) each represent a pair of diagrams that are related by symmetry and thus give equal contributions to the self-energy. In our first- and second-order calculations, we have seen that the graphene effective field theory renormalizes through the renormalization of the Fermi velocity and electron field strength. In this section, we will test whether the theory continues to be renormalizable at third order by computing the leading ultraviolet divergences at this order. In particular, we saw in Sec.~\ref{sec:recursionrels} that the coefficient of the $\log^2$ divergence at third order is fully determined by the coefficients of the $\log$ divergences arising at first- and second-order. Therefore, on general grounds it must be the case that $\log^2$ ultraviolet divergences arise at third order, while higher powers of $\log$ should not arise at this order. We will confirm this expectation by calculating the leading divergences in several of the diagrams shown in Fig.~\ref{fig:3loopdiagrams}.

Our first- and second-order results show that, at least up to those orders, ultraviolet divergences only arise in corrections to the electron self-energy. Therefore, among the 16 diagrams depicted in Fig.~\ref{fig:3loopdiagrams}, we expect that the strongest ultraviolet divergences will come from those diagrams which contain divergent self-energy subdiagrams. Nominally, there are eight such diagrams, corresponding to those labeled (a), (b), (c), (e), (g), (i), (j), (l) in Fig.~\ref{fig:3loopdiagrams}. However, it is clear that diagram (a) vanishes identically since it contains the two-loop rainbow diagram as a subdiagram, and we have already shown that this diagram vanishes identically in Sec.~\ref{sec:twolooprainbow}:
\beq
\Sigma_{3a}(q)=0.
\eeq
Also notice that diagram (g) would appear to have a $\log^3$ divergence since it contains two logarithmically divergent subdiagrams, potentially violating the recursion relations derived in Sec.~\ref{sec:recursionrels}, which would in turn endanger the renormalizability of the theory. However, we show in Appendix~\ref{app:threelooprainbowSE} that diagram (g) vanishes identically,
\beq
\Sigma_{3g}(q)=0,
\eeq
thus avoiding this potential danger and leaving only six non-trivial diagrams containing divergent subdiagrams at third order. (Also note that diagrams (n) and (o) vanish identically due to Furry's theorem.\cite{Furry_PR37}) The leading divergences of diagrams (b), (c), (e), (i), (j), (l) are computed explicitly in Appendix~\ref{app:threeloop}, and we summarize the results here:
\bea
\Sigma_{3b}(q)\!\!\!&{=}&\!\!\!i\left[{-}\tfrac{1}{480}q_0\gamma^0{+}\tfrac{1}{128}\left({-}\tfrac{29}{5}{+}8\log2\right)v_F\vec{q}{{\cdot}}\vec{\gamma}\right]\nn\\
&&\times\alpha^3\log^2(\Lambda/|\vec{q}|){+}\ldots,
\nn\\ \Sigma_{3c}(q)\!\!\!&{=}&\!\!\!{-}\tfrac{i}{16}\left (\log{2}{{-}}\tfrac{2}{3}\right )\alpha^3v_F\bq{\cdot}\vec{\gamma}\log^2\left (\Lambda/|\bq|\right ){+}\ldots,
\nn\\ \Sigma_{3e}(q)\!\!\!&{=}&\!\!\!i\left[\tfrac{1}{8}\left(\tfrac{41}{60}{-}\log2\right)q_0\gamma^0{+}\tfrac{1}{128}\left(\tfrac{247}{15}{-}24\log2\right)v_F\vec{q}{\cdot}\vec\gamma\right]\nn\\
&&\times\alpha^3\log^2(\Lambda/|\vec{q}|){+}\ldots,
\nn\\ \Sigma_{3i}(q)\!\!\!&{=}&\!\!\!{-}\tfrac{iN}{480}(4q_0\gamma^0{-}3v_F\bq{\cdot}\vec{\gamma})\alpha^3\log^2\left (\Lambda/|\bq|\right ){+}\ldots,
\nn\\ \Sigma_{3j}(q)\!\!\!&{=}&\!\!\!{-}\tfrac{iN}{96}v_F\bq{\cdot}\vec{\gamma}\alpha^3\log^2\left (\Lambda/|\bq|\right ){+}\ldots,
\nn\\ \Sigma_{3l}(q)\!\!\!&{=}&\!\!\!{-}\tfrac{iN}{240}\left(3q_0\gamma^0{{-}}v_F\vec{q}{\cdot}\vec\gamma\right)\alpha^3\log^2(\Lambda/|\vec{q}|){+}\ldots,
\nn\\\label{logsquaredterms}
\eea
where the ellipsis in each case represents subleading $\log\left (\Lambda/|\bq|\right )$ terms. These results demonstrate explicitly that the expected $\log^2$ divergences are indeed present at third order, and they are strongly indicative that no $\log^3$ divergences arise at this order, as is fully consistent with the analysis presented in Sec.~\ref{sec:recursionrels} and with the overall renormalizability of the graphene effective field theory at higher orders.

We can test whether we have obtained all of the $\log^2$ contributions at third order by making use of the second relation given in Eq.~(\ref{FrelsfromG2}): $F_{23}=-\tfrac{1}{2}F_{11}F_{12}$. Defining the quantities $\xi_{3t}$ and $\xi_{3x}$ in analogy with the second-order analysis of Sec.~\ref{sec:velrenorm}, it is straightforward to show that
\beq
F_{23}\alpha^3\log^2(\Lambda/|\vec{q}|)=\xi_{3x}-\xi_{3t}-\xi_1\xi_{2t},
\eeq
where $\xi_1$, $\xi_{2t}$, and $\xi_{2x}$ were defined in Eq.~(\ref{defofxi1xi2}). Recalling the values of $F_{11}$ and $F_{12}$ from Sec.~\ref{sec:velrenorm}, namely $F_{11}=\tfrac{1}{4}$ and $F_{12}=\tfrac{1}{2}\log2-\tfrac{2}{3}$, we find
\beq
F_{23}=\tfrac{1}{12}-\tfrac{1}{16}\log2,
\eeq
which implies
\beq
\xi_{3x}-\xi_{3t}=(\tfrac{1}{24}+\tfrac{1}{16}\log2)\alpha^3\log^2(\Lambda/|\vec{q}|).\label{correctxi3xminusxi3t}
\eeq
On the other hand, summing up the results shown in Eq.~(\ref{logsquaredterms}) for $N=2$, we obtain the following net difference between the spatial and temporal components
\beq
\xi_{3x}-\xi_{3t}=(\tfrac{1}{12}-\tfrac{1}{16}\log2)\alpha^3\log^2(\Lambda/|\vec{q}|).\label{xi3xminusxi3t}
\eeq
The fact that this last result differs from the correct answer, Eq.~(\ref{correctxi3xminusxi3t}), indicates that additional $\log^2$ terms must arise from some of the other diagrams in Fig.~\ref{fig:3loopdiagrams}, in particular from diagrams which do not contain divergent subdiagrams. These additional terms must be parametrically small given that Eqs.~(\ref{correctxi3xminusxi3t}) and (\ref{xi3xminusxi3t}) differ by only a few percent. It is most likely the case that these additional contributions arise from diagrams containing higher-order vertex corrections, namely diagrams (h) and (k) in Fig.~\ref{fig:3loopdiagrams}. We leave the explicit verification that additional $\log^2$ contributions exist to future work.

To give an example of a linear-logarithmic divergence, we consider the simplest such contribution, which is provided by diagram (p). This diagram can easily be calculated from the full RPA self-energy correction, as shown in Appendix~\ref{app:threeloop}. We quote the result here for convenience:
\beq
\Sigma_{3p}(q)=\frac{i\pi^2\alpha^3}{32}\left[-q_0\gamma^0+2v_F\vec{q}\cdot\vec\gamma\right]\log(\Lambda/|\vec{q}|).
\eeq
It is clear that this contribution is in no sense small when $\alpha\sim1$, suggesting that the third-order corrections to the electron self-energy and velocity are again likely to be comparable to the first-order results, as were the second-order corrections. It is possible that the additional contributions to the linear-log divergence coming from the remaining diagrams in Fig.~\ref{fig:3loopdiagrams} will lead to some cancelations, but in the absence of special symmetries and given the lack of such cancelations at second order, we believe that $\Sigma_{3p}$ is indicative of the magnitude of the full third-order correction. Therefore, the full third-order contribution to the velocity renormalization could significantly alter the value of the coupling at the quantum critical point, $\alpha_c$, we obtained from the two-loop analysis above.

\begin{figure}
\includegraphics[width=\columnwidth]{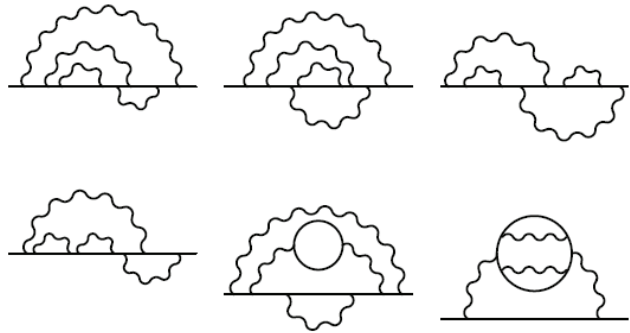}
\caption{Examples of diagrams contributing $\log^3$ divergent terms to the fourth-order electron self-energy.}\label{fig:4loopdiagrams}
\end{figure}
We conclude the perturbation theory analysis by making a few general comments regarding the structure of higher-order diagrams. First of all, it must be the case that at higher orders in perturbation theory, the divergence structure of the theory is such that at $n$th order, logarithmic divergences of the form $\log^m(\Lambda/|\vec{q}|)$ arise, where $m\le n-1$. This follows immediately from the recursion relation analysis developed in Sec.~\ref{sec:recursionrels} and is required by the renormalizability of the theory. Thus, at fourth order for example, the most divergent terms will scale as $\log^3(\Lambda/|\vec{q}|)$; some of the diagrams contributing to this leading divergence are shown in Fig.~\ref{fig:4loopdiagrams}. We further anticipate that the generalized rainbow or ``sunrise" diagrams, of which Figs.~\ref{fig:3loopdiagrams}a and \ref{fig:3loopdiagrams}g are examples, will continue to vanish at higher orders in perturbation theory. This class of diagrams consists of all graphs that do not contain fermion loops and are such that, if every Coulomb line is drawn above the main fermion line, then no two Coulomb lines cross. Of course, any diagram containing a sunrise diagram as a subdiagram will also vanish. Furthermore, Furry's theorem will continue to hold at higher orders, so that any diagram containing a fermion loop connected to an odd number of Coulomb lines will vanish as well.

\section{Comparison with experiments}\label{sec:experiment}

Although the coupling constant $\alpha$ in graphene being of order unity (e.g., $\alpha\approx0.4,0.8,2.2$ for graphene on BN, SiO${}_2$, vacuum respectively) in relevant experimental systems of interest indicates that perhaps a weak-coupling perturbative expansion in the coupling constant is not the appropriate theoretical tool to use for a quantitative understanding of the data, the fact is that many recent detailed experimental studies\cite{Elias_NP11,Yu_PNAS13,Siegel_PNAS11,Siegel_PRL13,Li_PRL09,Luican_PRB11,Chae_PRL12,Xia_NN09,Reed_Science10,Henriksen_PRL10,Bostwick_NP07,Bostwick_Science10,Li_NP08,Martin_NP09,Ju_NN11,Horng_PRB11,Wang_NP12,Nair_Science08,Mak_PRL08} have been carried out to study graphene many-body effects which are then (almost always) successfully compared with the leading-order perturbation theory. This presents a conundrum that, in spite of $\alpha$ being not particularly small, it appears that perturbative theoretic results are in good agreement with a wide variety of experimental results. In particular, there is absolutely no signature in any reported experimental data of a strong-coupling quantum phase transition or a gap opening near the Dirac point. Thus, the weak-coupling perturbative analysis appears to be at least in qualitative agreement with all existing experimental results on graphene, putting our current work in the proper context of graphene phenomenology.

In comparing theory and experiment, we first mention a key issue which is not always appreciated in the literature (particularly in experimental publications claiming agreement between theory and experiment). Many-body theories (or in general, any theory) restricted precisely to intrinsic graphene (i.e., undoped pristine graphene with the chemical potential at the Dirac point) are, by their very construction, about a hypothetical and idealized system which cannot exist in a laboratory. All real systems are extrinsic (i.e., doped) graphene with a finite chemical potential (and finite carrier density). Thus, experimental results should only be compared with theories for extrinsic graphene with a finite Fermi level. Of course, at finite temperatures, when $k_BT>E_F$, there is no essential difference between intrinsic and extrinsic graphene,\cite{DasSarma_PRB13b} but such a theory must incorporate qualitative finite temperature effects for it to be realistic.

\begin{figure}
\includegraphics[width=0.5\columnwidth]{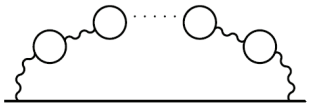}
\caption{The RPA self-energy is obtained by summing the infinite series of bubble diagrams, which have the structure depicted here. There is one such diagram in each order of the perturbative expansion, where the $n$th order diagram contains $n-1$ fermion loops (``bubbles"). The sum of all such diagrams (over all orders) is equivalent to a one-loop expansion in the dynamically screened Coulomb interaction. Details for the RPA self-energy calculation can be found in Refs.~[\onlinecite{DasSarma_PRB07},\onlinecite{Hwang_PRB07a},\onlinecite{DasSarma_PRB13}].}
\label{fig:RPAdiagram}
\end{figure}
Thus, the only existing theory enabling a direct quantitative comparison with experiment is RPA since the Hartree-Fock theory (i.e., the leading-order self-energy in the bare Coulomb interaction) has a pathological infrared divergence at the Fermi energy which is due to the long-range nature of the Coulomb interaction and which gets regularized by the screening effect in RPA (see Fig.~\ref{fig:RPAdiagram}), giving a meaningful result at the Fermi energy. This is easily seen by writing out the leading-order (in $\alpha$) results for the renormalized graphene velocity in various theories:
\bea
\frac{v_F^*(E)}{v_F}\!\!\!&{=}&\!\!\!1{+}\frac{\alpha}{4}\log(E_c/E),\label{vF1}\\
\frac{v_F^*(E_F)}{v_F}\!\!\!&{=}&\!\!\!1{-}\frac{\alpha}{\pi}\left(\frac{5}{3}{+}\log\alpha\right){+}\frac{\alpha}{4}\log(E_c/E_F),\label{vF2}\\
\frac{v_F^*(E)}{v_F}\!\!\!&{=}&\!\!\!1{+}\frac{\alpha}{4(1{+}\pi\alpha/2)}\log(E_c/E),\label{vF3}\\
\frac{v_F^*(E_F)}{v_F}\!\!\!&{=}&\!\!\!1{+}\alpha\left\{\frac{1}{4}\log(E_c/E_F){-}1{-}\frac{1}{4}\log\left(\frac{1{+}4\alpha}{4\alpha}\right)\right\}.\label{vF4}\nn\\&&
\eea
Equations (\ref{vF1})-(\ref{vF4}) above correspond respectively to the leading-order (in the bare interaction) single-loop result at the Dirac point,\cite{Gonzalez_NPB94} the analytical RPA result (obtained by summing the series of diagrams depicted in Fig.~\ref{fig:RPAdiagram} and expanding to $O(\alpha)$) at a finite Fermi energy $E_F$ (or at the finite Fermi momentum $k_F$),\cite{DasSarma_PRB07} the leading-order result in the statically screened Coulomb interaction at the Dirac point,\cite{Elias_NP11} and the RPA result obtained for a finite Fermi energy at $k=0$.\cite{DasSarma_PRB07} We emphasize that, whereas Eqs. (\ref{vF2}) and (\ref{vF4}) correspond to a realistic situation with a finite Fermi energy $E_F$, Eqs. (\ref{vF1}) and (\ref{vF3}) correspond manifestly to the Dirac point for the undoped intrinsic system where the concept of a Fermi energy does not apply. Therefore, any carrier density-dependent graphene Fermi velocity measurement, as, for example in Ref.~[\onlinecite{Elias_NP11}], where one simply substitutes $E_F\propto k_F\propto\sqrt{n}$ so that the $\log(E_c/E_F)$ term becomes $\frac{1}{2}\log(n_c/n)$, can {\it only} be described by Eqs. (\ref{vF2}) and (\ref{vF4}), and {\it not} by Eqs. (\ref{vF1}) and (\ref{vF3}). This makes complete sense because neither the concept of a Fermi energy, nor the concept of a carrier density applies to the Dirac point (i.e., Eqs. (\ref{vF1}), (\ref{vF3})), where the ultraviolet $\log(E_c/E)$ term only describes a flow where energy ({\it not} the density) is changing, but Eqs. (\ref{vF2}) and (\ref{vF4}) indeed describe the dependence of the renormalized Fermi velocity on the carrier density. This subtle point has not often been appreciated in the literature, and often pure Dirac point intrinsic theories which give $v_F^*(E)$ have been utilized in comparing theory with experiment simply by interpreting $v_F^*\equiv v_F^*(E_F)\equiv v_F^*(n)$, which has no a priori theoretical justification at the Dirac point where the concept of a changing carrier density or Fermi energy does not apply since $E_F\propto n\equiv0$ at the Dirac point! We emphasize, as has already been discussed in Ref.~[\onlinecite{Hwang_PRL07b}], the simple leading-order formula, Eq.~(\ref{vF1}), being equivalent to the Hartree-Fock self-energy expression, does not apply at the Fermi energy (i.e., one cannot just put $E=E_F$ in Eq.~(\ref{vF1})) because of the intrinsic infrared divergence of the Coulomb interaction.

We should also emphasize that RPA, being the infinite order sum of the bubble diagrams (and thus the leading-order expansion in the dynamically-screened effective interaction, see Fig.~\ref{fig:RPAdiagram}), represents the correct leading-order result in an expansion in the coupling constant $\alpha$, whereas the loop expansion in the bare interaction represents the leading-order result only in the most ultraviolet divergent term, as represented by the product, $\alpha \log (E_c/E)$, in Eq.~(\ref{vF1}). There are non-divergent corrections in $\alpha$ which are missed by the loop expansion, but which are correctly captured in RPA, as is evident in Eqs.~(\ref{vF2}) and (\ref{vF4}). Such non-divergent corrections are of course not important at (or close to) the Dirac point (i.e., for hypothetical intrinsic graphene), which is the (experimentally inaccessible) quantum-critical infrared fixed point in the problem, but away from the Dirac point (i.e., for any realistic experimental situations where the system is extrinsic, and has a finite chemical potential and finite doping), such non-divergent many-body corrections could very well be quantitatively important depending on the details of the situation (e.g., carrier density, the value of $\alpha$ itself, etc.).  In particular, in the very high-energy (or high-density) limit, the divergent contribution vanishes logarithmically, but the non-divergent RPA contributions still provide a many-body correction to the graphene velocity renormalization.

With the above comments in mind we note, however, that the leading ultraviolet divergence in all the theoretical formulas for the interacting Fermi velocity is, of course, exactly the same $\log(E_c/E)$ term, as necessarily follows from the fundamental renormalizability of the underlying graphene effective field theory. This indicates that perhaps purely on a heuristic level (and perhaps without any rigorous theoretical justification), one can replace $\log(E_c/E)$ by $\frac{1}{2}\log(n_c/n)$ in all of the above formulas, noting that $E_F\propto\sqrt{n}$. Such a heuristic procedure may not be unreasonable when the ultraviolet divergence dominates the velocity renormalization over all the subleading terms (for example, the other terms in the RPA expansion of Eq.~(\ref{vF2})). In comparing experiments with theories, this is the procedure we adopt below when we use pure intrinsic graphene Dirac point theories with the experimental data giving the velocity as a function of carrier density.

\begin{figure}
\begin{center}
\includegraphics[width=\columnwidth]{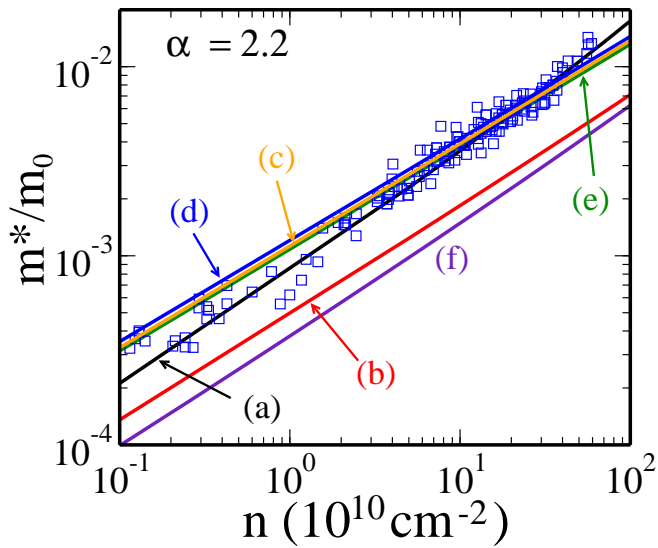} \\
\end{center}
\caption{Comparison of the experimental results (squares) from Ref.~[\onlinecite{Elias_NP11}] for the effective cyclotron mass versus carrier density for graphene suspended in vacuum ($\alpha=2.2$) with six different theoretical predictions: (a) The one-loop RPA result for extrinsic graphene given in Eq.~(\ref{vF2}) and first computed in Ref.~[\onlinecite{DasSarma_PRB07}]; (b) The one-loop (bare Coulomb interaction) result for intrinsic graphene given in Eq.~(\ref{vF1}), which was first calculated in Ref.~[\onlinecite{Gonzalez_NPB94}] and rederived in the current work in Sec.~\ref{sec:oneloopSE}; (c) The two-loop with geometric resummation result for intrinsic graphene first derived in the current work and given in Eq.~(\ref{vF5}); (d) The one-loop result with the statically screened Coulomb interaction for intrinsic graphene quoted in Eq.~(\ref{vF3}) and used in Ref.~[\onlinecite{Elias_NP11}]; (e) The full RPA off-shell solution obtained in Ref.~[\onlinecite{DasSarma_PRB13}] by solving Dyson's equation for the self-energy self-consistently; (f) The full RPA on-shell solution obtained in Ref.~[\onlinecite{DasSarma_PRB13}] by solving Dyson's equation for the self-energy (with the non-interacting energy inserted) numerically. In all cases, we have used that the effective cyclotron mass is related to the renormalized Fermi velocity via $m^*=\sqrt{\pi n}/v_F^*$ and have made the replacement $\log(E_c/E)$, $\log(E_c/E_F)\to\frac{1}{2}\log(n_c/n)$ as explained in the text, with $n_c=10^{15}\hbox{cm}^{-2}$. $m_0$ is the free electron mass, and the bare velocity is taken to be $v_F=10^8\hbox{cm/s}$.}\label{fig:exptheorycomp}
\end{figure}
In Fig. \ref{fig:exptheorycomp} we show a comparison between several theoretical results and the rather impressive experimental data of Ref.~[\onlinecite{Elias_NP11}] on the renormalized Fermi velocity as a function of carrier density for graphene suspended in vacuum. We choose this particular experiment because this experiment reports a large velocity renormalization covering a large (exceeding two orders of magnitude) density range, and thus this is the experiment which is most suitable for the possible observation of the ultraviolet divergence associated with the logarithmic term in the theory.

In Fig. \ref{fig:exptheorycomp} we compare the experimental data with six independent theoretical results. Three of them are given in Eqs. (\ref{vF1})-(\ref{vF3}) above, respectively, and the other three are: full RPA on-shell theory, full RPA off-shell theory, and our two-loop theory given in Sec.~\ref{sec:perttheory} of this paper. We leave out Eq.~(\ref{vF4}) above in the experimental comparison since this formula provides the RPA velocity renormalization in extrinsic graphene at $k=0$, where quasiparticles are ill-defined. The full RPA theory for extrinsic graphene has been described elsewhere by two of the authors,\cite{DasSarma_PRB07,Hwang_PRB07a,DasSarma_PRB13} where the word ``full" implies an exact numerical calculation of the RPA many-body self-energy without any expansion in $\alpha$ involved. The leading-order analytic expansion in $\alpha$ of the full RPA theory gives Eq.~(\ref{vF2}) for both on-shell and off-shell approximations, which respectively refer to using just the non-interacting energy in the Dyson equation for the self-energy (``on-shell") or the full numerical solution of the RPA self-energy with the Dyson equation solved self-consistently. We refer to Refs. [\onlinecite{DasSarma_PRB07}], [\onlinecite{Hwang_PRB07a}], [\onlinecite{DasSarma_PRB13}] for the full details of the extrinsic graphene RPA many-body theory. Finally, we mention that the two-loop results used in Fig. \ref{fig:exptheorycomp} for experimental comparison requires a geometric resummation of our direct intrinsic two-loop results given in Sec.~\ref{sec:perttheory} since the direct two-loop self-energy has a strong-coupling divergence for $\alpha>0.78$ (as discussed in Sec.~\ref{sec:perttheory}), and the experimental system under consideration here\cite{Elias_NP11} is graphene suspended in vacuum where $\alpha=2.2$ ($>0.78$). The relevant two-loop formula with a geometric series resummation gives:
\bea
&&\!\!\!\!\!\!v_F^*(E){=}v_F\left[1{+}\frac{\alpha}{4}\left\{1{+}\left(\frac{8}{3}{-}2\log2\right)\alpha\right\}^{{-}1}\log(E_c/E)\right].\nn\\\label{vF5}&&
\eea
It is easy to see that the leading-order expansion in $\alpha$ of the denominator on the right-hand-side of Eq.~(\ref{vF5}) leads to our two-loop velocity renormalization (Eq.~(\ref{velexp})) derived in Sec.~\ref{sec:perttheory}. Here, we have assumed a geometric resummation in order to avoid the unphysical negative velocity result (for $\alpha=2.2$) that the naive two-loop result would imply.

To compare theory and experiment, we do the heuristic replacement of $E\to\sqrt{n}$ for all the intrinsic theories (Eqs. (\ref{vF1}), (\ref{vF3}), (\ref{vF5})) and use the approximate $E_F\sim\sqrt{n}$ replacement in the extrinsic RPA theories. The most important features of the comparison between theories and the experimental data in Fig. \ref{fig:exptheorycomp} are the following:

(i) In general, the RPA results provide the best quantitative agreement with the experimental data,\cite{Geim_PC} as is expected since it is indeed the best available theoretical formalism for extrinsic graphene at finite carrier density.

(ii) Including two-loop corrections substantially improves the agreement between experiment and the intrinsic theory compared with the corresponding one-loop result although, given the number of approximations involved in the comparison (e.g., replacing energy by the square root of the density and the geometric resummation of the two-loop self-energy result), one cannot quite be sure that this improvement at the two-loop level is not a mere coincidence.

(iii) The on-shell numerical RPA approximation appears to be decisively in disagreement with the experimental data, indicating the necessity for the full self-consistent solution of the Dyson integral equation for extrinsic graphene.

(iv) In general, the logarithmic increase in the experimental Fermi velocity seems to be apparent in the data as predicted by all graphene theories by virtue of the underlying ultraviolet divergence in the Dirac-Weyl massless chiral effective field theories, but the current data cannot definitively confirm or rule out the presence of additional terms such as the subleading contributions which are present in the extrinsic RPA theories.

(v) More accurate quantitative data over a much broader range of carrier density would be necessary for the decisive observation of the log-divergent terms and our calculated higher-loop self-energy results presented in the current work, although the results shown in Fig. \ref{fig:exptheorycomp} indicate that the existing data of Ref.~[\onlinecite{Elias_NP11}] are indeed consistent with the higher-loop perturbative corrections discussed in the current work.

We make three final comments about our comparison between theory and the experiment of Ref.~[\onlinecite{Elias_NP11}]. First, the experiment of Ref.~[\onlinecite{Elias_NP11}] actually measures the low-field cyclotron effective mass, and not the graphene Fermi velocity, and we have uncritically assumed that the operational procedure used in Ref.~[\onlinecite{Elias_NP11}] to convert the cyclotron effective mass to an effective Fermi velocity by assuming the non-interacting relationship between cyclotron frequency and Fermi velocity remains valid (which is probably a reasonable assumption for extrinsic graphene since it is a Landau Fermi liquid at all carrier densities). Second, the experimental data (particularly in the context of its comparison with the various perturbation theoretic results shown in Fig.~\ref{fig:exptheorycomp}) show no indication of any strong-coupling behavior in terms of the opening of a gap or any other quantum phase transition which would be the hallmark of the strong-coupling behavior. This indicates that there is strong empirical evidence in the existing experimental data for monolayer graphene being a weak-coupling perturbative system. This is particularly true in the context of the fact that the data of Ref.~[\onlinecite{Elias_NP11}] shown in Fig. \ref{fig:exptheorycomp} not only represent the measured many-body corrections in graphene at the lowest carrier density (i.e., closest to the Dirac point), but also for the most strongly interacting graphene system (i.e., graphene in vacuum with $\kappa=1$) with $\alpha=2.2$ being the maximum allowed value of the bare coupling strength in graphene. The final comment is that one can substantially improve upon the comparison shown in Fig. \ref{fig:exptheorycomp} by eliminating the unknown (and somewhat arbitrary) cutoff density $n_c$ in favor of two distinct carrier densities, $n_1$ and $n_2$, at which the measured Fermi velocities, $v_F^*(n_1)$ and $v_F^*(n_2)$, can be compared with respect to the theoretical results. One can then vary $n_1$ or $n_2$ (or both) over the range of experimentally available densities to obtain detailed statistics about how various theories compare quantitatively with the measured data. Such a detailed statistical quantitative analysis of experimental data compared with the theory is clearly beyond the scope of the current theoretical work, but we have actually carried out some of this statistical analysis using the limited set of data points shown in Fig.~\ref{fig:exptheorycomp}, finding good agreement with the RPA off-shell theory, showing at least tentatively that in the current experimental situation, the logarithmic ultraviolet term is present, but the subleading non-divergent terms are quantitatively important.

\begin{figure}
\begin{center}
\includegraphics[width=\columnwidth]{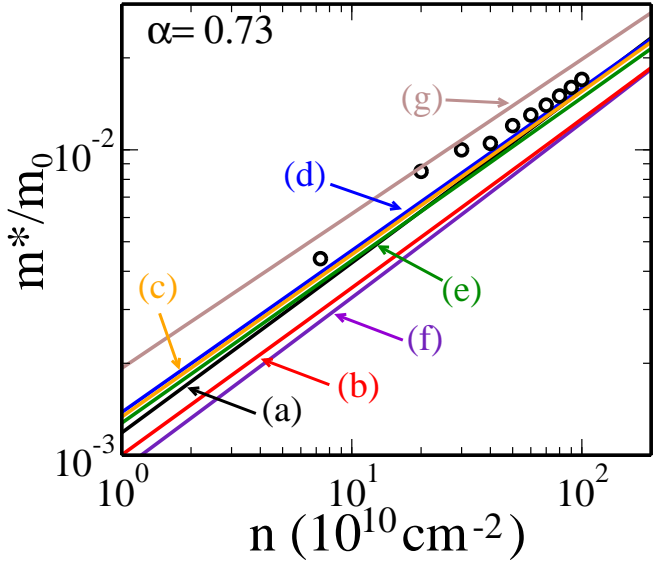} \\
\end{center}
\caption{Comparison of the experimental results (circles) from the Supplementary Information of Ref.~[\onlinecite{Elias_NP11}] for the effective cyclotron mass versus carrier density for graphene on a boron nitride substrate ($\alpha=0.73$) with seven different theoretical predictions: (a) The one-loop RPA result for extrinsic graphene given in Eq.~(\ref{vF2}) and first computed in Ref.~[\onlinecite{DasSarma_PRB07}]; (b) The one-loop (bare Coulomb interaction) result for intrinsic graphene given in Eq.~(\ref{vF1}), which was first calculated in Ref.~[\onlinecite{Gonzalez_NPB94}] and rederived in the current work in Sec.~\ref{sec:oneloopSE}; (c) The two-loop with geometric resummation result for intrinsic graphene first derived in the current work and given in Eq.~(\ref{vF5}); (d) The one-loop result with the statically screened Coulomb interaction for intrinsic graphene quoted in Eq.~(\ref{vF3}) and used in Ref.~[\onlinecite{Elias_NP11}]; (e) The full RPA off-shell solution obtained in Ref.~[\onlinecite{DasSarma_PRB13}] by solving Dyson's equation for the self-energy self-consistently; (f) The full RPA on-shell solution obtained in Ref.~[\onlinecite{DasSarma_PRB13}] by solving Dyson's equation for the self-energy (with the non-interacting energy inserted) numerically; (g) The two-loop bare Coulomb interaction result first derived in the current work and given in Eq.~(\ref{velexp}). In all cases, we have used that the effective cyclotron mass is related to the renormalized Fermi velocity via $m^*=\sqrt{\pi n}/v_F^*$ and have made the replacement $\log(E_c/E)$, $\log(E_c/E_F)\to\frac{1}{2}\log(n_c/n)$ as explained in the text, with $n_c=10^{15}\hbox{cm}^{-2}$. $m_0$ is the free electron mass, and the bare velocity is taken to be $v_F=10^8\hbox{cm/s}$.}\label{fig:exptheorycompBN}
\end{figure}
Similar conclusions can also be drawn in the context of graphene on a BN substrate, for which experimental data is available in the Supplemental Information of Ref.~[\onlinecite{Elias_NP11}]. A comparison of this data with the same six theory curves as in Fig.~\ref{fig:exptheorycomp} is shown in Fig.~\ref{fig:exptheorycompBN}, where it is again apparent that the one-loop bare Coulomb interaction result, Eq.~(\ref{vF1}), and the full RPA on-shell solution do not agree as well with the data as do the other four theory curves. Since $\alpha<\alpha_c=0.78$ in this case, we have also included in Fig.~\ref{fig:exptheorycompBN} our two-loop result for the bare Coulomb interaction, Eq.~(\ref{velexp}), which lies above the experimental data points as well as all the other theory curves. This is due to the fact that in this case $\alpha=0.73$ is sufficiently close to the critical value $\alpha_c=0.78$ that the suppression of renormalization effects that occurs at $\alpha_c$ is also manifest (albeit to a weaker degree) for $\alpha=0.73$, leading to a smaller reduction in the cyclotron mass. We further analyze the effect of proximity to the critical point below and discuss how, like in the case of $\alpha>\alpha_c$, a perturbative expansion in the bare Coulomb interaction appears inadequate even for $\alpha\lesssim\alpha_c$.

\begin{figure}
\includegraphics[width=\columnwidth]{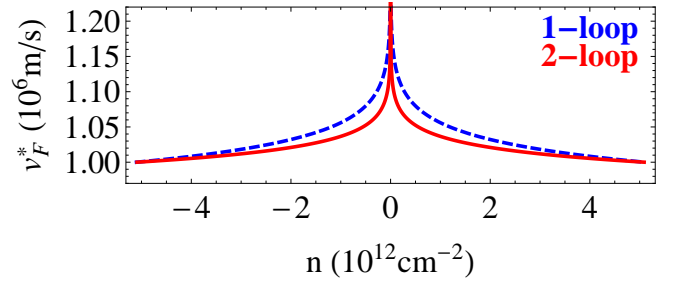}
\caption{One-loop (blue, dashed) and two-loop (red, solid) results for the renormalized Fermi velocity versus carrier density taken from Eq.~(\ref{twoloopvq}) (for the one-loop result we set $f_2=0$ in that equation), where we have made the replacements $|\vec{k}|\to n_k$, $|\vec{q}|\to n$, and where we have taken the dielectric constant to be $\kappa=8$. The figure is to be compared with Fig.~2B of Ref.~[\onlinecite{Yu_PNAS13}], which shows Fermi velocity measurements for graphene on an h-BN substrate. The RG flow reference point specified by $n_k=5\times10^{12}\hbox{cm}^{-2}$ and $v_F^*(n_k)=v_k=10^6\hbox{m/s}$ is taken from the right-most data point in Fig.~2B of Ref.~[\onlinecite{Yu_PNAS13}]. Note that there are no fitting parameters in this comparison.}\label{fig:vF1and2loop}
\end{figure}
A smaller yet still unmistakeable renormalization of the Fermi velocity was observed in quantum capacitance measurements of graphene on an h-BN substrate, as reported in Ref.~[\onlinecite{Yu_PNAS13}]. In this case, the dielectric constant is quite large ($\kappa=8$), leading to an effective coupling strength of $\alpha\approx0.28$, almost an order of magnitude smaller than in the case of graphene suspended in vacuum. Since this value is well below the critical value $\alpha_c=0.78$, no critical point arises in the RG flow, and the system remains weakly interacting as the Dirac point is approached by reducing the carrier density. Thus, no qualitative change in the Fermi velocity renormalization arises from the inclusion of the perturbative second-order correction, so that the second-order interaction effects are purely quantitative. This is illustrated in Fig.~\ref{fig:vF1and2loop}, where we plot both the one-loop and our two-loop perturbation theory results for the renormalized velocity as a function of carrier density (following the same procedure of replacing intrinsic graphene momenta with densities as explained above); it is apparent from the figure that the second-order correction results in a narrowing of the cusp near the Dirac point, but otherwise maintains the same basic behavior of the velocity relative to the one-loop case. Fig.~\ref{fig:vF1and2loop} should be compared with Fig.~2B of Ref.~[\onlinecite{Yu_PNAS13}], from which it is clear that both the one-loop and two-loop perturbation theory results are in reasonably close agreement with the experimental data. It is important to stress that, unlike in Ref.~[\onlinecite{Yu_PNAS13}], we are comparing experiment and theory without any fitting parameters. Whereas in that reference (and elsewhere in the literature) the ultraviolet cutoff $\Lambda$ (equivalently $n_c$) was used as a fitting parameter, our result for the renormalized Fermi velocity, Eq.~(\ref{twoloopvq}),
does not depend on $\Lambda$, as should be the case for a properly renormalized physical observable. The renormalized velocity at density $n$ instead depends on the value $v_k$ of the velocity at a separate, reference density $n_k$, where for Fig.~\ref{fig:vF1and2loop}, $v_k=10^6\hbox{m/s}$ and $n_k=5\times10^{12}\hbox{cm}^{-2}$ were taken from the data shown in Fig.~2B of Ref.~[\onlinecite{Yu_PNAS13}].

\begin{figure}
\includegraphics[width=\columnwidth]{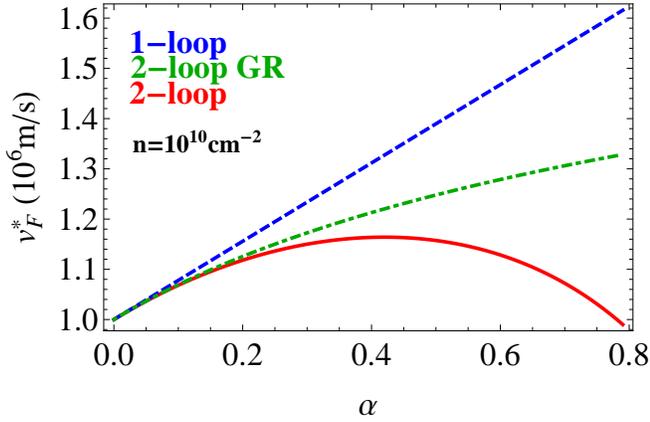}
\caption{One-loop (blue, dashed), two-loop (red, solid), and two-loop with geometric resummation (green, dot-dashed) results for the renormalized Fermi velocity versus bare fine structure constant $\alpha$ taken from Eq.~(\ref{twoloopvq}) (for the one-loop result we set $f_2=0$ throughout that equation, while for the two-loop with geometric resummation, we set $f_2=0$ in the argument of the logarithm in that equation), where we have made the replacements $|\vec{k}|\to n_k$, $|\vec{q}|\to n$, and where we have taken the density to be $n=10^{10}\hbox{cm}^{-2}$. The RG flow reference point was taken to be $n_k=5\times10^{12}\hbox{cm}^{-2}$ and $v_F^*(n_k)=v_k=10^6\hbox{m/s}$. It is clear from the figure that the two-loop velocity renormalization is strongly suppressed relative to the one-loop result for $\alpha\gtrsim0.4$, corresponding to substrates with dielectric constant $\kappa\lesssim5.5$.}\label{fig:vF1and2loopvsalpha}
\end{figure}
\begin{figure}
\includegraphics[width=\columnwidth]{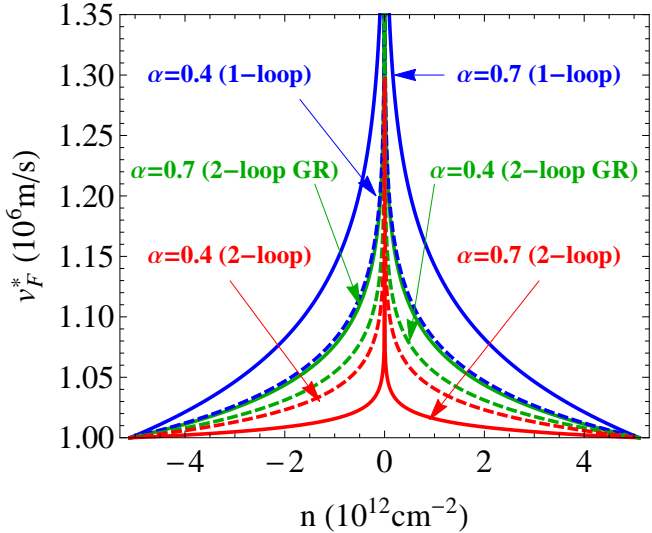}
\caption{One-loop and two-loop results for the renormalized Fermi velocity versus carrier density taken from Eq.~(\ref{twoloopvq}) (for the one-loop result we set $f_2=0$ throughout that equation, while for the two-loop with geometric resummation, we set $f_2=0$ in the argument of the logarithm in that equation), where we have made the replacements $|\vec{k}|\to n_k$, $|\vec{q}|\to n$. The RG flow reference point was taken to be $n_k=5\times10^{12}\hbox{cm}^{-2}$ and $v_F^*(n_k)=v_k=10^6\hbox{m/s}$. Results are shown for two different values of the bare fine structure constant: $\alpha=0.4$ and $\alpha=0.7$, corresponding to substrate dielectric constants of $\kappa=5.5$ and $\kappa=3.14$, respectively.}\label{fig:vF1and2loopVariousAlpha}
\end{figure}
The fact that both the one-loop and two-loop results in Fig.~\ref{fig:vF1and2loop} agree reasonably well with the experimental data in Ref.~[\onlinecite{Yu_PNAS13}] highlights the need for more quantitatively accurate data in order for a definitive conclusion to be reached regarding whether the higher-order many-body effects we predict are present in the experiment. More broadly, in order to truly ascertain the validity of higher-order perturbative graphene effective field theory, improved experimental accuracy would be needed not only for graphene on h-BN, but for graphene on a range of different substrates for which the effective coupling satisfies $\alpha<\alpha_c$, i.e., substrates with moderate to large dielectric constants. As illustrated in Figs.~\ref{fig:vF1and2loopvsalpha} and \ref{fig:vF1and2loopVariousAlpha}, the experimental visibility of higher-order effects should improve at more moderate values of the dielectric constant for which $\alpha\lesssim\alpha_c$ since the presence of the critical point at $\alpha_c$ leads to a suppression of the Fermi velocity relative to the leading-order, one-loop result so long as $\alpha$ is not too small. It is clear from these figures that even if the precise value of the substrate dielectric constant is not known, the observation of a 40-50\% enhancement of the Fermi velocity can automatically rule out the presence of higher-order corrections. This endeavor of searching for higher-order effects in graphene would be analogous to the on-going and increasingly accurate g-2 measurements carried out for QED and compared with very high-order perturbative calculations, a sixty-year enterprise that has established QED as the most precise theory in all of physics. However, unlike in QED, the asymptotic limit of graphene perturbation theory may already be reached at second order when $\alpha\lesssim\alpha_c$, meaning that comparing such measurements with our results for $\alpha<\alpha_c$ is critical for a fundamental understanding of graphene effective field theory. Also, it must be kept in mind that, unlike in QED, the finite carrier density of real graphene systems plays an important role, making it essential to perform the measurements as a function of density in order to eliminate the chemical potential effect by doing the RG connection between different carrier densities as we propose in this paper.

More accurate measurements would also be needed for a quantitative test of the second-order vacuum polarization results found in Sec.~\ref{sec:perttheory}. In particular, the vacuum polarization results lead to the following expressions for the effective dielectric constant and optical conductivity of graphene electrons: $\epsilon\approx1+\tfrac{\pi}{2}\alpha+0.78\alpha^2$ and $\sigma\approx e^2(1+0.01\alpha)/4$, respectively.\cite{Sodemann_PRB12} If we ignore the $O(\alpha^2)$ term in the expression for $\epsilon$, then we obtain $\epsilon\approx4.6(2.4)$ for $\alpha\approx2.2(0.9)$, whereas with the second-order correction, these numbers are substantially altered to the values $\epsilon\approx8.2(3.0)$. In Ref.~[\onlinecite{Reed_Science10}], a dielectric constant of $\epsilon=15.4_{-6.4}^{+39.6}$ was reported for graphene in vacuum ($\alpha\approx2.2$), while in Ref.~[\onlinecite{Wang_NP12}], a value of $\epsilon=3.0\pm1.0$ was measured for graphene on a BN substrate ($\alpha\approx0.9$). If we compare the theoretical results to both experimental results for $\epsilon$, then it is evident that the second-order perturbative corrections lead to an improvement in the agreement between theory and experiment, although further experiments and greater measurement accuracy would be needed to demonstrate that this apparent improvement is not a coincidence. Experiments have also found reasonable agreement with the universal, leading-order value of the optical conductivity, $\sigma=e^2/4$,\cite{Li_NP08,Nair_Science08,Mak_PRL08} but again greater accuracy is needed to confirm the presence of the small $O(\alpha)$ correction stemming from the second-order vacuum polarization.

We conclude our discussion of the experimental status of many-body effects with respect to graphene effective field theory by mentioning that a recent independent analysis by Geim\cite{Geim_PC} comparing the experimental data of Elias et al.\cite{Elias_NP11} on suspended graphene with the RPA theory of Refs.~[\onlinecite{DasSarma_PRB07},\onlinecite{DasSarma_PRB13}] comes to the same conclusion as ours that RPA indeed provides a better quantitative fit to the experimentally observed density-dependent velocity renormalization.  The same conclusion was also reached by Chae et al.,\cite{Chae_PRL12} who carried out a quantitative comparison between the RPA results for the density-dependent velocity renormalization\cite{DasSarma_PRB07,DasSarma_PRB13} with scanning tunneling spectroscopic data for doped graphene on BN substrates, obtaining impressive quantitative agreement between theory and experiment although the actual quantitative velocity renormalization in this experiment was rather small (around 20\% compared with around 200\% in Ref.~[\onlinecite{Elias_NP11}]).  A recent series of experiments by Siegel et al.\cite{Siegel_PNAS11,Siegel_PRL13} attempted to compare various theoretical results directly with ARPES data on the graphene velocity renormalization by changing the substrate dielectric constants.  Although not completely decisive, these experiments also conclude that RPA\cite{DasSarma_PRB07,Hwang_PRB08a,DasSarma_PRB13} seems to provide the best agreement with the experimental data regarding the substrate dependence of the graphene velocity renormalization.

\section{Discussion}\label{sec:discussion}

\subsection{General remarks}

We have in the current work investigated four distinct questions regarding quantum many-body corrections to the electronic properties of graphene:

I. How important are the higher-loop perturbative many-body corrections to the graphene self-energy and polarizability?

II. Is perturbative graphene effective field theory valid at these higher orders?

III. Does the divergence structure of the effective field theory remain consistent with renormalizability at higher orders?

IV. How do theory and experiment compare in graphene many-body electronic properties?

To answer questions I and II above, we carry out a complete two-loop analysis of the intrinsic graphene self-energy and polarizability diagrams by explicitly calculating all the relevant diagrams up to $O(\alpha^2)$ perturbatively. Although the fact that $\alpha$ is of order unity for real experiments immediately leads to the expectation that the higher-order corrections will be comparable in magnitude to the leading one-loop results, it is necessary to check this explicitly since the coefficients of higher-order terms could potentially be parametrically small. In the case of the two-loop Fermi velocity, this turns out not to be the case; we find in Eq.~(\ref{velexp}) that the numerical coefficient of the second-order correction is not small: $\frac{1}{2}\log2-\frac{2}{3}\approx-0.32$. Even in the case of graphene on a BN substrate where $\alpha\approx0.4$ is relatively small, the second-order contribution amounts to a 50\% correction to the first-order result, illustrating the fact that higher-order corrections are significant. Similar conclusions were also reached in the context of the dielectric function, as we discussed in the previous section. Furthermore, we find that for larger values of $\alpha$, specifically for $\alpha>\alpha_c\approx0.78$, the second-order correction reverses the RG flow of the Fermi velocity and running coupling such that the theory is now asymptotically free and flows to a new strong-coupling fixed point in the infrared, where the Fermi velocity is renormalized to zero and the renormalized coupling runs to infinity. Whether this strong-coupling fixed point is physical or a signature of the breakdown of the perturbation theory is unknown except that experimentally the weak-coupling physics seems to hold for all values of $\alpha$ ($\lesssim2.2$), including suspended graphene ($\alpha=2.2$). Our best estimate, based on an extension of Dyson's original argument for the eventual breakdown of QED perturbation theory,\cite{Dyson_PR52} is that the asymptotic expansion most likely starts diverging at very low orders, perhaps already beyond $O(\alpha)$, which may explain why theory and experiment agree in the first-order calculation, but the agreement is not quantitatively accurate. The issue of the asymptotic nature of weak-coupling graphene perturbation theory requires further analysis in future works.

To address question III, we also calculate representative third-order ($O(\alpha^3)$) corrections to some of the three-loop graphene self-energy diagrams. In these three-loop corrections, we find the expected $\log^2(E_c/E)$ ultraviolet divergence in the self-energy, whereas the only divergence in the first two loops, i.e., up to $O(\alpha^2)$, is the $\log(E_c/E)$ divergence, which is consistent with the standard theorem on the renormalization scheme independence of the lowest-order terms in the beta function for the effective coupling. We derive a set of recursion relations which relate the coefficients of the third-order $\log^2(E_c/E)$ terms to those of the first- and second-order $\log(E_c/E)$ divergences. We find that our explicit $O(\alpha^3)$ results for the electron self-energy are fully consistent with these recursion relations and with the higher-order renormalizability of the effective field theory.

Although we have not discussed the role of the Landau pole\cite{Landau_55} thus far, it may be worth commenting on it briefly at this point in relation to our analysis of the self-consistency of graphene perturbation theory. It is well known that when a renormalization group analysis is applied to generic quantum field theories which are not asymptotically free, there can exist a finite renormalization scale, the Landau scale, at which the effective coupling diverges, signifying that the theory is ill defined at energy scales above the Landau scale. In QED, the Landau scale is estimated to be astronomically large and thus merely a matter of theoretical interest. However, since the effective fine structure constant of graphene is much larger than that of QED, it is possible that the graphene effective field theory exhibits a Landau pole at experimentally relevant energy scales. To calculate the Landau scale, one generally needs full knowledge of the beta function for the effective coupling to all orders in perturbation theory, especially when the coupling is not small as in the case of graphene. We can at least give a rough estimate using first-order perturbation theory, from which it is straightforward to show that the Landau scale is $\Lambda_L\sim\hbar v_F\sqrt{n}e^{4/\alpha}$. For typical values of the Fermi velocity and electron density ($v_F\sim10^6\hbox{m/s}$, $n\sim10^{12}\hbox{cm}^{-2}$), $\Lambda_L$ is already comparable to the inverse lattice constant for graphene in vacuum ($\alpha=2.2$), while for graphene on a BN substrate with $\alpha=0.4$, $\Lambda_L$ is several orders of magnitude larger than the scale of the lattice. Since we already know that the theory does not apply at such high energy scales, we do not expect the Landau pole to hold any practical importance for the graphene effective field theory. Note that here we are assuming that the critical point $\alpha_c=0.78$ does not arise in a full, non-perturbative treatment of the theory; if the critical point is a real, physical feature of the theory, then no Landau pole arises at all.

Finally, with respect to item IV above, we have analyzed a recent experiment\cite{Elias_NP11} in depth using various graphene many-body theories, finding that the data are consistent with the existence of the ultraviolet logarithmic divergence of the Fermi velocity, but not completely quantitatively decisive yet. We find that the non-singular subleading many-body corrections to the self-energy are quantitatively significant, and the best theory for the comparison with experiment is the RPA theory for extrinsic graphene, where both the ultraviolet and the infrared divergence are properly accounted for by using RG and dynamical screening, respectively. The agreement between RPA and experimental data is quite encouraging,\cite{Elias_NP11,Chae_PRL12,Siegel_PNAS11} although more accurate measurements of the Fermi velocity over a larger density range than is available at the current time would be necessary for a definitive conclusion on the agreement between theory and experiment.

Given our conclusion that RPA, i.e., the leading-order theory in the dynamically screened Coulomb interaction (or equivalently, the sum of the infinite geometric series of bubble diagrams), is the appropriate quantitative theory for comparing theory and experiment in extrinsic graphene with finite carrier density (and by definition, all experiments are carried out in the extrinsic graphene with undoped intrinsic graphene being a purely theoretical abstraction), the natural question is whether one can go to higher orders in the dynamically screened Coulomb interaction in calculating the graphene self-energy in a systematic manner. This is a formidable task which has not been carried out in any system ever, but is worth considering in the context of graphene perturbation theory given the large value of the interaction strength ($\alpha\sim1$). Of course, the equations for the velocity renormalization suggest that the real perturbation parameter may be $\alpha/4$, where $4$ is the ground state spin-valley degeneracy factor, but even $\alpha/4$ is not necessarily small in graphene.

An interesting question in this context is whether the full RPA confirms the strong-coupling $1/N$ expansion approach,\cite{Son_PRB07} which yields a nontrivial critical point in the infinite coupling limit with a dynamical exponent $z<1$ for graphene, i.e., the graphene dispersion changes from $E=v_Fk$ to $E\sim k^z$, with $z<1$. Of course, the systematic weak-coupling perturbation theory maintains $z=1$ to all orders, with only an interaction-induced renormalization of the Fermi velocity, $v_F$. We have carried out a full and exhaustive numerical calculation with RPA to calculate the graphene quasiparticle dispersion, $E^*(k)$, finding that within the numerical error bars, $E^*(k_F)=v_F^*k_F$, i.e., $z=1$ always for all values of the finite carrier density. Thus, the nontrivial dynamical exponent $z<1$ is an artifact purely of the $1/N$ expansion around the infinite-coupling fixed point, and is not experimentally germane in spite of its theoretical curiosity.

The interaction-induced logarithmic renormalization exhibited by the Fermi velocity should also arise in other physical quantities, as is generally the case for a renormalizable field theory. To demonstrate that additional observables renormalize via the same logarithmic divergence, we explicitly calculate the renormalization of the spin susceptibility. Given the importance of RPA as a theoretical tool in understanding the electronic properties of real graphene, we give below a new theoretical RPA result for the interacting spin susceptibility in graphene, whose analytic RPA form has not been theoretically calculated before in this literature. Future experiments should be able to test our predictions for the RPA spin susceptibility result given below.

\subsection{Spin susceptibility}

In a normal Fermi liquid, the spin susceptibility is given by $\chi^* =n
d\xi/dB = g^*\mu_B N_F/2$, where $g^*$ is the effective g-factor,
$N_F$ is the density of states (DOS) at
the Fermi level, and $\xi=(n_{\uparrow}-n_{\downarrow})/n$ is the
spin polarization parameter. In graphene, the DOS is given by $N_F =
g_v g_s k_F/(2\pi \hbar v_F^*)$, therefore
$\chi^* \propto g^*/v_F^*$.
We express it as a relative spin susceptibility
\begin{equation}
\frac{\chi^*}{\chi_0}  =  \frac{g^*}{g} \frac{v_F}{v_F^*} ,
\end{equation}
where $\chi_0$ is the spin susceptibility of the noninteracting system.

In the presence of the Coulomb interaction, the quasiparticle energy of
graphene can be found by solving Dyson's equation
\begin{equation}
E(k) = \xi(k) + \Sigma[k,E(k)],
\end{equation}
where $\xi(k) = \hbar v_F k - \mu$ is the noninteracting energy
relative to the chemical
potential, and $\Sigma(k,\omega)$ is the self-energy.

In the presence of a weak magnetic field $B$, the quasiparticle
energy for two spin states can be written as
\begin{equation}
E_{\uparrow}(k)= \xi(k) + \frac{1}{2}g \mu_B B + \Sigma_{\uparrow}[k,E_{\uparrow}(k)],
\label{eup}
\end{equation}
\begin{equation}
E_{\downarrow}(k)= \xi(k) - \frac{1}{2}g \mu_B B + \Sigma_{\downarrow}[k,E_{\downarrow}(k)],
\label{edown}
\end{equation}
where $g$ is the free electron $g$-factor and $\mu_B$ the Bohr magneton.
Then the effective $g$-factor, $g^*$, can be found to be
\begin{eqnarray}
\!\!\!\!\!\!\!\!g^*\mu_B B & \equiv &  \eup -\edo \nonumber \\
\!\!\!\!\!\!\!\!           &  =     &  g \mu_B B + \Sigma_{\uparrow}[k,\eup] -
           \Sigma_{\uparrow}[k,\edo].
\label{eg}
\end{eqnarray}
The RPA self-energy correction due to the Coulomb interaction $V$ is given by\cite{DasSarma_PRB07}
\begin{eqnarray}
\Sigma_s[k,\omega] = - \sum_{s'} \int \frac{d^2k'}{(2\pi)^2} \int
\frac{d\nu}{2\pi i}
\frac{V_{k-k'}}{\epsilon(k-k',\nu)} \nonumber \\
F_{ss'}(k,k') G(q-k',\nu+\omega),
\label{self}
\end{eqnarray}
where $s,s'=\pm1$ are the band indices, $n_F(E)$ is the Fermi
function, and the chiral term $F_{ss'}(k,k')=(1+ss'\cos
\theta_{kk'})/2$  arises from the wave function overlap factor, where
$\theta_{kk'}$ is the angle between k and $k'$.
With Eqs. (\ref{eup})--(\ref{self}) we have the effective $g$-factor, $g^*$,
\begin{equation}
\frac{g}{g^*} = 1-\frac{k_F}{2\pi v_F^*} \int \frac{d\phi}{2\pi}
\frac{V_{k-k'}} {\epsilon(k-k')}F_{++}(k,k'),
\label{gstar}
\end{equation}
where $v_F^*$ is the renormalized quasiparticle velocity due to many-body interactions, and
within RPA, we have the
normalized quasiparticle velocity\cite{DasSarma_PRB07}
\begin{equation}
\frac{v_F^*}{v_F} = 1 + \frac{\alpha}{4} \log \frac{k_c}{k_F} -
\frac{\alpha}{\pi} \left [ \frac{5}{3} + \log \alpha \right ],
\label{vf}
\end{equation}
where $k_c \sim 1/a$ is the ultraviolet momentum cutoff, and the second (third)
term comes from exchange (correlation)
self-energy corrections.

The spin susceptibility is now given by
\begin{eqnarray}
\frac{\chi_0}{\chi^*} & = & \frac{g}{g^*}\frac{v_F^*}{v_F} \nonumber \\
& = & \frac{v_F^*}{v_F} - \frac{k_F}{2\pi v_F} \int \frac{d\phi}{2\pi}
\frac{V_{k-k'}} {\epsilon(k-k')}F_{++}(k,k')|_{k,k'=k_F}.\label{spinsusc9}\nn\\&&
\end{eqnarray}
Note that this equation can be compared with the corresponding non-chiral parabolic 2D version, Eq.~(10) in
Ref.~[\onlinecite{Zhang_PRB05}].

With Eq.~(\ref{spinsusc9}), we finally find the spin susceptibility up to $O(\alpha\log \alpha)$ to be
\begin{equation}
\frac{\chi}{\chi^*} = 1-\frac{\alpha}{\pi} \left [ \frac{5}{3} +
  \frac{\pi}{8} + \frac{3}{4} \log \alpha \right ] + \frac{\alpha}{4} \log
\frac{E_c}{E_F}.\label{spinsusc10}
\end{equation}

Eq.~(\ref{spinsusc10}) shows that the leading-order RPA susceptibility has exactly the same ultraviolet divergence
given by the $\log(E_c/E_F)$ term, and is thus renormalized by the velocity renormalization (or equivalently, by
the running coupling constant). We note, however, that the non-singular subleading terms could be quantitatively
significant unless the carrier density is very small.

\subsection{Zero-range electron-electron interactions}

Finally, we briefly discuss the structure of ultraviolet divergences for a hypothetical theory in which the Coulomb interaction is replaced by a zero-range electron-electron contact interaction. This theory should be relevant for experimental studies of artificial graphene realized in cold atomic gases confined in optical lattices. Recently, much progress has been made in constructing optical honeycomb lattices in the laboratory,\cite{SoltanPanahi_NP11,SoltanPanahi_NP12} and in controllably creating and moving Dirac cones with cold Fermi gases.\cite{Tarruell_Nature12} Although one might initially expect that a theory with contact interactions would be comparatively simpler in the absence of the long-range Coulomb force, it turns out that this theory is rather subtle and exhibits a significantly more complicated divergence structure. A detailed description of this theory, along with results for one-loop, two-loop, and certain $n$-loop corrections to the electron self-energy are presented in Appendix~\ref{app:zerorange}. In particular, we show that power-law divergences arise at every order of perturbation theory, and that the $n$th-order (RPA-type) ring diagram diverges with momentum $k$ like $k^{n+1}$, indicating that if the theory can be renormalized, the procedure would be quite different from the Coulomb-interaction case, and would not simply involve the renormalization of the Fermi velocity. Thus, it seems that the contact-interaction theory does not provide insight into the Coulomb-interaction problem, and moreover it may be unphysical.

\section{Conclusions}\label{sec:conclusions}

In this very long article, we have theoretically studied graphene many-body effects from a number of different (but closely related) perspectives.

Graphene, having an interaction coupling constant of order unity, can be construed to be a 2+1-dimensional strong-coupling version of massless chiral QED. Unfortunately, $\alpha\approx0.4-2.2$ in graphene is really an intermediate coupling situation where purely strong-coupling theories, which assume infinite interaction strengths, are inappropriate. We have investigated the applicability of the weak-coupling perturbation theory at the Dirac critical point (undoped with the Fermi level at the Dirac point and with zero carrier density) by going to three loops, i.e., $O(\alpha^3)$, with a complete analytical calculation up to all $O(\alpha^2)$ second-order terms. Our second-order results indicate that the perturbation theoretic weak-coupling series is asymptotic and well behaved for $\alpha<0.78$, while a strong-coupling critical point appears for $\alpha>0.78$, questioning the validity of a perturbative approach for more strongly-coupled systems such as graphene in vacuum. In the three-loop order, we explicitly showed the emergence of higher-order, $O(\log^2)$, ultraviolet singular terms, as we anticipated from simple RG arguments that rely on the renormalizability of the theory. However, the serious issue of the asymptotic nature of the perturbative expansion remains open; our best estimate indicates that the perturbative expansion may start diverging after only the first few terms.

We have also compared experimental results with various graphene many-body theories, finding that RPA, i.e., the infinite sum of bubble diagrams, provides the best available quantitative description of the experimental data since RPA regularizes both the short-wavelength ultraviolet divergence and the long-wavelength infrared divergence appropriately. Going beyond RPA in a systematic manner remains a great open challenge for future graphene theories. But, this intermediate-coupling nature of graphene (i.e. $\alpha\sim1$) is of course quite common in solid state electronic materials where, for example, in metals and semiconductors the dimensionless electron-electron interaction parameter $r_s$ (which for graphene is precisely the fine structure constant $\alpha$ as already mentioned in the Introduction) is invariably larger than unity, with $r_s\sim5-6$ for simple metals (Na, K, Li, etc.)\cite{Mahan} and $r_s>10$ for 2d semiconductor systems such as Si inversion layers and GaAs quantum wells at low carrier densities.\cite{Ando_RMP82} In general, RPA-based many-body theories\cite{Rice_AP65,Chaplik_SPJETP71,Ting_PRL75,DasSarma_PRB79,Giuliani_PRB82,Jalabert_PRB89,Hu_PRB93a,Hu_PRB93b,Zheng_PRB96,Zhang_PRB04} work quite well for such strong-coupling metallic systems, perhaps because the Fermi liquid ground state is stable under RG flow, and the RPA-expansion in terms of the dynamically screened Coulomb interaction turns out to be a reasonable technique, often referred to as the `GW' technique (where `W' is the dynamically-screened Coulomb interaction, and `G' is the electron Green's function).  Graphene is, however, qualitatively different from interacting 2d Fermi liquids because of the ultraviolet divergence (and the existence of the Dirac point as a critical point which has no analog in ordinary Fermi liquids), which could lead to a strong-coupling fixed point associated with chiral symmetry breaking.

The graphene interacting many-body problem is conceptually and technically difficult precisely because it suffers from the difficulties of both the ordinary interacting electron liquids (i.e., an infrared Coulomb divergence which must be regularized away from the Dirac point by using RPA) and the ultraviolet divergence of strong-coupling chiral QED at the Dirac point (i.e. the possibility of a flow toward a strong-coupling fixed point with a chiral-symmetry-breaking gap opening at the Dirac point).  As we have emphasized, the effective field theory for graphene is perfectly well defined and is renormalizable (and the strong-coupling problem is asymptotically free), but there is no effective theoretical technique available for dealing with the intermediate-coupling situation (as graphene is, since $\alpha\sim0.4-2.2$ is neither too large nor too small, and since the number of fermion flavors is $N=2$, which is again not large, thus making $1/N$ expansion-type RG theories ineffective).  Such intermediate-coupling problems are notoriously difficult to tackle in theoretical physics, and one must use experiments as the guide.

Our approach, based on a weak-coupling perturbative expansion of graphene effective field theory, is rooted entirely on the empirical evidence that the leading-order one-loop perturbation theory seems to work very well in graphene, and there is no experimental signature anywhere of any strong-coupling behavior (either a spontaneous gap opening at the Dirac point or the appearance of a dynamical exponent different from unity).  Given the empirical success of the weak-coupling theory in the leading order and the fact that the actual coupling is not very weak, we believe that a higher-order perturbative calculation, as we have done in this work, is absolutely necessary to establish the domain of validity of the weak-coupling theory. Our finding that, for $\alpha<0.78$, the weak-coupling theory gives moderate corrections to graphene properties, bringing theory and experiment slightly closer together for the measured graphene velocity and dielectric function, provides some confidence and justification for the applicability of the weak-coupling theory to study graphene many-body effects.  More experimental data using substrates with large dielectric constants so that $\alpha< 0.78$ is strictly satisfied would be necessary for future progress in the field.

It is important to emphasize that QED is touted as one of the greatest triumphs of theoretical physics simply because higher-order calculations (beyond $O(\alpha^4)$) have been done, obtaining an astonishing better than twelve decimal place agreement between theory and experiments.  This has, however, been possible purely because of luck, not because of any particular cleverness or theoretical breakthrough.  It just so happens that the smallness ($\sim1/137$) of the QED coupling constant allows accurate perturbative calculations up to 137 decimal places before the asymptotic perturbative series breaks down.  In graphene, the perturbative series may be breaking down already at the one- or two-loop level, and so perhaps, we should not expect better than simple qualitative agreement (i.e., just one significant digit in the dimensionless velocity!) between theory and experiment.  The situation may actually be worse than it is even for QCD (which is inherently strongly coupled) because in graphene, experiments give no hints at all of any strong-coupling behavior!

Thus, graphene many-body theory shares features of calculational difficulties with QED (in the sense that the perturbative series in $\alpha$ is asymptotic, but may already be diverging after the first few terms, unlike in QED, because $\alpha$ for graphene is not very small), QCD (in the sense that the strong-coupling problem is not amenable to analytical tools except perhaps in the trivial infinite-coupling problem, which is useless for graphene since $\alpha$ is not very large), and the metallic many-body problem in ordinary strongly-interacting Fermi liquids (in the sense that $r_s=\alpha$ is not small, and therefore standard diagrammatic techniques may not be accurate).  The problem is therefore interesting and highly nontrivial, which is the reason for our rather comprehensive analysis in the current work.

Actually, the many-body problem for real graphene (not intrinsic graphene with the Fermi level at the Dirac point, which is an idealization) is even more complex than what is discussed above.  In particular, all experimental graphene samples are doped, and the Fermi level is shifted away from the Dirac point, which has no analog in QED or QCD. The finite Fermi level is irrelevant as far as the RG flow is concerned since at low enough densities, the ultraviolet divergent logarithmic term dominates all other contributions, but for quantitative purposes the finite carrier density matters very much, and analytical calculations must go beyond the single-loop order in the dynamically screened Coulomb interaction, which is a problem of great difficulty.  In addition, at low carrier density near the Dirac point, even ultrapure graphene would have disorder effects which could potentially overwhelm many-body effects since disorder becomes important as the Fermi energy decreases.  Disorder is likely to be relevant in the RG sense, and of course, the problem of disorder and interaction together is a formidable unsolved problem in many-body physics. Again, disorder is not a complication that arises in QED or QCD.  One hand-waiving way of handling disorder is to cut off the RG flow at the disorder energy scale, but again, this would complicate any quantitative comparison between experiment and theory.  Obviously, much more work will be needed, both theoretically and experimentally, in order to understand the quantitative many-body effects in graphene; ours is just the first step in this journey, where we have brought the crucial questions into sharp focus and provided plausible partial answers in a few situations.

One important (albeit somewhat tentative) conclusion following from our work is that the best possible many-body theory for graphene self-energy and velocity renormalization may very well be RPA, i.e., the leading-order loop expansion in the dynamically screened Coulomb interaction, which, being an expansion in the screened interaction, does not have the artifact of the infrared bare Coulomb Hartree-Fock divergence in the self-energy at the Fermi level that the usual loop expansion in the bare Coulomb interaction suffers from.  Since all experiments are carried out, by definition, in doped graphene, RPA serves the crucial role of validating the nature of the ultraviolet log divergence that the loop expansion in the bare interaction for undoped graphene has uncovered. Both RPA and the one-loop theory give exactly the same ultraviolet divergence, justifying the RG approach for the running coupling in doped graphene.  Thus, as long as one is interested only in the  log divergence part of the self-energy (and not the sub-leading non-divergent terms), RPA and the one-loop expansion give exactly the same results, with RPA allowing the use of the theory at finite doping.  This result is still somewhat tentative for a number of reasons.  First, the asymptotic convergence of the series is unknown for the expansion in the dynamically screened Coulomb interaction since only the leading-order term (i.e. RPA) has so far been calculated.  It is important to calculate the higher-order self-energy diagrams in the dynamically screened Coulomb interaction to establish that they are small for a full justification of the theoretical framework.  This is  a formidable task well beyond the scope of the current work.  Second, if the graphene self-energy problem is indeed a strong-coupling problem, then even an apparently numerically convergent perturbative expansion in the dynamically screened interaction would not lead to the correct result since there is a new strong-coupling fixed point not accessible to any perturbative expansion.  We believe that such a strong-coupling behavior is unlikely in view of the existing experimental data, but the theoretical possibility of the existence of a strong-coupling fixed point cannot be ruled out given the large value of the graphene bare coupling constant.  As emphasized in this paper, the rather disappointing possibility that the graphene self-energy expansion ceases to be asymptotic already at the two-loop level (in contrast to QED which remains asymptotic at least up to 137 loops, possibly to much higher orders) would explain the observed approximate agreement between the one-loop theory and experiment, with the unfortunate implication that the agreement between experiment and theory in graphene may not improve in the future beyond what we have today.

{\it Note added}. A very recent work has just appeared\cite{Hofmann_arXiv14} which uses a perturbative expansion in the dynamically screened Coulomb interaction rather than the loop expansion in the bare Coulomb coupling used in this work, finding only a small correction to the leading-order RPA result,\cite{DasSarma_PRB07,Son_PRB07} which is consistent with our conclusion that RPA may very well be the quantitatively accurate theory for graphene many-body effects.

\bigskip
\centerline{\bf Acknowledgments}
\bigskip

We thank Andre Geim for providing us with the experimental data points. We also thank Johannes Hofmann for helpful discussions. This work is supported by LPS-CMTC and US-ONR.

\appendix

\section{Leading ultraviolet divergence at three loops}\label{app:threeloop}

In this appendix, we show that one of the three-loop corrections to the electron self-energy, that shown in Fig.~\ref{fig:3loopdiagrams}g, vanishes identically, thus avoiding a potential $\log^3$ divergence at third order that would violate renormalizability constraints. We also compute the leading $\log^2$ ultraviolet divergence for several of the remaining three-loop diagrams, in particular those which contain divergent self-energy subdiagrams. The results confirm the expectation that such $\log^2$ divergences arise at third order, while higher powers of $\log$ do not occur, as required by renormalizability and as discussed in Sec.~\ref{sec:threeloop}. We also calculate explicitly the diagram shown in Fig.~\ref{fig:3loopdiagrams}p, which exhibits a simple $\log$ divergence.

\subsection{First vertex correction to two-loop rainbow diagram}
\begin{figure}
\includegraphics[width=0.5\columnwidth]{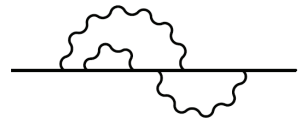}
\caption{First vertex correction to two-loop rainbow diagram.}
\label{fig:selfenergy3b}
\end{figure}

We begin by considering the diagram shown in Fig.~\ref{fig:selfenergy3b}.  We will denote
this diagram by $\Sigma_{3b}(q)$.  We note that it contains the sole first{-}order self{-}energy
correction as a subdiagram. The correction
is given by
\begin{eqnarray}
\Sigma_{3b}(q)\!\!\!&{=}&\!\!\!2{\int}\frac{d^3k}{(2\pi)^3}\,{\int}\frac{d^3p}{(2\pi)^3}\,\frac{g^2}{2|\bq{-}\bk|}\frac{g^2}{2|\bq{-}\bp|}i\gamma^0\frac{i\slashed{k}}{k^2}\frac{ig^2}{16\pi}\bk{\cdot}\vec{\gamma} \nn\\&\times&\log\left (\frac{\Lambda}{|\bk|}\right)\frac{i\slashed{k}}{k^2}i\gamma^0\frac{i(\slashed{k}{+}\slashed{p}{-}\slashed{q})}{(k{+}p{-}q)^2}
i\gamma^0\frac{i\slashed{p}}{p^2}i\gamma^0.
\end{eqnarray}
Note the overall factor of $2$; this is due to the presence of a second diagram with the self{-}energy
correction on the right{-}hand side.  This will give us an identical contribution.  With the aid
of the identity,
\begin{equation}
\gamma^0\slashed{k}\bk{\cdot}\vec{\gamma}\slashed{k}\gamma^0{=}(k_0^2{-}v_F^2|\bk|^2)\bk{\cdot}\vec{\gamma}{+}2v_F|\bk|^2k_0\gamma^0,
\end{equation}
we may rewrite this as
\begin{eqnarray}
\Sigma_{3b}(q)\!\!\!&{=}&\!\!\!i\frac{g^6}{32\pi}{\int}\frac{d^3k}{(2\pi)^3}\,{\int}\frac{d^3p}{(2\pi)^3}\,\frac{1}{|\bq{-}\bk||\bq{-}\bp|}\frac{1}{k^4p^2(k{+}p{-}q)^2}\nn\\
&\times&[(k_0^2{-}v_F^2|\bk|^2)\bk{\cdot}\vec{\gamma}{+}2v_F|\bk|^2k_0\gamma^0]\nn\\
&\times&[(k_0{+}p_0{-}q_0)\gamma^0{+}v_F(\bk{+}\bp{-}\bq){\cdot}\vec{\gamma}]\nn\\
&\times&(p_0\gamma^0{-}v_F\bp{\cdot}\vec{\gamma})\log\left (\frac{\Lambda}{|\bk|}\right ).
\end{eqnarray}

\subsubsection{Evaluation of integrals over temporal components}
We will now do the intergrals over $k_0$ and $p_0$.  If we multiply out the above expression, we
obtain
\begin{eqnarray}
&&\!\!\!\!\!\!\Sigma_{3b}(q){=}i\frac{g^6}{32\pi}{\int}\frac{d^2k}{(2\pi)^2}\,{\int}\frac{d^2p}{(2\pi)^2}\,\frac{1}{|\bq{-}\bk||\bq{-}\bp|}\log\left(\frac{\Lambda}{|\vec{k}|}\right)\nn\\
&&\!\!\!\!\!\!\times\Big[2v_F|\bk|^2B_1\gamma^0{-}2v_F^2|\bk|^2B_2\bp{\cdot}\vec{\gamma}{-}2v_F^2|\bk|^2B_3(\bk{+}\bp{-}\bq){\cdot}\vec{\gamma}\nn\\
&&\!\!\!\!\!\!{-}2v_F^3|\bk|^2B_4\gamma^0(\bk{+}\bp{-}\bq){\cdot}\vec{\gamma}\bp{\cdot}\vec{\gamma}{+}B_5\bk{\cdot}\vec{\gamma}{+}v_FB_6\gamma^0\bk{\cdot}\vec{\gamma}\bp{\cdot}\vec{\gamma}\nn\\
&&\!\!\!\!\!\!{+}v_FB_7\gamma^0\bk{\cdot}\vec{\gamma}(\bk{+}\bp{-}\bq){\cdot}\vec{\gamma}{-}v_F^2B_8\bk{\cdot}\vec{\gamma}(\bk{+}\bp{-}\bq){\cdot}\vec{\gamma}\bp{\cdot}\vec{\gamma}\Big],
\end{eqnarray}
where the eight integrals $B_k$ are given by
\bea
B_1\!\!\!&{=}&\!\!\!{\int}\frac{dk_0}{2\pi}\,{\int}\frac{dp_0}{2\pi}\,\frac{k_0p_0(k_0{+}p_0{-}q_0)}{k^4p^2(k{+}p{-}q)^2}\nn\\ &&\!\!\!\!\!\!{=}\frac{(|\bk|{+}|\bp|{+}|\bk{+}\bp{-}\bq|)q_0}{4|\bk|[q_0^2{+}v_F^2(|\bk|{+}|\bp|{+}|\bk{+}\bp{-}\bq|)^2]^2},
\eea
\bea
B_2\!\!\!&{=}&\!\!\!{\int}\frac{dk_0}{2\pi}\,{\int}\frac{dp_0}{2\pi}\,\frac{k_0(k_0{+}p_0{-}q_0)}{k^4p^2(k{+}p{-}q)^2}\nn\\
&&\!\!\!\!\!\!{=}\frac{v_F^2(|\bk|{+}|\bp|{+}|\bk{+}\bp{-}\bq|)^2{-}q_0^2}{8v_F^2|\bk||\bp|[q_0^2{+}v_F^2(|\bk|{+}|\bp|{+}|\bk{+}\bp{-}\bq|)^2]^2},
\eea
\bea
B_3\!\!\!&{=}&\!\!\!{\int}\frac{dk_0}{2\pi}\,{\int}\frac{dp_0}{2\pi}\,\frac{k_0p_0}{k^4p^2(k{+}p{-}q)^2}\nn\\ &&\!\!\!\!\!\!{=}\frac{q_0^2{-}v_F^2(|\bk|{+}|\bp|{+}|\bk{+}\bp{-}\bq|)^2}{8v_F^2|\bk||\bk{+}\bp{-}\bq|[q_0^2{+}v_F^2(|\bk|{+}|\bp|{+}|\bk{+}\bp{-}\bq|)^2]^2},\nn\\ \eea
\bea
B_4\!\!\!&{=}&\!\!\!{\int}\frac{dk_0}{2\pi}\,{\int}\frac{dp_0}{2\pi}\,\frac{k_0}{k^4p^2(k{+}p{-}q)^2}\nn\\
&&\!\!\!\!\!\!{=}\frac{(|\bk|{+}|\bp|{+}|\bk{+}\bp{-}\bq|)q_0}{4v_F^2|\bk||\bp||\bk{+}\bp{-}\bq|[q_0^2{+}v_F^2(|\bk|{+}|\bp|{+}|\bk{+}\bp{-}\bq|)^2]^2},\nn\\ \eea
\bea
B_5\!\!\!&{=}&\!\!\!{\int}\frac{dk_0}{2\pi}\,{\int}\frac{dp_0}{2\pi}\,\frac{(k_0^2{-}v_F^2|\bk|^2)p_0(k_0{+}p_0{-}q_0)}{k^4p^2(k{+}p{-}q)^2}\nn\\ &&\!\!\!\!\!\!{=}\frac{q_0^2{-}v_F^2(|\bk|{+}|\bp|{+}|\bk{+}\bp{-}\bq|)^2}{4[q_0^2{+}v_F^2(|\bk|{+}|\bp|{+}|\bk{+}\bp{-}\bq|)^2]^2},
\eea
\bea
B_6\!\!\!&{=}&\!\!\!{\int}\frac{dk_0}{2\pi}\,{\int}\frac{dp_0}{2\pi}\,\frac{(k_0^2{-}v_F^2|\bk|^2)(k_0{+}p_0{-}q_0)}{k^4p^2(k{+}p{-}q)^2}\nn\\ &&\!\!\!\!\!\!{=}\frac{(|\bk|{+}|\bp|{+}|\bk{+}\bp{-}\bq|)q_0}{2|\bp|[q_0^2{+}v_F^2(|\bk|{+}|\bp|{+}|\bk{+}\bp{-}\bq|)^2]^2},
\eea
\bea
B_7\!\!\!&{=}&\!\!\!{\int}\frac{dk_0}{2\pi}\,{\int}\frac{dp_0}{2\pi}\,\frac{(k_0^2{-}v_F^2|\bk|^2)p_0}{k^4p^2(k{+}p{-}q)^2}\nn\\ &&\!\!\!\!\!\!{=}{-}\frac{(|\bk|{+}|\bp|{+}|\bk{+}\bp{-}\bq|)q_0}{2|\bk{+}\bp{-}\bq|[q_0^2{+}v_F^2(|\bk|{+}|\bp|{+}|\bk{+}\bp{-}\bq|)^2]^2},\nn\\
\eea
\bea
B_8\!\!\!&{=}&\!\!\!{\int}\frac{dk_0}{2\pi}\,{\int}\frac{dp_0}{2\pi}\,\frac{k_0^2{-}v_F^2|\bk|^2}{k^4p^2(k{+}p{-}q)^2}\nn\\ &&\!\!\!\!\!\!{=}\frac{q_0^2{-}v_F^2(|\bk|{+}|\bp|{+}|\bk{+}\bp{-}\bq|)^2}{4v_F^2|\bp||\bk{+}\bp{-}\bq|[q_0^2{+}v_F^2(|\bk|{+}|\bp|{+}|\bk{+}\bp{-}\bq|)^2]^2}.\nn\\
\eea
We will now extract the temporal and spatial components of this self{-}energy correction.  To do so,
we simply multiply by the appropriate matrix and take the trace, making use of the anticommutation
relations for the Dirac matrices to simplify the result.  The temporal component is
\begin{eqnarray}
&&\!\!\!\!\!\!\Sigma_{3b,0}(q){=}\tfrac{1}{4}\mbox{Tr}[\gamma^0\Sigma_{3b}(q)]{=}i\frac{g^6}{32\pi}{\int}\frac{d^2k}{(2\pi)^2}\,{\int}\frac{d^2p}{(2\pi)^2}\,\nn\\
&&\!\!\!\!\!\!\times\frac{1}{|\bq{-}\bk||\bq{-}\bp|}\log\left(\frac{\Lambda}{|\vec{k}|}\right)\Big[2v_F|\bk|^2B_1\nn\\
&&\!\!\!\!\!\!{-}2v_F^3|\bk|^2B_4(\bk{+}\bp{-}\bq){\cdot}\bp{+}v_FB_6\bk{\cdot}\bp{+}v_FB_7\bk{\cdot}(\bk{+}\bp{-}\bq)\Big]\nn\\
&&\!\!\!\!\!\!{=}i\frac{g^6}{64\pi}q_0{\int}\frac{d^2k}{(2\pi)^2}\,{\int}\frac{d^2p}{(2\pi)^2}\,\frac{1}{|\bq{-}\bk||\bq{-}\bp|}\nn\\
&&\!\!\!\!\!\!\times\frac{v_F|\bk|(|\bk|{+}|\bp|{+}|\bk{+}\bp{-}\bq|)}{[q_0^2{+}v_F^2(|\bk|{+}|\bp|{+}|\bk{+}\bp{-}\bq|)^2]^2}\nn\\
&&\!\!\!\!\!\!\times\left [1{-}\frac{\bp{\cdot}(\bk{+}\bp{-}\bq)}{|\bp||\bk{+}\bp{-}\bq|}{+}\frac{\bk{\cdot}\bp}{|\bk||\bp|}{-}\frac{\bk{\cdot}(\bk{+}\bp{-}\bq)}{|\bk||\bk{+}\bp{-}\bq|}\right ]\log\left (\frac{\Lambda}{|\bk|}\right ),\nn\\
\end{eqnarray}
and the spatial component is
\begin{eqnarray}
&&\!\!\!\!\!\!\Sigma_{3b,i}(q){=}\tfrac{1}{4}\mbox{Tr}[\gamma^i\Sigma_{3b}(q)]{=}i\frac{g^6}{32\pi}{\int}\frac{d^2k}{(2\pi)^2}\,{\int}\frac{d^2p}{(2\pi)^2}\,\frac{1}{|\bq{-}\bk|}\nn\\
&&\!\!\!\!\!\!\times\frac{1}{|\bq{-}\bp|}\Big\{{-}2v_F^2|\bk|^2B_2p_i{-}2v_F^2|\bk|^2B_3(k_i{+}p_i{-}q_i){+}B_5k_i\nn\\
&&\!\!\!\!\!\!{-}v_F^2B_8\Big[\bp{\cdot}(\bk{+}\bp{-}\bq)k_i{-}\bk{\cdot}\bp(k_i{+}p_i{-}q_i){+}\bk{\cdot}(\bk{+}\bp{-}\bq)p_i\Big]\Big\}\nn\\
&&\!\!\!\!\!\!\times\log\left (\frac{\Lambda}{|\bk|}\right )\nn\\
&&\!\!\!\!\!\!{=}i\frac{g^6}{32\pi}\frac{q_i}{|\bq|^2}{\int}\frac{d^2k}{(2\pi)^2}\,{\int}\frac{d^2p}{(2\pi)^2}\,\frac{1}{|\bq{-}\bk||\bq{-}\bp|}\nn\\
&&\!\!\!\!\!\!\times\frac{|\bk|[q_0^2{-}v_F^2(|\bk|{+}|\bp|{+}|\bk{+}\bp{-}\bq|)^2]}{4[q_0^2{+}v_F^2(|\bk|{+}|\bp|{+}|\bk{+}\bp{-}\bq|)^2]^2}\bigg [\frac{\bp{\cdot}\bq}{|\bp|}{-}\frac{(\bk{+}\bp{-}\bq){\cdot}\bq}{|\bk{+}\bp{-}\bq|}{+}\frac{\bk{\cdot}\bq}{|\bk|}\nn\\
&&\!\!\!\!\!\!{-}\frac{|\bp|^2\bk{\cdot}\bq{+}|\bq|^2\bk{\cdot}\bp{+}|\bk|^2\bp{\cdot}\bq{-}2(\bk{\cdot}\bq)(\bp{\cdot}\bq)}{|\bk||\bp||\bk{+}\bp{-}\bq|}\bigg ]\log\left (\frac{\Lambda}{|\bk|}\right ).
\end{eqnarray}

\subsubsection{Extracting the divergence of the temporal component}
We first find the leading divergence of the temporal component.  The integrand is non{-}zero for
$\bq{=}0$, so that the leading divergence is found by simply setting $\bq{=}0$ in the integrand:
\begin{eqnarray}
\Sigma_{3b,0}(q)&{\approx}&i\frac{g^6}{64\pi}q_0{\int}\frac{d^2k}{(2\pi)^2}{\int}\frac{d^2p}{(2\pi)^2}\frac{1}{|\bp|}\nn\\
&\times&\frac{v_F(|\bk|{+}|\bp|{+}|\bk{+}\bp|)}{[q_0^2{+}v_F^2(|\bk|{+}|\bp|{+}|\bk{+}\bp)^2]^2}\nn\\
&\times&\left [1{-}\frac{\bp{\cdot}(\bk{+}\bp)}{|\bp||\bk{+}\bp|}{+}\frac{\bk{\cdot}\bp}{|\bk||\bp|}{-}\frac{\bk{\cdot}(\bk{+}\bp)}{|\bk||\bk{+}\bp}\right ]\log\left (\frac{\Lambda}{|\bk|}\right )\nn\\
&{=}&i\frac{g^6}{64\pi}q_0(\Psi_1{-}\Psi_2{-}\Psi_3{+}\Psi_4),
\end{eqnarray}
where
\begin{eqnarray}
\Psi_1\!\!\!&{=}&\!\!\!{\int}\frac{d^2k}{(2\pi)^2}{\int}\frac{d^2p}{(2\pi)^2}\frac{1}{|\bp|}S(\vec{k},\vec{p})\log\left (\frac{\Lambda}{|\bk|}\right ), \nn\\
\Psi_2\!\!\!&{=}&\!\!\!{\int}\frac{d^2k}{(2\pi)^2}{\int}\frac{d^2p}{(2\pi)^2}\frac{1}{|\bp|}S(\vec{k},\vec{p})\frac{\bp{\cdot}(\bk{+}\bp)}{|\bp||\bk{+}\bp|}\log\left (\frac{\Lambda}{|\bk|}\right ),\nn\\
\Psi_3\!\!\!&{=}&\!\!\!{\int}\frac{d^2k}{(2\pi)^2}{\int}\frac{d^2p}{(2\pi)^2}\frac{1}{|\bp|}S(\vec{k},\vec{p})\frac{\bk{\cdot}(\bk{+}\bp)}{|\bk||\bk{+}\bp|}\log\left (\frac{\Lambda}{|\bk|}\right ),\nn\\
\Psi_4\!\!\!&{=}&\!\!\!{\int}\frac{d^2k}{(2\pi)^2}{\int}\frac{d^2p}{(2\pi)^2}\frac{1}{|\bp|}S(\vec{k},\vec{p})\frac{\bk{\cdot}\bp}{|\bk||\bp|}\log\left (\frac{\Lambda}{|\bk|}\right ),
\end{eqnarray}
with
\beq
S(\vec{k},\vec{p})\equiv\frac{v_F(|\bk|{+}|\bp|{+}|\bk{+}\bp|)}{[q_0^2{+}v_F^2(|\bk|{+}|\bp|{+}|\bk{+}\bp)^2]^2}.
\eeq
To evaluate the $\bp$ integrals, we use elliptic coordinates.  Let $p_{||}$ and $p_{\perp}$ be
the components of $\bp$ parallel and perpendicular to $\bk$, respectively.  These components are
written in terms of the elliptic coordinates $\mu$ and $\nu$ as
\begin{equation}
p_{||}{=}\tfrac{1}{2}|\bk|(\cosh{\mu}\cos{\nu}{-}1),\,p_{\perp}{=}\tfrac{1}{2}|\bk|\sinh{\mu}\sin{\nu}.
\end{equation}
These coordinates have the ranges, $0\leq\mu<\infty$ and $0\leq\nu<2\pi$.  Some useful identities are
\begin{eqnarray}
&|\bp|{=}\tfrac{1}{2}|\bk|(\cosh{\mu}{-}\cos{\nu}),\,|\bk{+}\bp|{=}\tfrac{1}{2}|\bk|(\cosh{\mu}{+}\cos{\nu}),\cr
&|\bk|{+}|\bp|{+}|\bk{+}\bp|{=}|\bk|(\cosh{\mu}{+}1).
\end{eqnarray}
The integration measure is
\begin{equation}
d^2p{=}\tfrac{1}{4}|\bk|^2(\cosh^2{\mu}{-}\cos^2{\nu})\,d\mu\,d\nu{=}|\bp||\bk{+}\bp|\,d\mu\,d\nu.
\end{equation}
In terms of these coordinates, the $\Psi_k$ integrals above become
\begin{eqnarray}
\Psi_1\!\!\!&{=}&\!\!\!\frac{1}{4\pi^2}{\int}_{0}^{\infty}d\mu\,{\int}_{0}^{2\pi}d\nu\,f(\mu)(\cosh{\mu}{+}\cos{\nu}),\nn\\
\Psi_2\!\!\!&{=}&\!\!\!\frac{1}{4\pi^2}{\int}_{0}^{\infty}d\mu\,{\int}_{0}^{2\pi}d\nu\,f(\mu)\frac{\cosh^2{\mu}{+}\cos^2{\nu}{-}2}{\cosh{\mu}{-}\cos{\nu}},\nn\\
\Psi_3\!\!\!&{=}&\!\!\!\frac{1}{4\pi^2}{\int}_{0}^{\infty}d\mu\,{\int}_{0}^{2\pi}d\nu\,f(\mu)(1{+}\cosh{\mu}\cos{\nu}),\nn\\
\Psi_4\!\!\!&{=}&\!\!\!\frac{1}{4\pi^2}{\int}_{0}^{\infty}d\mu\,{\int}_{0}^{2\pi}d\nu\,f(\mu)\times\nn\\
&&\times\frac{(\cosh{\mu}\cos{\nu}{-}1)(\cosh{\mu}{+}\cos{\nu})}{\cosh{\mu}{-}\cos{\nu}},\nn\\
\end{eqnarray}
where
\begin{equation}
f(\mu){=}{\int}\frac{d^2k}{(2\pi)^2}\,\frac{v_F|\bk|(1{+}\cosh{\mu})}{2[q_0^2{+}v_F^2|\bk|^2(1{+}\cosh{\mu})^2]^2}|\bk|\log\left (\frac{\Lambda}{|\bk|}\right ).\label{defoffofmu}
\end{equation}
We first do the $\nu$ integrals, obtaining
\begin{eqnarray}
\Psi_1&{=}&\frac{1}{2\pi}{\int}_{0}^{\infty}d\mu\,f(\mu)\cosh{\mu},\nn\\
\Psi_2&{=}&{-}\frac{1}{2\pi}{\int}_{0}^{\infty}d\mu\,f(\mu)(\cosh{\mu}{-}2\sinh{\mu}),\nn\\
\Psi_3&{=}&\frac{1}{2\pi}{\int}_{0}^{\infty}d\mu\,f(\mu),\nn\\
\Psi_4&{=}&-\frac{1}{2\pi}{\int}_{0}^{\infty}d\mu\,f(\mu)e^{{-}2\mu}.
\end{eqnarray}
We now evaluate $f(\mu)$ in closed form; doing so, we obtain
\bea
f(\mu)&{=}&{-}\frac{1}{16\pi v_F^3(1{+}\cosh{\mu})^3}\times\nn\\
&&\times\bigg \{\log\left [1{+}\left (\frac{v_F\Lambda}{q_0}\right )^2(1{+}\cosh{\mu})^2\right ]\nn\\
&&{+}\mbox{Li}_2\left [{-}\left (\frac{v_F\Lambda}{q_0}\right )^2(1{+}\cosh{\mu})^2\right ]\bigg \},
\eea
where $\mbox{Li}_n(z)$ is the polylogarithm function,
\begin{equation}
\mbox{Li}_n(z){=}\sum_{k{=}1}^{\infty}\frac{z^k}{k^n}.
\end{equation}
For $z\to \infty$, we may approximate $\mbox{Li}_2({-}z^2)$ as
\begin{equation}
\mbox{Li}_2({-}z^2)\approx {-}2\log^2z.
\end{equation}
Since we assume that the cutoff $\Lambda$ is large, this approximation is
valid, and thus we may approximate $f(\mu)$ as
\begin{equation}
f(\mu)\approx\frac{1}{8\pi v_F^3(1{+}\cosh{\mu})^3}\log^2\left [\frac{v_F\Lambda}{|q_0|}(1{+}\cosh{\mu})\right ].
\end{equation}
If we now substitute this result into the $\Psi_k$ integrals and drop the
$1{+}\cosh{\mu}$ factor in the logarithm (which will only give sub{-}leading
terms if retained), then we may evaluate the remaining integrals over $\mu$,
obtaining
\begin{eqnarray}
\Psi_1\!\!\!&{=}&\!\!\!\frac{1}{16\pi^2 v_F^3}\left [{\int}_{0}^{\infty}d\mu\,\frac{\cosh{\mu}}{(1{+}\cosh{\mu})^3}\right ]\log^2\left (\frac{v_F\Lambda}{|q_0|}\right )\nn\\
&&\!\!\!\!\!\!{=}\frac{1}{80\pi^2v_F^3}\log^2\left (\frac{v_F\Lambda}{|q_0|}\right ),\nn\\
\Psi_2\!\!\!&{=}&\!\!\!{-}\frac{1}{16\pi^2 v_F^3}\left [{\int}_{0}^{\infty}d\mu\,\frac{\cosh{\mu}{-}2\sinh{\mu}}{(1{+}\cosh{\mu})^3}\right ]\log^2\left (\frac{v_F\Lambda}{|q_0|}\right )\nn\\
&&\!\!\!\!\!\!{=}\frac{1}{320\pi^2v_F^3}\log^2\left (\frac{v_F\Lambda}{|q_0|}\right ),\nn\\
\Psi_3\!\!\!&{=}&\!\!\!\frac{1}{16\pi^2 v_F^3}\left [{\int}_{0}^{\infty}d\mu\,\frac{1}{(1{+}\cosh{\mu})^3}\right ]\log^2\left (\frac{v_F\Lambda}{|q_0|}\right )\nn\\
&&\!\!\!\!\!\!{=}\frac{1}{120\pi^2v_F^3}\log^2\left (\frac{v_F\Lambda}{|q_0|}\right ),\nn\\
\Psi_4\!\!\!&{=}&\!\!\!-\frac{1}{16\pi^2 v_F^3}\left [{\int}_{0}^{\infty}d\mu\,\frac{e^{{-}2\mu}}{(1{+}\cosh{\mu})^3}\right ]\log^2\left (\frac{v_F\Lambda}{|q_0|}\right )\nn\\
&&\!\!\!\!\!\!{=}-\frac{1}{320\pi^2v_F^3}\log^2\left (\frac{v_F\Lambda}{|q_0|}\right ).
\end{eqnarray}
Putting everything together, we find that the leading divergence of the temporal
component of $\Sigma_{3b}(q)$ is
\bea
\Sigma_{3b,0}(q)&{=}&-\tfrac{1}{480}i\alpha^3 q_0\log^2\left (\frac{v_F\Lambda}{|q_0|}\right )\nn\\
&{\to}&-\tfrac{1}{480}i\alpha^3 q_0\log^2\left (\frac{\Lambda}{|\bq|}\right ).
\eea
We make the replacement, $\frac{|q_0|}{v_F}\to|\bq|$, for reasons similar to those for
making the same replacement in determining the leading divergence of $\Sigma_{2a}(q)$.

\subsubsection{Extracting the divergence of the spatial component}
We now turn our attention to the spatial component.  Note that, in this case, the integrand
is zero when $\bq{=}0$.  Therefore, in order to obtain the leading divergence of this component,
we will need to expand the integrand to first order in $\bq$.  We will find it easier to do
this if we shift $\bp$ by $\bq$ and interchange $\bk$ and $\bp$:
\begin{eqnarray}
&&\!\!\!\!\!\!\Sigma_{3b,i}(q){=}i\frac{g^6}{128\pi}\frac{q_i}{|\bq|^2}{\int}\frac{d^2k}{(2\pi)^2}\,{\int}\frac{d^2p}{(2\pi)^2}\,\frac{|\bp|}{|\bk||\bq{-}\bp|}\nn\\
&&\!\!\!\!\!\!\times\frac{q_0^2{-}v_F^2(|\bk{+}\bq|{+}|\bp|{+}|\bk{+}\bp|)^2}{[q_0^2{+}v_F^2(|\bk{+}\bq|{+}|\bp|{+}|\bk{+}\bp|)^2]^2}\nn\\
&&\!\!\!\!\!\!\times\left [\frac{\bp{\cdot}\bq}{|\bp|}{-}\frac{(\bk{+}\bp){\cdot}\bq}{|\bk{+}\bp|}{+}\frac{(\bk{+}\bq){\cdot}\bq}{|\bk{+}\bq|}{-}\frac{|\bp|(\bk{+}\bq){\cdot}\bq}{|\bk{+}\bq||\bk{+}\bp|}{-}\frac{|\bq|^2(\bk{+}\bq){\cdot}\bp}{|\bk{+}\bq||\bp||\bk{+}\bp|}\right.\nn\\
&&\!\!\!\!\!\!{-}\left.\frac{|\bk{+}\bq|\bp{\cdot}\bq}{|\bp||\bk{+}\bp|}{+}\frac{2[(\bk{+}\bq){\cdot}\bq](\bp{\cdot}\bq)}{|\bk{+}\bq||\bp||\bk{+}\bp|}\right ]\log\left (\frac{\Lambda}{|\bp|}\right ).
\end{eqnarray}
In doing the expansion, the following approximations will prove useful:
\begin{equation}
|\bk{+}\bq|\approx|\bk|{+}\frac{\bk{\cdot}\bq}{|\bk|},\,\frac{1}{|\bk{+}\bq|}\approx\frac{1}{|\bk|}{-}\frac{\bk{\cdot}\bq}{|\bk|^3},\,\frac{1}{|\bp{-}\bq|}\approx\frac{1}{|\bp|}{+}\frac{\bp{\cdot}\bq}{|\bp|^3},
\end{equation}
and
\begin{equation}
\frac{[q_0^2{-}v_F^2(|\bk{+}\bq|{+}|\bp|{+}|\bk{+}\bp|)^2]}{[q_0^2{+}v_F^2(|\bk{+}\bq|{+}|\bp|{+}|\bk{+}\bp|)^2]^2}\approx Q(\bk,\bp){+}R(\bk,\bp)\frac{\bk{\cdot}\bq}{|\bk|},
\end{equation}
where
\begin{eqnarray}
Q(\bk,\bp)\!\!\!&{=}&\!\!\!\frac{q_0^2{-}v_F^2(|\bk|{+}|\bp|{+}|\bk{+}\bp|)^2}{[q_0^2{+}v_F^2(|\bk|{+}|\bp|{+}|\bk{+}\bp|)^2]^2},\nn\\
R(\bk,\bp)\!\!\!&{=}&\!\!\!{-}2v_F^2(|\bk|{+}|\bp|{+}|\bk{+}\bp|)\times\nn\\
&&\times\frac{3q_0^2{-}v_F^2(|\bk|{+}|\bp|{+}|\bk{+}\bp|)^2}{[q_0^2{+}v_F^2(|\bk|{+}|\bp|{+}|\bk{+}\bp|)^2]^3}.\label{defofQandR}
\end{eqnarray}
Using these formulas, we may now write
\begin{eqnarray}
&&\!\!\!\!\!\!\Sigma_{3b,i}(q)\approx\frac{g^6}{128\pi}\frac{q_i}{|\bq|^2}{\int}\frac{d^2k}{(2\pi)^2}\,{\int}\frac{d^2p}{(2\pi)^2}\,\frac{1}{|\bk|}\left (1{+}\frac{\bp{\cdot}\bq}{|\bp|^2}\right )\nn\\
&&\!\!\!\!\!\!\times\left [Q(\bk,\bp){+}R(\bk,\bp)\frac{\bk{\cdot}\bq}{|\bk|}\right ]\log\left (\frac{\Lambda}{|\bp|}\right )\nn\\
&&\!\!\!\!\!\!\times\left \{\frac{\bp{\cdot}\bq}{|\bp|}{-}\frac{(\bk{+}\bp){\cdot}\bq}{|\bk{+}\bp|}{+}\left [\frac{(\bk{+}\bq){\cdot}\bq}{|\bk|}{-}\frac{|\bp|(\bk{+}\bq){\cdot}\bq}{|\bk||\bk{+}\bp|}{-}\frac{|\bq|^2(\bk{+}\bq){\cdot}\bp}{|\bk||\bp||\bk{+}\bp|}\right.\right.\nn\\
&&\!\!\!\!\!\!{+}\left.\left.\frac{2[(\bk{+}\bq){\cdot}\bq](\bp{\cdot}\bq)}{|\bk||\bp||\bk{+}\bp|}\right ]\left (1{-}\frac{\bk{\cdot}\bq}{|\bk|^2}\right ){-}\frac{|\bk|\bp{\cdot}\bq}{|\bp||\bk{+}\bp|}\left (1{+}\frac{\bk{\cdot}\bq}{|\bk|^2}\right )\right \}.\nn\\
\end{eqnarray}
When we multiply out the integrand, we find that some of the terms na\"ively appear to produce
a linear divergence, but these terms turn out to be zero.  We will therefore enumerate all of
the terms that give us a logarithmic divergence, of which there are the following fifteen:
\bea
\Xi_1\!\!\!&{=}&\!\!\!{\int}\frac{d^2k}{(2\pi)^2}\,{\int}\frac{d^2p}{(2\pi)^2}\,\frac{Q(\bk,\bp)}{|\bk|}\left [\frac{|\bq|^2}{|\bk|}{-}\frac{(\bk{\cdot}\bq)}{|\bk|^3}\right ]\log\left (\frac{\Lambda}{|\bp|}\right ),\nn\\
\eea
\bea
\Xi_2\!\!\!&{=}&\!\!\!{-}{\int}\frac{d^2k}{(2\pi)^2}\,{\int}\frac{d^2p}{(2\pi)^2}\,\frac{Q(\bk,\bp)}{|\bk|}\frac{|\bp|}{|\bk{+}\bp|}\left [\frac{|\bq|^2}{|\bk|}{-}\frac{(\bk{\cdot}\bq)}{|\bk|^3}\right ]\nn\\&&\times\log\left (\frac{\Lambda}{|\bp|}\right ),
\eea
\bea
\Xi_3\!\!\!&{=}&\!\!\!{-}{\int}\frac{d^2k}{(2\pi)^2}\,{\int}\frac{d^2p}{(2\pi)^2}\,\frac{Q(\bk,\bp)}{|\bk|}\frac{|\bq|^2\bk{\cdot}\bp}{|\bk||\bp||\bk{+}\bp|}\log\left (\frac{\Lambda}{|\bp|}\right ),\nn\\
\eea
\bea
\Xi_4\!\!\!&{=}&\!\!\!2{\int}\frac{d^2k}{(2\pi)^2}\,{\int}\frac{d^2p}{(2\pi)^2}\,\frac{Q(\bk,\bp)}{|\bk|}\frac{(\bk{\cdot}\bq)(\bp{\cdot}\bq)}{|\bk||\bp||\bk{+}\bp|}\log\left (\frac{\Lambda}{|\bp|}\right ),\nn\\
\eea
\bea
\Xi_5\!\!\!&{=}&\!\!\!{-}\tfrac{1}{2}\Xi_4,
\eea
\bea
\Xi_6\!\!\!&{=}&\!\!\!{\int}\frac{d^2k}{(2\pi)^2}\,{\int}\frac{d^2p}{(2\pi)^2}\,\frac{Q(\bk,\bp)}{|\bk|}\frac{(\bp{\cdot}\bq)^2}{|\bp|^3}\log\left (\frac{\Lambda}{|\bp|}\right ),\nn\\
\eea
\bea
\Xi_7\!\!\!&{=}&\!\!\!-{\int}\frac{d^2k}{(2\pi)^2}\,{\int}\frac{d^2p}{(2\pi)^2}\,\frac{Q(\bk,\bp)}{|\bk|}\frac{(\bp{\cdot}\bq)[(\bk{+}\bp){\cdot}\bq]}{|\bp|^2|\bk{+}\bp|}\log\left (\frac{\Lambda}{|\bp|}\right ),\nn\\
\eea
\bea
\Xi_8\!\!\!&{=}&\!\!\!{\int}\frac{d^2k}{(2\pi)^2}\,{\int}\frac{d^2p}{(2\pi)^2}\,\frac{Q(\bk,\bp)}{|\bk|}\frac{(\bk{\cdot}\bq)(\bp{\cdot}\bq)}{|\bk||\bp|^2}\log\left (\frac{\Lambda}{|\bp|}\right ),\nn\\
\eea
\bea
\Xi_9\!\!\!&{=}&\!\!\!{-}\tfrac{1}{2}\Xi_4,
\eea
\bea
\Xi_{10}\!\!\!&{=}&\!\!\!{-}{\int}\frac{d^2k}{(2\pi)^2}\,{\int}\frac{d^2p}{(2\pi)^2}\,\frac{Q(\bk,\bp)}{|\bk|}\frac{|\bk|(\bp{\cdot}\bq)^2}{|\bp|^3|\bk{+}\bp|}\log\left (\frac{\Lambda}{|\bp|}\right ),\nn\\
\eea
\bea
\Xi_{11}\!\!\!&{=}&\!\!\!{\int}\frac{d^2k}{(2\pi)^2}\,{\int}\frac{d^2p}{(2\pi)^2}\,\frac{R(\bk,\bp)}{|\bk|}\frac{(\bk{\cdot}\bq)(\bp{\cdot}\bq)}{|\bk||\bp|}\log\left (\frac{\Lambda}{|\bp|}\right ),\nn\\
\eea
\bea
\Xi_{12}\!\!\!&{=}&\!\!\!{-}{\int}\frac{d^2k}{(2\pi)^2}\,{\int}\frac{d^2p}{(2\pi)^2}\,\frac{R(\bk,\bp)}{|\bk|}\frac{(\bk{\cdot}\bq)[(\bk{+}\bp){\cdot}\bq]}{|\bk||\bk{+}\bp|}\log\left (\frac{\Lambda}{|\bp|}\right ),\nn\\
\eea
\bea
\Xi_{13}\!\!\!&{=}&\!\!\!{\int}\frac{d^2k}{(2\pi)^2}\,{\int}\frac{d^2p}{(2\pi)^2}\,\frac{R(\bk,\bp)}{|\bk|}\left (\frac{\bk{\cdot}\bq}{|\bk|}\right )^2\log\left (\frac{\Lambda}{|\bp|}\right ),\nn\\
\eea
\bea
\Xi_{14}\!\!\!&{=}&\!\!\!{-}{\int}\frac{d^2k}{(2\pi)^2}\,{\int}\frac{d^2p}{(2\pi)^2}\,\frac{R(\bk,\bp)}{|\bk|}\frac{|\bp|}{|\bk{+}\bp|}\left (\frac{\bk{\cdot}\bq}{|\bk|}\right )^2\log\left (\frac{\Lambda}{|\bp|}\right ),\nn\\
\eea
\bea
\Xi_{15}\!\!\!&{=}&\!\!\!{-}{\int}\frac{d^2k}{(2\pi)^2}\,{\int}\frac{d^2p}{(2\pi)^2}\,\frac{R(\bk,\bp)}{|\bk|}\frac{(\bk{\cdot}\bq)(\bp{\cdot}\bq)}{|\bp||\bk{+}\bp|}\log\left (\frac{\Lambda}{|\bp|}\right ).\nn\\
\eea
We will once again use elliptic coordinates to evaluate these integrals.  In these coordinates,
the functions $Q$ and $R$ become
\begin{eqnarray}
Q(\bk,\bp)\!\!\!&\to&\!\!\!\tilde{Q}(\mu,|\bk|){=}\frac{q_0^2{-}v_F^2|\bk|^2(1{+}\cosh{\mu})^2}{[q_0^2{+}v_F^2|\bk|^2(1{+}\cosh{\mu})^2]^2},\nn \\
R(\bk,\bp)\!\!\!&\to&\!\!\!\tilde{R}(\mu,|\bk|){=}\nn\\
&&\!\!\!\!\!\!{=}{-}\frac{2v_F^2|\bk|(1{+}\cosh{\mu})[3q_0^2{-}v_F^2|\bk|^2(1{+}\cosh{\mu})^2]}{[q_0^2{+}v_F^2|\bk|^2(1{+}\cosh{\mu})^2]^3}.\nn\\\label{defofQtandRt}
\end{eqnarray}
It will be helpful to write all scalar products in terms of components of the vectors parallel
and perpendicular to $\bk$; in this case, the components of $\bq$ will be $q_{||}{=}|\bq|\cos{\theta}$
and $q_\perp{=}{-}|\bq|\sin{\theta}$, where $\theta$ is the angle variable in the $\bk$ integration.
We will also replace the $|\bp|$ in the logarithms with $|\bk|$, since the terms that are dropped in
making this replacement will only contribute to the sub{-}leading divergence.  Upon performing the $\nu$
and $\theta$ integrals, we obtain
\bea
\Xi_1\!\!\!&{=}&\!\!\!\frac{|\bq|^2}{32\pi^2}{\int}_{0}^{\infty}d\mu\,{\int}_{0}^{\Lambda}d|\bk|\,\tilde{Q}(\mu,|\bk|)(\cosh^2{\mu}{-}\tfrac{1}{2})|\bk|\nn\\
&&\times\log\left (\frac{\Lambda}{|\bk|}\right ),
\eea
\bea
\Xi_2\!\!\!&{=}&\!\!\!{-}\frac{|\bq|^2}{32\pi^2}{\int}_{0}^{\infty}d\mu\,{\int}_{0}^{\Lambda}d|\bk|\,\tilde{Q}(\mu,|\bk|)(\cosh^2{\mu}{+}\tfrac{1}{2})|\bk|\nn\\
&&\times\log\left (\frac{\Lambda}{|\bk|}\right ),
\eea
\bea
\Xi_3\!\!\!&{=}&\!\!\!\frac{|\bq|^2}{8\pi^2}{\int}_{0}^{\infty}d\mu\,{\int}_{0}^{\Lambda}d|\bk|\,\tilde{Q}(\mu,|\bk|)|\bk|\log\left (\frac{\Lambda}{|\bk|}\right ),\nn\\
\eea
\bea
\Xi_4\!\!\!&{=}&\!\!\!{-}\Xi_3,
\eea
\bea
\Xi_5\!\!\!&{=}&\!\!\!\tfrac{1}{2}\Xi_3,
\eea
\bea
\Xi_6\!\!\!&{=}&\!\!\!\frac{|\bq|^2}{16\pi^2}{\int}_{0}^{\infty}d\mu\,{\int}_{0}^{\Lambda}d|\bk|\,\tilde{Q}(\mu,|\bk|)\cosh{\mu}|\bk|\log\left (\frac{\Lambda}{|\bk|}\right ),\nn\\
\eea
\bea
\Xi_7\!\!\!&{=}&\!\!\!\frac{|\bq|^2}{16\pi^2}{\int}_{0}^{\infty}d\mu\,{\int}_{0}^{\Lambda}d|\bk|\,\tilde{Q}(\mu,|\bk|)(\cosh{\mu}{-}2\sinh{\mu})|\bk|\nn\\
&&\times\log\left (\frac{\Lambda}{|\bk|}\right ),
\eea
\bea
\Xi_8\!\!\!&{=}&\!\!\!{-}\frac{|\bq|^2}{16\pi^2}{\int}_{0}^{\infty}d\mu\,{\int}_{0}^{\Lambda}d|\bk|\,\tilde{Q}(\mu,|\bk|)e^{{-}2\mu}|\bk|\log\left (\frac{\Lambda}{|\bk|}\right ),\nn\\
\eea
\bea
\Xi_9\!\!\!&{=}&\!\!\!\tfrac{1}{2}\Xi_3,
\eea
\bea
\Xi_{10}\!\!\!&{=}&\!\!\!{-}\Xi_3,
\eea
\bea
\Xi_{11}\!\!\!&{=}&\!\!\!{-}\frac{|\bq|^2}{64\pi^2}{\int}_{0}^{\infty}d\mu\,{\int}_{0}^{\Lambda}d|\bk|\,\tilde{R}(\mu,|\bk|)\cosh{\mu}|\bk|^2\nn\\
&&\times\log\left (\frac{\Lambda}{|\bk|}\right ),
\eea
\bea
\Xi_{12}\!\!\!&{=}&\!\!\!\Xi_{11},
\eea
\bea
\Xi_{13}\!\!\!&{=}&\!\!\!\frac{|\bq|^2}{32\pi^2}{\int}_{0}^{\infty}d\mu\,{\int}_{0}^{\Lambda}d|\bk|\,\tilde{R}(\mu,|\bk|)(\cosh^2{\mu}{-}\tfrac{1}{2})|\bk|^2\nn\\
&&\times\log\left (\frac{\Lambda}{|\bk|}\right ),
\eea
\bea
\Xi_{14}\!\!\!&{=}&\!\!\!{-}\frac{|\bq|^2}{32\pi^2}{\int}_{0}^{\infty}d\mu\,{\int}_{0}^{\Lambda}d|\bk|\,\tilde{R}(\mu,|\bk|)(\cosh^2{\mu}{+}\tfrac{1}{2})|\bk|^2\nn\\
&&\times\log\left (\frac{\Lambda}{|\bk|}\right ),
\eea
\bea
\Xi_{15}\!\!\!&{=}&\!\!\!\frac{|\bq|^2}{16\pi^2}{\int}_{0}^{\infty}d\mu\,{\int}_{0}^{\Lambda}d|\bk|\,\tilde{R}(\mu,|\bk|)|\bk|^2\log\left (\frac{\Lambda}{|\bk|}\right ).\nn\\
\eea
We now combine all of these results together, writing $\Sigma_{3b,i}(q)$ as a sum of an integral
involving $\tilde{Q}$ and another involving $\tilde{R}$, obtaining
\begin{eqnarray}
\Sigma_{3b,i}(q)&{=}&i\frac{g^6}{64\pi}\frac{q_i}{64\pi^2}{\int}_{0}^{\infty}d\mu{\int}_{0}^{\Lambda}d|\bk|\log\left (\frac{\Lambda}{|\bk|}\right )\nn\\
&\times&\Big[\tilde{Q}(\mu,|\bk|)({-}1{+}4e^{{-}\mu}{-}2e^{{-}2\mu})|\bk|\nn\\
&&{+}\tilde{R}(\mu,|\bk|)(1{-}\cosh{\mu})|\bk|^2\Big].
\end{eqnarray}
We first evaluate the $|\bk|$ integrals, obtaining
\begin{eqnarray}
&&{\int}_{0}^{\Lambda}d|\bk|\,\tilde{Q}(\mu,|\bk|)|\bk|\log\left (\frac{\Lambda}{|\bk|}\right )\nn\\
&&{=}\frac{1}{2v_F^2(1{+}\cosh{\mu})^2}\bigg\{\log\left [1{+}\left (\frac{v_F\Lambda}{q_0}\right )^2(1{+}\cosh{\mu})^2\right ]\nn\\
&&{+}\tfrac{1}{2}\mbox{Li}_2\left [{-}\left (\frac{v_F\Lambda}{q_0}\right )^2(1{+}\cosh{\mu})^2\right ]\bigg \}\nn\\
&&\approx{-}\frac{1}{2v_F^2(1{+}\cosh{\mu})^2}\log^2\left [\frac{v_F\Lambda}{|q_0|}(1{+}\cosh{\mu})\right ],\label{intofQt}
\end{eqnarray}
\begin{eqnarray}
&&{\int}_{0}^{\Lambda}d|\bk|\,\tilde{R}(\mu,|\bk|)|\bk|^2\log\left (\frac{\Lambda}{|\bk|}\right )\nn\\
&&{=}\frac{\Lambda^2}{(1{+}\cosh{\mu})[q_0^2{+}v_F^2\Lambda^2(1{+}\cosh{\mu})^2]}\nn\\
&&{-}\frac{3}{2v_F^2(1{+}\cosh{\mu})^3}\log\left [1{+}\left (\frac{v_F\Lambda}{q_0}\right )^2(1{+}\cosh{\mu})^2\right ]\nn\\
&&{-}\frac{1}{2v_F^2(1{+}\cosh{\mu})^3}\mbox{Li}_2\left [{-}\left (\frac{v_F\Lambda}{q_0}\right )^2(1{+}\cosh{\mu})^2\right ]\nn\\
&&\approx\frac{1}{v_F^2(1{+}\cosh{\mu})^3}\log^2\left [\frac{v_F\Lambda}{|q_0|}(1{+}\cosh{\mu})\right ].\label{intofRt}
\end{eqnarray}
We now substitute in these results and drop the $1{+}\cosh{\mu}$ term in the logarithm because it gives us
contributions to the sub{-}leading divergence, and evaluate the remaining integral on $\mu$,
obtaining
\bea
\Sigma_{3b,i}(q)&{=}&\tfrac{1}{128}\left ({-}\tfrac{29}{5}{+}8\log{2}\right )i\alpha^3v_Fq_i\log^2\left (\frac{v_F\Lambda}{|q_0|}\right )\nn\\
&\to&\tfrac{1}{128}\left ({-}\tfrac{29}{5}{+}8\log{2}\right )i\alpha^3v_Fq_i\log^2\left (\frac{\Lambda}{|\bq|}\right ).\nn\\
\eea
Note that, as with the temporal component, we made the replacement, $\frac{|q_0|}{v_F}\to|\bq|$.

Combining all of our results, the full leading divergence of $\Sigma_{3b}(q)$ is
\bea
&&\Sigma_{3b}(q){=}\nn\\
&&i\alpha^3\left [-\tfrac{1}{480}q_0\gamma^0{+}\tfrac{1}{128}\left ({-}\tfrac{29}{5}{+}8\log{2}\right )v_F\bq{\cdot}\vec{\gamma}\right ]\log^2\left (\frac{\Lambda}{|\bq|}\right ).\nn\\
\eea

\subsection{Second vertex correction to two-loop rainbow diagram}
\begin{figure}
\includegraphics[width=0.5\columnwidth]{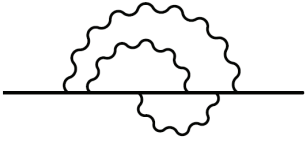}
\caption{Second vertex correction to two-loop rainbow diagram.}
\label{fig:selfenergy3c}
\end{figure}

We consider the diagram shown in Fig.~\ref{fig:selfenergy3c}.  We will
denote the value of this diagram by $\Sigma_{3c}(q)$.  We note that it contains
the second-order diagram shown in Fig.~\ref{fig:selfenergy2a} as a sub-diagram;
its value was denoted by $\Sigma_{2a}(q)$ and given in Eq.~(\ref{selfenergy2a}).
We are interested only in the leading divergence of $\Sigma_{3c}(q)$. We may express $\Sigma_{3c}(q)$
as
\bea
\Sigma_{3c}(q)\!\!\!&{=}&\!\!\!{\int}\frac{d^3k}{(2\pi)^3}\,\frac{g^2}{2|\bq{-}\bk|}i\gamma^0\frac{i\slashed{k}}{k^2}i\left (\log{2}{-}\tfrac{2}{3}\right )\alpha^2
\nn\\&&\!\!\!\times\left (\tfrac{1}{2}k_0\gamma^0{+}v_F\bk{\cdot}\vec{\gamma}\right )\log\left (\frac{\Lambda}{|\bk|}\right )\frac{i\slashed{k}}{k^2}i\gamma^0.
\eea
Using the anticommutation relations for the $\gamma$ matrices, we may rewrite this
as
\bea
\Sigma_{3c}(q)\!\!\!&{=}&\!\!\!2\pi i\left (\log{2}{-}\tfrac{2}{3}\right )\alpha^3v_F{\int}\frac{d^3k}{(2\pi)^3}\,\frac{1}{|\bq{-}\bk|}
\nn\\&&\!\!\!\!\!\!\times\gamma^0\frac{\slashed{k}}{k^2}\left (\tfrac{1}{2}k_0\gamma^0{+}v_F\bk{\cdot}\vec{\gamma}\right )\log\left (\frac{\Lambda}{|\bk|}\right )\frac{\slashed{k}}{k^2}\gamma^0.\nn\\
\eea
We now note that the denominator of the integrand is an even function of $k_0$, so
that terms in the numerator that are odd functions of $k_0$ vanish under integration.
Keeping only the non-vanishing terms, this becomes
\bea
\Sigma_{3c}(q)\!\!\!&{=}&\!\!\!-2\pi i\left (\log{2}{-}\tfrac{2}{3}\right )\alpha^3v_F{\int}\frac{d^3k}{(2\pi)^3}\,\frac{1}{|\bq{-}\bk|}
\nn\\&&\!\!\!\times\left (\frac{v_F^2|\bk|^2}{k^4}\right )v_F\bk{\cdot}\vec{\gamma}\log\left (\frac{\Lambda}{|\bk|}\right ).
\eea
We may now evaluate the integral over $k_0$, obtaining
\bea
\Sigma_{3c}(q)&{=}&-\tfrac{1}{2}\pi i\left (\log{2}{-}\tfrac{2}{3}\right )\alpha^3v_F{\int}\frac{d^2k}{(2\pi)^2}\,\frac{1}{|\bk||\bq{-}\bk|}
\nn\\&&\times\log\left (\frac{\Lambda}{|\bk|}\right )\bk{\cdot}\vec{\gamma}.
\eea
If we were to now choose our coordinate system such that $\bq$ lies along, say, the
$x$ axis, then one may see that the integrand is odd in $k_y$.  Therefore, we may make
the replacement,
\begin{equation}
\bk\cdot\vec{\gamma}\rightarrow\frac{\bk\cdot\bq}{|\bq|^2}\bq\cdot\vec{\gamma},
\end{equation}
thus obtaining
\bea
\!\!\!\Sigma_{3c}(q)\!\!\!&=&\!\!\!-\tfrac{1}{2}\pi i\left (\log{2}-\tfrac{2}{3}\right )\alpha^3v_F\times
\nn\\&&\!\!\!\times\int\frac{d^2k}{(2\pi)^2}\,\frac{1}{|\bk||\bq-\bk|}\frac{\bk\cdot\bq}{|\bq|^2}\log\left (\frac{\Lambda}{|\bk|}\right )\bq\cdot\vec{\gamma}.\nn\\
\eea
Because we are interested in the leading ultraviolet divergence, we now expand $\frac{1}{|\bq-\bk|}$
for large $|\bk|$:
\begin{equation}
\frac{1}{|\bq-\bk|}\approx\frac{1}{|\bk|}\left (1+\frac{\bk\cdot\bq}{|\bk|^2}\right )
\end{equation}
Na\"ively, we would expect that the leading term in this expansion will give us a linearly
divergent contribution to $\Sigma_{3c}(q)$.  However, this term is an odd function
of both components of $\bk$, so that it drops out.  The leading non-zero correction is
thus given by the {\it sub}-leading term, which only gives us a logarithmic divergence.
All higher terms are ultraviolet convergent.  The resulting integral for the leading
divergence is easily evaluated:
\bea
\Sigma_{3c}(q)\!\!\!&{=}&\!\!\!-\tfrac{1}{2}\pi i\left (\log{2}{-}\tfrac{2}{3}\right )\alpha^3v_F{\int}\frac{d^2k}{(2\pi)^2}\,\frac{1}{|\bk|^4}\frac{(\bk{\cdot}\bq)^2}{|\bq|^2}
\nn\\&&\qquad\times\log\left (\frac{\Lambda}{|\bk|}\right )\bq{\cdot}\vec{\gamma}
\nn\\\!\!\!&{=}&\!\!\!-\tfrac{1}{16}i\left (\log{2}{-}\tfrac{2}{3}\right )\alpha^3v_F\log^2\left (\frac{\Lambda}{|\bq|}\right )\bq{\cdot}\vec{\gamma}.
\eea
In evaluating this integral, we assume that the {\it infrared} divergence of this integral
is regularized by imposing a lower limit of $|\bq|$ on $|\bk|$.

\subsection{Self-energy correction to two-loop vertex correction}
\begin{figure}
\includegraphics[width=0.5\columnwidth]{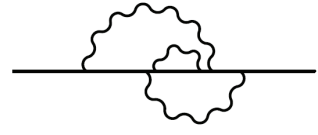}
\caption{Self-energy correction to two-loop vertex correction.}
\label{fig:selfenergy3e}
\end{figure}

We now turn our attention to the diagram in Fig.~\ref{fig:selfenergy3e}, the value of which we will denote by $\Sigma_{3e}(q)$. This diagram is
similar in appearance to Fig.~\ref{fig:selfenergy3b}, and indeed the evaluation of $\Sigma_{3e}(q)$ will be very similar to the
calculation of $\Sigma_{3b}(q)$ given above. The expression for this correction is
\begin{eqnarray}
&&\!\!\!\!\!\!\Sigma_{3e}(q){=}\nn\\&&\!\!\!\!\!\!{\int}\frac{d^3k}{(2\pi)^3}{\int}\frac{d^3p}{(2\pi)^3}\frac{g^2}{2|\bq{-}\bk|}\frac{g^2}{2|\bq{-}\bp|}i\gamma^0\frac{i\slashed{k}}{k^2}i\gamma^0\frac{i(\slashed{k}{+}\slashed{p}{-}\slashed{q})}{(k{+}p{-}q)^2}\cr
&&\!\!\!\!\!\!\times\frac{ig^2}{16\pi}(\bk{+}\bp{-}\bq){\cdot}\vec{\gamma}\log\left (\frac{\Lambda}{|\bk{+}\bp{-}\bq|}\right )\frac{i(\slashed{k}{+}\slashed{p}{-}\slashed{q})}{(k{+}p{-}q)^2}i\gamma^0\frac{i\slashed{p}}{p^2}i\gamma^0.\nn\\
\end{eqnarray}
To simplify the subsequent calculations, we make the substitution, $p\to {-}p{-}k{+}q$, obtaining
\begin{eqnarray}
\Sigma_{3e}(q)\!\!\!&{=}&\!\!\!i\frac{g^6}{64\pi}{\int}\frac{d^3k}{(2\pi)^3}{\int}\frac{d^3p}{(2\pi)^3}\frac{1}{|\bq{-}\bk||\bk{+}\bp|}\gamma^0\frac{\slashed{k}}{k^2}\gamma^0\frac{\slashed{p}}{p^2}\bp{\cdot}\vec{\gamma}\nn\\
&&\!\!\!\times\log\left (\frac{\Lambda}{|\bp|}\right )\frac{\slashed{p}}{p^2}\gamma^0\frac{\slashed{k}{+}\slashed{p}{-}\slashed{q}}{(k{+}p{-}q)^2}\gamma^0.
\end{eqnarray}
We may now use the identity,
\begin{equation}
\slashed{k}\bk{\cdot}\vec{\gamma}\slashed{k}{=}{-}(k_0^2{-}v_F^2|\bk|^2)\bk{\cdot}\vec{\gamma}{+}2v_F|\bk|^2k_0\gamma^0,
\end{equation}
to express the above as
\begin{eqnarray}
\Sigma_{3e}(q)\!\!\!&{=}&\!\!\!i\frac{g^6}{64\pi}{\int}\frac{d^3k}{(2\pi)^3}{\int}\frac{d^3p}{(2\pi)^3}\frac{1}{|\bq{-}\bk||\bk{+}\bp|}\frac{1}{k^2p^4(k{+}p{-}q)^2}\cr
&&\!\!\!\!\!\!\times(k_0\gamma^0{-}v_F\bk{\cdot}\vec{\gamma})[{-}(p_0^2{-}v_F^2|\bp|^2)\bp{\cdot}\vec{\gamma}{+}2v_F|\bp|^2p_0\gamma^0]\nn\\
&&\!\!\!\!\!\!\times[(k_0{+}p_0{-}q_0)\gamma^0{-}v_F(\bk{+}\bp{-}\bq){\cdot}\vec{\gamma}]\log\left (\frac{\Lambda}{|\bp|}\right ).
\end{eqnarray}
Evaluation of the $k_0$ and $p_0$ integrals is similar to the calculation of $\Sigma_{3b}$; the result is
\begin{eqnarray}
&&\!\!\!\!\!\!\Sigma_{3e}(q){=}\nn\\&&\!\!\!\!\!\!i\frac{g^6}{64\pi}{\int}\frac{d^2k}{(2\pi)^2}\,{\int}\frac{d^2p}{(2\pi)^2}\,\frac{1}{|\bq{-}\bk||\bk{+}\bp|}\log\left (\frac{\Lambda}{|\bp|}\right )\cr
&&\!\!\!\!\!\!\times[2v_F|\bp|^2\tilde{B}_1\gamma^0{-}2v_F^2|\bp|^2\tilde{B}_2\bk{\cdot}\vec{\gamma}{-}2v_F^2|\bp|^2\tilde{B}_3(\bk{+}\bp{-}\bq){\cdot}\vec{\gamma}\cr
&&\!\!\!\!\!\!{-}2v_F^3|\bp|^2\tilde{B}_4\gamma^0\bk{\cdot}\vec{\gamma}(\bk{+}\bp{-}\bq){\cdot}\vec{\gamma}{+}\tilde{B}_5\bp{\cdot}\vec{\gamma}{+}v_F\tilde{B}_6\gamma^0\bk{\cdot}\vec{\gamma}\bp{\cdot}\vec{\gamma}\cr
&&\!\!\!\!\!\!{+}v_F\tilde{B}_7\gamma^0\bp{\cdot}\vec{\gamma}(\bk{+}\bp{-}\bq){\cdot}\vec{\gamma}{-}v_F^2\tilde{B}_8\bk{\cdot}\vec{\gamma}\bp{\cdot}\vec{\gamma}(\bk{+}\bp{-}\bq){\cdot}\vec{\gamma}],\nn\\
\end{eqnarray}
where the $\tilde{B}_i$ are identical to the corresponding $B_i$ in the calculation of $\Sigma_{3b}$, except
that $\bk$ and $\bp$ are interchanged.  We now extract the temporal and spatial components, obtaining
\begin{eqnarray}
\Sigma_{3e,0}(q)\!\!\!&{=}&\!\!\!i\frac{g^6}{64\pi}q_0{\int}\frac{d^2k}{(2\pi)^2}{\int}\frac{d^2p}{(2\pi)^2} \frac{1}{|\bq{-}\bk||\bk{+}\bp|}\log\left (\frac{\Lambda}{|\bp|}\right ) \nn\\&&\!\!\!\!\!\!\times\frac{v_F|\bp|(|\bk|{+}|\bp|{+}|\bk{+}\bp{-}\bq|)}{2[q_0^2{+}v_F^2(|\bk|{+}|\bp|{+}|\bk{+}\bp{-}\bq|)^2]^2}\cr
&&\!\!\!\!\!\!\times\left [1{-}\frac{\bp{\cdot}(\bk{+}\bp{-}\bq)}{|\bp||\bk{+}\bp{-}\bq|}{+}\frac{\bk{\cdot}\bp}{|\bk||\bp|}{-}\frac{\bk{\cdot}(\bk{+}\bp{-}\bq)}{|\bk||\bk{+}\bp{-}\bq|}\right ],
\end{eqnarray}
and
\begin{eqnarray}
\Sigma_{3e,i}(q)\!\!\!&{=}&\!\!\!i\frac{g^6}{64\pi}\frac{q_i}{|\bq|^2}{\int}\frac{d^2k}{(2\pi)^2}{\int}\frac{d^2p}{(2\pi)^2}\frac{1}{|\bq{-}\bk||\bk{+}\bp|} \nn\\&&\!\!\!\!\!\!\times\frac{|\bp|[q_0^2{-}v_F^2(|\bk|{+}|\bp|{+}|\bk{+}\bp{-}\bq|)^2]}{4[q_0^2{+}v_F^2(|\bk|{+}|\bp|{+}|\bk{+}\bp{-}\bq|)^2]^2}\cr
&&\!\!\!\!\!\!\times\bigg [\frac{\bp{\cdot}\bq}{|\bp|}{-}\frac{(\bk{+}\bp{-}\bq){\cdot}\bq}{|\bk{+}\bp{-}\bq|}{+}\frac{\bk{\cdot}\bq}{|\bk|}\nn\\&&\!\!\!\!\!\! {-}\frac{|\bp|^2\bk{\cdot}\bq{-}|\bq|^2\bk{\cdot}\bp{-}|\bk|^2\bp{\cdot}\bq{+}2(\bk{\cdot}\bp)(\bk{\cdot}\bq)}{|\bk||\bp||\bk{+}\bp{-}\bq|}\bigg ]\cr
&&\!\!\!\!\!\!\times\log\left (\frac{\Lambda}{|\bp|}\right ).
\end{eqnarray}

\subsubsection{Extracting the divergence of the temporal component}
As before, we begin by extracting the divergence of the temporal component.  In this case, we
simply set $\bq{=}0$ in the integrand to do so:
\begin{eqnarray}
\Sigma_{3e,0}(q)\!\!\!&\approx&\!\!\!i\frac{g^6}{128\pi}q_0{\int}\frac{d^2k}{(2\pi)^2}{\int}\frac{d^2p}{(2\pi)^2}\frac{|\bp|}{|\bk||\bk{+}\bp|} \nn\\&&\!\!\!\!\!\!\times\frac{v_F(|\bk|{+}|\bp|{+}|\bk{+}\bp|)}{[q_0^2{+}v_F^2(|\bk|{+}|\bp|{+}|\bk{+}\bp|)^2]^2}\cr
&&\!\!\!\!\!\!\times\left [1{-}\frac{\bp{\cdot}(\bk{+}\bp)}{|\bp||\bk{+}\bp|}{+}\frac{\bk{\cdot}\bp}{|\bk||\bp|}{-}\frac{\bk{\cdot}(\bk{+}\bp)}{|\bk||\bk{+}\bp|}\right ]\log\left (\frac{\Lambda}{|\bp|}\right )\cr
&&\!\!\!\!\!\!{=}i\frac{g^6}{128\pi}q_0(\tilde{\Psi}_1{-}\tilde{\Psi}_2{-}\tilde{\Psi}_3{+}\tilde{\Psi}_4),
\end{eqnarray}
where the $\tilde{\Psi}_i$ are given by
\begin{eqnarray}
\tilde{\Psi}_1\!\!\!&{=}&\!\!\!{\int}\frac{d^2k}{(2\pi)^2}{\int}\frac{d^2p}{(2\pi)^2}\frac{|\bp|}{|\bk||\bk{+}\bp|}S(\vec{k},\vec{p})\log\left (\frac{\Lambda}{|\bp|}\right ), \\
\tilde{\Psi}_2\!\!\!&{=}&\!\!\!{\int}\frac{d^2k}{(2\pi)^2}{\int}\frac{d^2p}{(2\pi)^2}\frac{|\bp|}{|\bk||\bk{+}\bp|}S(\vec{k},\vec{p})\frac{\bk{\cdot}(\bk{+}\bp)}{|\bk||\bk{+}\bp|}\log\left (\frac{\Lambda}{|\bp|}\right ), \nonumber\\ \\
\tilde{\Psi}_3\!\!\!&{=}&\!\!\!{\int}\frac{d^2k}{(2\pi)^2}{\int}\frac{d^2p}{(2\pi)^2}\frac{|\bp|}{|\bk||\bk{+}\bp|}S(\vec{k},\vec{p})\frac{\bp{\cdot}(\bk{+}\bp)}{|\bp||\bk{+}\bp|}\log\left (\frac{\Lambda}{|\bp|}\right ), \nonumber\\ \\
\tilde{\Psi}_4\!\!\!&{=}&\!\!\!{\int}\frac{d^2k}{(2\pi)^2}{\int}\frac{d^2p}{(2\pi)^2}\frac{|\bp|}{|\bk||\bk{+}\bp|}S(\vec{k},\vec{p})\frac{\bk{\cdot}\bp}{|\bk||\bp|}\log\left (\frac{\Lambda}{|\bp|}\right ),\nn\\
\end{eqnarray}
with
\beq
S(\vec{k},\vec{p})=\frac{v_F(|\bk|{+}|\bp|{+}|\bk{+}\bp|)}{[q_0^2{+}v_F^2(|\bk|{+}|\bp|{+}|\bk{+}\bp|)^2]^2}.
\eeq
We now rewrite the integral over $\bk$ in terms of elliptic coordinates, similarly to how
we did before; the result is
\begin{eqnarray}
\tilde{\Psi}_1\!\!\!&{=}&\!\!\!\frac{1}{2\pi^2}{\int}_{0}^{\infty}d\mu\,{\int}_{0}^{2\pi}d\nu\,f(\mu),\nn \\
\tilde{\Psi}_2\!\!\!&{=}&\!\!\!\frac{1}{2\pi^2}{\int}_{0}^{\infty}d\mu\,{\int}_{0}^{2\pi}d\nu\,f(\mu)\frac{\cosh^2{\mu}{+}\cos^2{\nu}{-}2}{\cosh^2{\mu}{-}\cos^2{\nu}},\nn \\
\tilde{\Psi}_3\!\!\!&{=}&\!\!\!\frac{1}{2\pi^2}{\int}_{0}^{\infty}d\mu\,{\int}_{0}^{2\pi}d\nu\,f(\mu)\frac{1{+}\cosh{\mu}\cos{\nu}}{\cosh{\mu}{+}\cos{\nu}},\nn \\
\tilde{\Psi}_4\!\!\!&{=}&\!\!\!\frac{1}{2\pi^2}{\int}_{0}^{\infty}d\mu\,{\int}_{0}^{2\pi}d\nu\,f(\mu)\frac{\cosh{\mu}\cos{\nu}{-}1}{\cosh{\mu}{-}\cos{\nu}},
\end{eqnarray}
where $f(\mu)$ is defined in Eq.~(\ref{defoffofmu}).  If we now evaluate the $\nu$ integrals,
we obtain
\begin{eqnarray}
\tilde{\Psi}_1&{=}&\frac{1}{\pi}{\int}_{0}^{\infty}d\mu f(\mu), \cr
\tilde{\Psi}_2&{=}&\frac{1}{\pi}{\int}_{0}^{\infty}d\mu f(\mu)(2\tanh{\mu}{-}1), \cr
\tilde{\Psi}_3&{=}&\frac{1}{\pi}{\int}_{0}^{\infty}d\mu f(\mu)e^{{-}\mu}, \cr
\tilde{\Psi}_4&{=}&{-}\tilde{\Psi}_3.
\end{eqnarray}
Combining these results together, and using the asymptotic form of $f(\mu)$ derived earlier,
we obtain
\begin{eqnarray}
&&\!\!\!\!\!\!\Sigma_{3e,0}(q){=} \nn\\&&\!\!\!\!\!\!i\frac{g^6}{128\pi}\frac{q_0}{8\pi^2v_F^3}{\int}_0^{\infty}d\mu \frac{2(1{-}\tanh{\mu}{-}e^{{-}\mu})}{(1{+}\cosh{\mu})^3}\log^2\left (\frac{v_F\Lambda}{|q_0|}\right ) \cr
&&\!\!\!\!\!\!{=}\tfrac{1}{8}\left (\tfrac{41}{60}{-}\log{2}\right )i\alpha^3 q_0\log^2\left (\frac{v_F\Lambda}{|q_0|}\right )\nn\\&&\!\!\!\!\!\!\to\tfrac{1}{8}\left (\tfrac{41}{60}{-}\log{2}\right )i\alpha^3 q_0\log^2\left (\frac{\Lambda}{|\bq|}\right ).
\end{eqnarray}

\subsubsection{Extracting the divergence of the spatial component}
We now look at the spatial component.  If we once again shift $\bk$ by $\bq$ and then expand
the integrand to linear order in $\bq$, we obtain
\begin{eqnarray}
&&\!\!\!\!\!\!\Sigma_{3e,i}(q)\approx \nn\\&&\!\!\!\!\!\!i\frac{g^6}{256\pi}\frac{q_i}{|\bq|^2}{\int}\frac{d^2k}{(2\pi)^2}{\int}\frac{d^2p}{(2\pi)^2}\frac{|\bp|}{|\bk||\bk{+}\bp|}\left[1{-}\frac{(\bk{+}\bp){\cdot}\bq}{|\bk{+}\bp|^2}\right] \nn\\&&\!\!\!\!\!\!\times\left[Q(\bk,\bp){+}R(\bk,\bp)\frac{\bk{\cdot}\bq}{|\bk|}\right]\bigg\{\frac{\bp{\cdot}\bq}{|\bp|}{-}\frac{(\bk{+}\bp){\cdot}\bq}{|\bk{+}\bp|} \nn\\&&\!\!\!\!\!\!{+}\bigg[\frac{(\bk{+}\bq){\cdot}\bq}{|\bk|}{-}\frac{|\bp|(\bk{+}\bq){\cdot}\bq}{|\bk||\bk{+}\bp|}{+}\frac{|\bq|^2(\bk{+}\bq){\cdot}\bp}{|\bk||\bp||\bk{+}\bp|}\cr
&&\!\!\!\!\!\!{-}\frac{2[(\bk{+}\bq){\cdot}\bp][(\bk{+}\bq){\cdot}\bq)}{|\bk||\bp||\bk{+}\bp|}\bigg]\left(1{-}\frac{\bk{\cdot}\bq}{|\bk|^2}\right) \cr &&\!\!\!\!\!\!{+}\frac{|\bk|\bp{\cdot}\bq}{|\bp||\bk{+}\bp|}\left(1{+}\frac{\bk{\cdot}\bq}{|\bk|^2}\right)\bigg\}\log\left(\frac{\Lambda}{|\bp|}\right).\nn\\
\end{eqnarray}
The functions $Q(\bk,\bp)$ and $R(\bk,\bp)$ were defined in Eq.~(\ref{defofQandR}). Upon multiplying the integrand out, we again obtain terms that appear to be linearly divergent, but turn out to be zero.  We now enumerate the logarithmically divergent terms, of which there
are now seventeen:
\bea
\tilde{\Xi}_1\!\!\!&{=}&\!\!\!{\int}\frac{d^2k}{(2\pi)^2}\,{\int}\frac{d^2p}{(2\pi)^2}\,\frac{|\bp|Q(\bk,\bp)}{|\bk||\bk{+}\bp|}\left [\frac{|\bq|^2}{|\bk|}{-}\frac{(\bk{\cdot}\bq)^2}{|\bk|^3}\right ]\nn\\&&\!\!\!\!\!\!\times\log\left (\frac{\Lambda}{|\bp|}\right ),
\eea
\bea
\tilde{\Xi}_2\!\!\!&{=}&\!\!\!{-}{\int}\frac{d^2k}{(2\pi)^2}\,{\int}\frac{d^2p}{(2\pi)^2}\,\frac{|\bp|Q(\bk,\bp)}{|\bk||\bk{+}\bp|}\frac{|\bp|}{|\bk{+}\bp|}\left [\frac{|\bq|^2}{|\bk|}{-}\frac{(\bk{\cdot}\bq)^2}{|\bk|^3}\right ]\nn\\&&\!\!\!\!\!\!\times\log\left (\frac{\Lambda}{|\bp|}\right ),
\eea
\bea
\tilde{\Xi}_3\!\!\!&{=}&\!\!\!{\int}\frac{d^2k}{(2\pi)^2}\,{\int}\frac{d^2p}{(2\pi)^2}\,\frac{|\bp|Q(\bk,\bp)}{|\bk||\bk{+}\bp|}\frac{|\bq|^2\bk{\cdot}\bp}{|\bk||\bp||\bk{+}\bp|}\log\left (\frac{\Lambda}{|\bp|}\right ),\nn\\
\eea
\bea
\tilde{\Xi}_4\!\!\!&{=}&\!\!\!{-}2{\int}\frac{d^2k}{(2\pi)^2}\,{\int}\frac{d^2p}{(2\pi)^2}\,\frac{|\bp|Q(\bk,\bp)}{|\bk||\bk{+}\bp|}\frac{(\bk{\cdot}\bq)(\bp{\cdot}\bq)}{|\bk||\bp||\bk{+}\bp|}\log\left (\frac{\Lambda}{|\bp|}\right ) \cr
&&\!\!\!\!\!\!{+}2{\int}\frac{d^2k}{(2\pi)^2}\,{\int}\frac{d^2p}{(2\pi)^2}\,\frac{|\bp|Q(\bk,\bp)}{|\bk||\bk{+}\bp|}\frac{(\bk{\cdot}\bp)(\bk{\cdot}\bq)^2}{|\bk|^3|\bp||\bk{+}\bp|}\log\left (\frac{\Lambda}{|\bp|}\right )\nn\\&&\!\!\!\!\!\!{-}2\tilde{\Xi}_3,
\eea
\bea
\tilde{\Xi}_5\!\!\!&{=}&\!\!\!{\int}\frac{d^2k}{(2\pi)^2}\,{\int}\frac{d^2p}{(2\pi)^2}\,\frac{|\bp|Q(\bk,\bp)}{|\bk||\bk{+}\bp|}\frac{(\bk{\cdot}\bq)(\bp{\cdot}\bq)}{|\bk||\bp||\bk{+}\bp|}\log\left (\frac{\Lambda}{|\bp|}\right ),\nn\\
\eea
\bea
\tilde{\Xi}_6\!\!\!&{=}&\!\!\!{-}{\int}\frac{d^2k}{(2\pi)^2}\,{\int}\frac{d^2p}{(2\pi)^2}\,\frac{|\bp|Q(\bk,\bp)}{|\bk||\bk{+}\bp|}\frac{(\bp{\cdot}\bq)[(\bk{+}\bp){\cdot}\bq]}{|\bp||\bk{+}\bp|^2} \nn\\&&\!\!\!\!\!\!\times\log\left (\frac{\Lambda}{|\bp|}\right ),
\eea
\bea
\tilde{\Xi}_7\!\!\!&{=}&\!\!\!{\int}\frac{d^2k}{(2\pi)^2}\,{\int}\frac{d^2p}{(2\pi)^2}\,\frac{|\bp|Q(\bk,\bp)}{|\bk||\bk{+}\bp|}\frac{[(\bk{+}\bp){\cdot}\bq]^2}{|\bk{+}\bp|^3}\log\left (\frac{\Lambda}{|\bp|}\right ),\nn\\
\eea
\bea
\tilde{\Xi}_8\!\!\!&{=}&\!\!\!{-}{\int}\frac{d^2k}{(2\pi)^2}\,{\int}\frac{d^2p}{(2\pi)^2}\,\frac{|\bp|Q(\bk,\bp)}{|\bk||\bk{+}\bp|}\frac{(\bk{\cdot}\bq)[(\bk{+}\bp){\cdot}\bq]}{|\bk||\bk{+}\bp|^2} \nn\\&&\!\!\!\!\!\!\times\log\left (\frac{\Lambda}{|\bp|}\right ),
\eea
\bea
\tilde{\Xi}_9\!\!\!&{=}&\!\!\!{\int}\frac{d^2k}{(2\pi)^2}\,{\int}\frac{d^2p}{(2\pi)^2}\,\frac{|\bp|Q(\bk,\bp)}{|\bk||\bk{+}\bp|}\frac{|\bp|(\bk{\cdot}\bq)[(\bk{+}\bp){\cdot}\bq]}{|\bk||\bk{+}\bp|^3} \nn\\&&\!\!\!\!\!\!\times\log\left (\frac{\Lambda}{|\bp|}\right ),
\eea
\bea
\tilde{\Xi}_{10}\!\!\!&{=}&\!\!\!{-}{\int}\frac{d^2k}{(2\pi)^2}\,{\int}\frac{d^2p}{(2\pi)^2}\,\frac{|\bp|Q(\bk,\bp)}{|\bk||\bk{+}\bp|}\frac{|\bk|(\bp{\cdot}\bq)[(\bk{+}\bp){\cdot}\bq]}{|\bp||\bk{+}\bp|^3} \nn\\&&\!\!\!\!\!\!\times\log\left (\frac{\Lambda}{|\bp|}\right ),
\eea
\bea
\tilde{\Xi}_{11}\!\!\!&{=}&\!\!\!{\int}\frac{d^2k}{(2\pi)^2}\,{\int}\frac{d^2p}{(2\pi)^2}\,\frac{|\bp|R(\bk,\bp)}{|\bk||\bk{+}\bp|}\frac{(\bk{\cdot}\bq)(\bp{\cdot}\bq)}{|\bk||\bp|}\log\left (\frac{\Lambda}{|\bp|}\right ),\nn\\
\eea
\bea
\tilde{\Xi}_{12}\!\!\!&{=}&\!\!\!{-}{\int}\frac{d^2k}{(2\pi)^2}\,{\int}\frac{d^2p}{(2\pi)^2}\,\frac{|\bp|R(\bk,\bp)}{|\bk||\bk{+}\bp|}\frac{(\bk{\cdot}\bq)[(\bk{+}\bp){\cdot}\bq]}{|\bk||\bk{+}\bp|} \nn\\&&\!\!\!\!\!\!\times\log\left (\frac{\Lambda}{|\bp|}\right ),
\eea
\bea
\tilde{\Xi}_{13}\!\!\!&{=}&\!\!\!{\int}\frac{d^2k}{(2\pi)^2}\,{\int}\frac{d^2p}{(2\pi)^2}\,\frac{|\bp|R(\bk,\bp)}{|\bk||\bk{+}\bp|}\left (\frac{\bk{\cdot}\bq}{|\bk|}\right )^2\log\left (\frac{\Lambda}{|\bp|}\right ),\nn\\
\eea
\bea
\tilde{\Xi}_{14}\!\!\!&{=}&\!\!\!{-}{\int}\frac{d^2k}{(2\pi)^2}\,{\int}\frac{d^2p}{(2\pi)^2}\,\frac{|\bp|R(\bk,\bp)}{|\bk||\bk{+}\bp|}\frac{|\bp|}{|\bk{+}\bp|}\left (\frac{\bk{\cdot}\bq}{|\bk|}\right )^2\nn\\&&\!\!\!\!\!\!\times\log\left (\frac{\Lambda}{|\bp|}\right ),
\eea
\bea
\tilde{\Xi}_{15}\!\!\!&{=}&\!\!\!{\int}\frac{d^2k}{(2\pi)^2}\,{\int}\frac{d^2p}{(2\pi)^2}\,\frac{|\bp|R(\bk,\bp)}{|\bk||\bk{+}\bp|}\frac{(\bk{\cdot}\bq)(\bp{\cdot}\bq)}{|\bp||\bk{+}\bp|}\log\left (\frac{\Lambda}{|\bp|}\right ),\nn\\
\eea
\bea
\tilde{\Xi}_{16}\!\!\!&{=}&\!\!\!2{\int}\frac{d^2k}{(2\pi)^2}\,{\int}\frac{d^2p}{(2\pi)^2}\,\frac{|\bp|Q(\bk,\bp)}{|\bk||\bk{+}\bp|}\frac{(\bk{\cdot}\bp)(\bk{\cdot}\bq)[(\bk{+}\bp){\cdot}\bq]}{|\bk||\bp||\bk{+}\bp|^3} \nn\\&&\!\!\!\!\!\!\times\log\left (\frac{\Lambda}{|\bp|}\right ),
\eea
\bea
\tilde{\Xi}_{17}\!\!\!&{=}&\!\!\!{-}2{\int}\frac{d^2k}{(2\pi)^2}\,{\int}\frac{d^2p}{(2\pi)^2}\,\frac{|\bp|R(\bk,\bp)}{|\bk||\bk{+}\bp|}\frac{(\bk{\cdot}\bp)(\bk{\cdot}\bq)^2}{|\bk|^2|\bp||\bk{+}\bp|} \nn\\&&\!\!\!\!\!\!\times\log\left (\frac{\Lambda}{|\bp|}\right ).
\eea
We now rewrite the $\bp$ integral in terms of elliptic coordinates.  Upon
performing the $\nu$ and $\theta$ integrals, we obtain
\bea
\tilde{\Xi}_1\!\!\!&{=}&\!\!\!\frac{|\bq|^2}{32\pi^2}{\int}_{0}^{\infty}d\mu{\int}_{0}^{\Lambda}d|\bk|\tilde{Q}(\mu,|\bk|)|\bk|\log\left (\frac{\Lambda}{|\bk|}\right )\nn\\&&\!\!\!\!\!\!\times(\cosh^2{\mu}{+}\tfrac{1}{2}),
\eea
\bea
\tilde{\Xi}_2\!\!\!&{=}&\!\!\!{-}\frac{|\bq|^2}{32\pi^2}{\int}_{0}^{\infty}d\mu\,{\int}_{0}^{\Lambda}d|\bk|\,\tilde{Q}(\mu,|\bk|)|\bk|\log\left (\frac{\Lambda}{|\bk|}\right )\nn\\&&\!\!\!\!\!\!\times(8\coth{\mu}{-}4e^{{-}2\mu}{+}\tfrac{1}{2}\cosh{2\mu}{-}4),
\eea
\bea
\tilde{\Xi}_3\!\!\!&{=}&\!\!\!\frac{|\bq|^2}{8\pi^2}{\int}_{0}^{\infty}d\mu\,{\int}_{0}^{\Lambda}d|\bk|\,\tilde{Q}(\mu,|\bk|)|\bk|\log\left (\frac{\Lambda}{|\bk|}\right )\nn\\&&\!\!\!\!\!\!\times(2{+}e^{{-}2\mu}{-}4\coth{\mu}),
\eea
\bea
\tilde{\Xi}_4\!\!\!&{=}&\!\!\!{-}2\tilde{\Xi}_3,
\eea
\bea
\tilde{\Xi}_5\!\!\!&{=}&\!\!\!\tfrac{1}{2}\tilde{\Xi}_3,
\eea
\bea
\tilde{\Xi}_6\!\!\!&{=}&\!\!\!{-}\frac{|\bq|^2}{16\pi^2}{\int}_{0}^{\infty}d\mu\,{\int}_{0}^{\Lambda}d|\bk|\,\tilde{Q}(\mu,|\bk|)|\bk|\log\left (\frac{\Lambda}{|\bk|}\right )\nn\\&&\!\!\!\!\!\!\times(2e^{{-}\mu}{+}\cosh{\mu}),
\eea
\bea
\tilde{\Xi}_7\!\!\!&{=}&\!\!\!\frac{|\bq|^2}{16\pi^2}{\int}_{0}^{\infty}d\mu\,{\int}_{0}^{\Lambda}d|\bk|\,\tilde{Q}(\mu,|\bk|)|\bk|\log\left (\frac{\Lambda}{|\bk|}\right )\nn\\&&\!\!\!\!\!\!\times(4\coth{\mu}{-}3)\cosh{\mu},
\eea
\bea
\tilde{\Xi}_8\!\!\!&{=}&\!\!\!\frac{|\bq|^2}{16\pi^2}{\int}_{0}^{\infty}d\mu\,{\int}_{0}^{\Lambda}d|\bk|\,\tilde{Q}(\mu,|\bk|)|\bk|\log\left (\frac{\Lambda}{|\bk|}\right )\nn\\&&\!\!\!\!\!\!\times(1{+}2e^{{-}2\mu}),
\eea
\bea
\tilde{\Xi}_9\!\!\!&{=}&\!\!\!\frac{|\bq|^2}{16\pi^2}{\int}_{0}^{\infty}d\mu\,{\int}_{0}^{\Lambda}d|\bk|\,\tilde{Q}(\mu,|\bk|)|\bk|\log\left (\frac{\Lambda}{|\bk|}\right )\nn\\&&\!\!\!\!\!\!\times(2{+}3e^{{-}2\mu}{-}4\coth^3{\mu}),
\eea
\bea
\tilde{\Xi}_{10}\!\!\!&{=}&\!\!\!{-}\frac{|\bq|^2}{8\pi^2}{\int}_{0}^{\infty}d\mu\,{\int}_{0}^{\Lambda}d|\bk|\,\tilde{Q}(\mu,|\bk|)|\bk|\log\left (\frac{\Lambda}{|\bk|}\right )\nn\\&&\!\!\!\!\!\!\times(2\coth^3{\mu}{-}1),
\eea
\bea
\tilde{\Xi}_{11}\!\!\!&{=}&\!\!\!{-}\frac{3|\bq|^2}{64\pi^2}{\int}_{0}^{\infty}d\mu\,{\int}_{0}^{\Lambda}d|\bk|\,\tilde{R}(\mu,|\bk|)|\bk|^2\log\left (\frac{\Lambda}{|\bk|}\right )\nn\\&&\!\!\!\!\!\!\times\cosh{\mu},
\eea
\bea
\tilde{\Xi}_{12}\!\!\!&{=}&\!\!\!{-}\frac{|\bq|^2}{64\pi^2}{\int}_{0}^{\infty}d\mu\,{\int}_{0}^{\Lambda}d|\bk|\,\tilde{R}(\mu,|\bk|)|\bk|^2\log\left (\frac{\Lambda}{|\bk|}\right )\nn\\&&\!\!\!\!\!\!\times(2\cosh{3\mu}{-}8\cosh^2{\mu}\sinh{\mu}{+}\cosh{\mu}),
\eea
\bea
\tilde{\Xi}_{13}\!\!\!&{=}&\!\!\!\frac{|\bq|^2}{32\pi^2}{\int}_{0}^{\infty}d\mu\,{\int}_{0}^{\Lambda}d|\bk|\,\tilde{R}(\mu,|\bk|)|\bk|^2\log\left (\frac{\Lambda}{|\bk|}\right )\nn\\&&\!\!\!\!\!\!\times(\cosh^2{\mu}{+}\tfrac{1}{2}),
\eea
\bea
\tilde{\Xi}_{14}\!\!\!&{=}&\!\!\!{-}\frac{|\bq|^2}{32\pi^2}{\int}_{0}^{\infty}d\mu\,{\int}_{0}^{\Lambda}d|\bk|\,\tilde{R}(\mu,|\bk|)|\bk|^2\log\left (\frac{\Lambda}{|\bk|}\right )\nn\\&&\!\!\!\!\!\!\times(8\coth{\mu}{-}4e^{{-}2\mu}{+}\tfrac{1}{2}\cosh{2\mu}{-}4),
\eea
\bea
\tilde{\Xi}_{15}\!\!\!&{=}&\!\!\!\frac{|\bq|^2}{16\pi^2}{\int}_{0}^{\infty}d\mu\,{\int}_{0}^{\Lambda}d|\bk|\,\tilde{R}(\mu,|\bk|)|\bk|^2\log\left (\frac{\Lambda}{|\bk|}\right )\nn\\&&\!\!\!\!\!\!\times(2{+}e^{{-}2\mu}{-}4\coth{\mu}),
\eea
\bea
\tilde{\Xi}_{16}\!\!\!&{=}&\!\!\!\frac{|\bq|^2}{8\pi^2}{\int}_{0}^{\infty}d\mu\,{\int}_{0}^{\Lambda}d|\bk|\,\tilde{Q}(\mu,|\bk|)|\bk|\log\left (\frac{\Lambda}{|\bk|}\right )\nn\\&&\!\!\!\!\!\!\times[2\coth{\mu}(2\coth^2{\mu}{-}1){-}e^{{-}2\mu}{-}1],
\eea
\bea
\tilde{\Xi}_{17}\!\!\!&{=}&\!\!\!{-}2\tilde{\Xi}_{15},
\eea
where $\tilde{Q}(\mu,|\bk|)$ and $\tilde{R}(\mu,|\bk|)$ were defined in Eq.~(\ref{defofQtandRt}). We may now do the integrals on $\bk$ using Eqs.~(\ref{intofQt}) and (\ref{intofRt}). Doing so and collecting all of the results, we obtain
\begin{eqnarray}
&&\!\!\!\!\!\!\Sigma_{3e,i}(q){=}\tfrac{1}{128}i\alpha^3 v_F q_i{\int}_{0}^{\infty} d\mu\,\times
\nn\\&&\!\!\!\!\!\!\times \left [{-}\frac{-1+16/(1+e^\mu)-4e^{-2\mu}(-2+3e^{\mu})}{2(1{+}\cosh{\mu})^2}\right.\cr
&&\!\!\!\!\!\!{+}\left.\frac{1{-}e^{{-}\mu}{+}2e^{{-}2\mu}{-}e^{{-}3\mu}{-}\cosh{\mu}}{(1{+}\cosh{\mu})^3}\right ]\log^2\left (\frac{v_F\Lambda}{|q_0|}\right ) \cr
&&\!\!\!\!\!\!{=}\tfrac{1}{128}\left (\tfrac{247}{15}{-}24\log{2}\right )i\alpha^3 v_F q_i\log^2\left (\frac{v_F\Lambda}{|q_0|}\right )\nn\\&&\!\!\!\!\!\!\to\tfrac{1}{128}\left (\tfrac{247}{15}{-}24\log{2}\right )i\alpha^3 v_F q_i\log^2\left (\frac{\Lambda}{|\bq|}\right ). \nonumber\\
\end{eqnarray}
The full leading divergence of $\Sigma_{3e}(q)$ is thus
\begin{eqnarray}
\Sigma_{3e}(q)\!\!\!&{=}&\!\!\!\left [\tfrac{1}{8}\left (\tfrac{41}{60}{-}\log{2}\right )q_0\gamma^0{+}\tfrac{1}{128}\left (\tfrac{247}{15}{-}24\log{2}\right )v_F \bq{\cdot}\vec{\gamma}\right ]\nn\\&&\times i\alpha^3\log^2\left (\frac{\Lambda}{|\bq|}\right ).
\end{eqnarray}

\subsection{Self{-}energy correction to two{-}loop rainbow diagram}\label{app:threelooprainbowSE}
\begin{figure}
\includegraphics[width=0.5\columnwidth]{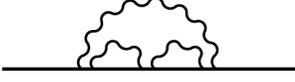}
\caption[Self-energy]{Self-energy correction to the two-loop rainbow diagram.}
\label{fig:selfenergy3g}
\end{figure}
The diagram shown in Fig.~\ref{fig:selfenergy3g} evaluates to
\bea
&&\!\!\!\!\!\!\!\Sigma_{3g}(q)=
\nn\\&&\!\!\!\!\!\!\!{-}{\int}\frac{d^3k}{(2\pi)^3}\gamma^0G_0(k)\Sigma_1(k)G_0(k)\Sigma_1(k)G_0(k)\gamma^0D_0(q{-}k)
\nn\\&&\!\!\!\!\!\!\!{=}\frac{{-}ig^6}{512\pi^2}{\int}\frac{d^3k}{(2\pi)^3}\gamma^0\frac{{\slashed k}\vec{k}{\cdot}\vec\gamma{\slashed k}\vec{k}{\cdot}\vec\gamma{\slashed k}}{k^6}\gamma^0\log^2(\Lambda/|\vec{k}|)\frac{1}{|\vec{q}{-}\vec{k}|}
\nn\\&&\!\!\!\!\!\!\!{=}\frac{ig^6}{512\pi^2}{\int}\frac{d^3k}{(2\pi)^3}[k_0\gamma^0{-}v_F\vec{k}{\cdot}\vec\gamma][k_0\gamma^0{+}v_F\vec{k}{\cdot}\vec\gamma] [k_0\gamma^0{-}v_F\vec{k}{\cdot}\vec\gamma]
\nn\\&&\qquad\qquad\times\frac{|\vec{k}|^2}{k^6}\log^2(\Lambda/|\vec{k}|)\frac{1}{|\vec{q}{-}\vec{k}|}
\nn\\&&\!\!\!\!\!\!\!{=}\frac{ig^6}{512\pi^2}{\int}\frac{d^3k}{(2\pi)^3}[k_0\gamma^0{-}v_F\vec{k}{\cdot}\vec\gamma][k_0^2{+}2v_F\vec{k}{\cdot}\vec\gamma k_0\gamma^0{-}v_F^2|\vec{k}|^2]
\nn\\&&\qquad\qquad\times\frac{|\vec{k}|^2}{k^6}\log^2(\Lambda/|\vec{k}|)\frac{1}{|\vec{q}{-}\vec{k}|}
\nn\\&&\!\!\!\!\!\!\!{=}\frac{ig^6}{512\pi^2}{\int}\frac{d^3k}{(2\pi)^3}[k_0^3\gamma^0{-}3v_Fk_0^2\vec{k}{\cdot}\vec\gamma{-}3v_F^2k_0|\vec{k}|^2\gamma^0 {+}v_F^3|\vec{k}|^2\vec{k}{\cdot}\vec\gamma]
\nn\\&&\qquad\qquad\times\frac{|\vec{k}|^2}{k^6}\log^2(\Lambda/|\vec{k}|)\frac{1}{|\vec{q}{-}\vec{k}|}
\nn\\&&\!\!\!\!\!\!\!{=}\frac{ig^6}{512\pi^2}{\int}\frac{d^3k}{(2\pi)^3}\frac{v_F^3|\vec{k}|^2{-}3v_Fk_0^2}{k^6}\vec{k}{\cdot}\vec\gamma|\vec{k}|^2 \log^2(\Lambda/|\vec{k}|)\frac{1}{|\vec{q}{-}\vec{k}|}.\nn\\&&
\eea
In the final step, we discarded the terms odd in $k_0$ since these terms will vanish identically under the $k_0$ integration. In fact, the terms even in $k_0$ also vanish under the integration:
\beq
{\int} dk_0\frac{v_F^3|\vec{k}|^2{-}3v_Fk_0^2}{k^6}{=}0.
\eeq
Therefore, the diagram vanishes identically:
\beq
\Sigma_{3g}(q){=}0.
\eeq

\subsection{Self-energy correction to one-loop bubble diagram.}
\begin{figure}
\includegraphics[width=0.5\columnwidth]{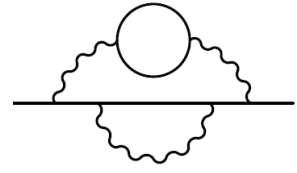}
\caption{Self-energy correction to bubble diagram.}
\label{fig:selfenergy3i}
\end{figure}

We now turn our attention to the diagram shown in Fig.~\ref{fig:selfenergy3i}, whose value
will be denoted by $\Sigma_{3i}(q)$.  Its value is given by
\bea
\Sigma_{3i}(q)&{=}&{\int}\frac{d^3k}{(2\pi)^3}\,i\gamma^0\frac{g^2}{2|\bq{-}\bk|}\Pi_B(q{-}k)\frac{g^2}{2|\bq{-}\bk|}
\nn\\&&\times\frac{i\slashed{k}}{k^2}\Sigma_1(k)\frac{i\slashed{k}}{k^2}i\gamma^0,
\eea
where $\Pi_B(q)$ is the value of the bubble diagram given in Eq.~(\ref{oneloopbubble}), and $\Sigma_1(q)$ is the one-loop self-energy given in Eq.~(\ref{selfenergy}). Substituting these into $\Sigma_{3i}(q)$, we obtain
\bea
\Sigma_{3i}(q)&=&-i\frac{g^6}{512\pi}N\int\frac{d^3k}{(2\pi)^3}\,\frac{1}{k^4\sqrt{(q-k)^2}}
\nn\\&&\times\gamma^0\slashed{k}\bk\cdot\vec{\gamma}\slashed{k}\gamma^0\log\left (\frac{\Lambda}{|\bk|}\right ).
\eea
After some algebra, we may show that
\begin{equation}
\gamma^0\slashed{k}\bk\cdot\vec{\gamma}\slashed{k}\gamma^0=(k_0^2-v_F^2|\bk|^2)\bk\cdot\vec{\gamma}+2v_F|\bk|^2k_0\gamma^0,
\end{equation}
so that
\bea
\Sigma_{3i}(q)&{=}&-i\frac{g^6}{512\pi}N{\int}\frac{d^3k}{(2\pi)^3}\,\frac{1}{k^4\sqrt{(q{-}k)^2}}
\nn\\&&\times[(k_0^2{-}v_F^2|\bk|^2)\bk{\cdot}\vec{\gamma}{+}2v_F|\bk|^2k_0\gamma^0]\log\left (\frac{\Lambda}{|\bk|}\right ).\nn\\&&
\eea
We are again interested in finding the leading divergence, so we expand $\frac{1}{\sqrt{(q-k)^2}}$
as follows:
\begin{equation}
\frac{1}{\sqrt{(q-k)^2}}=\frac{1}{k}\left (1+\frac{k\cdot q}{k^2}\right )
\end{equation}
Substituting this expansion into $\Sigma_{3i}(q)$ and keeping only terms that are
even in $k_0$, we obtain
\bea
\Sigma_{3i}(q)&=&-i\frac{g^6}{512\pi}N\int\frac{d^3k}{(2\pi)^3}\,\frac{1}{k^7}\Big[2v_F|\bk|^2k_0^2q_0\gamma^0
\nn\\&&+v_F^2(k_0^2-v_F^2|\bk|^2)(\bk\cdot\bq)(\bk\cdot\vec{\gamma})\Big]\log\left (\frac{\Lambda}{|\bk|}\right ).\nn\\&&
\eea
In the second term, we note that the cross terms in the product, $(\bk\cdot\bq)(\bk\cdot\vec{\gamma})$,
are odd functions of each component of $\bk$, and that the entire integral is symmetric
under the exchange of the two components of $\bk$, i.e. under $k_x\leftrightarrow k_y$.
As a result, we may make the replacement,
\begin{equation}
(\bk\cdot\bq)(\bk\cdot\vec{\gamma})=\tfrac{1}{2}|\bk|^2\bq\cdot\vec{\gamma},
\end{equation}
thus obtaining
\bea
\Sigma_{3i}(q)&=&-i\frac{g^6}{512\pi}N\int\frac{d^3k}{(2\pi)^3}\,\frac{1}{k^7}\Big[2v_F|\bk|^2k_0^2q_0\gamma^0
\nn\\&&+\tfrac{1}{2}v_F^2|\bk|^2(k_0^2-v_F^2|\bk|^2)\bq\cdot\vec{\gamma}\Big]\log\left (\frac{\Lambda}{|\bk|}\right ).\nn\\&&
\eea
We may now do the integrals over $k_0$, obtaining
\bea
\Sigma_{3i}(q)&=&-i\frac{g^6}{512\pi}\frac{1}{15\pi v_F}N(4q_0\gamma^0-3v_F\bq\cdot\vec{\gamma})
\nn\\&&\times\int\frac{d^2k}{(2\pi)^2}\,\frac{1}{(v_F|\bk|)^2}\log\left (\frac{\Lambda}{|\bk|}\right ).\nn\\&&
\eea
We may now easily evaluate the remaining integral, obtaining
\begin{equation}
\Sigma_{3i}(q)=-i\frac{N}{60}\frac{g^6}{512\pi^3v_F^3}(4q_0\gamma^0-3v_F\bq\cdot\vec{\gamma})\log^2\left (\frac{\Lambda}{|\bq|}\right ),
\end{equation}
where, once again, we assumed an infrared cutoff given by $|\bq|$.  Rewriting this
in terms of $\alpha$, we finally arrive at
\begin{equation}
\Sigma_{3i}(q)=-i\frac{N\alpha^3}{480}(4q_0\gamma^0-3v_F\bq\cdot\vec{\gamma})\log^2\left (\frac{\Lambda}{|\bq|}\right ).
\end{equation}

\subsection{Bubble diagram correction to one-loop electron self-energy.}
\begin{figure}
\includegraphics[width=0.5\columnwidth]{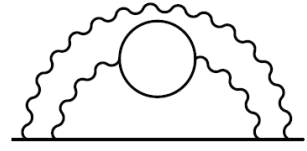}
\caption{Bubble diagram correction to one-loop electron self-energy.}
\label{fig:selfenergy3j}
\end{figure}

We now calculate the diagram shown in Fig.~\ref{fig:selfenergy3j}, which is similar in
appearance to the previous case.  We will denote this diagram by $\Sigma_{3j}(q)$.  It
contains the divergent sub-diagram shown in Fig.~\ref{fig:selfenergy2c}, which we denoted
by $\Sigma_{2c}(q)$ and expressed explicitly in Eq.~(\ref{selfenergy2c}). The value of $\Sigma_{3j}(q)$ is
\begin{equation}
\Sigma_{3j}(q)=\int \frac{d^3k}{(2\pi)^3}\,\gamma^0\frac{g^2}{2|\bq-\bk|}\frac{\slashed{k}}{k^2}\Sigma_{2c}(k)\frac{\slashed{k}}{k^2}\gamma^0.
\end{equation}
To evaluate this expression, we will find the following identities useful:
\bea
\!\!\!\!\!\!\gamma^0\slashed{k}k_0\gamma^0\slashed{k}\gamma^0\!\!\!&{=}&\!\!\!(k_0^2{-}v_F^2|\bk|^2)k_0\gamma^0{-}2v_Fk_0^2\bk{\cdot}\vec{\gamma}, \\
\!\!\!\!\!\!\gamma^0\slashed{k}\bk{\cdot}\vec{\gamma}\slashed{k}\gamma^0\!\!\!&{=}&\!\!\!2v_F|\bk|^2k_0\gamma^0{+}(k_0^2{-}v_F^2|\bk|^2)\bk{\cdot}\vec{\gamma}.
\eea
Using these identities, along with the above expression for $\Sigma_{2c}(q)$, we obtain,
after dropping terms that will integrate to zero by virtue of being odd functions of $k_0$,
\bea
\Sigma_{3j}(q)&{=}&\tfrac{1}{6}N\pi i\alpha^3 v_F{\int}\frac{d^3k}{(2\pi)^3}\,\frac{1}{k^4|\bq{-}\bk|}({-}3k_0^2{+}v_F^2|\bk|^2)
\nn\\&&\times v_F\bk{\cdot}\vec{\gamma}\log\left (\frac{\Lambda}{|\bk|}\right ).
\eea
Evaluating the $k_0$ integral, we obtain
\begin{equation}
\Sigma_{3j}(q)=-\tfrac{1}{12}N\pi i\alpha^3\int\frac{d^2k}{(2\pi)^2}\,\frac{1}{|\bk||\bq-\bk|}v_F\bk\cdot\vec{\gamma}\log\left (\frac{\Lambda}{|\bk|}\right ).
\end{equation}
As with the first diagram considered here, we note that the integrand is odd in the component
of $\bk$ perpendicular to $\bq$, and thus make the replacement,
\begin{equation}
\bk\cdot\vec{\gamma}\to\frac{\bk\cdot\bq}{|\bq|^2}\bq\cdot\vec{\gamma},
\end{equation}
as before.  The integral now becomes
\begin{equation}
\Sigma_{3j}(q){=}-\tfrac{1}{12}N\pi i\alpha^3{\int}\frac{d^2k}{(2\pi)^2}\,\frac{1}{|\bk||\bq{-}\bk|}\frac{\bk{\cdot}\bq}{|\bq|^2}v_F\bq{\cdot}\vec{\gamma}\log\left (\frac{\Lambda}{|\bk|}\right ).
\end{equation}
We now expand $\frac{1}{|\bq-\bk|}$ in powers of $\bq$.  As before, the leading term drops out
because it gives us a term that is odd in $\bk$, so that the sub-leading term is the lowest-order
non-zero contribution:
\begin{equation}
\Sigma_{3j}(q){=}-\tfrac{1}{12}N\pi i\alpha^3{\int}\frac{d^2k}{(2\pi)^2}\,\frac{(\bk{\cdot}\bq)^2}{|\bk|^4|\bq|^2}\log\left (\frac{\Lambda}{|\bk|}\right )v_F\bq{\cdot}\vec{\gamma}.
\end{equation}
We now evaluate the remaining integral on $\bk$, obtaining
\begin{equation}
\Sigma_{3j}(q)=-\tfrac{i}{96}N\alpha^3\log^2\left (\frac{\Lambda}{|\bq|}\right )v_F\bq\cdot\vec{\gamma}.
\end{equation}
We once again assumed a lower cutoff of $|\bq|$ on the magnitude of $\bk$ to regularize
the infrared divergence of the integral.

\subsection{Self{-}energy vacuum polarization correction to electron self{-}energy}

\begin{figure}
\includegraphics[width=0.5\columnwidth]{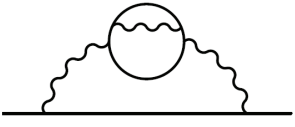}
\caption{Three{-}loop self{-}energy vacuum polarization correction to the electron self{-}energy.}
\label{fig:selfenergy3l}
\end{figure}

The three{-}loop self{-}energy vacuum polarization correction to the electron self{-}energy is shown in Fig.~\ref{fig:selfenergy3l}. This diagram evaluates to
\beq
\Sigma_{3l}(q){=}{-}2{\int}\frac{d^3k}{(2\pi)^3}\gamma^0G_0(q{-}k)\gamma^0D_0(k)^2\Pi_{SE}(k),
\eeq
where we have included an overall factor of 2 to account for the symmetry of the diagram. Recall that the self{-}energy correction to the vacuum polarization was found to be (Eq.~(\ref{fullPa}))
\beq
\Pi_{SE}(k){=}\frac{N\alpha |\vec{k}|}{32\pi v_F}\left[\frac{\pi}{2}\frac{v_F^3|\vec{k}|^3}{k^3}\log(\Lambda/|\vec{k}|){+}I_a(ik_0/v_F|\vec{k}|)\right],
\eeq
where the function $I_a(y)$ was given in Eq.~(\ref{Iaexplicit}). We begin by focusing on the term involving the logarithmic divergence in $\Pi_{SE}(q)$:
\bea
&&\Sigma_{3l,div}(q){=}\nn\\&&{-}\frac{iN\alpha g^4v_F^2}{128}{\int}\frac{d^3k}{(2\pi)^3}\gamma^0\frac{\slashed q{-}\slashed k}{(q{-}k)^2}\gamma^0\frac{1}{|\vec{k}|^2}\frac{|\vec{k}|^4}{k^3}\log(\Lambda/|\vec{k}|).\nn\\&&
\eea
We are only interested in the ultraviolet divergence of this integral, so we expand the denominator:
\beq
\frac{1}{(q{-}k)^2}\approx\frac{1}{k^2}\left(1{+}2\frac{k{\cdot} q}{k^2}\right).
\eeq
As usual, a potential linear divergence cancels identically due to the antisymmetry under $k\to{-}k$. The leading divergence is thus logarithmic:
\bea
&&\Sigma_{3l,div}(q){=}\nn\\&&
{-}\frac{iN\alpha g^4v_F^2}{128}{\int}\frac{d^3k}{(2\pi)^3}\gamma^0\left[\slashed q{-}2\frac{k_0^2q_0\gamma^0{+}v_F^3(\vec{k}{\cdot}\vec\gamma)(\vec{k}{\cdot}\vec{q})}{k^2}\right]
\nn\\&&\qquad\qquad\times\gamma^0 \frac{|\vec{k}|^2}{k^5}\log(\Lambda/|\vec{k}|).
\eea
In the second term inside the square brackets, we have kept only those terms which are symmetric under $k_0\to {-}k_0$. The $k_0$ integration can be done straightforwardly, with the result
\bea
&&\Sigma_{3l,div}(q){=}
\nn\\&&{-}\frac{iN\pi\alpha^3}{16}{\int}\frac{d^2k}{(2\pi)^2}\gamma^0\left[\frac{4}{3}\slashed q{-}\frac{8}{15}q_0\gamma^0{-}\frac{32}{15}v_F\frac{(\vec{k}{\cdot}\vec\gamma)(\vec{k}{\cdot}\vec{q})}{|\vec k|^2}\right]
\nn\\&&\qquad\qquad\times\gamma^0\frac{\log(\Lambda/|\vec{k}|)}{|\vec{k}|^2}.
\eea
In the third term inside the square brackets, we may keep only the terms symmetric under $k_1\to{-}k_1$. This, in combination with the symmetry under swapping $k_1\Leftrightarrow k_2$, allows us to make the replacement $(\vec{k}{\cdot}\vec\gamma)(\vec{k}{\cdot}\vec{q})\to1/2|\vec{k}|^2\vec{q}{\cdot}\vec\gamma$, with the result:
\bea
&&\!\!\!\!\!\!\!\Sigma_{3l,div}(q)=
\nn\\&&\!\!\!\!\!\!\!\frac{{-}iN\pi\alpha^3}{16}{\int}\frac{d^2k}{(2\pi)^2}\gamma^0\left[\frac{4}{3}\slashed q{-}\frac{8}{15}q_0\gamma^0{-}\frac{16}{15}v_F\vec{q}{\cdot}\vec\gamma\right]\gamma^0\frac{\log(\Lambda/|\vec{k}|)}{|\vec{k}|^2}
\nn\\&&\!\!\!\!\!\!\!{=}{-}\frac{iN\pi\alpha^3}{16}\left(\frac{4}{5}q_0\gamma^0{-}\frac{4}{15}v_F\vec{q}{\cdot}\vec\gamma\right){\int}\frac{d^2k}{(2\pi)^2}\frac{\log(\Lambda/|\vec{k}|)}{|\vec{k}|^2}
\nn\\&&\!\!\!\!\!\!\!{=}{-}\frac{iN\alpha^3}{240}\left(3q_0\gamma^0{-}v_F\vec{q}{\cdot}\vec\gamma\right)\log^2(\Lambda/|\vec{q}|).
\eea
In the final step, we made the assumption that the infrared divergence of the final integration is regulated by $|\vec{q}|$.

\subsection{Two-bubble RPA correction to electron self-energy}

\begin{figure}
\includegraphics[width=0.5\columnwidth]{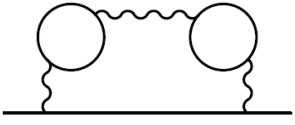}
\caption{Two-bubble RPA correction to electron self-energy.}
\label{fig:selfenergy3p}
\end{figure}

To give an example of a third-order simple logarithmic (as opposed to double logarithmic) divergence, we will calculate the two-bubble RPA diagram shown in Fig.~\ref{fig:selfenergy3p}, the value of which we denote by $\Sigma_{3p}(q)$. This diagram can be easily evaluated by starting from the full RPA result, which was originally obtained in Refs.~[\onlinecite{DasSarma_PRB07}], [\onlinecite{Son_PRB07}]:
\beq
\Sigma_{RPA}(q)=\frac{4i}{N\pi^2}\left[f_0(\lambda)q_0\gamma^0+f_1(\lambda)v_F\vec{q}\cdot\vec\gamma\right]\log(\Lambda/|\vec{q}|),
\eeq
where for $\lambda<1$, the functions $f_0$ and $f_1$ are given by
\bea
f_0(\lambda)&\equiv&-\frac{2-\lambda^2}{\lambda\sqrt{1-\lambda^2}}\arccos\lambda-2+\frac{\pi}{\lambda},\nn\\
f_1(\lambda)&\equiv&-\frac{\sqrt{1-\lambda^2}}{\lambda}\arccos\lambda-1+\frac{\pi}{2\lambda},
\eea
and we have defined an effective coupling
\beq
\lambda\equiv\frac{g^2N}{16v_F}=\frac{\pi N\alpha}{4}.
\eeq
We can extract the third-order contribution to $\Sigma_{RPA}(q)$ by expanding $f_0$ and $f_1$ to third order in $\lambda$:
\bea
f_0(\lambda)=\frac{\lambda^2}{3}-\frac{\pi\lambda^3}{8}+O(\lambda^4),\nn\\
f_1(\lambda)=\frac{\pi\lambda}{4}-\frac{\lambda^2}{3}+\frac{\pi\lambda^3}{16}+O(\lambda^4).
\eea
The third-order terms then yield
\beq
\Sigma_{3p}(q)=\frac{i\pi^2\alpha^3}{32}\left[-q_0\gamma^0+2v_F\vec{q}\cdot\vec\gamma\right]\log(\Lambda/|\vec{q}|).
\eeq

\section{electron self-energy with zero-range interaction}\label{app:zerorange}

Given the complexity of graphene perturbation theory with the Coulomb interaction, it is tempting to try replacing this long-range interaction with an effective short-range one in an attempt to arrive at a simpler theory. In particular, we will consider 2+1-dimensional Dirac electrons interacting via a delta-function contact interaction instead of the long-range Coulomb repulsion. In this case, the Feynman rules remain the same as before, except that we now use a dashed line to denote the propagator associated with the contact interaction, with the value of this propagator given by
\beq
D_0^{zr}=\frac{g^2r_0}{2}.
\eeq
Here, $r_0$ is a scale factor which controls the strength of the interaction and which preserves the dimensionality of the propagator relative to the 2d Coulomb interaction case. We will see that while inserting this contact interaction in place of the Coulomb force does offer some algebraic simplification in the course of computing diagrams, it does so at the expense of introducing new conceptual complexities related to the physicality and renormalizability of the resulting theory.

\subsection{One-loop electron self-energy for zero-range interaction and an ambiguity}

\begin{figure}
\includegraphics[width=0.5\columnwidth]{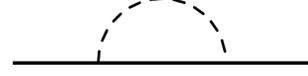}
\caption[Self-energy]{One-loop correction to the electron self-energy for zero-range interaction.}
\label{fig:oneloopSEzerorange}
\end{figure}

The diagram for the one-loop electron self-energy in the case of a zero-range contact interaction is shown in Fig.~\ref{fig:oneloopSEzerorange}. This diagram evaluates to
\bea
\Sigma_1^{zr}(q)&{=}&{-}{\int}\frac{d^3k}{(2\pi)^3}\gamma^0G_0(k)\gamma^0D_0^{zr}
\nn\\&{=}&\frac{{-}ig^2r_0}{2}{\int}\frac{d^3k}{(2\pi)^3}\gamma^0\frac{\slashed k}{k^2}\gamma^0{=}0.
\eea
In the last step, we determined that the integral over $k$ vanishes because the integrand is an odd function of $k$.

This argument may be too fast, however. Consider the following expression:
\bea
\Sigma_1^{zr}(q)&{=}&{-}{\int}\frac{d^3k}{(2\pi)^3}\gamma^0G_0(q{-}k)\gamma^0D_0^{zr}
\nn\\&{=}&\frac{{-}ig^2r_0}{2}{\int}\frac{d^3k}{(2\pi)^3}\gamma^0\frac{\slashed q{-}\slashed k}{(q{-}k)^2}\gamma^0.
\eea
This expression should be equally valid as it arises simply from a different definition of the loop momentum $k$. Performing the integration over $k_0$, we find
\beq
\Sigma_1^{zr}(q){=}\frac{ig^2r_0}{4}{\int}\frac{d^2k}{(2\pi)^2}\frac{(\vec{q}{-}\vec{k}){\cdot}\vec\gamma}{|\vec{q}{-}\vec{k}|}.
\eeq
Note that if at this point we changed variables, $\vec{k}\to\vec{q}{-}\vec{k}$, we would again conclude that the integral vanishes because the integrand would become odd. There is a problem here though because under such a coordinate transformation, the limits of integration (which have so far been kept implicit) will also change because of the implicit momentum cutoff. Indeed, if we do not make this coordinate change and proceed to evaluate the integral, we find
\bea
\Sigma_1^{zr}(q)&{=}&\frac{ig^2r_0}{4}{\int}\frac{d^2k}{(2\pi)^2}\frac{(\vec{q}{-}\vec{k}){\cdot}\vec\gamma}{|\vec{q}{-}\vec{k}|}
\nn\\&\approx&\frac{ig^2r_0}{4}{\int}\frac{d^2k}{(2\pi)^2}\frac{1}{|\vec{k}|}\left[\vec{q}{\cdot}\vec\gamma{-}\frac{(\vec{k}{\cdot}\vec\gamma)(\vec{k}{\cdot}\vec{q})}{|\vec{k}|^2}\right] \nn\\&{=}&\frac{i\alpha r_0k_c}{4}v_F\vec{q}{\cdot}\vec\gamma.
\eea
This result is not only nonzero, it is linearly divergent.

This discrepancy appears to be due to the momentum-independence of the photon propagator and can be understood from an even simpler toy model. Suppose that we have a one-dimensional theory where the photon propagator is again momentum-independent, but now the fermion propagator scales linearly with momentum: $G(k)\sim k$. In this case, the one-loop self-energy would be given by
\beq
\Sigma\sim{\int} dk G(k)\sim {\int_{{-}k_c}^{k_c}} dk k=0.
\eeq
However, we could just as well have defined the fermion propagator to carry momentum $q{-}k$ to obtain
\beq
\Sigma\sim{\int} dk G(q{-}k)\sim {\int_{{-}k_c}^{k_c}} dk (q{-}k)=2qk_c.
\eeq
Note that these two integrals are not related by a coordinate transformation because the integration limits remained the same. In each case, we chose the cutoff to be the maximal value of $k$, but the definition of $k$ is different in each case, and it is unclear which definition should be used. The same sort of behavior arises in the 2+1 dimensional theory, suggesting that this theory may be unphysical.

One possible way to lift this ambiguity is to modify the interaction slightly. We do this by re-interpreting $k_c$ as a cutoff scale associated with the interaction itself and by redefining the photon propagator as
\beq
D_0^{zr}(q){=}\frac{g^2r_0}{2}\theta(k_c{-}|\vec{q}|),
\eeq
so that the interaction strictly vanishes about the momentum scale $k_c$. Using this new definition of the photon propagator, for the one choice of loop momentum $k$ we now find (after performing the integration over $k_0$)
\beq
\Sigma_1^{zr}(q){=}\frac{ig^2r_0}{16\pi^2}{\int} d^2k\frac{(\vec{q}{-}\vec{k}){\cdot}\vec\gamma}{|\vec{q}{-}\vec{k}|}\theta(k_c{-}|\vec{k}|),
\eeq
while for the other choice of $k$ we obtain
\beq
\Sigma_1^{zr}(q){=}\frac{ig^2r_0}{16\pi^2}{\int} d^2k\frac{\vec{k}{\cdot}\vec\gamma}{|\vec{k}|}\theta(k_c{-}|\vec{q}{-}\vec{k}|).
\eeq
Since we have a cutoff coming from the interaction itself, it is no longer necessary to include a cutoff in the integration limits. The two results are then manifestly equivalent since we may perform the coordinate transformation $\vec{k}\to\vec{q}{-}\vec{k}$ in the latter integral to obtain the former. Either integral can then be performed by expanding the denominator in the large $|\vec{k}|$ limit as before, yielding
\beq
\Sigma_1^{zr}(q){=}\frac{i\alpha r_0k_c}{4}v_F\vec{q}{\cdot}\vec\gamma.
\eeq
Therefore, for this modified contact interaction, we unambiguously find a linear divergence in the one-loop correction to the electron self-energy. We will continue to use the modified contact interaction throughout the remainder of this section.

\subsection{Two-loop rainbow correction to self-energy for zero-range interaction}

\begin{figure}
\includegraphics[width=0.5\columnwidth]{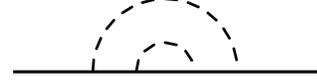}
\caption[Self-energy]{Two-loop rainbow correction to the electron self-energy for zero-range interaction.}
\label{fig:SE2bzerorange}
\end{figure}
The two-loop rainbow diagram correction to the electron self-energy is shown in Fig.~\ref{fig:SE2bzerorange}. The contribution from this diagram is given by
\bea
\Sigma_{2b}^{zr}(q)\!\!\!&{=}&\!\!\!{-}{\int}\frac{d^3k}{(2\pi)^3}\gamma^0G_0(q{-}k)\Sigma_1(q{-}k)G_0(q{-}k)\gamma^0D_0^{zr}(k)
\nn\\\!\!\!&{=}&\!\!\!\frac{i\pi\alpha^2r_0^2k_cv_F^2}{2}{\int}\frac{d^3k}{(2\pi)^3}\gamma^0\frac{\slashed q{-}\slashed k}{(q{-}k)^2}(\vec{q}{-}\vec{k}){\cdot}\vec\gamma\frac{\slashed q{-}\slashed k}{(q{-}k)^2}\gamma^0
\nn\\&&\qquad\qquad\times\theta(k_c{-}|\vec{k}|)
\nn\\\!\!\!&{=}&\!\!\!\frac{i\pi\alpha^2r_0^2k_cv_F^2}{2}{\int}\frac{d^3k}{(2\pi)^3}\frac{1}{k^4}\gamma^0\left[{-}k_0^2{+}v_F^2|\vec{k}|^2\right]\vec{k}{\cdot}\vec\gamma\gamma^0
\nn\\&&\qquad\qquad\times\theta(k_c{-}|\vec{q}{-}\vec{k}|)
\nn\\\!\!\!&{=}&\!\!\!0.
\eea
The contribution vanishes identically due to the integration over $k_0$.

\subsection{Two-loop bubble correction to electron self-energy for zero-range interaction}

\begin{figure}
\includegraphics[width=0.5\columnwidth]{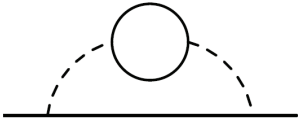}
\caption[Bubble]{Two-loop bubble correction to the electron self-energy for zero-range interaction.}
\label{fig:SE2czerorange}
\end{figure}
The diagram for the two-loop bubble correction to the electron self-energy in the case of the zero-range interaction is shown in Fig.~\ref{fig:SE2czerorange}. This diagram evaluates to
\bea
\Sigma_{2c}^{zr}(q)&{=}&{-}{\int}\frac{d^3k}{(2\pi)^3}\gamma^0G_0(q{-}k)\gamma^0\Pi_B(k)(D_0^{zr}(k))^2\nn\\
&{=}&\frac{iNg^4r_0^2}{32}{\int}\frac{d^3k}{(2\pi)^3}\gamma^0\frac{\slashed q{-}\slashed k}{(q{-}k)^2}\gamma^0\frac{|\vec{k}|^2}{\sqrt{k^2}}\theta(k_c{-}|\vec{k}|).\nn\\&&
\eea
Here, we have inserted the value for the one{-}loop vacuum polarization function, $\Pi_B(k)$, computed earlier in Eq.~(\ref{oneloopbubble}). As in the case of the two{-}loop vertex correction considered above, we may extract the divergent terms by expanding the denominator:
\bea
\frac{1}{(q{-}k)^2}\!\!\!&{=}&\!\!\!\frac{1}{k^2}\bigg\{1{+}2\frac{k{\cdot} q}{k^2}{+}4\frac{(k{\cdot} q)^2}{k^4}{-}\frac{q^2}{k^2}
\nn\\&&\!\!\!{+}4\frac{k{\cdot} q}{k^6}\left[2(k{\cdot} q)^2{-}k^2q^2\right]\bigg\}{+}O(1/k^6).\label{denomexp}
\eea
The first two terms of this expansion produce a quadratic divergence:
\bea
&&\!\!\!\!\!\!\Sigma_{2c,quad}^{zr}(q)=
\nn\\&&\!\!\!\!\!\!\frac{{-}iNg^4r_0^2}{32}{\int}\frac{d^3k}{(2\pi)^3}\frac{|\vec{k}|^2}{k^3}\left(1{+}2\frac{k{\cdot} q}{k^2}\right)\Big(k_0\gamma^0{-}v_F\vec{k}{\cdot}\vec\gamma
\nn\\&&\qquad\qquad{-}q_0\gamma^0{+}v_F\vec{q}{\cdot}\vec\gamma\Big)\theta(k_c{-}|\vec{k}|)
\nn\\&&\!\!\!\!\!\!{=}\frac{{-}iNg^4r_0^2}{32}{\int}\frac{d^3k}{(2\pi)^3}\frac{|\vec{k}|^2}{k^3}\bigg\{{-}q_0\gamma^0{+}v_F\vec{q}{\cdot}\vec\gamma{+}2\frac{k_0^2}{k^2}q_0\gamma^0 \nn\\&&\qquad\qquad{-}\frac{2v_F^3}{k^2}(\vec{k}{\cdot}\vec{q})(\vec{k}{\cdot}\vec\gamma)\bigg\}\theta(k_c{-}|\vec{k}|)
\nn\\&&\!\!\!\!\!\!{=}\frac{{-}iNg^4r_0^2}{32}{\int}\frac{d^3k}{(2\pi)^3}\frac{|\vec{k}|^2}{k^3}\bigg\{{-}q_0\gamma^0{+}v_F\vec{q}{\cdot}\vec\gamma{+}2\frac{k_0^2}{k^2}q_0\gamma^0 \nn\\&&\qquad\qquad{-}\frac{v_F^2|\vec{k}|^2}{k^2}v_F\vec{q}{\cdot}\vec\gamma\bigg\}\theta(k_c{-}|\vec{k}|).
\eea
The $k$ integrals were computed above, and we quote them again here for convenience:
\bea
{\int}\frac{d^3k}{(2\pi)^3}\frac{v_F^2|\vec{k}|^2}{k^3}\theta(k_c{-}|\vec{k}|)&{=}&\frac{1}{(2\pi)^3}2{\int} d^2k{=}\frac{k_c^2}{4\pi^2},\nn\\
{\int}\frac{d^3k}{(2\pi)^3}\frac{v_F^2|\vec{k}|^2k_0^2}{k^5}\theta(k_c{-}|\vec{k}|)&{=}&\frac{1}{(2\pi)^3}\frac{2}{3}{\int} d^2k{=}\frac{k_c^2}{12\pi^2},\nn\\
{\int}\frac{d^3k}{(2\pi)^3}\frac{v_F^4|\vec{k}|^4}{k^5}\theta(k_c{-}|\vec{k}|)&{=}&\frac{1}{(2\pi)^3}\frac{4}{3}{\int} d^2k{=}\frac{k_c^2}{6\pi^2}.\nn\\&&
\eea
Plugging in these results, we obtain
\bea
&&\Sigma_{2c,quad}^{zr}(q)=
\nn\\&&\frac{{-}iNg^4r_0^2k_c^2}{384\pi^2v_F^2}\left\{3({-}q_0\gamma^0{+}v_F\vec{q}{\cdot}\vec\gamma){+}2q_0\gamma^0 {-}2v_F\vec{q}{\cdot}\vec\gamma\right\}
\nn\\&&\qquad{=}\frac{i\alpha^2r_0^2k_c^2}{12}(q_0\gamma^0{-}v_F\vec{q}{\cdot}\vec\gamma).
\eea
Combining this result with the other contributions to the quadratic divergence of the two{-}loop self{-}energy, we have
\bea
\Sigma_{2,quad}^{zr}(q)&{=}&\Sigma_{2a,quad}^{zr}(q){+}\Sigma_{2c,quad}^{zr}(q)
\nn\\&{=}&\frac{i\alpha^2r_0^2k_c^2}{32}(3q_0\gamma^0{-}2v_F\vec{q}{\cdot}\vec\gamma).
\eea

The remaining terms in the expansion shown in Eq.~(\ref{denomexp}) give rise to a logarithmic divergence:
\bea
&&\!\!\!\!\!\!\!\Sigma_{2c,log}^{zr}(q)=
\nn\\&&\!\!\!\!\!\!\!\frac{{-}iNg^4r_0^2}{32}{\int}\frac{d^3k}{(2\pi)^3}\frac{|\vec{k}|^2}{k^3}\bigg\{\left(4\frac{(k{\cdot} q)^2}{k^4}{-}\frac{q^2}{k^2}\right)\left({-}q_0\gamma^0{+}v_F\vec{q}{\cdot}\vec\gamma\right)
\nn\\&&\!\!\!\!\!\!\!{+}4\frac{k{\cdot} q}{k^6}\left(2(k{\cdot} q)^2{-}k^2q^2\right)\left(k_0\gamma^0{-}v_F\vec{k}{\cdot}\vec\gamma\right)\bigg\}\theta(k_c{-}|\vec{k}|).
\eea
The first half of the integral has been computed already in the course of computing $\Sigma_{2a,log}^{zr}(q)$:
\bea
&&\!\!\!\!\!\!\!\frac{{-}iNg^4r_0^2}{32}{\int}\frac{d^3k}{(2\pi)^3}\frac{|\vec{k}|^2}{k^3}\left(4\frac{(k{\cdot} q)^2}{k^4}{-}\frac{q^2}{k^2}\right)\left({-}q_0\gamma^0{+}v_F\vec{q}{\cdot}\vec\gamma\right)
\nn\\&&\qquad\qquad\qquad\times\theta(k_c{-}|\vec{k}|)
\nn\\&&\!\!\!\!\!\!\!{=}{-}4NT_1
\nn\\&&\!\!\!\!\!\!\!{=}\frac{{-}iN\alpha^2r_0^2}{30v_F^2}(q_0\gamma^0{-}v_F\vec{q}{\cdot}\vec\gamma)(q_0^2{-}3v_F^2|\vec{q}|^2)\log(k_c/|\vec{q}|).
\eea
The remainder of the integral can be computed with the help of the following results:
\bea
\!\!\!\!\!\!I_1\!\!\!&{=}&\!\!\!8{\int}\frac{d^3k}{(2\pi)^3}\frac{|\vec{k}|^2}{k^9}(k{\cdot} q)^3k_0\gamma^0\theta(k_c{-}|\vec{k}|)
\nn\\\!\!\!&{=}&\!\!\!\frac{2}{35\pi^2v_F^4}(4q_0^2{+}v_F^2|\vec{q}|^2)q_0\gamma^0\log(k_c/|\vec{q}|),
\eea
\bea
\!\!\!\!\!\!I_2\!\!\!&{=}&\!\!\!{-}8v_F{\int}\frac{d^3k}{(2\pi)^3}\frac{|\vec{k}|^2}{k^9}(k{\cdot} q)^3\vec{k}{\cdot}\vec\gamma\theta(k_c{-}|\vec{k}|)
\nn\\\!\!\!&{=}&\!\!\!{-}\frac{8}{35\pi^2v_F^4}(2q_0^2{+}3v_F^2|\vec{q}|^2)v_F\vec{q}{\cdot}\vec\gamma\log(k_c/|\vec{q}|),
\eea
\bea
\!\!\!\!\!\!I_3\!\!\!&{=}&\!\!\!{-}4q^2{\int}\frac{d^3k}{(2\pi)^3}\frac{|\vec{k}|^2}{k^7}(k{\cdot} q)k_0\gamma^0\theta(k_c{-}|\vec{k}|)
\nn\\\!\!\!&{=}&\!\!\!{-}\frac{4}{15\pi^2v_F^4}q^2q_0\gamma^0\log(k_c/|\vec{q}|),
\eea
\bea
\!\!\!\!\!\!I_4\!\!\!&{=}&\!\!\!4v_Fq^2{\int}\frac{d^3k}{(2\pi)^3}\frac{|\vec{k}|^2}{k^7}(k{\cdot} q)(\vec{k}{\cdot}\vec\gamma)\theta(k_c{-}|\vec{k}|)
\nn\\\!\!\!&{=}&\!\!\!\frac{8}{15\pi^2v_F^4}q^2v_F\vec{q}{\cdot}\vec\gamma\log(k_c/|\vec{q}|).
\eea
The logarithmic divergence is then
\bea
\Sigma_{2c,log}^{zr}(q)\!\!\!&{=}&\!\!\!{-}4NT_1{-}\frac{iNg^4r_0^2}{32}\sum_iI_i
\nn\\\!\!\!&{=}&\!\!\!\frac{{-}iN\alpha^2r_0^2}{210v_F^2}\Big[(3q_0^2{-}43v_F^2|\vec{q}|^2)q_0\gamma^0
\nn\\&&{+}(q_0^2{+}5v_F^2|\vec{q}|^2)v_F\vec{q}{\cdot}\vec\gamma\Big]\log(k_c/|\vec{q}|).\nn\\&&
\eea
Combining the results from the different diagrams then gives for the full logarithmic divergence at second order
\bea
\Sigma_{2,log}^{zr}(q)&{=}&\Sigma_{2a,log}^{zr}(q){+}\Sigma_{2c,log}^{zr}(q)
\nn\\&{=}&\frac{i\alpha^2r_0^2}{840v_F^2}\Big[({-}25q_0^2{+}335v_F^2|\vec{q}|^2)q_0\gamma^0 \nn\\&&{-}(13q_0^2{+}37v_F^2|\vec{q}|^2)v_F\vec{q}{\cdot}\vec\gamma\Big]\log(k_c/|\vec{q}|).\nn\\&&
\eea
The complete divergence of the self{-}energy for the zero{-}range interaction up to two loops is then
\bea
\!\!\!\Sigma_{2,div}^{zr}(q)\!\!\!&{=}&\!\!\!\frac{i\alpha^2r_0^2k_c^2}{32}(3q_0\gamma^0{-}2v_F\vec{q}{\cdot}\vec\gamma)
\nn\\&&\!\!\!{+}\frac{i\alpha^2r_0^2}{840v_F^2}\Big[({-}25q_0^2{+}335v_F^2|\vec{q}|^2)q_0\gamma^0 \nn\\&&\!\!\!{-}(13q_0^2{+}37v_F^2|\vec{q}|^2)v_F\vec{q}{\cdot}\vec\gamma\Big]\log(k_c/|\vec{q}|).\nn\\&&
\eea

\subsection{Many-bubble correction to electron self-energy for zero-range interaction}

\begin{figure}
\includegraphics[width=0.5\columnwidth]{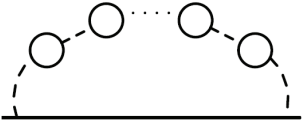}
\caption[Bubble]{$n$-bubble correction to the electron self-energy for zero-range interaction.}
\label{fig:SEnczerorange}
\end{figure}
The diagram corresponding to the $n${-}bubble correction to the electron self{-}energy in the case of the zero{-}range interaction is shown in Fig.~\ref{fig:SEnczerorange}. This diagram evaluates to
\bea
\Sigma_{n,c}^{zr}(q)\!\!\!&{=}&\!\!\!{-}{\int}\frac{d^3k}{(2\pi)^3}\gamma^0G_0(q{-}k)\gamma^0(\Pi_B(k))^n(D_0^{zr}(k))^{n{+}1}
\nn\\\!\!\!&{=}&\!\!\!\frac{({-}1)^{n{+}1}iN^ng^{2(n{+}1)}r_0^{n{+}1}}{2^{4n{+}1}}{\int}\frac{d^3k}{(2\pi)^3}\gamma^0\frac{\slashed q{-}\slashed k}{(q{-}k)^2}\gamma^0
\nn\\&&\qquad\qquad\times\frac{|\vec{k}|^{2n}}{k^n}\theta(k_c{-}|\vec{k}|).
\eea
The $(q{-}k)^2$ factor in the denominator can be expanded in the large $k$ limit precisely as in the case of the single{-}bubble correction considered in the previous section, and we obtain
\bea
&&\!\!\!\!\!\!\!\Sigma_{n,c}^{zr}(q)\approx\frac{({-}1)^{n}iN^ng^{2(n{+}1)}r_0^{n{+}1}}{2^{4n{+}1}}{\int}\frac{d^3k}{(2\pi)^3}\frac{|\vec{k}|^{2n}}{k^{n{+}2}}
\nn\\&&\!\!\!\!\!\!\!\times\left\{{-}q_0\gamma^0{+}v_F\vec{q}{\cdot}\vec\gamma{+}2\frac{k_0^2}{k^2}q_0\gamma^0 {-}\frac{v_F^2|\vec{k}|^2}{k^2}v_F\vec{q}{\cdot}\vec\gamma\right\}\theta(k_c{-}|\vec{k}|).\nn\\&&
\eea
Introducing a momentum cutoff $k_c$, the integrals over $k$ are easily computed:
\bea
&&\!\!\!\!\!\!\!\!\!\!\!{\int}\frac{d^3k}{(2\pi)^3}\frac{|\vec{k}|^{2n}}{k^{n{+}2}}\theta(k_c{-}|\vec{k}|)=
\nn\\&&\frac{1}{(2\pi)^3} \frac{\sqrt{\pi}\Gamma\left(\frac{1{+}n}{2}\right)}{\Gamma\left(1{+}\frac{n}{2}\right)v_F^{n{+}1}}{\int} d^2k|\vec{k}|^{n{-}1}\theta(k_c{-}|\vec{k}|)
\nn\\&&{=}\frac{\Gamma\left(\frac{1{+}n}{2}\right)k_c^{n{+}1}}{4\pi^{3/2}(n{+}1)\Gamma\left(1{+}\frac{n}{2}\right)v_F^{n{+}1}},
\eea
\bea
&&\!\!\!\!\!\!\!\!\!\!\!{\int}\frac{d^3k}{(2\pi)^3}\frac{|\vec{k}|^{2n}k_0^2}{k^{n{+}4}}\theta(k_c{-}|\vec{k}|)=
\nn\\&&\frac{1}{(2\pi)^3} \frac{\sqrt{\pi}\Gamma\left(\frac{1{+}n}{2}\right)}{2\Gamma\left(2{+}\frac{n}{2}\right)v_F^{n{+}1}}{\int} d^2k|\vec{k}|^{n{-}1}\theta(k_c{-}|\vec{k}|)
\nn\\&&{=}\frac{\Gamma\left(\frac{1{+}n}{2}\right)k_c^{n{+}1}}{8\pi^{3/2}(n{+}1)\Gamma\left(2{+}\frac{n}{2}\right)v_F^{n{+}1}},
\eea
\bea
&&\!\!\!\!\!\!\!\!\!\!\!{\int}\frac{d^3k}{(2\pi)^3}\frac{v_F^4|\vec{k}|^4}{k^5}\theta(k_c{-}|\vec{k}|)=
\nn\\&&\frac{1}{(2\pi)^3} \frac{\sqrt{\pi}\Gamma\left(\frac{3{+}n}{2}\right)}{\Gamma\left(2{+}\frac{n}{2}\right)v_F^{n{+}3}}{\int} d^2k|\vec{k}|^{n{-}1}\theta(k_c{-}|\vec{k}|)
\nn\\&&{=}\frac{\Gamma\left(\frac{3{+}n}{2}\right)k_c^{n{+}1}}{4\pi^{3/2}(n{+}1)\Gamma\left(2{+}\frac{n}{2}\right)v_F^{n{+}3}}.
\eea
Plugging these in and simplifying, we find the following result for the leading{-}order divergence of the $n${-}bubble self{-}energy correction:
\bea
&&\Sigma_{n,c}^{zr}(q)\approx
\nn\\&&\frac{({-}1)^{n{+}1}i\pi^{n{-}\frac{1}{2}}\Gamma\left(\frac{1{+}n}{2}\right)\alpha^{n{+}1}(r_0k_c)^{n{+}1}}{2^{n{+}2}(n{+}1)\Gamma\left(2{+}\frac{n}{2}\right)} [nq_0\gamma^0{-}v_F\vec{q}{\cdot}\vec\gamma].\nn\\&&
\eea
It is clear that the order of the divergence increases with the number of bubbles in the diagram. Unless there is a magical cancelation occurring at every order (we have seen in previous sections that this does not seem to occur at second order), the divergence structure of the theory appears to be dramatically different from that of the Coulomb-interaction theory.



\end{document}